\newcommand{\thesisTitleFrontmatter}{IMPROVING DRAM PERFORMANCE, RELIABILITY, AND SECURITY \\ BY \hhm{RIGOROUSLY} UNDERSTANDING INTRINSIC DRAM OPERATION}

\newcommand{\thesisTitlePlain}{Improving DRAM Performance, Reliability, and Security by Rigorously Understanding Intrinsic DRAM Operation}
\newcommand{\thesisDissNumber}{28621}
\newcommand{\thesisAuthor}{Hasan Hassan}
\newcommand{\thesisAuthorTitle}{B.Sc. \& M.Sc., TOBB University of Economics and Technology}
\newcommand{\thesisUni}{\protect{ETH Z\"urich}}

\newcommand{\thesisYear}{2022}
\newcommand{\thesisDOI}{TBD}

\newif\ifcameraready
\camerareadytrue

\newif\ifisthesis
\isthesistrue

\newcommand{\versionnum}[0]{3.1}

\ifcameraready
    \documentclass[12pt,oneside,a4paper]{ethzthesis}
\else
    \documentclass[12pt,oneside,a4paper,version]{ethzthesis}
\fi

\definecolor{darkamber}{rgb}{0.9, 0.49, 0.0}
\definecolor{amber}{rgb}{0.0, 0.0, 0.0}
\definecolor{extradarkamber}{rgb}{0.0, 0.0, 0.0}

\newcommand*\circled[1]{\tikz[baseline=(char.base)]{
    \node[shape=circle,fill,inner sep=1pt] (char) {\textcolor{white}{\textbf{#1}}};}}

\usepackage[binary-units=true]{siunitx}
\usepackage{algorithm2e}

\usepackage{listings}
\usepackage{multirow}
\usepackage{float}
\usepackage{bm}
\usepackage{amsmath}

\usepackage{import}

\usepackage{makecell}
\usepackage{multirow}
\usepackage{multicol}
\usepackage{colortbl}
\usepackage{booktabs}
\usepackage{threeparttable}
\usepackage{lscape}
\usepackage{afterpage}

\usepackage[colorinlistoftodos,size=small]{todonotes}

\usepackage{cleveref}
\crefformat{chapter}{Chapter #2#1#3}
\crefformat{section}{Section #2#1#3}
\crefformat{subsection}{Section #2#1#3}
\crefformat{subsubsection}{Section #2#1#3}
\crefformat{appendix}{Appendix #2#1#3}

\usepackage{soul}
\soulregister\cite7
\soulregister\ref7
\soulregister\pageref7

\colorlet{LightYellow}{yellow!60!white}
\colorlet{LightBlue}{yellow!60!white}

\ifcameraready
    \DeclareRobustCommand{\bgyellow}[1]{{#1}}
    
\else
    \DeclareRobustCommand{\bgyellow}[1]{{#1}}
    
\fi

\usepackage[most]{tcolorbox}
\newtcolorbox{yellowbox}{
    colframe=yellow!60!white,
    colback =yellow!60!white,
        grow to left by=4mm,
        grow to right by=4mm,
        boxrule=0mm,
        before skip =0mm,
        after skip =0mm,
    top=0mm, bottom=0mm, right=3.25mm, left=3.25mm,
}

\newtcolorbox{bluebox}{
    colframe=LightBlue,
    colback =LightBlue,
        grow to left by=4mm,
        grow to right by=4mm,
        boxrule=0mm,
        before skip =0mm,
        after skip =0mm,
    top=0mm, bottom=0mm, right=3.25mm, left=3.25mm,
}

\newtcolorbox{yellowbox_break}{
    breakable,
    colframe=yellow!60!white,
    colback =yellow!60!white,
        grow to left by=4mm,
        grow to right by=4mm,
        boxrule=0mm,
        before skip =0mm,
        after skip =0mm,
    top=0mm, bottom=0mm, right=3.25mm, left=3.25mm,
}

\newlength\MyIndent
\ifcameraready
    \newenvironment{yellowb}{}{}
    \newenvironment{yellowb_break}{}{}
\else
    \newenvironment{yellowb}{}{}
    \newenvironment{yellowb_break}{}{}
\fi

\ifcameraready
  \newcommand{\hh}[1]{{#1}\xspace}
  \newcommand{\hhm}[1]{{#1}\xspace}
  \newcommand{\hhmii}[1]{{#1}\xspace}
  \newcommand{\hhmiii}[1]{{#1}\xspace}
  \newcommand{\hhmiv}[1]{{#1}\xspace}
  \newcommand{\hhmv}[1]{{#1}\xspace}
  \newcommand{\hhms}[1]{{#1}\xspace}
\else
  \newcommand{\hh}[1]{{#1}\xspace}
  \newcommand{\hhm}[1]{{#1}\xspace}
  \newcommand{\hhmii}[1]{{#1}\xspace}
  \newcommand{\hhmiii}[1]{{#1}\xspace}
  \newcommand{\hhmiv}[1]{{#1}\xspace}
  \newcommand{\hhmv}[1]{{#1}\xspace}
  \newcommand{\hhms}[1]{{\color{blue}{#1}}\xspace}
\fi

\makeatletter
\newcommand\requiredelimiter[2][########]{%
  \ifdefined#2%
    \def\@temp{\def#2#1}%
    \expandafter\@temp\expandafter{#2}%
  \else
    \@latex@error{\noexpand#2undefined}\@ehc
  \fi
}
\@onlypreamble\requiredelimiter
\makeatother

\newcommand{\squishlist} {
    \begin{list}{$\bullet$} {
        \setlength{\itemsep}{-2pt}
        \setlength{\parsep}{2pt}
        \setlength{\topsep}{0pt}
        \setlength{\partopsep}{0pt}
        \setlength{\leftmargin}{1.0em}
        \setlength{\labelwidth}{1em}
        \setlength{\labelsep}{0.5em}
    }
}

\newcommand{\squishend} {
    \end{list}
}

\newcommand{\mr}[2]{\multicolumn{1}{c}{\multirow{#1}{*}{\makecell{#2}}}}
\newcommand{\stripe}{\rowcolor{blue!5}}

\newcommand{\trcd}{\texttt{{tRCD}}\xspace}
\newcommand{\tras}{\texttt{{tRAS}}\xspace}
\newcommand{\trp}{\texttt{{tRP}}\xspace}
\newcommand{\twr}{\texttt{{tWR}}\xspace}
\newcommand{\trfc}{\texttt{{tRFC}}\xspace}
\newcommand{\trefi}{\texttt{{tREFI}}\xspace}
\newcommand{\tcl}{\texttt{{tCL}}\xspace}

\newcommand{\twtr}{\texttt{{tWTR}}\xspace}
\newcommand{\trtw}{\texttt{{tRTW}}\xspace}
\newcommand{\tfaw}{\texttt{{tFAW}}\xspace}
\newcommand{\trefw}{\texttt{{tREFW}}\xspace}

\newcommand{\cmdact}{\texttt{{ACT}}\xspace}
\newcommand{\cmdread}{\texttt{{RD}}\xspace}
\newcommand{\cmdwrite}{\texttt{{WR}}\xspace}
\newcommand{\cmdprech}{\texttt{{PRE}}\xspace}
\newcommand{\cmdpre}{\texttt{{PRE}}\xspace}
\newcommand{\cmdrefresh}{\texttt{{REF}}\xspace}

\newcommand{\cmdras}{\texttt{{RAS}}\xspace}
\newcommand{\cmdcas}{\texttt{{CAS}}\xspace}
\newcommand{\cmdwe}{\texttt{{WE}}\xspace}
\newcommand{\cmdcke}{\texttt{{CKE}}\xspace}
\newcommand{\cmdcs}{\texttt{{CS}}\xspace}

\newcommand{\cmark}{\ding{51}}%
\newcommand{\xmark}{\ding{55}}%

\begin{document}
\frenchspacing
\raggedbottom
\selectlanguage{english}
\pagenumbering{roman}
\pagestyle{plain}


\begin{titlepage}
    \begin{center}
        \begingroup
        \MakeUppercase{Diss. ETH No. \thesisDissNumber{}}
        \endgroup
    
        \hfill

        \vfill

        \begingroup
            \textbf{\thesisTitleFrontmatter}
        \endgroup

        \vfill

        \begingroup
            A thesis submitted to attain the degree of\\
            \vspace{0.5em}
            \MakeUppercase{Doctor of Sciences} of \MakeUppercase{\thesisUni} \\
            \vspace{0.5em}
            (Dr. sc. \thesisUni) \\
            
        \endgroup

        \vfill

        \begingroup
            presented by\\
            \vspace{0.5em}
            \MakeUppercase{\thesisAuthor}\\
            \thesisAuthorTitle\\
            \vspace{0.5em}
            born on 4 July 1991\\
        \endgroup

        \vfill

        \begingroup
            accepted on the recommendation of\\
            \vspace{0.5em}
            Prof.\ Dr.\ Onur Mutlu, examiner\\
            \vspace{0.5em}
            Prof. Dr. Derek Chiou, co-examiner \\
            \vspace{0.5em}
            Prof. Dr. Mattan Erez, co-examiner \\
            \vspace{0.5em}
            Dr. Mike O'Connor, co-examiner \\
            \vspace{0.5em}
            Prof. Dr. Moinuddin Qureshi, co-examiner \\
            \vspace{0.5em}
            Dr. Christian Weis, co-examiner
        \endgroup

        \vfill

        \thesisYear%

        \vfill
    \end{center}
\end{titlepage}

\thispagestyle{empty}

\hfill

\vfill

\noindent\thesisAuthor: \textit{\thesisTitlePlain,} 
\textcopyright\ \thesisYear

\bigskip

\noindent\MakeUppercase{DOI}: \thesisDOI

%
%
%
%
%

\cleardoublepage
\thispagestyle{empty}

\vspace*{3cm}

\begin{center}
    \textit{To my wife, Seray.}
\end{center}

\medskip

\clearpage
\chapter*{\vspace{-30pt}Acknowledgments}
\addcontentsline{toc}{chapter}{Acknowledgments}

I received tremendous help from many people throughout the six years I spent at ETH
Z\"urich.

Foremost, I extend my deepest gratitude to my advisor, Onur Mutlu, for his continuous support and guidance during my PhD journey.  
Back when I visited his research group at CMU as an intern, his passion in teaching and research was what
steered me towards pursuing a PhD degree.
As my PhD advisor, he has always been patient with me and provided insightful
criticism and advice on my research.
Without him, I may not have even attempted to pursue a PhD degree and could not have written this thesis.

I would like to express my sincere gratitude to my PhD committee members, Derek Chiou, Mattan Erez, Mike O'Connor, Moinuddin Qureshi, and Christian Weis, for their time and efforts to review my thesis and provide valuable feedback.

I am grateful to all SAFARI group members for providing stimulating intellectual environment and their friendship.
I thank Jeremie Kim and Minesh Patel for their company since my early days at ETH as friends and excellent researchers. Surviving PhD would not be possible without them.
I thank the other members of SAFARI, including Giray Ya\u{g}l{\i}k\c{c}{\i}, Can
F{\i}rt{\i}na, Juan G\'{o}mez Luna, Ataberk Olgun, Haocong Luo, Geraldo De
Oliveira, Lois Orosa, Nisa Bostanc{\i}, Rahul Bera, Yahya Can Tu\u{g}rul,
Konstantinos Kanellopoulos, Jisung Park, Nika Mansouri Ghiasi, Mohammed Alser,
Jawad Haj-Yahya, Arash Tavakkol, Yaohua Wang, Jo\"el Lindegger, Mohammad
Sadrosadati, Roknoddin Azizibarzoki, Jo\~ao Ferreira, Christian Rossi, Tracy
Ewen, and many others for their collaboration and support.

I am also thankful to wonderful collaborators I had the chance to work with during my PhD. I thank Saugata Ghose for the great assistance he provided during my time at CMU and after. I also thank many others that I met at CMU, including Nandita Vijaykumar, Vivek Seshadri, Samira Khan, Donghyuk Lee, Gennady Pekhimenko, Amirali Boroumand, Kevin Chang, Yixin Luo, Lavanya Subramanian, Rachata Ausavarungnirun, Damla Senol, and Nastaran Hajinazar. 
I thank Kaveh Razavi for fascinating discussions on various security topics. I
thank Victor van der Veen, Pietro Frigo, and Emanuele Vannacci for their
contributions.
%
I thank my internship mentors Stephan Meier and Tyler Huberty for their guidance during my time at Apple.

Finally, I would like to thank my family for their endless support, encouragement, and love. I thank my parents, Meryem and Ibrahim, and my sister, Ay\c{s}eg\"ul, for always standing by me and enabling all this. Above all I would like to thank my wife Seray, who kept me motivated to write this thesis with her unwavering support and love.

\clearpage
\chapter*{Abstract}
\addcontentsline{toc}{chapter}{Abstract}

DRAM is the primary technology used for main memory in modern systems.
Unfortunately, as DRAM scales down to smaller technology nodes, it faces key
challenges in both data integrity and latency, which strongly affect overall
system reliability, security, and performance. To develop reliable, secure, and
high-performance DRAM-based main memory for future systems, it is critical to
\hhm{rigorously} characterize, analyze, and understand various aspects (e.g.,
reliability, retention, latency\hhm{, RowHammer vulnerability}) of existing DRAM
chips and their architecture.
\hhm{The goal of this dissertation is to 1) develop techniques and infrastructures to enable such rigorous characterization, analysis, and understanding, and 2) enable new mechanisms to improve DRAM performance, reliability, and security based on the developed understanding.}

\hhm{To this end, in} this dissertation, we 1) design, implement, and prototype
a \hhm{new} practical-to-use and flexible FPGA-based DRAM characterization
infrastructure \hhm{(called SoftMC)}, 2) use the DRAM characterization
infrastructure to \hhmiv{develop a new experimental methodology} \hhm{(called
U-TRR)} to uncover the operation of existing \hhm{proprietary} in-DRAM RowHammer
protection mechanisms and craft new RowHammer access patterns to efficiently
circumvent these RowHammer protection mechanisms, 3) propose a new DRAM
architecture\hhm{, called Self-Managing DRAM,} for enabling autonomous and
efficient in-DRAM maintenance operations \hhm{that enable not only better
performance, efficiency, and reliability but also faster and easier adoption of
changes to DRAM chips}, and 4) propose a versatile DRAM substrate\hhm{, called
the Copy-Row (CROW) substrate,} that enables new mechanisms for improving DRAM
performance, energy consumption, and reliability.

\textbf{SoftMC.}  To develop reliable and high-performance DRAM-based main
memory in future systems, it is critical to \hhm{experimentally} characterize,
understand, and analyze various aspects (e.g., reliability, latency) of existing
DRAM chips. To enable this, there is a strong need for a publicly-available DRAM
testing infrastructure that can flexibly and efficiently test DRAM chips in a
manner accessible to both software and hardware developers. To this end, we
design and prototype SoftMC: a flexible and practical FPGA-based DRAM testing
infrastructure. SoftMC implements all low-level DRAM operations (i.e., DDR
commands) available in a typical memory controller (e.g., opening a row in a
bank, reading a specific column address, performing a refresh operation,
enforcing various timing constraints between commands). Using these low-level
operations, SoftMC can test and characterize any (existing or new) DRAM
mechanism that uses the existing DDR interface. SoftMC provides its users with a
simple and intuitive high-level programming interface that completely hides the
low-level details of the FPGA. SoftMC is freely available as an open-source tool
and it has enabled many research projects since its release\hhm{, leading to new understanding and new techniques}.

\textbf{U-TRR.} \hhm{RowHammer is a critical vulnerability in modern DRAM chips
that can lead to reliability, safety, and security problems in computing
systems. As such, DRAM vendors have been implementing techniques to protect DRAM
chips against RowHammer.} We challenge the claim of DRAM vendors that their DRAM
chips are completely protected against RowHammer using proprietary,
undocumented, and obscure on-die Target Row Refresh (TRR) mechanisms. To assess
the security guarantees of recent DRAM chips, we develop Uncovering TRR (U-TRR),
\hhm{a new} experimental methodology to analyze in-DRAM TRR implementations.
U-TRR is based on the new observation that data retention failures in DRAM
enable a side channel that leaks information on how TRR refreshes potential
victim rows. U-TRR allows us to (i) understand how logical DRAM rows are laid
out physically in silicon; (ii) study undocumented on-die TRR mechanisms; and
(iii) combine (i) and (ii) to evaluate the RowHammer security guarantees of
modern DRAM chips. We show how U-TRR allows us to craft RowHammer access
patterns that successfully circumvent the TRR mechanisms employed in 45 DRAM
modules of the three major DRAM vendors. We find that the DRAM modules we
analyze are vulnerable to RowHammer, having bit flips in up to 99.9\% of all
DRAM rows \hhm{and that simple error-correcting codes cannot prevent bit flips.
As such, more robust techniques to \hhmiv{protect against} the RowHammer
vulnerability are necessary}. \hhmiv{We publicly release the source code of our
implementation of the U-TRR methodology.}

\textbf{Self-Managing DRAM.} 
\hhm{To ensure reliable and secure DRAM operation, three types of maintenance
operations are typically required: 1) DRAM refresh, 2) RowHammer protection, and
3) memory scrubbing. The reliability and security of DRAM chips continuously
worsen as DRAM technology node scales to smaller sizes. Consequently, new DRAM
chip generations necessitate making existing maintenance operations more
aggressive (e.g., lowering the refresh period) and introducing new types of
maintenance operations (e.g., targeted refresh for mitigating RowHammer) while
keeping the overheads of maintenance operations minimal. Unfortunately,
modifying the existing DRAM maintenance operations is difficult due to the
current rigid DRAM interface that places the memory controller completely in
charge of DRAM control. Implementing new or modifying existing maintenance
operations often require difficult-to-realize changes in the DRAM interface, the
memory controller, and \hhmiv{potentially} other system components (e.g., system
software). Our goal is to 1) ease, and thus accelerate, the process of
implementing new DRAM maintenance operations and 2) enable more efficient
in-DRAM maintenance operations.} To this end, we propose Self-Managing DRAM
(SMD), a new low-cost DRAM architecture that enables implementing new in-DRAM
maintenance mechanisms with no further changes in the DRAM interface, memory
controller, or other system components. We use SMD to implement six maintenance
mechanisms for three use cases: 1) \hhm{DRAM} refresh, 2) RowHammer protection,
and 3) memory scrubbing. Our evaluations show that SMD-based maintenance
operations have significantly lower system performance and energy overheads
compared to conventional DDR4 DRAM. \hhm{A combination of SMD-based maintenance
mechanisms that perform refresh, RowHammer protection, and memory scrubbing
achieve \hhmiv{significant speedup and lower DRAM energy across a wide variety
of system configurations. SMD's benefits increase as DRAM chips become denser.}
We publicly release all SMD source code and data.}

\textbf{CROW.} \hhm{Three} major challenges to DRAM scaling \hhm{(i.e., high
access latencies, high refresh overheads, and increasing reliability problems
like RowHammer)} are difficult to solve efficiently by directly modifying the
underlying cell array structure. This is because commodity DRAM implements an
extremely dense DRAM cell array that is optimized for low area-per-bit. Because
of its density, even a small change in the DRAM cell array structure may incur
non-negligible area overhead. \hhm{Thus, we would like to} lower the DRAM access
latency, reduce the refresh overhead, and improve DRAM reliability with no
changes to the DRAM cell architecture, and with only minimal changes to the DRAM
chip. \hhm{To this end, we} propose Copy-Row DRAM (CROW), a flexible substrate
that enables new mechanisms for improving DRAM performance, energy efficiency,
and reliability. We use the CROW substrate to implement 1) a low-cost in-DRAM
caching mechanism that lowers DRAM activation latency to frequently-accessed
rows by 38\% and 2) a mechanism that avoids the use of short-retention-time rows
to mitigate the performance and energy overhead of DRAM refresh operations.
CROW's flexibility allows the implementation of both mechanisms at the same
time. Our evaluations show that the two \hhm{CROW-based} mechanisms
synergistically improve system performance by 20.0\% and reduce DRAM energy by
22.3\% for memory-intensive four-core workloads. \hhm{We publicly release the
source code of CROW.}

\hhm{Holistically, via these four major contributions, this} dissertation shows
the importance of \hhm{rigorously} characterizing the \hhm{reliability, latency,
and RowHammer vulnerability} of existing DRAM chips and understanding their
architecture for developing practical and low-overhead mechanisms for
efficiently improving DRAM reliability, security, and performance. We believe
and hope that\hhm{, via the new infrastructure, understanding, and techniques we
develop and enable,} this dissertation encourages \hhm{similar experimental
understanding-driven} innovation in the design of future memories.

\clearpage
\chapter*{Zusammenfassung}
\addcontentsline{toc}{chapter}{Zusammenfassung}

DRAM ist die wichtigste Hauptspeicher-Technologie in modernen Systemen. Leider
steht DRAM bei der Verkleinerung auf kleinere Technologieknoten vor grossen
Herausforderungen in Bezug auf Datenintegrität und Latenz, die die
Zuverlässigkeit, Sicherheit und Leistung des Gesamtsystems stark
beeinträchtigen. Um zuverlässige, sichere und leistungsstarke DRAM-basierte
Hauptspeicher für künftige Systeme zu entwickeln, müssen verschiedene Aspekte
(z. B. Zuverlässigkeit, Datenerhaltung, Latenz, RowHammer-Anfälligkeit) der
vorhandenen DRAM-Chips und ihrer Architektur genau charakterisiert, analysiert
und verstanden werden. Ziel dieser Dissertation ist es, 1) Techniken und
Infrastrukturen zu entwickeln, die eine solche rigorose Charakterisierung,
Analyse und ein solches rigoroses Verständnis ermöglichen, und 2) neue
Mechanismen zur Verbesserung der DRAM-Leistung, -Zuverlässigkeit und -Sicherheit
auf der Grundlage des entwickelten Verständnisses zu ermöglichen.

Zu diesem Zweck entwickeln, implementieren und prototypisieren wir in dieser
Dissertation 1) eine neue praktisch einsetzbare und flexible FPGA-basierte
DRAM-Charakterisierungs\-infrastruktur (genannt SoftMC), 2) \hhmiv{die
DRAM-Charakterisierungs\-infrastruktur zur Entwicklung einer neuen
experimentellen Methodik (genannt U-TRR) verwendet, um die Funktionsweise
bestehender proprietärer In-DRAM RowHammer-Schutzmechanismen aufzudecken und
neue RowHammer-Zugriffsmuster zu entwickeln, um diese
RowHammer-Schutzmechanismen effizient zu umgehen,} 3) eine neue
DRAM-Architektur, genannt Self-Managing DRAM, vorschlagen, um autonome und
effiziente In-DRAM Wartungsvorgänge zu ermöglichen, die nicht nur eine bessere
Leistung, Effizienz und Zuverlässigkeit, sondern auch eine schnellere und
einfachere Annahme von Änderungen an DRAM-Chips ermöglichen, und 4) ein
vielseitiges DRAM-Substrat, genannt Copy-Row (CROW)-Substrat, vorschlagen, das
neue Mechanismen zur Verbesserung der DRAM-Leistung, des Energieverbrauchs und
der Zuverlässigkeit ermöglicht.

\textbf{SoftMC.} Für die Entwicklung zuverlässiger und leistungsfähiger
DRAM-basierter Hauptspeicher in zukünftigen Systemen ist es von entscheidender
Bedeutung, verschiedene Aspekte (z. B. Zuverlässigkeit, Latenz) der vorhandenen
DRAM-Chips experimentell zu charakterisieren, zu verstehen und zu analysieren.
Um dies zu ermöglichen, besteht ein dringender Bedarf an einer öffentlich
zugänglichen DRAM-Testinfrastruktur, mit der DRAM-Chips flexibel und effizient
getestet werden können und die sowohl für Software- als auch für
Hardware-Entwickler zugänglich ist. Zu diesem Zweck entwickeln und
prototypisieren wir SoftMC: eine flexible und praktische FPGA-basierte
DRAM-Testinfrastruktur. SoftMC implementiert alle Low-Level-DRAM-Operationen
(d.h. DDR-Befehle), die in einem typischen Speicher-Controller verfügbar sind
(z.B. Öffnen einer Zeile in einer Bank, Lesen einer bestimmten Spaltenadresse,
Durchführen einer Refresh-Operation, Durchsetzen verschiedener
Timing-Beschränkungen zwischen Befehlen). Mit diesen Low-Level-Operationen kann
SoftMC jeden (existierenden oder neuen) DRAM-Mechanismus, der die existierende
DDR-Schnittstelle nutzt, testen und charakterisieren. SoftMC bietet seinen
Benutzern eine einfache und intuitive High-Level-Programmierschnittstelle, die
die Low-Level-Details des FPGAs vollständig verbirgt. SoftMC ist als
Open-Source-Tool frei verfügbar und hat seit seiner Veröffentlichung viele
Forschungsprojekte ermöglicht, die zu neuen Erkenntnissen und neuen Techniken
geführt haben.

\textbf{U-TRR.} RowHammer ist eine kritische Schwachstelle in modernen
DRAM-Chips, die zu Zuverlässigkeits-, Sicherheits- und Schutzproblemen in
Computersystemen führen kann. Aus diesem Grund haben DRAM-Hersteller Techniken
zum Schutz von DRAM-Chips vor RowHammer implementiert. Wir hinterfragen die
Behauptung von DRAM-Anbietern, ihre DRAM-Chips seien durch proprietäre,
undokumentierte und undurchsichtige on-die Target Row Refresh (TRR)-Mechanismen
vollständig gegen RowHammer geschützt. Um die Sicherheitsgarantien aktueller
DRAM-Chips zu bewerten, entwickeln wir Uncovering TRR (U-TRR), eine neue
experimentelle Methode zur Analyse von TRR-Implementierungen im DRAM. U-TRR
basiert auf der neuen Beobachtung, dass Fehler bei der Datenerhaltung in DRAM
einen Seitenkanal ermöglichen, der Informationen darüber preisgibt, wie TRR
potenzielle Opfer-Zeilen refresht. U-TRR ermöglicht es uns, (i) zu verstehen,
wie logische DRAM-Zeilen physisch in auf dem chip angeordnet sind; (ii)
undokumentierte on-die TRR-Mechanismen zu untersuchen; und (iii) Punkte (i) und
(ii) zu kombinieren, um die RowHammer-Sicherheitsgarantien moderner DRAM-Chips
zu bewerten. Wir zeigen, wie U-TRR es uns ermöglicht, RowHammer-Zugriffsmuster
zu erstellen, die erfolgreich die TRR-Mechanismen umgehen, die in 45
DRAM-Modulen der drei großen DRAM-Hersteller eingesetzt werden. Wir stellen
fest, dass die von uns analysierten DRAM-Module anfällig für RowHammer sind, da
in bis zu 99,9\% aller DRAM-Zeilen Bitflips auftreten, und dass einfache
Fehlerkorrekturcodes Bitflips nicht verhindern können. Daher werden robustere
Techniken zur Verhinderung der RowHammer-Schwachstelle benötigt. \hhmiv{Wir
veröffentlichen den Quellcode unserer Implementierung der U-TRR-Methode.}

\textbf{Self-Managing DRAM}. Um einen zuverlässigen und sicheren DRAM-Betrieb zu
gewährleisten, sind in der Regel drei Arten von Wartungsvorgängen erforderlich:
1) DRAM-Refresh, 2) RowHammer-Schutz und 3) Memory-Scrubbing. Die
Zuverlässigkeit und Sicherheit von DRAM-Chips verschlechtert sich ständig, da
die DRAM-Technologieknoten immer kleiner werden. Folglich müssen bei neuen
DRAM-Chip-Generationen die bestehenden Wartungsvorgänge aggressiver gestaltet
werden (z. B. Verringerung des Refresh-Periode) und neue Arten von
Wartungsvorgängen eingeführt werden (z. B. gezielte Refreshes zur Milderung von
RowHammer), während der Overhead der Wartungsvorgänge minimal gehalten wird.
Leider ist die Änderung der bestehenden DRAM-Wartungsvorgänge aufgrund der
derzeitigen starren DRAM-Schnittstelle, die die DRAM-Steuerung vollständig dem
Speicher-Controller überlässt, schwierig. Die Implementierung neuer oder die
Änderung bestehender Wartungsvorgänge erfordert häufig schwer zu realisierende
Änderungen an der DRAM-Schnittstelle, dem Speicher-Controller und anderen
Systemkomponenten (z. B. Systemsoftware). Unser Ziel ist es, 1) den Prozess der
Implementierung neuer DRAM-Wartungsvorgänge zu vereinfachen und damit zu
beschleunigen und 2) effizientere In-DRAM-Wartungsvorgänge zu ermöglichen. Zu
diesem Zweck schlagen wir Self-Managing DRAM (SMD) vor, eine neue,
kostengünstige DRAM-Architektur, die die Implementierung neuer In-DRAM
Wartungsmechanismen ohne weitere Änderungen an der DRAM-Schnittstelle, dem
Speicher-Controller oder anderen Systemkomponenten ermöglicht. Wir verwenden SMD
zur Implementierung von sechs Wartungsmechanismen für drei Anwendungsfälle: 1)
DRAM-Refresh, 2) RowHammer-Schutz und 3) Memory-Scrubbing. Unsere Auswertungen
zeigen, dass SMD-basierte Wartungsvorgänge im Vergleich zu konventionellem
DDR4-DRAM eine deutlich geringere Systemleistungs- und Energiekosten haben. 
\hhmiv{Eine Kombination aus SMD-basierten Wartungsmechanismen, die Refresh-, RowHammer-Schutz- und Memory-Scrubbing-Operationen durchführen, führt zu einer erheblichen Beschleunigung und einem geringeren DRAM-Energieverbrauch in einer Vielzahl von Systemkonfigurationen. Die Vorteile von SMD nehmen zu, je dichter die DRAM-Chips werden.}
Wir veröffentlichen den gesamten SMD-Quellcode und -Daten.

\textbf{CROW.} Die drei grössten Herausforderungen bei der Skalierung von DRAM
(d. h. hohe Zugriffslatenzen, hohe Refresh-Kostenund zunehmende
Zuverlässigkeitsprobleme wie RowHammer) lassen sich nur schwer durch direkte
Änderung der zugrunde liegenden Zell-Array-Struktur effizient lösen. Dies liegt
daran, dass handelsüblicher DRAM ein extrem dichtes DRAM-Zellen-Array
implementiert, das auf eine geringe Fläche pro Bit optimiert ist. Aufgrund
dieser Dichte kann selbst eine kleine Änderung der DRAM-Zellen-Array-Struktur zu
einem nicht zu vernachlässigenden Flächen-Overhead führen. Daher möchten wir die
DRAM-Zugriffslatenz verringern, die Refresh-Kosten- reduzieren und die
DRAM-Zuverlässigkeit verbessern, ohne die DRAM-Zellenarchitektur zu verändern
und mit nur minimalen Änderungen am DRAM-Chip. Zu diesem Zweck schlagen wir
Copy-Row DRAM (CROW) vor, ein flexibles Substrat, das neue Mechanismen zur
Verbesserung der DRAM-Leistung, Energieeffizienz und Zuverlässigkeit ermöglicht.
Wir verwenden das CROW-Substrat, um 1) einen kostengünstigen In-DRAM
Caching-Mechanismus zu implementieren, der die DRAM-Aktivierungslatenz für
häufig genutzte Zeilen um 38\% verringert, und 2) einen Mechanismus, der die
Verwendung von Zeilen mit kurzer Datenerhaltungsdauer vermeidet, um den
Leistungs- und Energie-Overhead von DRAM-Refresh-perationen zu verringern. Die
Flexibilität von CROW ermöglicht die gleichzeitige Implementierung beider
Mechanismen. Unsere Auswertungen zeigen, dass die beiden CROW-basierten
Mechanismen die Systemleistung um 20,0\% verbessern und den
DRAM-Energieverbrauch bei speicherintensiven Vier-Kern-Workloads um 22,3\%
senken. Wir veröffentlichen den Quellcode von CROW.

Mit diesen vier Hauptbeiträgen zeigt diese Dissertation, wie wichtig es ist, die
Zuverlässigkeit, Latenz und RowHammer-Anfälligkeit bestehender DRAM-Chips genau
zu charakterisieren und ihre Architektur zu verstehen, um praktische und
kostengünstige Mechanismen zur effizienten Verbesserung der Zuverlässigkeit,
Sicherheit und Leistung von DRAM zu entwickeln. Wir glauben und hoffen, dass
diese Dissertation durch die neue Infrastruktur, das Verständnis und die
Techniken, die wir entwickeln und ermöglichen, ähnliche, auf experimentellem
Verständnis beruhende Innovationen bei der Entwicklung zukünftiger Speicher
anregt.

\pagestyle{headings}
\cleardoublepage
\tableofcontents
\newpage
\listoffigures
\newpage
\listoftables

\cleardoublepage
\pagenumbering{arabic}%
\chapter{Introduction}

\section{Problem Discussion}

\hhm{Dynamic Random Access Memory (DRAM)}~\cite{dennard1968field} has long been
the dominant memory technology used in almost all computing systems due to its
low latency and low cost per bit. DRAM vendors \hhm{scale the} DRAM technology
\hhm{by shrinking} DRAM cells to \hhm{continuously} reduce the cost of
DRAM~\cite{mutlu2013memory}. Unfortunately, while the density of the DRAM chips
has been increasing as a result of \hhm{technology} scaling, high-density DRAM
chips face three critical challenges~\cite{mutlu2014research, mutlu2013memory}:
(1)~high access latencies, (2)~high refresh overheads, and (3)~increasing
reliability problems. \hhm{These} three \hhm{DRAM scaling challenges} have a
major impact on the \hhm{performance, energy, and robustness (i.e., reliability,
safety, and security)} of a system. \hhm{We examine each of these critical
challenges in turn.}

\noindent\textbf{Challenge 1: High Access Latency.} The high DRAM access latency
is a challenge to improving system performance and energy efficiency. While DRAM
capacity increased significantly over the last two decades~\cite{jedec2008ddr3,
jedec2012ddr4, son2013reducing, chang2016understanding, lee2013tiered,
lee2015adaptive, mutlu2014research, mutlu2013memory, lee2016reducing,
patel2022enabling, patel2022case, luo2020clr}, DRAM access latency decreased
only slightly~\cite{lee2013tiered, mutlu2014research, chang2016understanding,
mutlu2013memory, lee2016reducing, patel2022enabling, patel2022case,
lee2015adaptive, luo2020clr, borkar2011future, hassan2016chargecache}. The high
DRAM access latency significantly degrades the performance of many
workloads~\cite{ferdman2012clearing, huang2014moby, gutierrez2011full,
zhu2015microarchitectural, hestness2014comparative, lee2015adaptive,
lee2013tiered, ghose2019demystifying, mutlu2003runahead, bera2019dspatch,
boroumand2018google, ghose2019processing, kanev2015profiling, koppula2019eden,
liu2019binary, son2013reducing, wilkes2001memory, wulf1995hitting,
oliveira2021damov, boroumand2021google, boroumand2022polynesia,
boroumand2020thesis}. The performance impact is particularly large for
applications that 1)~have working sets exceeding the cache capacity of the
system, 2)~suffer from high instruction and data cache miss rates, and 3)~have
low memory-level parallelism. While \hhm{some DRAM vendors} offer
latency-optimized DRAM modules~\cite{micron2021rldram, sato1998fast}, these
modules have significantly lower capacity and higher cost compared to commodity
DRAM~\cite{lee2013tiered, chang2016low, kim2012case}. Thus, reducing the high
DRAM access latency \emph{without \hhm{significantly affecting} capacity and
cost} in commodity DRAM remains an important
challenge~\cite{ghose2019demystifying, lee2013tiered, mutlu2013memory,
mutlu2014research}.

\noindent\textbf{Challenge 2: High Refresh Overhead.} The high DRAM refresh
overhead is a challenge to improving system performance and energy consumption.
A DRAM cell stores data in a capacitor that leaks charge over time. To maintain
correctness, \emph{every} DRAM cell requires periodic \emph{refresh} operations
that restore the charge level in a cell. As the DRAM cell size decreases with
process technology scaling, newer DRAM devices contain \hhm{both} more DRAM
cells \hhm{and smaller DRAM cells} than older DRAM devices~\cite{itrs}. As a
result, the performance and energy overheads of refresh operations scale
unfavorably \hhm{as DRAM technology scales into the
future}~\cite{chang2014improving, kang2014co, liu2012raidr}. In modern
DDR5~\cite{jedec2021ddr5} devices, the memory controller refreshes \emph{every}
DRAM cell every \SI{32}{\milli\second} \hhm{at the nominal temperature range and
every \SI{16}{\milli\second} at the extended temperature range}. Previous
studies show that 1)~refresh operations incur large performance overheads, as
DRAM cells \emph{cannot} be accessed when the cells are being
refreshed~\cite{chang2014improving, liu2012raidr, mukundan2013understanding,
nair2014refresh, jafri2020refresh, das2018vrl, qureshi2015avatar,
baek2014refresh, bhati2013coordinated, bhati2015flexible, cui2014dtail,
emma2008rethinking, ghosh2007smart, kim2020charge, liu2012flikker, nair2013case,
stuecheli2010elastic, zhang2014cream}; and 2)~\hhm{a large \hhmiv{fraction}
(e.g., 50\% in \SI{64}{\giga\bit} DRAM chips)} of the total DRAM energy is
consumed by the refresh operations~\cite{liu2012raidr, chang2014improving,
qureshi2015avatar, jafri2020refresh, liu2012flikker, kim2020charge,
stuecheli2010elastic, zhang2014cream}. 

\noindent\textbf{Challenge 3: Increasing Reliability Problems.} The increasing
vulnerability of DRAM cells to various failure mechanisms is an important
challenge to maintaining DRAM reliability with a small system performance and
energy overhead. As the process technology \hhm{node size reduces}, DRAM cells
\hhm{get} smaller and closer to each other, and thus \hhm{they} become more
susceptible to failures~\cite{mandelman2002challenges, mutlu2013memory,
redeker2002investigation, yaney1987meta, konishi1989analysis, liu2012raidr,
liu2013experimental, khan2014efficacy, khan2016case, khan2016parbor,
khan2017detecting, mutlu2014research, kim2020revisiting, orosa2021deeper,
mutlu2017rowhammer, kim2014flipping, mutlu2019rowhammer,
yauglikcci2022understanding}. A \hhmiv{real} \hhm{and prevalent} example of such
a failure mechanism in modern DRAM is \hhm{the RowHammer
vulnerability}~\cite{kim2014flipping, mutlu2017rowhammer, mutlu2019rowhammer,
mutlu2023fundamentally}. RowHammer causes \emph{disturbance errors} (i.e., bit
flips in vulnerable DRAM cells that are not being accessed) in DRAM rows
physically adjacent to a row that is repeatedly activated many times. In
addition to being detrimental to DRAM reliability, RowHammer also poses a threat
to system security \hhm{and safety}. Prior works~\cite{kim2020revisiting,
cojocar2019exploiting,yaglikci2021blockhammer, kim2014flipping,
park2020graphene, apple2015about, brasser2016can, konoth2018zebram,
van2018guardion, aweke2016anvil, lee2019twice, seyedzadeh2017cbt, son2017making,
you2019mrloc, greenfield2016throttling, yauglikcci2021security, kim2022mithril,
taouil2021lightroad, devaux2021method, marazzi2023rega, woo2023scalable,
wi2023shadow, kim2023ddr5} propose various mechanisms that protect systems
against RowHammer at the cost of increased performance and energy overheads.
\hhm{DRAM \hhmiv{also exhibits} other reliability issues \hhmiv{(e.g., data
retention failures, soft errors)} that necessitate \hhmiv{expensive} solutions
(e.g., increasing refresh rate, memory scrubbing), \hhmiv{which lead} to high
performance and energy overheads.}

\noindent\textbf{Our Goal.} To combat the \hhm{system-level implications of the
technology} scaling challenges of DRAM, it is essential to \hhm{rigorously
understand} DRAM cell behavior and the architecture of existing DRAM chips.
\hhm{The goal of this dissertation is to 1) develop techniques and
infrastructures to enable such rigorous characterization, analysis, and
understanding, and 2) enable new mechanisms to improve DRAM performance,
reliability, and security based on the developed understanding.}

\hhm{To this end,} in this dissertation, we first design, implement, and
prototype SoftMC, a practical-to-use and flexible FPGA-based DRAM
characterization infrastructure. SoftMC enables detailed analyses on DRAM cell
characteristics of real DRAM chips via precise control on the low-level DRAM
interface.
Second, we use SoftMC to uncover the operation and weaknesses of RowHammer
protection mechanisms implemented in modern DRAM chips. We develop a methodology
\hhm{(called U-TRR)} that uses DRAM retention failures as a side-channel to
monitor when a RowHammer protection mechanism performs additional refresh
operation to protect DRAM rows susceptible to a RowHammer attack. Based on the
understanding we develop, we craft new RowHammer access patterns that
efficiently circumvent in-DRAM RowHammer protection mechanisms implemented in
modern DDR4 DRAM chips from three major DRAM vendors.
Third, we propose \hhm{Self-Managing DRAM (SMD),} a new DRAM architecture and
interface to enable in-DRAM maintenance operations to efficiently perform 1)
DRAM refresh, 2) RowHammer protection, and 3) memory scrubbing. Our proposal
eases the adoption of new DRAM maintenance mechanisms by setting the memory
controller free from managing such DRAM maintenance operations.
Finally, we propose \hhm{the Copy-Row (CROW) DRAM} substrate that exploits the
fast and efficient data movement within a DRAM subarray to enable new mechanisms
for improving DRAM performance, energy consumption, and reliability.



\section{Thesis Statement}

Our thesis statement is \hhm{as follows}:

\begin{center}
    \parbox{15cm}{
        \emph{Understanding DRAM characteristics and operation through
        rigorous experimentation\\using real DRAM chips pave the way for developing
        new mechanisms that significantly\\improve DRAM performance, energy consumption, reliability,
        and security.}
    }
\end{center}

\section{Our Approach}

\hhm{We first enable detailed characterization of modern DRAM chips by designing
and prototyping a flexible FPGA-based infrastructure, which \hhmiv{we then} use
to develop understanding and mechanisms for improving system performance,
efficiency, reliability, and security. In the remainder of this section, we
briefly introduce our DRAM characterization infrastructure, new characterization
\hhmiv{studies}, and new mechanisms.}

\subsection{\hhm{Infrastructure for Characterizing DRAM Chips}}

\hhm{To} easily study the reliability characteristics and the operation of real
DRAM chips, we design and prototype \emph{SoftMC (Soft Memory Controller)}, a
flexible and easy-to-use experimental DRAM testing infrastructure. SoftMC is an
FPGA-based DRAM testing infrastructure that exposes the low-level DRAM command
interface to its users via a high-level programming interface. The high-level
programming interface completely hides the low-level details of the FPGA from
users. Users implement their DRAM experiment routines or mechanisms in a
high-level language that automatically gets translated into the low-level SoftMC
operations in the FPGA. SoftMC offers a wide range of use cases, such as
characterizing the effects of variation within a DRAM chip and across DRAM
chips, verifying the correctness of new DRAM mechanisms on actual hardware, and
experimentally discovering the reliability, retention, and timing
characteristics of an unknown or newly-designed DRAM chip (or finding the best
specifications for a known DRAM chip). \hhm{SoftMC was published at HPCA
2017~\cite{hassan2017softmc} and released as an open source
tool~\cite{softmcsource} at the same time. Since then, many
works~\cite{koppula2019eden,chang2016understanding,
chang2017understanding,frigo2020trrespass,orosa2021deeper,gao2019computedram,kim2020revisiting,kim2019d,hassan2021uncovering,khan2017detecting,ghose2018your,yauglikcci2022understanding,olgun2021quac,orosa2021codic,
talukder2018exploiting, talukder2019prelatpuf, talukder2018ldpuf,
talukder2020towards, bepary2022dram,farmani2021rhat, yaglikci2022hira, gao2022fracdram} used
different versions of SoftMC in various DRAM characterization studies \hhmiv{as
well as studies that introduced new ideas}.}

\subsection{\hhm{Understanding in-DRAM RowHammer Protection Mechanisms}}
\hhm{We} use SoftMC to study the security guarantees of RowHammer protection
mechanisms that DRAM vendors implement in their chips. Different vendors
implement different RowHammer protection mechanisms\hhm{,} commonly referred to
as \emph{Target Row Refresh (TRR)}. The main idea of TRR is to detect an
aggressor row (i.e., a row that is \hhm{frequently} activated \hhm{within
a time frame}) and refresh its victim rows (i.e., neighboring rows that are
physically adjacent to the aggressor row). However, the exact operation of TRR
is unknown. Some of the major DRAM vendors advertise \emph{RowHammer-free} DDR4
DRAM chips~\cite{lee2014green, micronddr4, frigo2020trrespass}. However, none of
the DRAM vendors have so far disclosed the implementation details let alone
proved the protection guarantees of their TRR mechanisms. Thus, we need new
methods for identifying whether or not DRAM chips are fully secure against
RowHammer. We develop U-TRR, a practical methodology that uncovers the inner
workings of TRR mechanisms in modern DRAM chips. The goal of U-TRR is to enable
the observation (i.e., uncovering) of all \hhm{protective} refreshes generated
by TRR after inducing a carefully crafted sequence of DRAM accesses. To make
this possible, we make a key observation that retention failures that occur on a
DRAM row can be used as a side channel to detect \emph{when} the row is
refreshed, due to either TRR or periodic refresh operations. U-TRR uncovers
important details of the TRR designs of three major DRAM vendors. We demonstrate
the usefulness of these insights for developing effective RowHammer attacks on
DRAM chips from each vendor by crafting specialized DRAM access patterns that
hammer a row enough times to cause a RowHammer bit flip \emph{without} alerting
the TRR protection mechanism (i.e., by redirecting potential TRR refreshes
\emph{away from} the victim rows). In our evaluation, we find that all tested
DRAM modules with different manufacturing dates (from 2016 to 2020) are
vulnerable to the new access patterns we can craft via U-TRR. \hhm{We open
source our U-TRR methodology and make it freely available for further research
and development~\cite{utrrsource}.}

\subsection{\hhm{Enabling Easy Adoption of Efficient in-DRAM Maintenance
Operations}}
\hhm{We develop} a new DRAM architecture and interface, Self-Managing DRAM, to
enable autonomous in-DRAM maintenance operations. For reliable and secure
operation of modern and future DRAM chips, \emph{DRAM maintenance operations},
such as periodic refresh, RowHammer protection, and memory scrubbing are
critical. An ideal maintenance operation should efficiently improve DRAM
reliability and security with minimal system performance and energy consumption
overheads. As DRAM technology node scales to smaller sizes, the reliability and
security of DRAM chips worsen. As a result, new DRAM chip generations
necessitate making existing maintenance operations more intensive (e.g.,
\hhm{increasing the refresh rate}~\cite{jedec2021ddr5, jedec2020lpddr5,
apple2015about} \hhm{and memory scrubbing frequency~\cite{qureshi2015avatar}})
and introducing new types of maintenance operations (e.g., targeted
refresh~\cite{hassan2021uncovering,frigo2020trrespass,jattke2022blacksmith} and
DDR5 RFM~\cite{jedec2021ddr5} as RowHammer defenses). Unfortunately, modifying
the DRAM maintenance operations is difficult due to the current rigid DRAM
interface that places the memory controller completely in charge of DRAM
control. Implementing new or modifying existing maintenance operations often
require difficult-to-realize changes in the DRAM interface, the memory
controller, and other system components (e.g., system software). Self-Managing
DRAM 1) eases the process and thus speed of implementing new DRAM maintenance
operations and 2) enables more efficient in-DRAM maintenance operations. Using
Self-Managing DRAM, we develop maintenance operations for refresh, RowHammer
protection, and memory scrubbing that incur significantly lower system
performance and energy overhead compared to existing \hhm{state-of-the-art}
approaches. We believe that enabling autonomous in-DRAM maintenance operations
would encourage innovation, reduce time-to-market, and ease adoption of novel
mechanisms that improve DRAM efficiency, reliability, and security. \hhm{To
foster research and development in this direction, we open source our
Self-Managing DRAM framework~\cite{smdsource} \hhmiv{and early-release our
Self-Managing DRAM work on arXiv.org~\cite{hassan2022self}.}}

\subsection{\hhm{Improving DRAM Access Latency and Reducing DRAM Refresh
Overhead via Simultaneous Row Activation}}
\hhm{We} propose Copy-Row DRAM (CROW), a flexible in-DRAM substrate that can be
used in multiple different ways to address the performance, energy efficiency,
and reliability challenges of DRAM. The key idea of CROW is to provide a fast,
low-cost mechanism to duplicate select rows in DRAM that contain data that is
most sensitive or vulnerable to latency, refresh, or reliability issues. At a
high level, CROW partitions each DRAM subarray into two regions (\emph{regular}
rows and \emph{copy} rows) and enables independent control over the rows in each
region. Using CROW, we develop two novel mechanisms: 1) \emph{CROW-cache}, which
reduces the DRAM access latency and 2) \emph{CROW-ref}, which reduces the DRAM
refresh overhead. The key idea of \emph{CROW-cache} is to 1) duplicate data from
recently-accessed regular rows into a small cache made up of \emph{copy rows},
and 2) simultaneously activate a duplicated regular row with its corresponding
\emph{copy row} when the regular row needs to be activated again. By activating
both the regular row and its corresponding \emph{copy row}, \emph{CROW-cache}
reduces the time needed to open the row and begin performing read and/or write
requests to the row by 38\%. The key idea of \emph{CROW-ref} is to avoid storing
any data in DRAM rows that contain weak cells, such that the entire DRAM chip
can use a longer refresh interval. \emph{CROW-ref} uses an efficient profiling
mechanism~\cite{patel2017reaper} to identify weak DRAM cells, remaps the regular
rows containing weak cells to strong \emph{copy rows}, and records the remapping
information. Our evaluations show that \emph{CROW-cache} and \emph{CROW-ref}
significantly improve system performance and energy efficiency compared to
conventional DRAM and incur small hardware area overhead. \hhmiv{We also
describe how CROW can be used to mitigate the RowHammer vulnerability.} \hhm{To
aid future research and development, we open source
CROW~\cite{crow_spice_github}.}

\section{Contributions}

This dissertation makes the following \hhm{major} contributions:

\begin{enumerate}
\item We introduce SoftMC, the first open-source FPGA-based experimental memory
testing infrastructure. SoftMC implements all low-level DRAM operations in a
programmable memory controller that is exposed to the user with a flexible and
easy-to-use interface, and hence enables the efficient characterization of
modern DRAM chips and evaluation of mechanisms built on top of low-level DRAM
operations. To our knowledge, SoftMC is the first publicly-available
infrastructure that exposes a high-level programming interface to ease memory
testing and characterization.

    \begin{enumerate}
        \item We provide a prototype implementation of SoftMC with a
        high-level software interface for users and a low-level FPGA-based
        implementation of the memory controller. We have released the
        software interface and the implementation publicly as a
        freely-available open-source tool~\cite{softmcsource}.

        \item We demonstrate the capability, flexibility, and programming ease
        of SoftMC by implementing two example use cases. Our first use case
        demonstrates the ease of use of SoftMC by implementing a routine for
        retention time characterization of DRAM cells. Our second use case
        demonstrates the effectiveness of SoftMC as a new tool to test existing
        or new mechanisms on existing memory chips. Using SoftMC, we demonstrate
        that the expected effect (i.e., highly-charged DRAM rows can be accessed
        faster than others) of two recently-proposed mechanisms is not
        observable in 24 modern DRAM chips from three major \hhm{vendors}.
    \end{enumerate}

\item \hhm{Using SoftMC, we} extensively analyze 45 modern DDR4 modules from
three major DRAM vendors to understand the operation and security guarantess of
their in-DRAM RowHammer protection mechanisms called Target Row Refresh (TRR).
We develop U-TRR, a new methodology for reverse-engineering TRR mechanisms and
assessing their security \hhm{properties}. 

    \begin{enumerate}
        \item We use U-TRR to understand and uncover the TRR implementations of
        45 DDR4 modules from the three major DRAM vendors. This evaluation shows
        that our new methodology is broadly applicable to any DRAM chip.

        \item Leveraging the TRR implementation details uncovered by U-TRR, we
        craft specialized RowHammer access patterns that make existing TRR
        protections ineffective. 

        \item Our specialized U-TRR-discovered access patterns are
        significantly more effective than patterns from the
        state-of-the-art~\cite{frigo2020trrespass}: we show that our new RowHammer
        access patterns cause 1) bit flips in all 45 DDR4
        modules we comprehensively examine, 2) bit flips in up to
        99.9\% of the all rows in a DRAM bank, and 3) two and more (up
        to $7$) bit flips in a single 8-byte dataword, enabling practical RowHammer
        attacks in systems that employ ECC.

        \item \hhm{We release the source code of our U-TRR methodology to aid
        future research and development~\cite{utrrsource}.}
    \end{enumerate}

\item We \hhm{develop} Self-Managing DRAM (\texttt{SMD}), a new DRAM chip design
and interface to enable autonomous and efficient in-DRAM maintenance operations.
We implement \texttt{SMD} with small changes to modern DRAM chips and memory
controllers. \hhm{The core components of SMD incur area overhead of only 1.63\%
of a \SI{45.5}{\milli\meter\squared} DRAM chip and increase DRAM row activation
latency by only 0.4\%.}

    \begin{enumerate}
        \item We use \texttt{SMD} to implement efficient DRAM maintenance
        mechanisms for three use cases: periodic refresh, RowHammer protection,
        and memory scrubbing. 

        \item We rigorously evaluate the performance and energy of the new
        maintenance mechanisms. \texttt{SMD} provides large performance and
        energy benefits while also improving reliability \hhm{and security}
        across a variety of systems and workloads.

        \item We open source our Self-Managing DRAM framework to aid future research and development in this direction~\cite{smdsource}.
    \end{enumerate}

\item We \hhm{introduce} Copy-Row DRAM (CROW), a flexible and low-cost substrate
in commodity DRAM that enables mechanisms for improving DRAM performance, energy
efficiency, and reliability by providing two sets of rows that have independent
control in each subarray. CROW does not change the extremely-dense cell array
and has \emph{low} cost (0.48\% additional area overhead in the DRAM chip,
\SI{11.3}{\kibi\byte} storage overhead in the memory controller, and 1.6\% DRAM
storage capacity overhead).

    \begin{enumerate}
        \item We propose three mechanisms that exploit the CROW substrate:
        1)~CROW-cache, an in-DRAM cache to reduce the DRAM access latency,
        2)~CROW-ref, a remapping scheme for weak DRAM rows to reduce the DRAM
        refresh overhead, and 3)~a mechanism for mitigating the RowHammer
        vulnerability. We show that CROW allows these mechanisms to be employed
        at the same time.

        \item We evaluate the performance, energy savings, and overhead of
        CROW-cache and CROW-ref, showing significant performance and energy
        benefits over a state-of-the-art commodity DRAM chip. We also compare
        CROW-cache to two prior proposals to reduce DRAM
        latency~\cite{kim2012case, lee2013tiered} and show that CROW-cache is
        more area- and energy-efficient at reducing DRAM latency.

        \item We make the source code of CROW publicly available to aid future
        research and development~\cite{crow_spice_github}.
    \end{enumerate}

\end{enumerate}

\section{Outline}

This dissertation consists of 8 chapters. \cref{chap:bg} provides background on
DRAM organization, operation, and \hhmii{problems}, which are essential for
understanding our proposals in later chapters. In \cref{chap:softmc}, we
describe SoftMC, \hhm{our new} FPGA-based DRAM characterization infrastructure.
\cref{chap:utrr} presents our U-TRR methodology for uncovering the operation of
in-DRAM TRR mechanisms using SoftMC. \cref{chap:smd} introduces Self-Managing
DRAM (SMD), a new DRAM architecture and interface for enabling efficient and
autonomous in-DRAM maintenance operations. \cref{chap:crow} presents Copy-Row
DRAM (CROW), a substrate that makes slight modifications in DRAM subarray
architecture to enable \hhm{efficient} mechanisms for improving DRAM
performance, energy efficiency, and reliability. Finally, \cref{chap:conc}
summarizes the dissertation and provides future research directions that this
dissertation enables.

\chapter{Background}
\label{chap:bg}

In this chapter, we \hhm{provide the background on DRAM necessary for
understanding our \hhmiv{methods,} observations, and \hhmiv{techniques}
presented in later chapters. We first describe DRAM organization and operation.
Afterward, we discuss the \hhmiv{major} performance, reliability, and security
problems of DRAM} \hhmiv{that we tackle}. We refer the reader to \hhm{many}
prior works \hhm{in the area}~\cite{chang2016understanding, chang2014improving,
chang2016low, chang2017understanding, hassan2019crow, hassan2016chargecache,
kim2014flipping, kim2012case, lee2015adaptive, lee2013tiered,
lee2016simultaneous, lee2017design, lee2015decoupled, liu2012raidr,
liu2013experimental, seshadri2013rowclone, seshadri2017ambit,
seshadri2015gather, seshadri2020indram, zhang2014half, kim2020revisiting,
frigo2020trrespass, luo2020clr, yaglikci2021blockhammer, wang2020figaro,
cojocar2020are, patel2019understanding, bhati2015flexible, zhang2016restore,
kim2019d, kim2018dram, olgun2021quac, kim2015ramulator,
yauglikcci2022understanding, qureshi2015avatar, orosa2021deeper, orosa2021codic,
hassan2021uncovering, nair2013case, mutlu2013memory, mutlu2014research,
mutlu2017rowhammer, mutlu2019rowhammer, bostanci2022dr, hajinazar2021simdram,
patel2020bit, khan2017detecting, khan2016parbor, khan2016case, khan2014efficacy,
davis2001modern, rixner2000memory, keeth2007dram, itoh2013vlsi, jacob2010memory,
gong2015clean, gong2018duo, jeong2012balancing, jeong2012drsim, kim2015bamboo,
kim2015frugal, kim2016all, kim2016relaxfault, yoon2010virtualized,
nair2013archshield, nair2016xed, saileshwar2022randomized,
chandrasekar2014exploiting, jung2014optimized, jung2015dramsys,
jung2015omitting, jung2016reverse, jung2017platform, kraft2018improving,
weis2015retention, weis2017dramspec, chandrasekar2012drampower} for
\hhm{various} details on DRAM background.



\section{DRAM Organization}
\label{sec:dram_organization}

\begin{figure}[!ht]
    \centering
    \includegraphics[width=.95\linewidth]{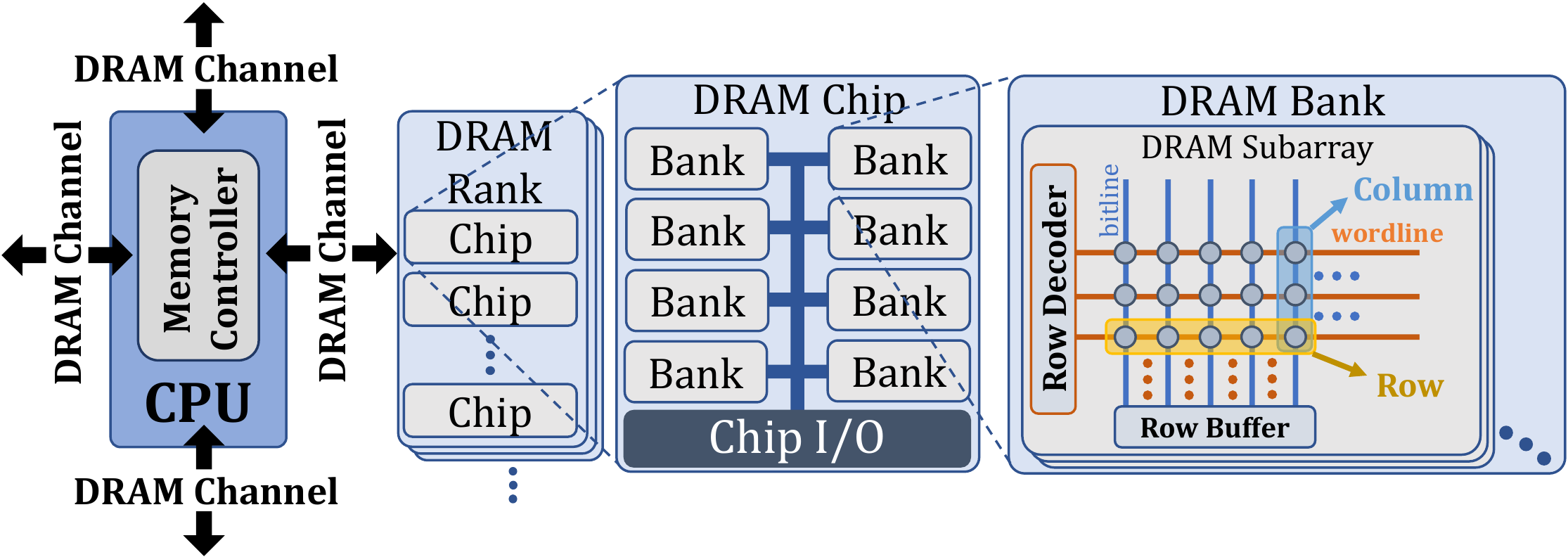}
    \caption{Typical DRAM system organization.}
    \label{bg_fig:dram_organization}
\end{figure}

Fig.~\ref{bg_fig:dram_organization} shows the typical organization of a modern
DRAM system. DRAM is organized into hierarchical \hhm{arrays consisting of}
billions of DRAM cells \hhmiv{in total, where} each \hhmiv{cell stores} one bit
of data. In modern systems, a CPU chip implements a set of memory controllers,
where each memory controller interfaces with a DRAM \emph{channel} to perform
read, write, and maintenance operations (e.g., refresh\hhm{, RowHammer
protection, memory scrubbing}) via a dedicated I/O bus that is independent of
other channels in the system. A DRAM channel can host one or more \emph{DRAM
modules}, where each module consists of one or more \emph{DRAM ranks.} A rank is
comprised of multiple \emph{DRAM chips} that operate in lock step and ranks in
the same channel time-share the channel's I/O bus.

A DRAM chip consists of multiple DRAM \emph{banks}, which \hhm{share an internal
bus that connects them to \hhmiv{the} chip's I/O circuitry}. Within a DRAM bank,
DRAM cells are organized into multiple (e.g., 128) dense two-dimensional arrays
of DRAM cells called
\emph{subarrays}~\cite{kim2012case,chang2014improving,seshadri2013rowclone} and
corresponding peripheral circuitry for manipulating the data within the
subarray. A row of cells (i.e., DRAM \emph{row}) within a subarray share a wire
(i.e., \emph{wordline}), which is driven by a \emph{row decoder} to \emph{open}
(i.e., select) the row of cells to be read or written. A column of cells (i.e.,
DRAM \emph{column}) within a subarray share a wire (i.e., \emph{bitline}), which
is used to read and write to the cells with the help of a \emph{row buffer}
(consisting of \emph{sense amplifiers}). This hierarchical layout of DRAM cells
enables any data in the DRAM system to be accessed and updated using unique
\hhm{channel,} rank, bank, row, and column addresses.


\section{DRAM Operation}
\label{sec:dram_operation}

The memory controller interfaces with DRAM using a series of commands sent over
the I/O bus. There are four major commands that are used to access DRAM:
\cmdact, \cmdwrite, \cmdread, and \cmdprech. DRAM command
scheduling~\cite{rixner2000memory, ipek2008self, kim2010atlas, kim2010thread,
mutlu2007stall, mutlu2008parallelism, subramanian2016bliss} is tightly regulated
by a set of \emph{timing parameters}, which guarantee that enough time
\hhm{passes} after a certain command such that DRAM provides or retains data
correctly. \hhm{DRAM commands and timing parameters are defined by DRAM
standards~\cite{jedec2012ddr4,jedec2021ddr5, jedec2021ddr5, jedec2016gddr6} and
they constitute part of the interface between the memory controller and the DRAM
chip.}

\begin{figure}[h] 
    \centering
    \includegraphics[width=0.9\linewidth]{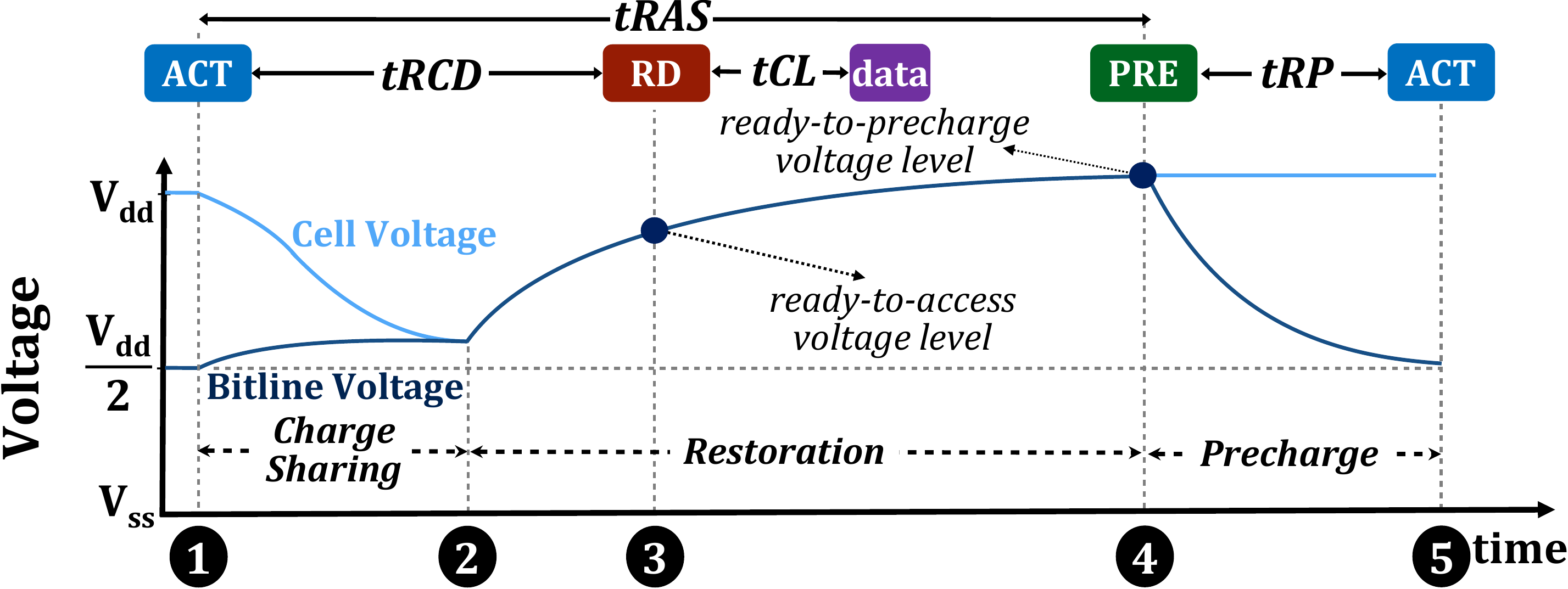} 
    \caption{Commands, timing parameters, and cell/bitline voltages during a DRAM read operation.}
    \label{bg_fig:latency}
\end{figure}

Figure~\ref{bg_fig:latency}\hhm{, which we explain in the next several
subsections,} illustrates the relationship between the commands issued to
perform a DRAM read, their governing timing parameters\hhm{, and their effect on
cell and bitline voltages}. The memory controller enforces \hhm{relevant} timing
parameters as it schedules each DRAM command. Aside from the DRAM access
commands, the memory controller also periodically issues a \hhm{refresh
(\cmdrefresh)} command to prevent data loss due to leakage of charge from the
cell capacitors over time.

\subsection{Activate (\cmdact)} 
The \cmdact command \emph{activates} (opens) a DRAM row by transferring the data
contained in the cell capacitors to the row buffer. \cmdact{} latency is
governed by the \trcd{} timing parameter, which ensures that enough time has
passed since the \cmdact is issued for the data to stabilize in the row buffer
(such that it can be read \hhm{by issuing a \cmdread{} command}).

\cmdact consists of two major steps: 1) capacitor-bitline \emph{charge sharing}
and 2) \emph{charge restoration}. Charge sharing begins by enabling the wordline
(\circled{1} in Figure~\ref{bg_fig:latency}), which allows the cell capacitor to
share charge with the bitline, and thus perturb the precharged bitline voltage.
Once the cell and bitline voltages equalize due to charge sharing, charge
restoration starts (\circled{2}). During charge restoration, the sense
amplifiers are enabled to first detect the bitline voltage shift, and later
restore the bitline to a full $V_{ss}$ or $V_{dd}$ depending on the direction of
the shift. Once the bitline is restored to a \emph{ready-to-access} voltage
level (\circled{3}), the other DRAM commands (e.g., \cmdread, \cmdwrite) can be
issued to the bank.

\subsection{Read (\cmdread)}
After a row activation, the memory controller reads data from the open row by
issuing \hhm{a} \cmdread command. The \cmdread{} \hhm{command} includes a column
address, which indicates the portion of the open row to be read. When a DRAM
chip receives \hhm{a} \cmdread command, it first loads the requested portion of
the open row into the global row buffer. After the data is in the global row
buffer, the DRAM chip sends the data across the data bus to the memory
controller. The \cmdread{} command is governed by the timing parameter \tcl,
after which the data appears on the data bus.

\subsection{Write (\cmdwrite)}
The \cmdwrite command (not shown in Figure~\ref{bg_fig:latency}) modifies data
in an open DRAM row. The operation of \cmdwrite is analogous to \cmdact in that
both commands require waiting enough time for the sense amplifiers to restore
the data in the DRAM cells.  Similar to how a sense amplifier restores a cell
capacitor during the second step of \cmdact{} \hhm{(i.e., charge restoration)},
in case of a \cmdwrite, the sense amplifier restores the capacitor with the new
data value that the \cmdwrite command provides. The restoration latency for
\cmdwrite is governed by the \twr timing parameter. For both \cmdact and
\cmdwrite commands, the restoration latency originates from the sense amplifier
driving a bitline to replenish the charge of the DRAM cell
capacitor~\cite{kim2012case,kang2014co,lee2015adaptive,zhang2015exploiting}.

\subsection{Precharge (\cmdprech)} 
\cmdprech is used to \emph{close} an open DRAM row and prepare the DRAM bank for
activation of another row. The memory controller can follow an \cmdact with
\cmdprech to the same bank after at least the time interval specified by
\hhm{the \tras timing parameter}. \tras ensures that enough time has passed to
fully restore the DRAM cells of the activated row to a \emph{ready-to-precharge
voltage} (\circled{4} in Figure~\ref{bg_fig:latency}). The latency of \cmdprech
is governed by the \hhm{\trp{} timing parameter}, which allows enough time to
set the bitline voltage back to \hhm{the} reference \hhm{voltage} level (e.g.,
$V_{dd}/2$). After \trp{} \hhm{(\circled{5} in Figure~\ref{bg_fig:latency})},
the memory controller can issue an \cmdact to open a new row in the same bank.

\subsection{\hhm{Refresh (\cmdrefresh{})}}

\hhm{A DRAM cell cannot store its data permanently, as the cell capacitor leaks
charge over time. The retention time of a DRAM cell is defined as the length of
time for which the data can still be correctly read out of the cell after data
is stored in the cell. To ensure data integrity, a DRAM cell must be
periodically refreshed. To enable such periodic refresh of all DRAM cells, the
memory controller periodically issues a \emph{refresh} (\cmdrefresh{}) command
(e.g., every \SI{7.8}{\micro\second} or \SI{3.9}{\micro\second} \hhm{in chips
implementing the DDR4 standard~\cite{jedec2012ddr4}}) to ensure that every DRAM
cell is refreshed once at a fixed \emph{refresh window} (i.e., typically 32 or
\SI{64}{\milli\second} in chips implementing the DDR4
standard~\cite{jedec2012ddr4})~\cite{jedec2014lpddr4, jedec2008ddr3,
liu2013experimental,
liu2012raidr,jedec2016gddr6,jedec2012ddr4,jedec2020lpddr5,jedec2021ddr5}. A DRAM
chip refreshes several (e.g., $16$) rows upon receiving a single \cmdrefresh{}
command\hhmiv{, which takes \trfc{} (e.g., $\approx\SI{350}{\nano\second}$) to
complete.} Typically $8192$ \cmdrefresh{} commands \hhm{are required to refresh
the entire DDR4} DRAM chip in a refresh window.} 
\hhms{The new DDR5 standard~\cite{jedec2021ddr5} similarly completes refreshing
the entire DRAM chip with $8192$ \cmdrefresh{} commands but requires
\SI{32}{\milli\second} or \SI{16}{\milli\second} refresh window, i.e., a
\cmdrefresh{} is issued every \SI{3.9}{\micro\second} when operating at up to
\SI{85}{\celsius} and \SI{1.95}{\micro\second} when operating between
\SI{85}{\celsius}-\SI{95}{\celsius}.}


\section{\hhm{DRAM Performance Issues}}
\label{bg:sec:dram_performance}

\hhm{In this section, we discuss the \hhmiv{major} performance issues of DRAM
\hhmiv{that we tackle in this dissertation}.}

\subsection{\hhm{DRAM Access Latency}}
\label{bg:subsec:dram_latency}

\hhm{ In contrast to tremendous improvement in DRAM chip density, DRAM access
latency has reduced \emph{only} slightly~\cite{lee2013tiered,
lee2015adaptive,hassan2016chargecache, chang2016low, borkar2011future,
chang2017thesis, chang2016understanding, choi2015multiple, hennessy2011computer,
lee2016reducing, nguyen2018nonblocking, son2013reducing,patel2022enabling}. Over
the last two decades, DRAM density increased by \hhmiv{more than} two orders of
magnitude, whereas \trcd{} and \tras latencies decreased by \emph{only} 0.81\%
and 1.33\% per year between 2000 and 2015~\cite{patel2022enabling}.

Even though technology node scaling would normally reduce latencies in the DRAM
circuit in conjunction with enabling higher DRAM capacity, DRAM vendors
typically design their DRAM chips to trade the latency reduction for further
improvements in DRAM capacity. For example, vendors amortize the large area cost
of a sense amplifier by increasing the number of DRAM cells connected to the
same sense amplifier, which increases the bitline
length~\cite{hassan2016chargecache,lee2013tiered}. Increasing the bitline length
leads to higher resistance and parasitic capacitance on the connections between
a DRAM cell and a sense amplifier. Consequently, a long bitline results in
higher access latency than a short bitline.

To reduce the long DRAM latencies, many prior works propose various techniques
that exploit temporal charge variation in DRAM
cells~\cite{hassan2016chargecache,
shin2014nuat,wang2018reducing,zhang2016restore}, exploit design-induced latency
variation~\cite{lee2017design, kim2018solar, chandrasekar2014exploiting}, use
in-DRAM caching~\cite{chang2016low,hassan2019crow, kim2012case, hidaka1990cache,
gulur2012multiple,wang2020figaro}, make static~\cite{lee2013tiered,
micron2021rldram, son2013reducing, takemura2007long} and
dynamic~\cite{luo2020clr, choi2015multiple} capacity-latency tradeoff, and
reduce timing parameters under certain operating
conditions~\cite{chang2016understanding, chang2017understanding,
lee2015adaptive}. }

\subsection{\hhm{DRAM Refresh Overhead}}
\label{bg:subsec:dram_refresh_performance}

\hhm{ Periodic DRAM refresh operations incur high system performance (and energy
consumption) \hhmiv{overheads}. As the DRAM cell size decreases with process
technology scaling, newer DRAM devices contain more DRAM cells than older DRAM
devices~\cite{itrs}. As a result, while DRAM capacity increases, the performance
and energy overheads of the refresh operations scale
unfavorably~\cite{chang2014improving, kang2014co, liu2012raidr}. In modern
LPDDR4~\cite{micron-lpddr4} devices, the memory controller refreshes
\emph{every} DRAM cell every \SI{32}{\milli\second} \hhmiv{(when temperature is
above \SI{85}{\celsius}, every \SI{16}{\milli\second})}. Previous studies show
that 1)~refresh operations incur large performance overheads, as DRAM cells
\emph{cannot} be accessed when the cells are being
refreshed~\cite{chang2014improving, liu2012raidr, mukundan2013understanding,
nair2014refresh}; and 2)~up to 50\% of the total DRAM energy is consumed by the
refresh operations~\cite{liu2012raidr, chang2014improving}.

To mitigate DRAM refresh overhead, prior works propose various techniques that
exploit retention time variation~\cite{liu2012raidr,liu2012flikker,
baek2014refresh,isen2009eskimo,jung2015omitting,kim2003block,mukundan2013understanding,
patel2017reaper, qureshi2015avatar,venkatesan2006retention}, simultaneously
activate multiple DRAM rows~\cite{riho2014partial,
hassan2019crow,choi2015multiple}, parallelize refreshes with
accesses~\cite{chang2014improving, yaglikci2022hira}, optimize scheduling of
refresh commands~\cite{bhati2013coordinated, cui2014dtail, emma2008rethinking,
ghosh2007smart, nair2014refresh, nair2013case,stuecheli2010elastic}, and exploit
data pattern
variation~\cite{khan2014efficacy,khan2016case,khan2016parbor,khan2017detecting,patel2005energy}.}



\section{DRAM Reliability \hhm{Issues}}
\label{bg:sec:dram_realiability}

\hhm{In this section, we discuss the \hhmiv{major} reliability issues of DRAM \hhmiv{that we tackle in this dissertation}.}

\subsection{Charge Leakage and \hhm{Data Retention}}
\label{bg:subsec:refresh}

A DRAM cell stores a data value in the form of charge in its capacitor (e.g., a
charged cell can represent 0 or 1 and vice versa). Since the capacitor naturally
loses charge over time, the capacitor charge must be actively and periodically
\emph{refreshed} to prevent information loss due to a data retention
failure~\cite{patel2017reaper, jedec2014lpddr4, jedec2008ddr3,
liu2013experimental, khan2014efficacy, khan2017detecting, khan2016parbor,
liu2012raidr,khan2016case,qureshi2015avatar,kim2009new,lee2015adaptive,chandrasekar2014exploiting,chang2017thesis,chang2016understanding,chang2017understanding,hassan2017softmc,jung2016reverse,jung2014optimized,jung2015omitting,kim2015architectural,kim2014flipping,lee2016reducing,lee2017design,meza2015revisiting,schroeder2009dram,sridharan2015memory,sridharan2012study,weis2015retention,weis2015thermal,patel2020bit}.
\hhm{As we explain in~\cref{bg:subsec:dram_refresh_performance}, DRAM refresh
incurs significant system performance and energy consumption overheads.}

Although all DRAM cell in a DRAM chip are refreshed at a fixed refresh interval,
the retention times of different cells can significantly
vary~\cite{patel2017reaper,liu2012raidr,qureshi2015avatar,khan2017detecting,khan2016case,khan2016parbor,khan2014efficacy,liu2013experimental,kim2009new,lee2015adaptive}.
Many prior
works~\cite{liu2013experimental,qureshi2015avatar,khan2017detecting,khan2016case,khan2016parbor,khan2014efficacy,
kim2009new, lee2015adaptive,
patel2017reaper,chandrasekar2014exploiting,chang2017thesis,chang2016understanding,chang2017understanding,hassan2017softmc,jung2016reverse,jung2014optimized,jung2015omitting,kim2015architectural,kim2014flipping,lee2016reducing,lee2017design,meza2015revisiting,schroeder2009dram,sridharan2015memory,sridharan2012study,weis2015retention,weis2015thermal,patel2020bit}
perform detailed experimental studies to analyze data retention and reliability
characteristics of real DRAM chips. Other prior
works~\cite{lin2012secret,liu2012raidr,nair2013archshield,ohsawa1998optimizing,venkatesan2006retention,wang2014proactivedram}
propose techniques to mitigate DRAM retention failures.









\subsection{Violating Access and Refresh Timings}

DRAM vendors specify timing parameters that a memory controller must satisfy
when accessing and refreshing their DRAM chip in order to ensure reliable
operation. \hhm{However, deliberately violating DRAM timing parameters is useful
for several cases.} \hhm{First, as we explain in~\cref{bg:subsec:dram_latency},
violating DRAM} timing parameters can improve DRAM latency and energy efficiency
at the cost of increased bit error rate\hhm{. Second, prior works show that
deliberately violating DRAM timing parameters provides new functions like
Physical Unclonable Functions (PUFs)\cite{kim2018dram, sutar2016d,
talukder2019prelatpuf, talukder2018ldpuf, tehranipoor2017investigation}, True
Random Number Generators
(TRNGs)\cite{kim2019d,bostanci2022dr,olgun2021quac,sutar2018d,talukder2018exploiting,tehranipoor2016robust},
and computation using
DRAM~\cite{gao2019computedram,seshadri2017ambit,seshadri2013rowclone,orosa2021codic,
olgun2021pidram}}.
%
%
\hhm{Third, DRAM errors caused by violating DRAM timing parameters are useful
for understanding various DRAM \hhmiv{properties and behavior under} different
operating
conditions~\cite{liu2013experimental,qureshi2015avatar,khan2017detecting,khan2016case,khan2016parbor,khan2014efficacy,
kim2009new, lee2015adaptive,
patel2017reaper,chandrasekar2014exploiting,chang2017thesis,chang2016understanding,chang2017understanding,hassan2017softmc,jung2016reverse,jung2014optimized,jung2015omitting,kim2015architectural,kim2014flipping,lee2016reducing,lee2017design,meza2015revisiting,schroeder2009dram,sridharan2015memory,sridharan2012study,weis2015retention,weis2015thermal,patel2020bit}
and reverse-engineering DRAM circuit
design~\cite{patel2022enabling,patel2020bit,patel2019understanding,patel2021harp,chang2016understanding,
hassan2021uncovering, jung2016reverse, kim2018solar,
kim2018dram,kraft2018improving, lee2015adaptive,mukhanov2020dstress,
orosa2021deeper}.}

\subsection{Variable Retention Time (VRT)}

Variable Retention Time (VRT)~\cite{bacchini2014characterization,
qureshi2015avatar, yaney1987meta, restle1992dram, shirley2014copula,
kim2011characterization, kim2011study, kumar2014detection, mori2005origin,
ohyu2006quantitative, khan2014efficacy, kang2014co, liu2013experimental} is a
phenomenon where certain DRAM cells switch between low and high retention time
states in an unpredictable manner. When in low retention time state, a DRAM cell
becomes more likely to experience a retention error. \hhm{Therefore, VRT effects
make accurately profiling the retention time of DRAM cells difficult.} With DRAM
technology scaling, VRT grows into an even more critical
problem~\cite{micron2017whitepaper, kang2014co, cha2017defect}. Despite the
efforts to identify and mitigate VRT-related DRAM
errors~\cite{qureshi2015avatar,
sharifi2017online,patel2017reaper,khan2014efficacy,khan2016case,khan2016parbor,khan2017detecting,lee2015adaptive,kim2019d},
VRT still remains as a key DRAM scaling challenge~\cite{kang2014co,
liu2013experimental}.





\subsection{Environmental Factors}

Various environmental factors degrade the reliability of a DRAM chip. First,
random external events such as particle strikes~\cite{borucki2008comparison,
may1979alpha, guenzer1979single, ziegler1996terrestrial,sridharan2012study,
schroeder2009dram, hwang2012cosmic} may cause transient bit flips in DRAM.
Second, prior experimental studies~\cite{liu2013experimental,
hamamoto1998retention, hou2013fpga, kong2008analysis, kim2010high,
patel2017reaper,khan2014efficacy,khan2016case,khan2016parbor,khan2017detecting,halderman2008lest}
show that increase in DRAM operation temperature significantly accelerates
charge leakage, leading to lower DRAM cell retention times. Thus, high operating
temperatures, especially temperatures above the range specified by DRAM vendors
(e.g., $>\SI{95}{\celsius}$~\cite{jedec2012ddr4,jedec2021ddr5, jedec2020lpddr5,
jedec2014lpddr4, jedec2016gddr6}), can severely impact the reliability of a DRAM
chip. Third, a DRAM chip may suffer from increased reliability issues when set
to operate at reduced supply voltage. Chang et al.~\cite{chang2017understanding}
analyzes the DRAM bit errors at reduces voltage levels and observes that such
bit errors can be avoided by increasing DRAM access timing parameters.
\hhm{Fourth, prior studies~\cite{schroeder2009dram,kim2016ecc,meza2015revisiting}
report that aging is an important factor affecting DRAM reliability and
lifetime. Only after 10-18 months of being in use, DRAM chips exhibit increased
error rates~\cite{schroeder2009dram}.}


\section{\hhm{DRAM Security Issues}}
\label{bg:sec:dram_security}

\hhm{In this section, we discuss \hhmiii{RowHammer and cold boot attacks, which
are major} security issues of DRAM.} \hhmiv{We tackle RowHammer in this
dissertation, but our findings and techniques can potentially be useful for
tackling cold boot attacks as well.}

\subsection{RowHammer} 
\label{bg:subsec:rowhammer}

Modern DRAM chips suffer from disturbance errors that occur when a high number
of activations (within a refresh window) to one DRAM row unintentionally affects
the values of cells in nearby rows~\cite{kim2014flipping}. This phenomenon,
popularly called \emph{RowHammer}~\cite{kim2014flipping,
mutlu2019rowhammer,mutlu2017rowhammer, mutlu2023fundamentally}, stems from
electromagnetic interference between circuit elements. RowHammer becomes
exacerbated as manufacturing process technology node size (and hence DRAM cell
size) shrinks and circuit elements are placed closer
together~\cite{kim2020revisiting,mutlu2017rowhammer}. As demonstrated in prior
\hhmiv{works}~\cite{kim2014flipping, kim2020revisiting,
orosa2021deeper,yauglikcci2022understanding,hassan2021uncovering,cojocar2020are,frigo2020trrespass,jattke2022blacksmith},
the RowHammer effect is strongest between immediately physically-adjacent rows.
RowHammer bit flips are most likely to appear in neighboring rows physically
adjacent to a \emph{hammered row} that is activated many times (e.g., $139K$ in
DDR3~\cite{kim2014flipping}, $10K$ in DDR4~\cite{kim2020revisiting}, and $4.8K$
in LPDDR4~\cite{kim2020revisiting})\footnote{\hhm{For DDR3
chips,~\cite{kim2014flipping} reports the minimum number of row activations on a
\emph{single} aggressor row (i.e., single-sided RowHammer) to cause a RowHammer
bit flip. For DDR4 and LPDDR4 chips,~\cite{kim2020revisiting} reports the
minimum number of row activations to \emph{each of the two} immediately-adjacent
aggressor rows (i.e., double-sided RowHammer).}}. A hammered row is also called
an \emph{aggressor row} and a nearby row that is affected by the hammered row is
called a \emph{victim row}, regardless of whether or not the victim row actually
experiences RowHammer bit flips.

\begin{figure}[!ht]
    \captionsetup[subfigure]{justification=centering, size=scriptsize}
    \centering
    \begin{subfigure}[b]{.48\linewidth}
        \centering
        \includegraphics[width=\linewidth]{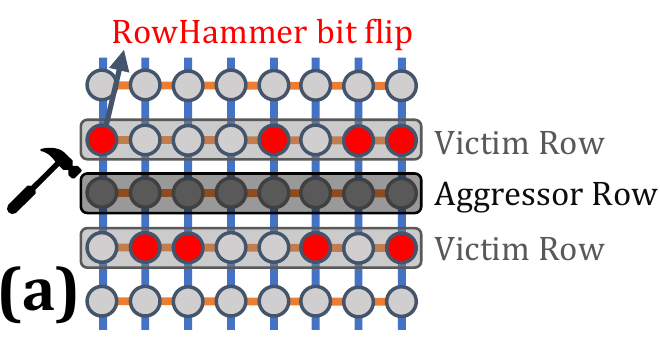}
    \end{subfigure}\quad
    \begin{subfigure}[b]{.48\linewidth}
        \centering
        \includegraphics[width=\linewidth]{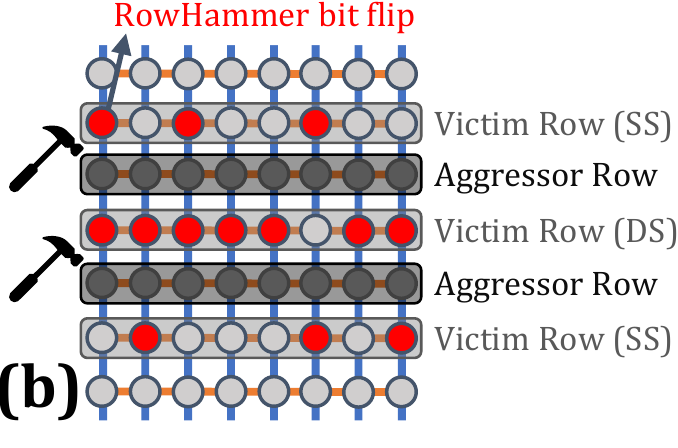}
    \end{subfigure}
    \caption{Typical Single-sided (SS) and Double-sided (DS) RowHammer access patterns.}
    \label{bg_fig:rowhammer_access}
\end{figure}

To most effectively exploit the RowHammer phenomenon, attackers typically
perform $i$) single-sided RowHammer (i.e., repeatedly activate one aggressor row
that is physically adjacent to the victim row, as we show in
Fig.~\ref{bg_fig:rowhammer_access}a)~\cite{kim2014flipping} or $ii$)
double-sided RowHammer (i.e., repeatedly activate in an alternating manner two
aggressor rows that are both physically adjacent to the victim row, as we show
in Fig.~\ref{bg_fig:rowhammer_access}b)~\cite{seaborn2015exploiting,
rh_project_zero}. Prior works have shown that double-sided RowHammer leads to
more bit flips and does so more quickly than single-sided
RowHammer~\cite{kim2014flipping,kim2020revisiting, mutlu2019rowhammer,
seaborn2015exploiting, rh_project_zero,
orosa2021deeper,yauglikcci2022understanding,hassan2021uncovering,cojocar2020are,frigo2020trrespass,jattke2022blacksmith}.

\hhm{Since the discovery of RowHammer, researchers have proposed many techniques
that take advantage of the RowHammer vulnerability to compromise operating
systems~\cite{seaborn2015exploiting, van2018guardion, veen2016drammer,
gruss2016rowhammer, cojocar2019exploiting, qiao2016new, kwong2020rambleed,
pessl2016drama, bhattacharya2016curious, jang2017sgx, zhang2020pthammer,
cojocar2020are, weissman2020jackhammer, ji2019pinpoint}, web
browsers~\cite{gruss2018another, bosman2016dedup, frigo2020trrespass,
deridder2021smash, frigo2018grand}, cloud virtual machines~\cite{razavi2016flip,
xiao2016one}, remote servers~\cite{tatar2018defeating, lipp2018nethammer}, and
deep neural networks~\cite{yao2020deephammer, hong2019terminal}. To mitigate
RowHammer, early solutions were immediate quick fixes such as \hhmiv{increasing}
the DRAM refresh rate~\cite{apple2015about, rh-cisco, lenovo2015row,
enterprise2015hp} and relying on ECC to fix RowHammer bit
flips~\cite{kim2014flipping,seaborn2015exploiting,razavi2016flip,veen2016drammer,gruss2018another},
which were shown to be ineffective~\cite{cojocar2019exploiting,aweke2016anvil,
kim2014flipping}. Other solutions include isolating sensitive data from DRAM
rows that an attacker can hammer~\cite{brasser2016can, konoth2018zebram,
van2018guardion}, keeping track of row activation counts and refreshing
potential victim rows~\cite{aweke2016anvil, kim2014flipping, lee2019twice,
park2020graphene, seyedzadeh2017cbt, son2017making, you2019mrloc,
yauglikcci2021security, kim2022mithril, taouil2021lightroad,
devaux2021method,marazzi2022protrr, marazzi2023rega, qureshi2022hydra,
woo2023scalable, wi2023shadow}, throttling row activations to prevent a row from
being activated more than a certain threshold~\cite{yaglikci2021blockhammer,
kim2014flipping, greenfield2016throttling}\hhmiv{, and reducing wordline
voltage~\cite{yauglikcci2022understanding}}. 

To this date, DRAM vendors implement proprietary in-DRAM RowHammer solutions.
Despite the claims of the DRAM vendors that their DRAM chips are
``RowHammer-free''~\cite{lee2014green}, prior works~\cite{frigo2020trrespass,
hassan2021uncovering, jattke2022blacksmith} demonstrate that certain
carefully-crafted access patterns can be used to bypass these in-DRAM protection
mechanisms. In~\cref{chap:utrr}, we further discuss the operation and security
properties of RowHammer protection mechanisms implemented in current DRAM
devices.

}


\subsection{\hhm{Cold Boot Attacks}} 
\label{bg:subsec:cold_boot}

\hhm{When a DRAM chip is powered off, the DRAM cells in the chip are expected to
lose their \hhmiv{values} as cells can no longer be refreshed. However, the
majority of DRAM cells are capable of correctly retaining their data for much
longer than the typical 64 or \SI{32}{\milli\second} refresh
period~\cite{liu2013experimental,patel2017reaper,khan2017detecting,khan2016parbor,khan2016case,khan2014efficacy,lee2015adaptive,halderman2008lest,kim2009new,qureshi2015avatar}.
Cooling the DRAM chip further extends the retention capability of DRAM cells
because their charge leakage rates \hhmiv{greatly} decrease as the operating
temperature reduces~\cite{liu2013experimental, hamamoto1998retention,
hou2013fpga, kong2008analysis, kim2010high,
patel2017reaper,khan2014efficacy,khan2016case,khan2016parbor,khan2017detecting,halderman2008lest}.
Halderman et al.~\cite{halderman2008lest} show that a DRAM chip can correctly
retain 99.9\% of its data for minutes when cooled down to \SI{-50}{\celsius}
using a compressed air can. The fact that a DRAM chip retains the majority of
its data after a power loss poses a critical security problem. An attacker can
retrieve sensitive information (e.g., encryption keys) from a DRAM chip by
plugging in the chip to a machine that the attacker owns. To mitigate DRAM cold
boot attacks, prior works propose techniques that obfuscate data written to
DRAM~\cite{mosalikanti2011high, yitbarek2017cold,gruhn2013practicability,
bauer2016lest}, store encryption keys only in CPU
registers~\cite{simmons2011security, muller2011tresor}, use full memory
encryption~\cite{henson2014memory, yang2005improving}, and erase DRAM
content~\cite{orosa2021codic, seol2019amnesiac}.}




\chapter[The SoftMC Infrastructure]{SoftMC: A Flexible and Practical Open-Source Infrastructure \\ for Enabling Experimental DRAM Studies}
\label{chap:softmc}

DRAM process technology node scaling leads to key challenges that critically
impact DRAM reliability and performance. To accurately understand the DRAM cell
behavior in terms of reliability and latency, it is critical to characterize and
analyze {\em real DRAM chips}.
In this chapter, we design, prototype, and demonstrate the basic capabilities of
a flexible and easy-to-use experimental DRAM characterization infrastructure,
called {\em SoftMC (Soft Memory Controller)}. SoftMC is an open-source
FPGA-based DRAM characterization infrastructure, consisting of a programmable
memory controller that can control and test memory modules designed for the
commonly-used DDR (Double Data Rate) interface. 
SoftMC provides a simple and intuitive high-level programming interface that
completely hides the low-level details of the FPGA from users. Users implement
their test routines or mechanisms in a high-level language that automatically
gets translated into the low-level SoftMC memory controller operations in the
FPGA.

SoftMC offers a wide range of use cases, such as characterizing the effects of
variation within a DRAM chip and across DRAM chips, verifying the correctness of
new DRAM mechanisms on actual hardware, and experimentally discovering the
reliability, retention, and timing characteristics of an unknown or
newly-designed DRAM chip (or finding the best specifications for a known DRAM
chip). We demonstrate the potential and ease of use of SoftMC by implementing
two use cases.
First, we demonstrate the ease of use of SoftMC's high-level interface by
implementing a simple experiment to characterize the retention time behavior of
cells in modern DRAM chips. Our test results match the prior experimental
studies that characterize DRAM retention time in modern DRAM
chips~\cite{liu2012raidr, liu2013experimental, khan2014efficacy,
qureshi2015avatar}, providing a validation of our infrastructure. Second, we
demonstrate the flexibility and capability of SoftMC by validating two
recently-proposed DRAM latency reduction mechanisms~\cite{hassan2016chargecache,
shin2014nuat}. These mechanisms exploit the idea that highly-charged DRAM cells
can be accessed with low latency. Our SoftMC-based experimental analysis of 24
real DRAM chips from three major DRAM vendors demonstrates that the expected
latency reduction effect of these mechanisms is \emph{not} observable in
existing DRAM chips. This experiment demonstrates {\em (i)} the importance of
experimentally characterizing real DRAM chips to understand the behavior of DRAM
cells, and designing mechanisms that are based on this experimental
understanding; and {\em (ii)} the effectiveness of SoftMC in testing (validating
or refuting) new ideas on existing memory modules.

We also discuss several other use cases of SoftMC, including the ability to
characterize emerging non-volatile memory modules that obey the DDR standard. We
hope that SoftMC inspires other new studies, ideas, and methodologies in memory
system design. \hhm{In fact, since the release of SoftMC to the public as part
of this research, many works~\cite{koppula2019eden,chang2016understanding,
chang2017understanding,frigo2020trrespass,orosa2021deeper,gao2019computedram,kim2020revisiting,kim2019d,hassan2021uncovering,khan2017detecting,ghose2018your,yauglikcci2022understanding,olgun2021quac,orosa2021codic, talukder2018exploiting, talukder2019prelatpuf, talukder2018ldpuf, talukder2020towards, bepary2022dram,farmani2021rhat, yaglikci2022hira, gao2022fracdram} used different
versions of SoftMC in various DRAM characterization studies. SoftMC is openly
and freely available at~\cite{softmcsource}.}

\section{Motivation and Goal}

At smaller technology nodes, it is becoming increasingly difficult to store and
retain enough charge in a DRAM cell, causing various reliability and performance
issues~\cite{liu2013experimental,mandelman2002challenges, mutlu2013memory,
redeker2002investigation, yaney1987meta, konishi1989analysis, liu2012raidr,
khan2014efficacy, khan2016case, khan2016parbor, khan2017detecting,
mutlu2014research, kim2020revisiting, orosa2021deeper, mutlu2017rowhammer,
kim2014flipping, mutlu2019rowhammer,
yauglikcci2022understanding,chang2014improving, qureshi2015avatar,
jafri2020refresh, liu2012flikker, kim2020charge, stuecheli2010elastic,
zhang2014cream}. Ensuring reliable operation of the DRAM cells is a key
challenge in future technology nodes~\cite{mutlu2013memory,
mandelman2002challenges, kim2005technology, mueller2005challenges, liu2012raidr,
kang2014co, liu2013experimental, redeker2002investigation, yaney1987meta,
konishi1989analysis, khan2014efficacy, khan2016case, khan2016parbor,
khan2017detecting, mutlu2014research, kim2020revisiting, orosa2021deeper,
mutlu2017rowhammer, kim2014flipping, mutlu2019rowhammer,
yauglikcci2022understanding}.

The fundamental problem of retaining data with less charge in smaller cells
directly impacts the reliability and performance of DRAM cells. First, smaller
cells placed in close proximity make cells more susceptible to various types of
interference. This potentially disrupts DRAM operation by flipping bits in DRAM,
resulting in major reliability issues~\cite{kim2014flipping, nakagome1988impact,
redeker2002investigation, schroeder2009dram, sridharan2013feng,
meza2015revisiting,mutlu2013memory, mandelman2002challenges, kim2005technology,
mueller2005challenges, liu2012raidr, kang2014co, liu2013experimental,
yaney1987meta, konishi1989analysis, khan2014efficacy, khan2016case,
khan2016parbor, khan2017detecting, mutlu2014research, kim2020revisiting,
orosa2021deeper, mutlu2017rowhammer, mutlu2019rowhammer,
yauglikcci2022understanding}, which can lead to system
failure~\cite{schroeder2009dram,meza2015revisiting,cojocar2019exploiting,mutlu2019rowhammer,cojocar2020are}
or security breaches~\cite{kim2014flipping, seaborn2015exploiting,
rh_project_zero, gruss2016rowhammer, veen2016drammer, xiao2016one,
razavi2016flip, van2018guardion, cojocar2019exploiting, qiao2016new,
kwong2020rambleed, pessl2016drama, bhattacharya2016curious, jang2017sgx,
zhang2020pthammer, cojocar2020are, weissman2020jackhammer,
ji2019pinpoint,gruss2018another, bosman2016dedup, frigo2020trrespass,
deridder2021smash, frigo2018grand, tatar2018defeating,
lipp2018nethammer,yao2020deephammer, hong2019terminal}. Second, it takes longer
time to access a cell with less charge~\cite{hassan2016chargecache,
lee2015adaptive, shin2014nuat, lee2017design, chang2016understanding,
chang2017understanding,wang2018reducing}, and write latency increases as the
access transistor size
reduces~\cite{kang2014co,lee2015adaptive,wang2018reducing}. Thus, smaller cells
directly impact DRAM latency, as DRAM access latency is determined by the
worst-case (i.e., {\em slowest}) cell in any
chip~\cite{chandrasekar2014exploiting,
lee2015adaptive,patel2017reaper,lee2017design}. DRAM access latency has not
improved with technology scaling in the past decade~\cite{lee2013tiered,
borkar2011future, mutlu2013memory, lee2015adaptive,hassan2016chargecache,
chang2016low, chang2017thesis, chang2016understanding, choi2015multiple,
hennessy2011computer, lee2016reducing, nguyen2018nonblocking,
son2013reducing,patel2022enabling}, and, in fact, some latencies are expected to
increase~\cite{kang2014co}, making memory latency an increasingly critical
system performance bottleneck.

As such, there is a significant need for new mechanisms that improve the
reliability and performance of DRAM-based main memory systems. In order to
design, evaluate, and validate many such mechanisms, it is important to
accurately characterize, analyze, and understand DRAM (cell) behavior in terms
of reliability and latency. For such an understanding to be accurate, it is
critical that the characterization and analysis be based on the
\emph{experimental} studies of {\em real DRAM chips}, since a large number of
factors (e.g., various types of cell-to-cell
interference~\cite{nakagome1988impact, redeker2002investigation,
kim2014flipping,kim2020revisiting,orosa2021deeper}, inter- and intra-die process
variation~\cite{chang2016understanding, lee2016reducing, nassif2000delay,
chandrasekar2014exploiting, lee2015adaptive,lee2017design}, random
effects~\cite{liu2013experimental, yaney1987meta, srinivasan1994accurate,
hazucha2000impact,qureshi2015avatar}, operating
conditions~\cite{liu2013experimental,
lee2010mechanism,chang2017understanding,yauglikcci2022understanding,orosa2021deeper},
internal organization~\cite{liu2013experimental, hidaka1989twisted,
khan2016parbor,lee2017design}, stored data patterns~\cite{khan2016case,
khan2016parbor, liu2013experimental, khan2017detecting,
khan2014efficacy,qureshi2015avatar,kim2020revisiting,orosa2021deeper})
concurrently impact the reliability and latency of cells. Many of these
phenomena and their interactions cannot be properly modeled (e.g., in simulation
or using analytical methods) without rigorous experimental characterization and
analysis of real DRAM chips.  The need for such experimental characterization
and analysis, with the goal of building the understanding necessary to improve
the reliability and performance of future DRAM-based main memories at various
levels (both software and hardware), motivates the need for a publicly-available
DRAM testing infrastructure that can enable system users and designers to
characterize real DRAM chips.


A publicly-available DRAM testing infrastructure that can characterize real DRAM
chips enables new mechanisms to improve DRAM reliability and latency. In this
work, we argue that such a testing infrastructure should have two key features
to ensure widespread adoption among architects and designers:
\emph{(i)}~flexibility and \emph{(ii)}~ease of use.

\textbf{Flexibility.} As discussed in \cref{sec:dram_operation}, a DRAM
module is accessed by issuing specific commands (e.g., \cmdact, \cmdprech) in a
particular sequence with a strict delay between the commands (specified by  the
timing parameters, e.g., \trp, \tras). A DRAM testing infrastructure should
implement all low-level DRAM operations (i.e., DDR commands) with tunable timing
parameters without any restriction on the ordering of DRAM commands. Such a
design enables flexibility at two levels. First, it enables comprehensive
testing of \emph{any} DRAM operation with the ability to customize the length of
each timing constraint. For example, we can implement a retention test with
different refresh intervals to characterize the distribution of retention time
in modern DRAM chips. Such a characterization can enable new mechanisms to
reduce the number of refresh operations in DRAM, leading to performance and
power efficiency improvements.  Second, it enables testing of DRAM chips with
high-level test programs, which can consist of \emph{any combination of DRAM
operations and timings}. Such flexibility is extremely powerful to test the
impact of existing or new DRAM mechanisms in real DRAM chips.

\textbf{Ease of Use.} A DRAM testing infrastructure should provide a simple and
intuitive programming interface that minimizes programming effort and time. An
interface that hides the details of the underlying implementation is accessible
to a wide range of users. With such a high-level abstraction, even users that
lack hardware design experience should be able to develop DRAM tests.

In this work, we propose and prototype a publicly-available, open-source DRAM
testing infrastructure that can enable system users and designers to easily
characterize real DRAM chips. Our experimental DRAM testing infrastructure,
called \emph{SoftMC (Soft Memory Controller)}, can test DDR-based memory modules
with a flexible and easy-to-use interface. In the next section, we discuss the
shortcomings of existing tools and platforms that can be used to test DRAM
chips, and explain how SoftMC is designed to avoid these shortcomings.
\section{Related Work}
\label{softmc_sec:related}

No prior DRAM testing infrastructure provides both \emph{flexibility} and
\emph{ease of use} properties, which are critical for enabling widespread
adoption of the infrastructure. Three different kinds of tools/infrastructure
are available today for characterizing DRAM behavior. As we will describe, each
kind of tool has some shortcomings. The goal of SoftMC is to eliminate
\emph{all} of these shortcomings.

\textbf{Commercial Testing Infrastructures.}
A large number of commercial DRAM testing platforms (e.g.,~\cite{advantest,
nickel, teradyne, futureplus}) are available in the market. Such platforms are
optimized for throughput (i.e., to test as many DRAM chips as possible in a
given time period), and generally apply a \emph{fixed test pattern} to the units
under test. Thus, since they lack support for flexibility in defining the test
routine, these infrastructures are not suitable for detailed DRAM
characterization where the goal is to investigate new issues and new ideas.
Furthermore, such testing equipment is usually quite expensive, which makes
these infrastructures an impractical option for research in academia. Industry
may also have internal DRAM development and testing tools, but, to our
knowledge, these are proprietary and are unlikely to be made openly available.

We aim for SoftMC to be a low-cost (i.e., free) and flexible open-source
alternative to commercial testing equipment that can enable new research
directions and mechanisms. For example, prior work~\cite{yang2015random}
recently proposed a random command pattern generator to validate DRAM chips
against uncommon yet supported (according to JEDEC specifications) DDR command
patterns. Using the test patterns on commercial test equipment, this work
demonstrates that specific sequences of commands introduce failures in current
DRAM chips. SoftMC flexibly supports the ability to issue an arbitrary command
sequence, and therefore can be used as a low-cost method for  validating DRAM
chips against problems that arise due to command ordering.

\textbf{FPGA-Based Testing Infrastructures.}
Several prior works proposed FPGA-based DRAM testing
infrastructures~\cite{huang2000fpga, hou2013fpga, keezer2015fpga}.
Unfortunately, all of them lack flexibility and/or a simple user interface, and
none are open-source. The FPGA-based infrastructure proposed by Huang et
al.~\cite{huang2000fpga} provides a high-level interface for developing DRAM
tests, but the interface is limited to defining only data patterns and march
algorithms for the tests. Hou et al.~\cite{hou2013fpga} propose an FPGA-based
test platform whose capability is limited to analyzing only the data retention
time of the DRAM cells. Another work~\cite{keezer2015fpga} develops a custom
memory testing board with an FPGA chip, specifically designed to test memories
at a very high data rate. However, it requires low-level knowledge to develop
FPGA programs, and even then offers only limited flexibility in defining a test
routine. On the other hand, SoftMC aims to provide \emph{full control over all
DRAM commands} using a high-level \emph{software interface}, and it is
open-source.

PARDIS~\cite{bojnordi2012pardis} is a reconfigurable logic (e.g., FPGA) based
programmable memory controller meant to be implemented inside microprocessor
chips. PARDIS is capable of optimizing memory scheduling algorithms, refresh
operations, etc. at run-time based on application characteristics, and can
improve system performance and efficiency. However, it does not provide
programmability for DRAM commands and timing parameters, and therefore cannot be
used for detailed DRAM characterization.

\textbf{Built-In Self Test (BIST).}
A BIST mechanism (e.g,~\cite{you1985self, querbach2014reusable,
querbach2016architecture, bernardi2010programmable, yang2015hybrid}) is implemented
inside the DRAM chip to enable fixed test patterns and algorithms. Using such an
approach, DRAM tests can be performed faster than with other testing platforms.
However, BIST has two major flexibility issues, since the testing logic is
hard-coded into the hardware: \emph{(i)} BIST offers only a limited number of
tests that are fixed at hardware design time. \emph{(ii)} A limited set of DRAM
chips, which come with BIST support, can be tested. In contrast, SoftMC allows
for the implementation of a wide range of DRAM test routines and supports any
off-the-shelf DRAM chip that is compatible with the DDR interface.

We conclude that prior work lacks either the flexibility or the ease-of-use
properties that are critical for performing detailed DRAM characterization. To
fill the gap left by current infrastructures, we introduce an open-source DRAM
testing infrastructure, SoftMC, that fulfills these two properties.
\section{SoftMC Design}

SoftMC is an FPGA-based open-source programmable memory controller that provides
a high-level software interface, which the users can use to initiate any DRAM
operation from a host machine. The FPGA component of SoftMC collects the DRAM
operation requests incoming from the host machine and executes them on real DRAM
chips that are attached to the FPGA board. In this section, we explain in detail
the major components of our infrastructure.

\subsection{High-Level Design}
\label{softmc_sec:highlevel}

\begin{figure}[!t] 
    \centering
    \includegraphics[width=0.53\linewidth]{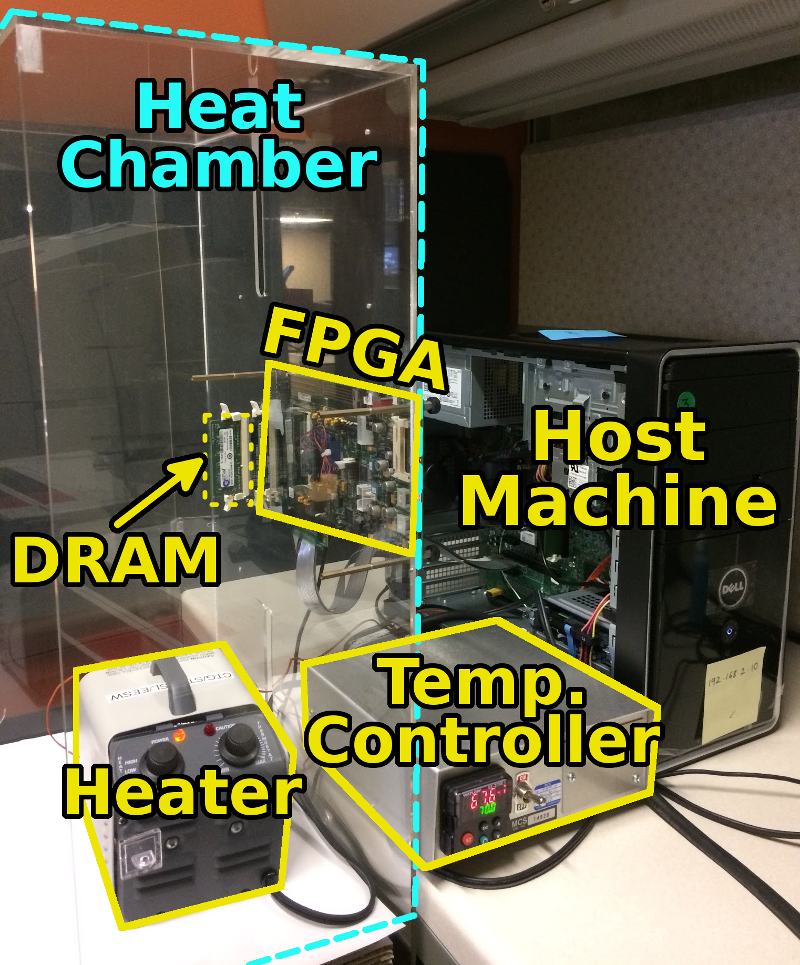}
    \caption{Our SoftMC infrastructure \hhm{(2017)}.}
    \label{softmc_fig:softmc_overview}
\end{figure}

Figure~\ref{softmc_fig:softmc_overview} shows our SoftMC infrastructure. It comprises
three major components:
\squishlist
\item  In the host machine, the \emph{SoftMC API} provides a high-level software
    interface (in C++) for users to communicate with the SoftMC hardware. The
    API provides user-level \emph{functions} to 1) send SoftMC
    \emph{instructions} from the host machine to the hardware and 2) receive
    data, which the hardware reads from the DRAM, back to the host machine. An
    \emph{instruction} encodes and specifies an operation that the hardware is
    capable of performing (see \cref{softmc_sec:softmc_design}). It is used to
    communicate the user-level \emph{function} to the SoftMC hardware such that
    the hardware can execute the necessary operations to satisfy the user-level
    \emph{function}. The SoftMC API provides several functions (See
    \cref{softmc_sec:interface}) that users can call to easily generate an
    instruction to perform any of the operations supported by the SoftMC
    infrastructure. For example, the API contains a \emph{genACT()} function,
    which users call to generate an instruction to activate a row in DRAM.

\item The \emph{driver} is responsible for transferring instructions and data
    between the host machine and the FPGA across a PCIe bus. To implement the
    driver, we use RIFFA~\cite{jacobsen2015riffa}. SoftMC execution is
    \emph{not} affected by the long transfer latency of the PCIe interface, as
    our design sends \emph{all of the instructions} over the PCIe bus to the
    FPGA \emph{before the test routine begins,} and buffers the commands within
    the FPGA. Therefore, SoftMC guarantees that PCIe-related delays do not
    affect the precise user-defined timings between each instruction.

\item Within the FPGA, the core \emph{SoftMC hardware} queues the instructions,
    and executes them in an appropriate hardware component. For example, if the
    hardware receives an instruction that indicates a DRAM command, it sends
    DDR-compatible signals to the DRAM to issue the command.  
\squishend

\subsection{\hhm{\hhm{DRAM} Command Interface}} 
\label{bg:subsec::dram_cmd}
\hhm{
\hhm{DRAM} commands are transmitted from the memory controller to the DRAM
module across a memory bus. On the memory bus, each command is encoded using
five output signals (\cmdcke, \cmdcs, \cmdras, \cmdcas, and \cmdwe).
Enabling/disabling these signals corresponds to specific commands (as specified
by the DDR standard).  First, the \cmdcke signal (\emph{clock enable})
determines whether the DRAM is in ``standby mode'' (ready to be accessed) or
``power-down mode''. Second, the \cmdcs (\emph{chip selection}) signal specifies
the rank that should receive the issued command.  Third, the \cmdras (\emph{row
address strobe})/\cmdcas (\emph{column address strobe}) signal  is used to
generate commands related to DRAM row/column operations. Fourth, the \cmdwe
signal (\emph{write enable}) in combination with \cmdras and \cmdcas, generates
the specific row/column command. For example, enabling \cmdcas and \cmdwe
together generates a \cmdwrite command, while enabling only \cmdcas indicates a
\cmdread command. }

\subsection{SoftMC API}
\label{softmc_sec:interface}
SoftMC offers users complete control over the DDR-based DRAM memory modules by
providing an API (Application Programming Interface) in a high-level programming
language (C++).  This API consists of several C++ functions that generate
\emph{instructions} for the SoftMC hardware, which resides in the FPGA. These
instructions provide support for \emph{(i)}~all possible DDR DRAM commands; and
\emph{(ii)}~managing the SoftMC hardware, e.g., inserting a delay between
commands, configuring the auto-refresh mechanism (see \cref{softmc_sec:impl}).
Using the SoftMC API, any sequence of DRAM commands can be easily and flexibly
implemented in SoftMC with little effort and time from the programmer.  

When users would like to issue a sequence of commands to the DRAM, they
\emph{only} need to make SoftMC API function calls. Users can easily generate
\emph{instructions} by calling the API \emph{functions}. Users can insert the
generated instructions to an \emph{InstructionSequence}, which is a data
structure that the SoftMC API provides to keep the instructions in order and
easily send them to the SoftMC hardware via the driver. There are three types of
API functions: \emph{(i)~command functions}, which provide the user the ability
to specify DDR-compatible DRAM commands (e.g., \cmdact, \cmdprech);
\emph{(ii)}~the \emph{wait function}, which provide the user the ability to
specify a time interval between two commands (thereby allowing the user to
determine the latencies of each timing parameter); and \emph{(iii)~aux
functions}, which provide the user with control over various auxiliary
operations. Programming the memory controller using such high-level C++ function
calls enables users who are inexperienced in hardware design to easily develop
sophisticated memory controllers.

\textbf{Command Functions.}
For every DDR command, we provide a command function that generates an
instruction, which instructs the SoftMC hardware to execute the command. For
example, \mbox{\emph{genACT(b, r)}} function generates an instruction that
informs the SoftMC controller to issue an \cmdact to DRAM row \emph{r} in bank
\emph{b}. Program~\ref{softmc_code:softmc_write} shows a short example program that
writes to a single cache line (i.e., DRAM column) in a closed row. In
Program~\ref{softmc_code:softmc_write}, there are three command functions:
\emph{genACT()}, \emph{genWR()}, and \emph{genPRE()}, in lines 2, 4, and 6,
which generate SoftMC hardware instructions that inform the SoftMC hardware to
issue \cmdact, \cmdwrite, and \cmdprech{} commands to the DRAM.

\renewcommand\lstlistingname{Program}
\lstset{
    basicstyle=\footnotesize\ttfamily, 
    numbers=left,              
    numberstyle=\tiny,          
    numbersep=5pt,              
    tabsize=2,                  
    extendedchars=true,
    breaklines=true,            
    frame=b,
    stringstyle=\color{white}\ttfamily, 
    showspaces=false,
    showtabs=false,
    xleftmargin=17pt,
    framexleftmargin=17pt,
    framexrightmargin=5pt,
    framexbottommargin=4pt,
    showstringspaces=false
}

\begin{lstlisting}[language=C++,
captionpos=b, caption=Performing a write operation to a closed row using
the SoftMC API.,
label=softmc_code:softmc_write]
InstructionSequence iseq;
iseq.insert(genACT(bank, row));
iseq.insert(genWAIT(tRCD));
iseq.insert(genWR(bank, col, data));
iseq.insert(genWAIT(tCL+tBL+tWR));
iseq.insert(genPRE(bank));
iseq.insert(genWAIT(tRP));
iseq.insert(genEND());
iseq.execute(fpga));
\end{lstlisting}

\textbf{Wait Function.}
We provide a single function, \emph{genWAIT(t)}, which generates an instruction
to inform the SoftMC hardware to wait \emph{t} DRAM cycles before executing the
next instruction. With this one function, the user can implement any timing
constraint. For example, in Program~\ref{softmc_code:softmc_write}, the controller
should wait for \trcd after the \cmdact before issuing a \cmdwrite to the DRAM.
We simply add a call to \emph{genWAIT()} on Line~3 to insert the \trcd delay.
Users can easily provide any value that they want to test for the time interval
(e.g., a reduced \trcd, to see if errors occur when the timing parameter is
reduced). 

\textbf{Aux Functions.}
There are a number of auxiliary functions that allow the user to configure the
SoftMC hardware. One such function, \emph{genBUSDIR(dir)}, allows the user to
switch (i.e., turn around) the DRAM bus direction. To reduce the number of IO
pins, the DDR interface uses a bi-directional data bus between DRAM and the
controller. The bus must be set up to move data from DRAM to the controller
before issuing a \cmdread, and from the controller to DRAM before issuing a
\cmdwrite. After \emph{genBUSDIR()} is called, the user should follow it with a
\emph{genWAIT()} function, to ensure that the bus is given enough time to
complete the turn around. The time interval for this wait corresponds to the
\twtr and \trtw timing parameters in the DDR interface.

We provide an auxiliary function to control auto-refresh behavior. The SoftMC
hardware includes logic to \emph{automatically} refresh the DRAM cells after a
given time interval (\trefi) has passed. Auto-refresh operations are not added
to the code by the user, and are instead handled directly within our hardware.
In order to adjust the interval at which refresh is performed, the user can
invoke the \emph{genREF\_CONFIG()} function. To disable the auto-refresh
mechanism, the user needs to set the refresh interval to 0 by calling that
function. The user can invoke the same function to adjust the number of cycles
that specify the refresh latency (\trfc). We describe the auto-refresh
functionality of the SoftMC hardware in more detail in \cref{softmc_sec:impl}.

Finally, we provide a function, \emph{genEND()}, which generates an \emph{END}
instruction to inform the SoftMC hardware that the end of the instruction
sequence has been reached, and that the SoftMC hardware can start executing the
sequence. Note that the programmer should call \emph{iseq.execute(fpga)} to send
the instruction sequence to the driver, which in turn sends the sequence to the
SoftMC hardware. Program~\ref{softmc_code:softmc_write} shows an example of this in
Lines 8--9.

\subsection{Instruction Types and Encoding}
\label{softmc_sec:softmc_design}

Figure~\ref{softmc_fig:softmc_ddr_fields} shows the encodings for the key SoftMC
instruction types.\footnote{We refer the reader to the source-code and manual of
SoftMC for the encoding of \emph{all} instruction types~\cite{softmcsource}.}
All SoftMC instructions are 32 bits wide, where the most-significant 4 bits
indicate the type of the instruction.

\begin{figure}[H]
\begin{center}

\begin{tabular}{|c|c|c|c|c|c|c|c|c|}
\multicolumn{1}{c}{\textbf{InstrType}} &  \multicolumn{8}{c}{}  \\
\hline
DDR (4) & \emph{unused} (3) & \multicolumn{5}{c|}{\cmdcke, \cmdcs(2), \cmdras,
\cmdcas, \cmdwe (6)} & Bank (3) & Addr (16) \\
\hline
WAIT (4) & \multicolumn{8}{c|}{cycles (28)} \\
\hline
BUSDIR (4) & \multicolumn{7}{c|}{\emph{unused} (27)} & dir (1) \\
\hline
END (4) & \multicolumn{8}{c|}{\emph{unused} (28)} \\
\hline
\end{tabular}
\end{center}
\caption{Key SoftMC instruction types.}
\label{softmc_fig:softmc_ddr_fields}
\end{figure}

\textbf{DDR Instruction Type.}
Recall from \cref{bg:subsec::dram_cmd} that the DDR commands are encoded using
several fields (e.g., \cmdras, \cmdcas, \cmdwe) that correspond to the signals
of the DDR interface, which sits between the memory controller and the DRAM. The
SoftMC instructions generated by our API functions encode these same fields, as
shown in the instruction encoding for the DDR type instructions in
Figure~\ref{softmc_fig:softmc_ddr_fields}. Thus, DDR type instructions can
represent \emph{any} DDR command from the DDR3
specification~\cite{jedec2008ddr3}. For example, the user needs to set \cmdcke =
1, \cmdcs = 0, \cmdras = 0, \cmdcas = 1, and \cmdwe = 1 to create an \cmdact
command to open the row in the bank specified by \emph{Address} and \emph{Bank},
respectively. Other DDR commands can easily be represented by setting the
appropriate instruction fields according to the DDR3 specification.

\textbf{Other Instruction Types.}
Other instructions carry information to the FPGA using the bottom 28 bits of
their encoding. For example, the \emph{WAIT} instruction uses these bits to
encode the wait latency, i.e., wait cycles. The \emph{BUSDIR} instruction uses
the least significant bit of the bottom 28~bits to encode the new direction of
the data bus. The \emph{END} instruction requires no additional parameters,
using only the \emph{InstrType} field.

\subsection{Hardware Architecture}
\label{softmc_sec:impl}

\begin{figure}[!ht] 
    \centering
    \includegraphics[width=0.90\linewidth]{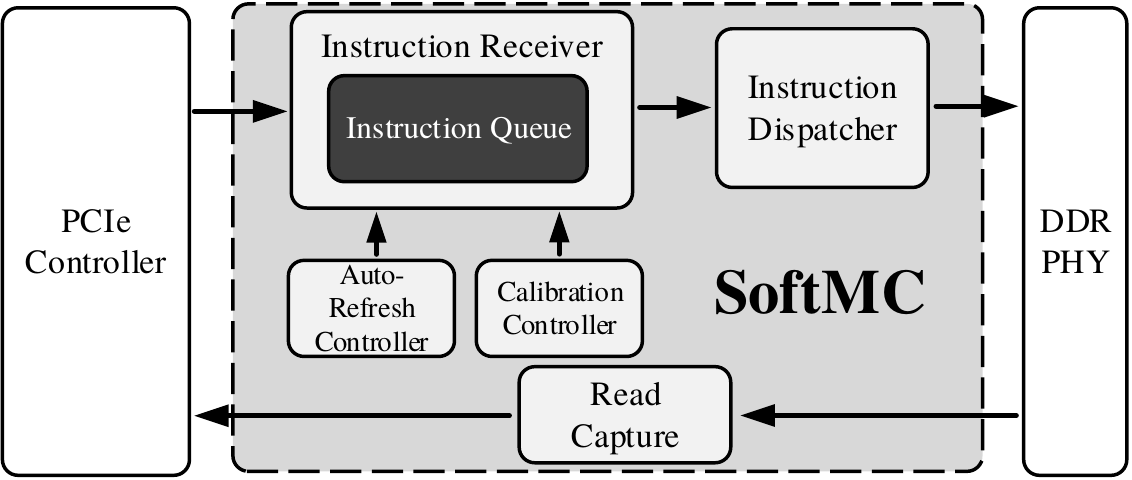}
    \caption{SoftMC hardware architecture.}
    \label{softmc_fig:softmc_arch}
\end{figure}

The driver sends the SoftMC instructions across the PCIe bus to the FPGA, where
the \emph{SoftMC hardware} resides. Figure~\ref{softmc_fig:softmc_arch} shows the five
components that make up the \emph{SoftMC hardware}: \emph{(i)} an
\emph{instruction receiver}, which accumulates the instructions sent over the
PCIe bus; \emph{(ii)} an \emph{instruction dispatcher}, which decodes each
instruction into DRAM control signals and address bits, and then sends the
decoded bits to the DRAM module through the memory bus; \emph{(iii)} a
\emph{read capture module}, which receives data read from the DRAM and sends it
to the host machine over PCIe; \emph{(iv)} an \emph{auto-refresh controller};
and \emph{(v)} a \emph{calibration controller}, responsible for ensuring data
integrity on the memory bus. SoftMC also includes two interfaces: a \emph{PCIe
bus controller} and a \emph{DDR PHY}. Next, we describe each of these components
in detail.

\textbf{Instruction Receiver.} 
When the driver sends SoftMC instructions across the PCIe bus, the instruction
receiver buffers them in an \emph{instruction queue} until the \emph{END}
instruction is received.  When the \emph{END} instruction arrives, the
instruction receiver notifies the instruction dispatcher that it can start
fetching instructions from the instruction queue. Execution ends when the
instruction queue is fully drained.

\textbf{Instruction Dispatcher.} 
The instruction dispatcher fetches an instruction, decodes it, and executes the
operation that the instruction indicates, at the corresponding component. For
example, if the type of the instruction is \emph{DDR}, the dispatcher decodes
the control signals and address bits from the instruction, and then sends them
to the DRAM using the DDR PHY. Since the FPGAs operate at a lower frequency than
the memory bus, the dispatcher supports fetching and issuing of multiple DDR
commands per cycle. As another example, if the instruction type is \emph{WAIT},
it is not sent to the DRAM, but is instead used to count down the number of
cycles before the instruction dispatcher can issue the next instruction to the
DRAM. Other instruction types (e.g., \emph{BUSDIR, END}) are handled in a
similar manner.

\textbf{Read Capture Module.} 
After a \cmdread command is sent to the DRAM module, the read capture module
receives the data emitted by the DRAM from the DDR PHY, and sends it to the host
machine via the PCIe controller. The read capture module is responsible for
ensuring data alignment and ordering, in case the DDR PHY operates on a
different clock domain or reorders the data output.

\textbf{Auto-Refresh Controller.}
The auto-refresh controller includes two registers, which store the refresh
interval (\trefi) and refresh latency (\trfc) values (See
\cref{sec:dram_operation}). It issues periodic refresh operations to DRAM, based
on the values stored in the registers.  If the instruction dispatcher is in the
middle of executing an instruction sequence that was received from the host
machine, at the time a refresh is scheduled to start, the auto-refresh
controller \emph{delays} the refresh operation until the dispatcher finishes the
execution of the instruction sequence. Note that a small delay does not pose any
problems, as the DDR standard allows for a small amount of flexibility for
scheduling refresh operations~\cite{chang2014improving, jedec2008ddr3,
mukundan2013understanding}. However, if the instruction sequence is too long,
and delays the refresh operation beyond a critical interval (i.e., more than 8X
the \trefi), it is the user's responsibility to redesign the test routine and
split up the long instruction sequence into smaller sequences.

\textbf{Calibration Module.}
This module ensures the data integrity of the memory bus. It periodically issues
a command to perform \emph{short ZQ calibration}, which is an operation
specified by the DDR standard to calibrate the output driver and on-die
termination circuits~\cite{jedec2014lpddr4, david2011memory}. SoftMC provides
hardware support for \emph{automatic} calibration, in order to ease programmer
effort.

\textbf{PCIe Controller and DDR PHY.}
The PCIe controller acts as an interface between the host machine and the rest
of the SoftMC hardware. The DDR PHY sits between the SoftMC hardware and the
DRAM module, and is responsible for initialization and clock synchronization
between the FPGA and the DRAM module.

\subsection{SoftMC Prototype}

We describe the key implementation details of our first prototype of SoftMC.

\textbf{FPGA Board.} The current SoftMC prototype targets a Xilinx ML605 FPGA
board~\cite{ml605}. It consists of \emph{(i)} a Virtex-6 FPGA chip, which we use
to implement our SoftMC design; and \emph{(ii)} a SO-DIMM slot, where we plug in
the real DRAM modules to be tested.

\textbf{PCIe Interface.} The PCIe controller uses \emph{(i)} the Xilinx PCIe
Endpoint IP Core~\cite{xilpcie}, which implements low-level operations related
to  the physical layer of the PCIe bus; and \emph{(ii)} the RIFFA IP
Core~\cite{jacobsen2015riffa}, which provides high-level protocol functions for
data communication. The communication between the host machine and the FPGA
board is established through a PCIe 2.0 link.

\textbf{DDR PHY.} SoftMC uses the Xilinx DDR PHY IP Core~\cite{xilphy} to
communicate with the DRAM module. This PHY implements the low-level operations
to transfer data over the memory channel to/from the DRAM module. The PHY
transmits two DDR commands each memory controller cycle. As a result, our SoftMC
implementation  fetches and dispatches two instructions from the instruction
queue every cycle.

\textbf{Performance.}
To evaluate the performance of our prototype, we time the execution of a test
routine that performs operations on a particular region of the DRAM and then
reads data back from this region, repeating the test on different DRAM regions
until the entire DRAM module has been tested. We model the expected overall
execution time of the test as:

\begin{equation}
\label{softmc_eq:1}
Expected~Execution~Time = (S + E + R) * \dfrac{Capacity}{Region~Size}
\end{equation}

where \emph{S (Send)}, \emph{E (Execute)}, and \emph{R (Receive)} are the
host-to-FPGA PCIe latency, execution time of the command sequence in the FPGA,
and FPGA-to-host PCIe latency, respectively, that is required for each DRAM
region. For our prototype, we measure \emph{S} and \emph{R} to both be
\SI{22}{\micro\second} on average (with $\pm$\SI{2}{\micro\second} variation).
\emph{E} varies based on the length of the command sequence. For the sequence
given in Program~\ref{softmc_code:write_row}, which we use in our first use case
(\cref{softmc_sec:usecase1}), we calculate \emph{E} as \SI{16}{\micro\second}
using typical DDR3 timing parameters. Since that command sequence tests only a
single 8KB row, the \emph{Region Size} is 8KB. For an experiment testing a full
4GB memory module completely, we find that our prototype runs
Program~\ref{softmc_code:write_row} on the entire module in approximately 31.5 seconds.
An at-speed DRAM controller, which, at best, has \emph{S} and \emph{R} equal to
zero, would run the same test in approximately 11.5 seconds, i.e., only 2.7x
faster. As \emph{E} increases, the performance of our SoftMC prototype gets
closer to the performance of an at-speed DRAM controller. We conclude that our
SoftMC prototype is fast enough to run test routines that cover an entire memory
module.

\section{Example Use Cases}
\label{softmc_sec:use_cases}
Using our SoftMC prototype, we perform two case studies on randomly-selected
real DRAM chips from three manufacturers. First, we discuss how a simple
retention test can be implemented using SoftMC, and present the experimental
results of that test (\cref{softmc_sec:usecase1}). Second, we demonstrate how
SoftMC can be leveraged to test the expected effect of two recently-proposed
mechanisms that aim to reduce DRAM access latency (\cref{softmc_sec:nuat_test}).
Both use cases demonstrate the flexibility and ease of use of SoftMC.

\subsection{Retention Time Distribution Study}
\label{softmc_sec:usecase1}

This test aims to characterize data retention time in different DRAM modules.
The retention time of a cell can be determined by testing the cell with
different refresh intervals. The cell fails at a refresh interval that is
greater than its retention time. In this test, we gradually increase the refresh
interval from the default \SI{64}{\milli\second} and count the number of bytes
that have an incorrect value at each refresh interval.

\subsubsection{Test Methodology}
\label{softmc_sec:ret_test_method}

We perform a simple test to measure the retention time of the cells in a DRAM
chip. Our test consists of three steps: \emph{(i)} We write a reference data
pattern (e.g. all zeros, or all ones) to an entire row. \emph{(ii)} We wait for
the specified refresh interval, so that the row is idle for that time and all
cells gradually leak charge. \emph{(iii)} We read data back from the same row
and compare it against the reference pattern that we wrote in the first step.
Any mismatch indicates that the cell could not hold its data for that duration,
which resulted in a bit flip. We count the number of bytes that have bit flips
for each test. 

We repeat this procedure for all rows in the DRAM module. The read and write
operations in the test are issued with the standard timing parameters, to make
sure that the only timing delay that affects the cells is the refresh interval.

\subsubsection{Implementation with SoftMC.}

\textbf{Writing Data to DRAM.} 
In Program~\ref{softmc_code:write_row}, we present the implementation of the
first part of our retention time test, where we write data to a row, using the
SoftMC API. First, to activate the row, we insert the instruction generated by
the \emph{genACT()} function to an instance of the \emph{InstructionSequence}
(Lines~1-2). This function is followed by a \emph{genWAIT()} function (Line~3)
that ensures that the activation completes with the standard timing parameter
\trcd. Second, we issue write instructions to write the data pattern in each
column of the row. This is implemented in a loop, where, in each iteration, we
call \emph{genWR()} (Line~5), followed by a call to \emph{genWAIT()} function
(Line~6) that ensures proper delay between two \cmdwrite{} operations. After
writing to all columns of the row, we insert another delay (Line~8) to account
for the \emph{write recovery} time (\cref{sec:dram_operation}). Third, once we
have written to all columns, we close the row by precharging it. This is done by
the \emph{genPRE()} function (Line~9), followed by a \emph{genWAIT()} function
with standard \trp timing. Finally, we call the \emph{genEND()} function to
indicate the end of the instruction sequence, and send the test program to the
FPGA by calling the \emph{execute()} function.

\textbf{Employing a Specific Refresh Interval.} 
Using SoftMC, we can implement  the target refresh interval in two ways. We can
use the auto-refresh support provided by the SoftMC hardware, by setting the
\trefi parameter to our target value, and letting the FPGA take care of the
refresh operations. Alternatively, we can disable auto-refresh, and manually
control the refresh operations from the software. In this case, the user is
responsible for issuing refresh operations at the right time. In this retention
test, we disable auto-refresh and use a software clock to determine when we
should read back data from the row (i.e., refresh the row).

\begin{lstlisting}[language=C++, captionpos=b, caption=Writing data
to a row using the SoftMC API.,
label=softmc_code:write_row]
InstructionSequence iseq;
iseq.insert(genACT(bank, row));
iseq.insert(genWAIT(tRCD));
for(int col = 0; col < COLUMNS; col++){
  iseq.insert(genWR(bank, col, data));
  iseq.insert(genWAIT(tBL));
}
iseq.insert(genWAIT(tCL+tWR));
iseq.insert(genPRE(bank));
iseq.insert(genWAIT(tRP));
iseq.insert(genEND());
iseq.execute(fpga));
\end{lstlisting}

\textbf{Reading Data from DRAM.} Reading data back from the DRAM requires steps
similar to DRAM writes (presented in Program~\ref{softmc_code:write_row}). The only
difference is that, instead of issuing a \cmdwrite command, we need to issue a
\cmdread command and enforce read-related timing parameters. In the SoftMC API,
this is done by calling the \emph{genRD()} function in place of the
\emph{genWR()} function, and specifying the appropriate read-related timing
parameters. After the read operation is done, the FPGA sends back the data read
from the DRAM module, and the user can access that data using the
\emph{fpga\_recv()} function provided by the driver.

Based on the intuitive code implementation of the retention test, we conclude
 that it requires minimal effort to write test programs using the SoftMC API.
 Our full test is provided in our open-source release~\cite{softmcsource}.

\subsubsection{Results}

\begin{figure}[!hb] 
    \centering
    \includegraphics[width=.92\linewidth]{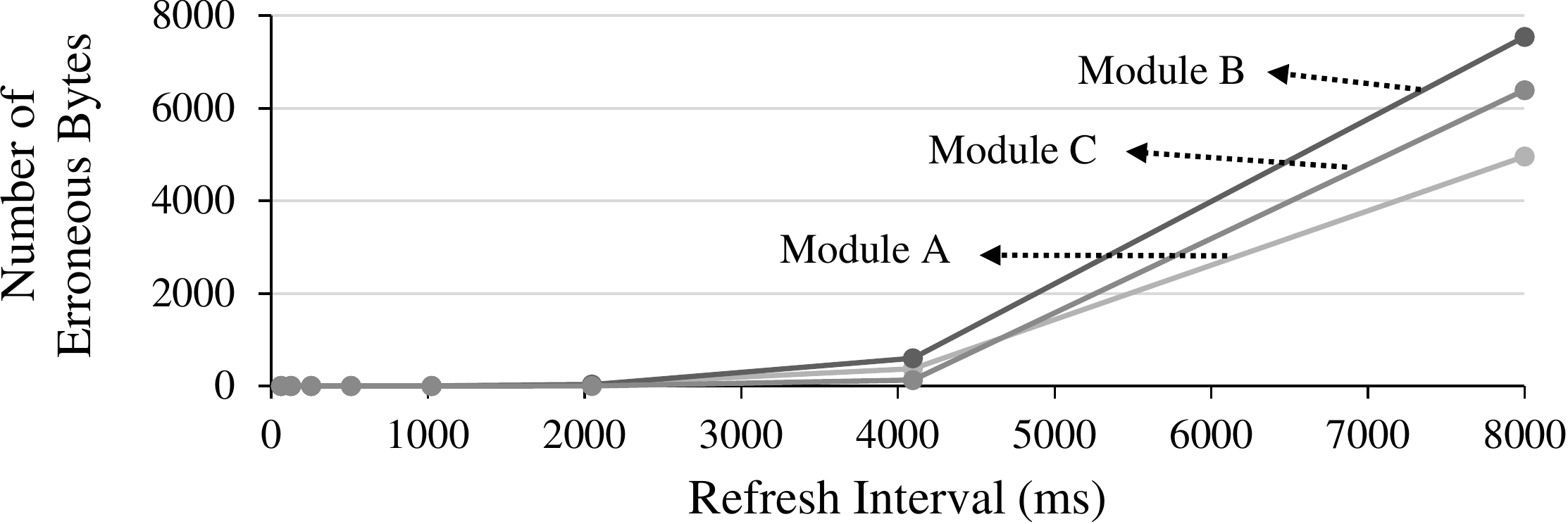}
    \caption{Number of erroneous bytes observed in retention time tests.}
    \label{softmc_fig:ret_time_test}
\end{figure}

We perform the retention time test at room temperature, using 24 chips from
three major manufacturers. We vary the refresh interval from
\SI{64}{\milli\second} to \SI{8192}{\milli\second} exponentially.
Figure~\ref{softmc_fig:ret_time_test} shows the results for the test, where the x-axis
shows the refresh interval in milliseconds, and the y-axis shows the number of
erroneous bytes found in each interval. We make two major observations.

\emph{(i)} We do not observe any retention failures until we test with a refresh
interval of \SI{1}{\second}. This shows that there is a large safety margin for
the refresh interval in modern DRAM chips, which is conservatively set to
\SI{64}{\milli\second} by the DDR standard.\footnote{DRAM manufacturers perform
retention tests that are similar to ours (but with proprietary in-house
infrastructures that are not disclosed). Their results are similar to
ours~\cite{liu2013experimental, lee2015adaptive, khan2014efficacy,
chang2016understanding}, showing significant margin for the refresh interval.
This margin is added to ensure reliable DRAM operation for the worst-case
operating conditions (i.e., worst case temperature) and for worst-case cells, as
has been shown by prior works~\cite{lee2015adaptive, liu2013experimental,
khan2014efficacy, chang2016understanding}.}

\emph{(ii)} We observe that the number of failures increases
\hhmiii{drastically} with the increase in refresh interval.

Prior experimental studies on retention time of DRAM cells have reported similar
observations as ours~\cite{khan2014efficacy, liu2013experimental,
lee2015adaptive, hou2013fpga, hamamoto1998retention}. We conclude that SoftMC
can easily reproduce prior known DRAM experimental results, validating the
correctness of our testing infrastructure and showing its flexibility and ease
of use. 

\subsection{Evaluating the Expected Effect of Two Recently-Proposed Mechanisms
in Existing DRAM Chips}
\label{softmc_sec:nuat_test}

Two recently-proposed mechanisms, ChargeCache~\cite{hassan2016chargecache} and
NUAT~\cite{shin2014nuat}, provide low-latency access to highly-charged DRAM
cells. They both are based on the key idea that a highly-charged cell can be
accessed faster than a cell with less charge~\cite{lee2015adaptive}. ChargeCache
observes that cells belonging to \emph{recently-accessed} DRAM rows are in a
highly-charged state and that such rows are likely to be accessed again in the
near future. ChargeCache exploits the highly-charged state of these
recently-accessed rows to lower the latency for later accesses to them. NUAT
observes that \emph{recently-refreshed} cells are in highly-charged state, and
thus it lowers the latency for accesses to recently-refreshed rows. Prior to
issuing an \cmdact command to DRAM, both ChargeCache and NUAT determine whether
the target row is in a highly-charged state. If so, the memory controller uses
reduced \trcd and \tras timing parameters to perform the access. 

In this section, we evaluate whether or not the expected latency reduction
effect of these two works is observable in existing DRAM modules, using SoftMC.
We first describe our methodology for evaluating the improvement in the \trcd
and \tras parameters. We then show the results we obtain using SoftMC, and
discuss our observations. 

\subsubsection{Evaluating DRAM Latency with SoftMC.} 
\label{softmc_sec:trcd_test}
\label{softmc_sec:tras_test}

\textbf{Methodology.} In our experiments, we use 24 DDR3 chips (i.e., three
SO-DIMMs) from three major vendors. To stress DRAM reliability and maximize the
amount of cell charge leakage, we raise the test temperature to 80$^{\circ}$C
(significantly higher than the common-case operating range of
35-55$^{\circ}$C~\cite{lee2015adaptive}) by enclosing our FPGA infrastructure in a
temperature-controlled heat chamber (see Figure~\ref{softmc_fig:softmc_overview}). For
all experiments, the temperature within the heat chamber was maintained within
0.5$^{\circ}$C of the target 80$^{\circ}$C temperature.

To study the impact of charge variation in cells on access latency in existing
DRAM chips, which is dominated by the \trcd and \tras timing
parameters~\cite{lee2013tiered, lee2015adaptive, chang2016understanding} (see
\cref{sec:dram_operation}), we perform experiments to test the headroom for
reducing these parameters. In our experiments, we vary one of the two timing
parameters, and test whether the original data can be read back correctly with
the reduced timing. If the data that is read out contains errors, this indicates
that the timing parameter cannot be reduced to the tested value without inducing
errors in the data. We perform the tests using a variety of data patterns (e.g.,
0x00, 0xFF, 0xAA, 0x55) because 1) different DRAM cells store information (i.e.,
0 or 1) in different states (i.e., charged or empty)~\cite{liu2013experimental}
and 2) we would like to stress DRAM reliability by increasing the interference
between adjacent bitlines~\cite{khan2016parbor, liu2013experimental,
khan2016case}. We also perform tests using different refresh intervals, to study
whether the variation in charge leakage increases significantly if the time
between refreshes increases.

\begin{figure}[ht]
        \centering
        \subfloat[\trcd Test] {
                \includegraphics[width=0.65\linewidth]{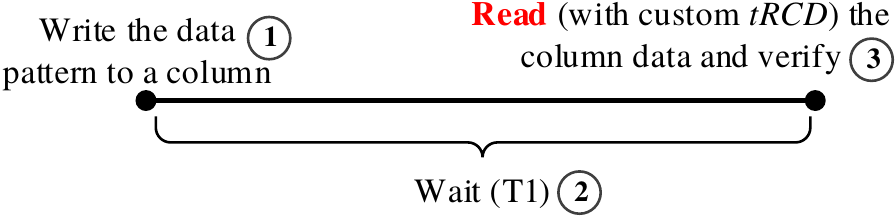}%
                \label{softmc_fig:trcd_test}
        }

        \subfloat[\tras Test] {
                \includegraphics[width=0.65\linewidth]{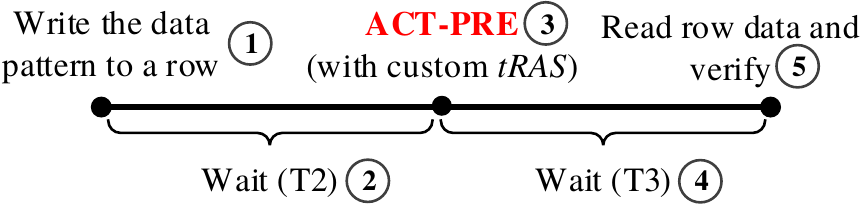}%
                \label{softmc_fig:tras_test}
        }
	\caption{Timelines that illustrate the methodology for testing the
improvement of (a) \trcd and (b) \tras on highly-charged DRAM cells.}
        \label{softmc_fig:nuat_tests}
\end{figure}

\textbf{tRCD Test.}
We measure how highly-charged cells affect the \trcd timing parameter, by using
a custom \trcd value to read data from a row to which we previously wrote a
reference data pattern. We adjust the time between writing a reference data
pattern and performing the read, to vary the amount of charge stored within the
cells of a row. In Figure~\ref{softmc_fig:trcd_test}, we show the command sequence that
we use to test whether recently-refreshed DRAM cells can be accessed with a
lower \trcd, compared to cells that are close to the end of the refresh
interval. We perform the write and read operations to each DRAM row one column
at a time, to ensure that each read incurs the \trcd latency. First (\circled{1}
in Figure~\ref{softmc_fig:trcd_test}), we perform a reference write to the DRAM column
under test by issuing \cmdact, \cmdwrite, and \cmdprech successively with the
\emph{default} DRAM timing parameters. Next (\circled{2}), we wait for the
duration of a time interval~(\textbf{T1}), which is the refresh interval in
practice, to vary the charge contained in the cells. When we wait longer, we
expect the target cells to have less charge at the end of the interval. We cover
a wide range of wait intervals, evaluating values between 1 and
\SI{512}{\milli\second}. Finally (\circled{3}), we read the data from the column
that we previously wrote to and compare it with the reference pattern. We
perform the read with the custom \trcd value for that specific test. We evaluate
\trcd values ranging from 3 to 6 (default) cycles. Since a \trcd of 3 cycles
produced errors in every run, we did not perform any experiments with a lower
\trcd.

We process multiple rows in an interleaved manner (i.e., we write to multiple
rows, wait, and then verify their data one after another) in order to further
stress the reliability of DRAM~\cite{lee2015adaptive}.  We repeat this process
for all DRAM rows to evaluate the entire memory module.

\textbf{tRAS Test.} 
We measure the effect of accessing highly-charged rows on the \tras timing
parameter by issuing the \cmdact and \cmdprech commands, with a custom \tras
value, to a row. We check if that row still contains the same data that it held
before the \cmdact-\cmdprech command pair was issued. Figure~\ref{softmc_fig:tras_test}
illustrates the methodology for testing the effect of the refresh interval on
\tras. First (\circled{1}), we write the reference data pattern to the selected
DRAM row with the default timing parameters. Different from the \trcd test, we
write to \emph{every column} in the open row (before switching to another row)
to save cycles by eliminating a significant amount of \cmdact and \cmdprech
commands, thereby reducing the testing time. Next (\circled{2}), we wait for the
duration of time interval \textbf{T2}, during which the DRAM cells lose a
certain amount of charge. To refresh the cells (\circled{3}), we issue an
\cmdact-\cmdprech command pair associated with a custom \tras value. When the
\cmdact-\cmdprech pair is issued, the charge in the cells of the target DRAM row
may not be fully restored if the wait time is too long or the \tras value is too
short, potentially leading to loss of data. Next (\circled{4}), we wait again
for a period of time \textbf{T3} to allow the cells to leak a portion of their
charge. Finally (\circled{5}), we read the row using the default timing
parameters and test whether it still retains the correct data. Similar to the
\trcd test, to stress the reliability of DRAM, we simultaneously perform the
\tras test on multiple DRAM rows.

We would expect, from this experiment, that the data is likely to maintain its
integrity when evaluating reduced \tras with shorter wait times (\textbf{T2}),
as the higher charge retained in a cell with a short wait time can enable a
reliable reduction on \tras. In contrast, we would expect failures to be more
likely when using a reduced \tras{} with a longer wait time, because the cells
would have a low amount of charge that is not enough to reliably reduce \tras.

\subsubsection{Results}

We analyze the results of the \trcd and \tras tests, for 24 real DRAM chips from
different vendors, using the test programs detailed in
\cref{softmc_sec:trcd_test}. We evaluate \trcd values ranging from 3 to 6
cycles, and \tras values ranging from 2 to 14 cycles, where the maximum number
for each is the default timing parameter value. For both tests, we evaluate
refresh intervals between 8 and \SI{512}{\milli\second} and measure the number
of observed errors during each experiment.

\begin{figure}[!b]
        \centering
        \subfloat[Module A] {
                \includegraphics[height=1.92in]{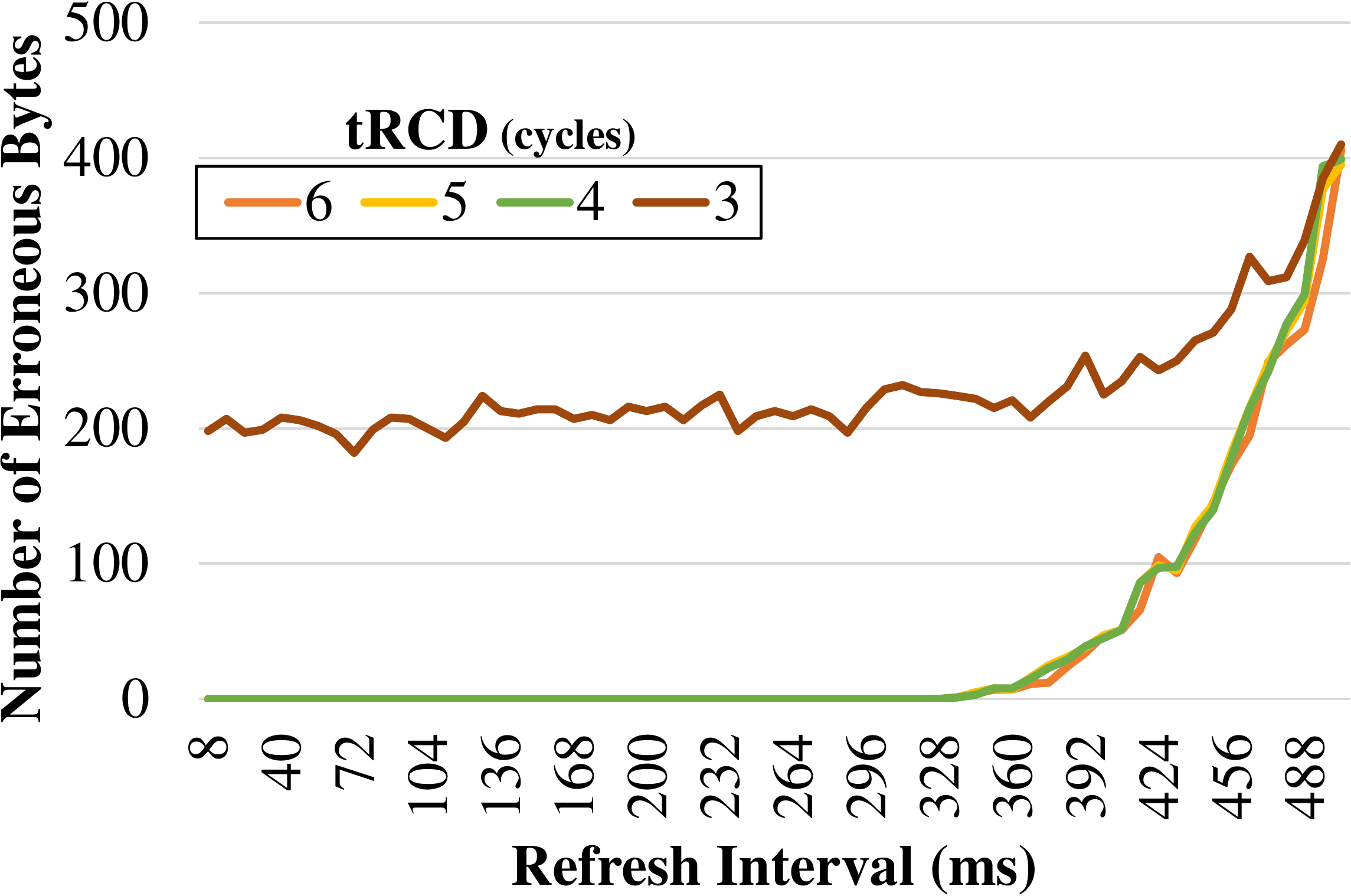}
                \label{softmc_fig:trcd_res1}
        }
        \subfloat[Module B] {
                \includegraphics[height=1.92in]{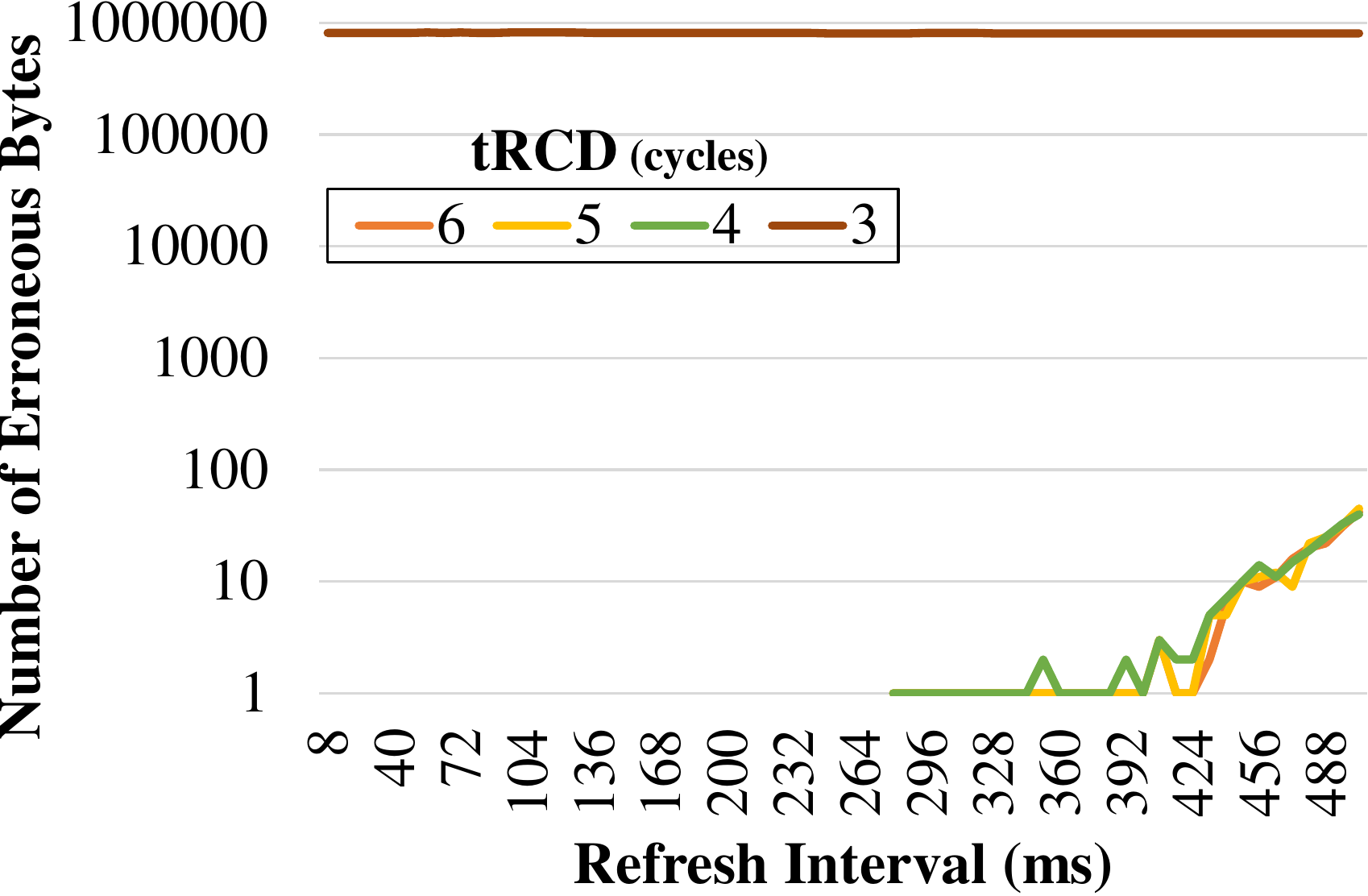}
                \label{softmc_fig:trcd_res2}
        }\\
        \subfloat[Module C] {
                \includegraphics[height=1.92in]{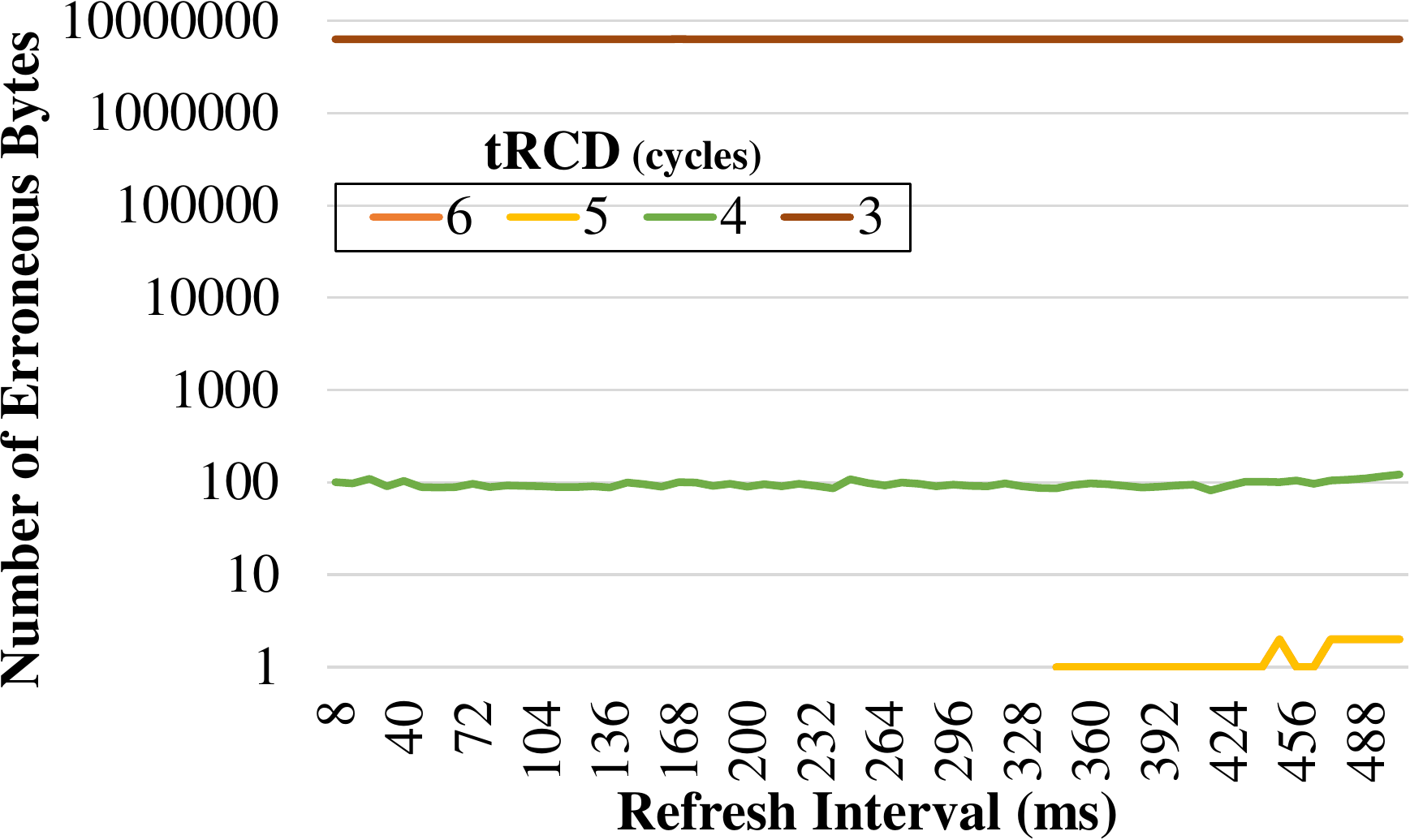}
                \label{softmc_fig:trcd_res3}
        }
	\caption{Effect of reducing \trcd{} on the number of errors at various
refresh intervals.} 
        \label{softmc_fig:trcd_test_results}
\end{figure}

\begin{figure}[!t]
        \centering
        \subfloat[Module A] {
                \includegraphics[height=1.92in]{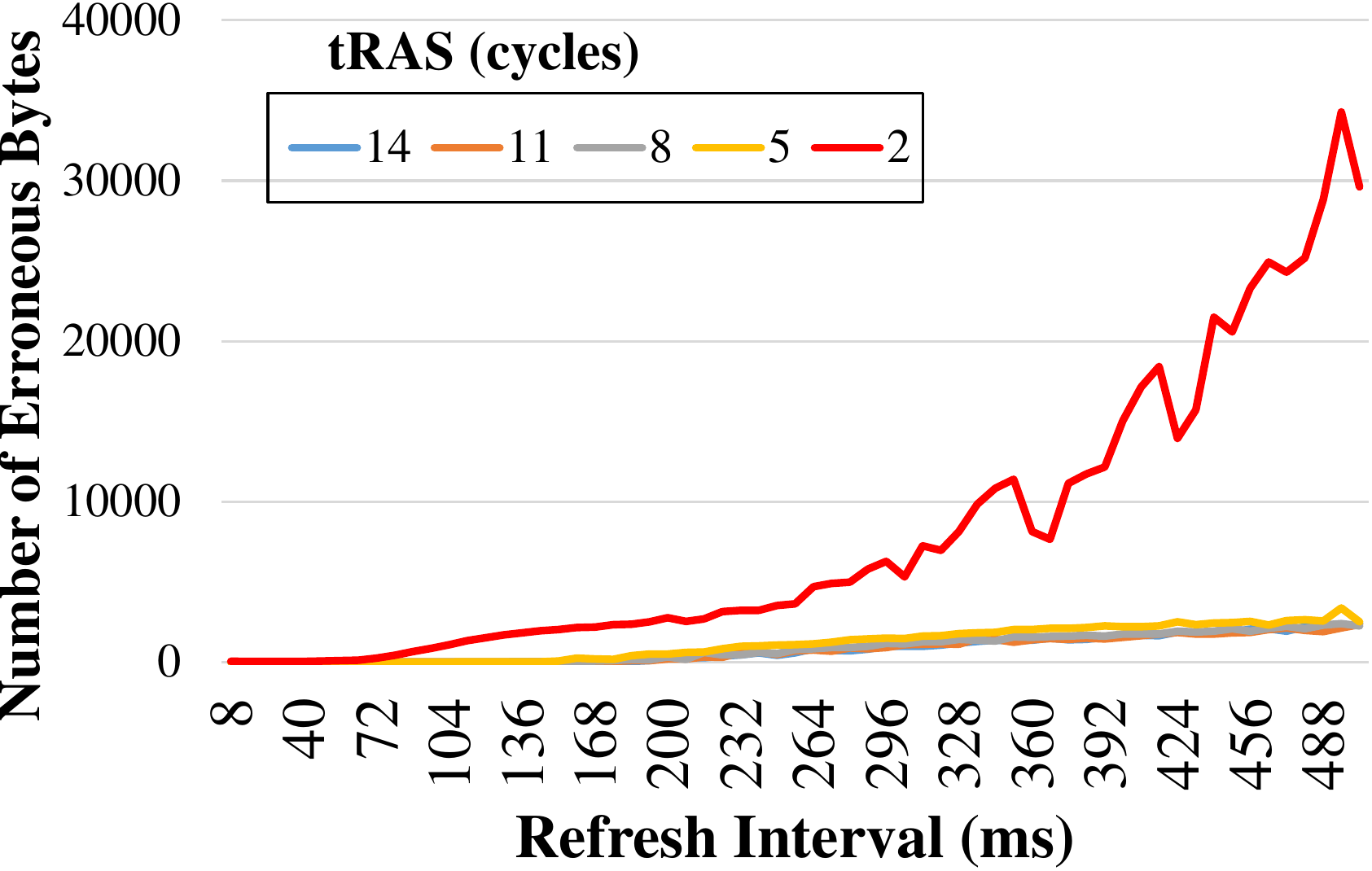}
                \label{softmc_fig:tras_res1}
        }
        \subfloat[Module B] {
                \includegraphics[height=1.92in]{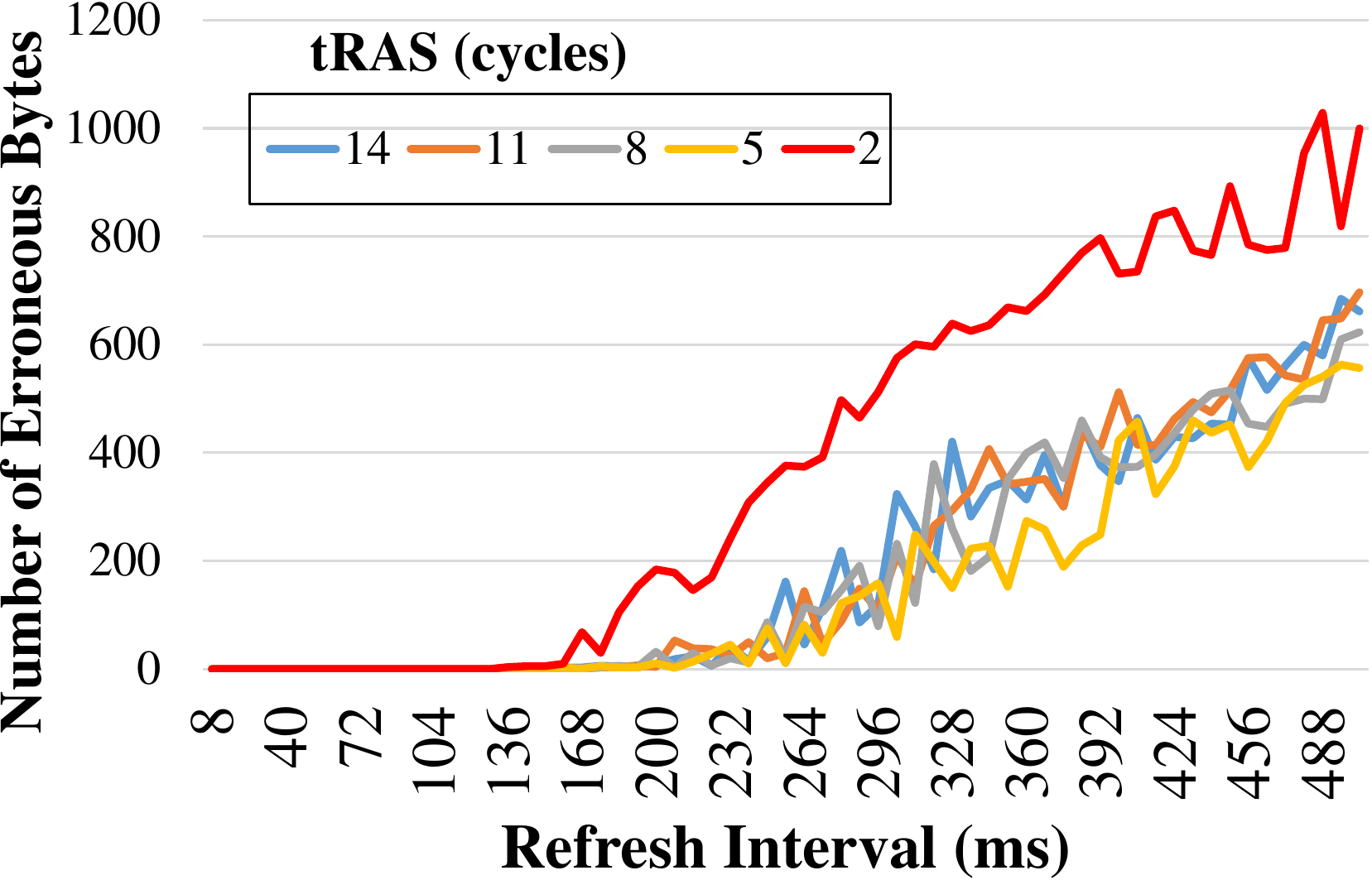}
                \label{softmc_fig:tras_res2}
        }\\
        \subfloat[Module C] {
                \includegraphics[height=1.92in]{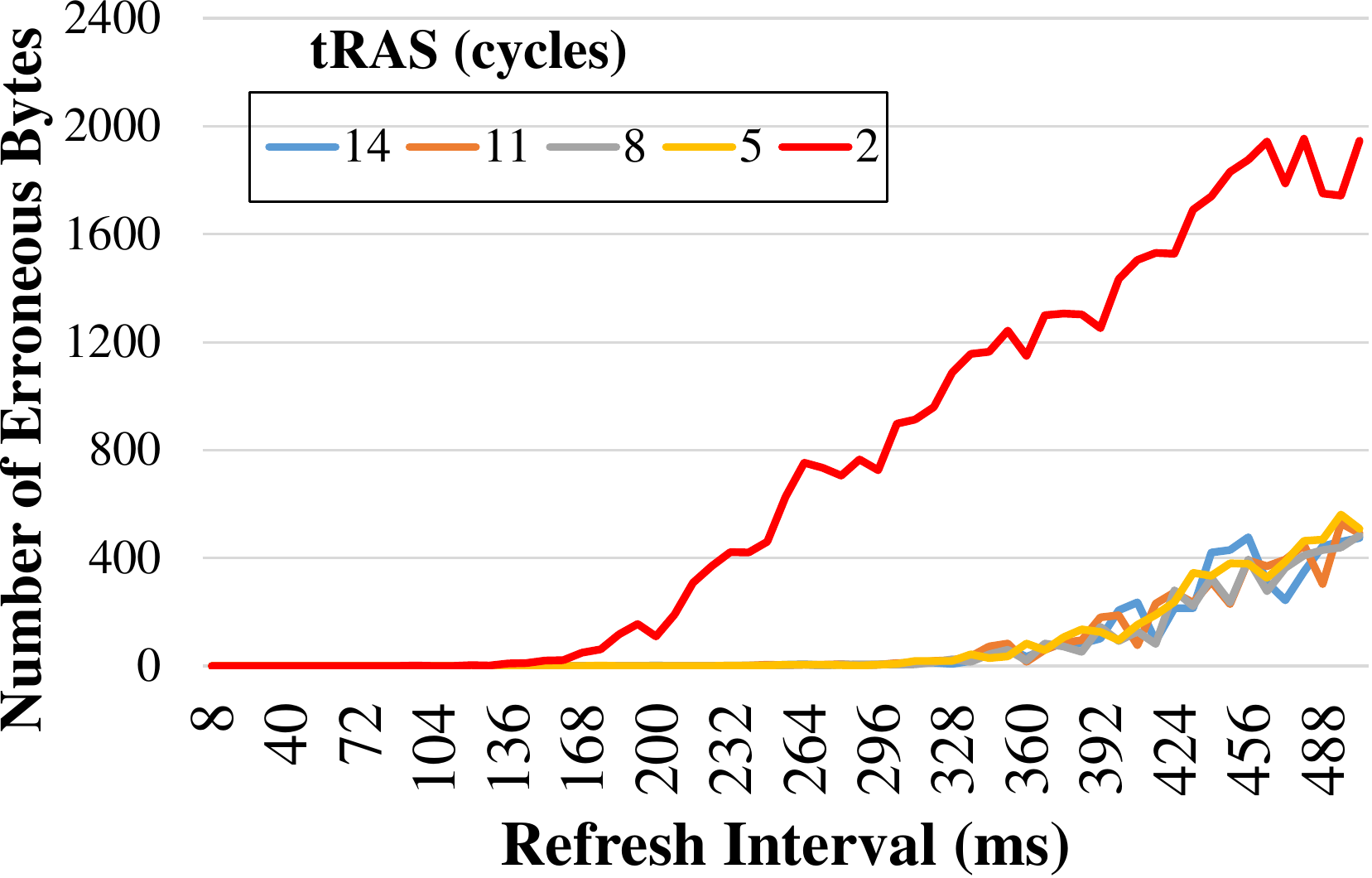}
                \label{softmc_fig:tras_res3}
        }
        \caption{Effect of reducing \tras{} on the number of errors at various refresh intervals.} 
        \label{softmc_fig:tras_test_results}
\end{figure}

Figures~\ref{softmc_fig:trcd_test_results} and~\ref{softmc_fig:tras_test_results} depict the
results for the \trcd test and the \tras test, respectively, for three DRAM
modules (each from a different DRAM vendor). We make three major observations:

\emph{(i)} \emph{Within the duration of the standard
refresh interval (\SI{64}{\milli\second}), DRAM cells do not leak
a sufficient amount of charge to have a negative impact on DRAM access
latency.\footnote{Other studies have shown methods to take advantage
of the fact that latencies can be reduced without incurring
errors~\cite{lee2015adaptive, chang2016understanding}.}} For refresh
intervals less than or equal to \SI{64}{\milli\second}, we
observe little to no variation in the number of errors induced.
Within this refresh interval range, depending on the \trcd or \tras value, the
errors generated are either zero or a constant number. We make
the same observation in both the \trcd and \tras tests for all three DRAM
modules.

For all the modules tested, using different data patterns and stressing DRAM
operation with temperatures significantly higher than the common-case operating
conditions, we can significantly reduce \trcd and \tras parameters, without
observing any errors. We observe errors only when \trcd and \tras parameters are
too small to correctly perform the DRAM access, regardless of the charge amount
of the accessed cells.

\emph{(ii)} \emph{The large safety margin employed by the manufacturers protects
DRAM against errors even when accessing DRAM cells with low latency.} We observe
no change in the number of induced errors for \trcd values less than the default
of 6 cycles (down to 4 cycles in modules A and B, and 5 cycles in module C). We
observe a similar trend in the \tras test: \tras can be reduced from the default
value of 14 cycles to 5 cycles without increasing the number of induced errors
for \emph{any} refresh interval. 

We conclude that even at temperatures much higher than typical operating
conditions, there exists a large safety margin for access latency in existing
DRAM chips. This demonstrates that DRAM cells are much \emph{stronger} than
their timing specifications indicate.\footnote{Similar observations were made by
prior work~\cite{lee2015adaptive, chandrasekar2014exploiting,
chang2016understanding}.} In other words, the timing margin in most DRAM cells
is very large, given the existing timing parameters.

\emph{(iii)} \emph{The expected effect of ChargeCache and NUAT, that
highly-charged cells can be accessed with lower latency, is slightly observable
only when very long refresh intervals are used.} For each of the tests, we
observe a significant increase in the number of errors at refresh intervals that
are much higher than the typical refresh interval of \SI{64}{\milli\second},
demonstrating the variation in charge held by each of the DRAM cells. Based on
the assumptions made by ChargeCache and NUAT, we expect that when lower values
of \trcd and \tras are employed, the error rate should increase more rapidly.
However, we find that for all but the minimum values of \trcd and \tras (and for
\mbox{\trcd= 4} for module C), the \trcd and \tras latencies have almost no
impact on the error rate.

We believe that the reason we cannot observe the expected latency reduction
effect of ChargeCache and NUAT on existing DRAM modules is due to the internal
behavior of existing DRAM chips that does not allow latencies to be reduced
beyond a certain point: we cannot \emph{externally control} when the sense
amplifier gets enabled, since this is dictated with a fixed latency internally,
regardless of the charge amount in the cell. The sense amplifiers are enabled
only after charge sharing, which starts by enabling the wordline and lasts until
sufficient amount of charge flows from the activated cell into the bitline (See
\cref{sec:dram_operation}), is expected to complete. Within existing DRAM
chips, the expected charge sharing latency (i.e., the time when the sense
amplifiers get enabled) is not represented by a timing parameter managed by the
memory controller. Instead, the latency is controlled internally within the DRAM
using a fixed value~\cite{tomishima2016dram, keeth2007dram}. ChargeCache and
NUAT require that charge sharing completes in less time, and the sense
amplifiers get enabled faster for a highly-charged cell. However, since existing
DRAMs provide no way to control the time it takes to enable the sense
amplifiers, we cannot harness the potential latency reduction possible for
highly-charged cells~\cite{tomishima2016dram}. Reducing \trcd affects the time
spent \emph{only after charge sharing}, at which point the bitline voltages
exhibit similar behavior regardless of the amount of charge initially stored
within the cell. Consequently, we are unable to observe the expected latency
reduction effect of ChargeCache and NUAT by simply reducing \trcd, even though
we believe that the mechanisms are sound and can reduce latency (assuming the
behavior of DRAM chips is modified). If the DDR interface exposes a method of
controlling the time it takes to enable the sense amplifiers in the future,
SoftMC can be easily modified to use the method and fully evaluate the latency
reduction effect of ChargeCache and NUAT.

\textbf{Summary.}
Overall, we make two major conclusions from the implementation and
experimental results of our DRAM latency experiments. First, SoftMC provides a
simple and easy-to-use interface to quickly implement tests that
characterize modern DRAM chips. Second, SoftMC is an effective
tool to validate or refute the expected effect of existing or new
mechanisms on existing DRAM chips.

\section{Limitations of SoftMC}

In the previous section, we presented and discussed some examples demonstrating
the benefits of SoftMC. These use cases illustrate the flexibility, ease of use,
and capability of SoftMC in characterizing DRAM operations and evaluating new
DRAM mechanisms (using the DDR interface). Although our SoftMC prototype already
caters for a wide range of use cases, it also has some limitations that arise
mainly from its current hardware implementation.

\sisetup{range-phrase=-}
\sisetup{range-units=single}

\textbf{Inability to Evaluate System Performance.} 
Studies where the applications run on the host machine and access memory within
the module under test can demonstrate the impact of memory reliability and
latency on system performance using  real applications.  However, such studies
are difficult to perform in any FPGA-based DRAM testing infrastructure that
connects  the host and the FPGA over the PCIe bus (including SoftMC). This is
due to the  the long latency associated with the PCIe bus (in
microseconds~\cite{koop2008performance, bittner2014direct, kadric2012fpga}; in
contrast to DRAM access latency that is within
\SIrange{15}{80}{\nano\second}~\cite{lee2013tiered, son2013reducing}). Note that
this long PCIe bus latency does \emph{not} affect our tests (as explained in
\cref{softmc_sec:highlevel}), but it would affect the performance of a system
that would use SoftMC as the main memory controller. One way to enable
performance studies with SoftMC is to add support for trace-based execution,
where the traces (i.e., memory access requests) are collected by executing
workloads on real systems or simulating them. The host PC can transform the
traces into SoftMC instructions and transmit them to SoftMC hardware, where we
can buffer the instructions and emulate microarchitectural behavior of a memory
controller developed using SoftMC while avoiding the PCIe bus latency. This
functionality requires coordination between the system/simulator and SoftMC, and
we expect to add such support by enabling coordination between our DRAM
simulator, Ramulator~\cite{kim2015ramulator, ramulatorgithub}, and a future
release of SoftMC.

\textbf{Coarse-Grained Timing Resolution.}
As FPGAs are significantly slower than the memory bus, the minimum time interval
between two consecutive commands that the FPGA can send to the memory bus is
limited by the FPGA frequency. In our current prototype, the DDR interface of
the ML605 operates at 400MHz, allowing SoftMC to issue two consecutive commands
with a minimum interval of \SI{2.5}{\nano\second}. Therefore, the timing
parameters in SoftMC can be changed only at the granularity of
\SI{2.5}{\nano\second}. However, it is possible to support finer-granularity
resolution with faster FPGA boards~\cite{vc709}.

\textbf{Limitation on the Number of Instructions Stored in the FPGA.}
The number of instructions in one test program (that is executed atomically in
SoftMC) is limited by the size of the \emph{Instruction Queue} in SoftMC. To
keep FPGA resource usage low, the size of the \emph{Instruction Queue} in the
current SoftMC prototype is limited to 8192 instructions. In the future, instead
of increasing the size of the \emph{Instruction Queue}, which would increase
resource utilization in the FPGA, we plan to extend the SoftMC hardware with
control flow instructions to support arbitrary-length tests. Adding control flow
instructions would enable loops that can iterate over large structures (e.g.,
DRAM rows, columns) in hardware with a small number of SoftMC instructions,
avoiding the need for the host machine to issue a large number of instructions
in a loop-unrolled manner. Such control flow support would further improve the
ease of use of SoftMC.

\section{Research Directions Enabled by SoftMC}

We believe SoftMC can enable many new studies of the behavior of DRAM and other
memories. \hhm{In fact, since the release of SoftMC to the public as part of
this research, many works~\cite{koppula2019eden,chang2016understanding,
chang2017understanding,frigo2020trrespass,orosa2021deeper,gao2019computedram,kim2020revisiting,kim2019d,hassan2021uncovering,khan2017detecting,ghose2018your,yauglikcci2022understanding,olgun2021quac,orosa2021codic, talukder2018exploiting, talukder2019prelatpuf, talukder2018ldpuf, talukder2020towards, bepary2022dram,farmani2021rhat, yaglikci2022hira, gao2022fracdram}
used different versions of SoftMC in various DRAM characterization studies. In
this section, we} briefly describe several \hhm{other example studies that
SoftMC can further enable}. 

\sloppypar{
\textbf{More Characterization of DRAM.}
The SoftMC DRAM testing infrastructure can  
 test any DRAM mechanism consisting of low-level DDR commands. Therefore, it
enables a wide range of characterization and analysis studies of real DRAM
modules that would otherwise not have been possible without such an
infrastructure. We discuss three such example research directions. }

First, as DRAM scales down to smaller technology nodes, it faces key challenges
in both reliability and latency~\cite{mutlu2013memory, mandelman2002challenges,
kim2005technology, mueller2005challenges, liu2012raidr, kang2014co,
liu2013experimental, khan2016case,
khan2016parbor,khan2014efficacy,khan2017detecting,mutlu2019rowhammer,kim2020revisiting,orosa2021deeper}.
Unfortunately, there is no comprehensive experimental study that characterizes
and analyzes the trends in DRAM cell operations and behavior with technology
scaling across various DRAM generations. The SoftMC infrastructure can help us
answer various questions to this end: How are the cell characteristics,
reliability, and latency changing with different generations of technology
nodes? Do all DRAM operations and cells get affected by scaling at the same
rate? Which DRAM operations are getting worse?  

Second, aging-related failures in DRAM can potentially affect the reliability
and availability of systems in the field~\cite{meza2015revisiting,
schroeder2009dram}. However, the causes, characteristics, and impact of
\emph{aging} have remained largely unstudied. Using SoftMC, it is possible to
devise controlled experiments to analyze and characterize aging. The SoftMC
infrastructure can help us answer questions such as: How prevalent are
aging-related failures? What types of usage accelerate aging? How can we design
architectural techniques that can slow down the aging process?

Third, prior works show that the failure rate of DRAM modules in large data
centers is significant, largely affecting the cost and downtime in data
centers~\cite{schroeder2009dram, sridharan2013feng, meza2015revisiting,
luo2014characterizing}. Unfortunately, there is no study that analyzes DRAM
modules that have failed in the field to determine the common causes of failure.
Our SoftMC infrastructure can test faulty DRAM modules and help answer various
research questions: What are the dominant types of DRAM failures at runtime? Are
failures correlated to any location or specific structure in DRAM? Do all chips
from the same generation exhibit the same failure characteristics?

\sloppypar{
\textbf{Characterization of Non-Volatile Memory.}
The SoftMC infrastructure can test any chip compatible with the DDR interface.
Such a design makes the scope of the chips that can be tested by SoftMC go well
beyond just DRAM. With the emergence of byte addressable non-volatile  memories
(e.g., PCM~\cite{raoux2008phare,lee2010phase, lee2009architecting,
lee2010phase}, STT-RAM~\cite{kawahara2008spram,kultursay2013evaluating},
ReRAM~\cite{akinaga2010resistive, wong2012metal}), several vendors are working
towards manufacturing DDR-compatible non-volatile memory chips at a large
scale~\cite{micron2016xpoint,everspin2021sttmram}. When these chips become
commercially available, it will be critical to characterize and analyze them in
order to understand, exploit, and/or correct their behavior. We believe that
SoftMC can be seamlessly used to characterize these chips, and can help enable
future mechanisms for NVM.}

SoftMC will hopefully enable other works that build on it in various ways. For
example, future work can extend the infrastructure to enable researchers to
analyze memory scheduling (e.g.,~\cite{mutlu2008parallelism, mutlu2007stall,
subramanian2014blacklisting, kim2010atlas, kim2010thread,
muralidhara2011reducing, ebrahimi2011parallel}) and memory power
management~\cite{david2011memory, deng2011memscale} mechanisms, and allow them
to develop new mechanisms using a programmable memory controller and real
workloads.  We conclude that characterization with SoftMC enables a wide range
of research directions in DDR-compatible memory chips (DRAM or NVM), leading to
better understanding of these technologies and helping to develop mechanisms
that improve the reliability and performance of future memory systems.

\section{Other Related Work}

Although no prior work provides an open-source DRAM testing infrastructure
similar to SoftMC, infrastructures for testing other types of memories have been
developed. Cai et al.~\cite{cai2011fpga} developed a platform for characterizing
NAND flash memory. They propose a flash controller, implemented on an FPGA, to
quickly characterize error patterns of existing flash memory chips. They expose
the functions of the flash translation layer (i.e., the flash chip interface) to
the software developer via the host machine connected to the FPGA board, similar
to how we expose the DDR interface to the user in SoftMC. Many
works~\cite{luo2016enabling, cai2015read, luo2015warm, cai2015data,
cai2014neighbor, cai2013program, cai2013error, cai2013threshold, cai2012flash,
cai2012error, cai2017vulnerabilities} use this flash memory testing
infrastructure to study various aspects of flash chips.

Our prior works~\cite{chang2016understanding, khan2016parbor, khan2016case,
kim2014flipping, lee2016reducing, khan2014efficacy, liu2013experimental,
lee2015adaptive} developed and used FPGA-based infrastructures for a wide range
of DRAM studies. Liu et al.~\cite{liu2013experimental} and Khan et
al.~\cite{khan2014efficacy} analyzed the data retention behavior of modern DRAM
chips and proposed mechanisms for mitigating retention failures. Khan et
al.~\cite{khan2016parbor, khan2016case} studied data-dependent failures in DRAM,
and developed techniques for efficiently detecting and handling them. Lee et
al.~\cite{lee2015adaptive, lee2016reducing} analyzed latency characteristics of
modern DRAM chips and proposed mechanisms for latency reduction. Kim et
al.~\cite{kim2014flipping} discovered a new reliability issue in existing DRAM,
called \emph{RowHammer}, which can lead to security
breaches~\cite{seaborn2015exploiting, rh_project_zero, gruss2016rowhammer,
veen2016drammer, xiao2016one, razavi2016flip}. Chang et
al.~\cite{chang2016understanding} used SoftMC to characterize latency variation
across DRAM cells for fundamental DRAM operations (e.g., activation, precharge).
SoftMC evolved out of these previous infrastructures, to address the need to
make the infrastructure flexible and easy to use.

\section{Summary}
\label{softmc_sec:conclusion}

This \hhm{chapter} introduces the first publicly-available FPGA-based DRAM
testing infrastructure, \emph{SoftMC} (Soft Memory Controller), which provides a
programmable memory controller with a flexible and easy-to-use software
interface. SoftMC enables the flexibility to test any standard DRAM operation
and any (existing or new) mechanism comprising of such operations. It provides
an intuitive high-level software interface for the user to invoke low-level DRAM
operations, in order to minimize programming effort and time. We provide a
prototype implementation of SoftMC, and we have released it publicly as a
freely-available open-source tool~\cite{softmcsource}.

We demonstrate the capability, flexibility, and programming ease of SoftMC by
implementing two example use cases.  Our experimental analyses demonstrate the
effectiveness of SoftMC as a new tool to \emph{(i)} perform detailed
characterization of various DRAM parameters (e.g., refresh interval and access
latency) as well as the relationships between them, and \emph{(ii)} test the
expected effects of existing or new mechanisms (e.g., whether or not
highly-charged cells can be accessed faster in existing DRAM chips). We believe
and hope that SoftMC, with its flexibility and ease of use, can enable many
other studies, ideas and methodologies in the design of future memory systems,
by making memory control and characterization easily accessible to a wide range
of software and hardware developers.

\chapter[Uncovering Target Row Refresh]{Uncovering In-DRAM RowHammer Protection Mechanisms: A New Methodology, Custom RowHammer Patterns, and Implications}
\label{chap:utrr}

\newcommand{\valARI}{62.5} 

The RowHammer vulnerability in DRAM is a critical threat to system security. To
protect against RowHammer, vendors commit to security-through-obscurity: modern
DRAM chips rely on undocumented, proprietary, on-die mitigations, commonly known
as \emph{Target Row Refresh} (TRR). At a high level, TRR detects and refreshes
potential RowHammer-victim rows, but its exact implementations are not openly
disclosed. Security guarantees of TRR mechanisms cannot be easily studied due to
their proprietary nature.

To assess the security guarantees of recent DRAM chips, we present
\emph{Uncovering TRR} (\method{}), an experimental methodology to analyze
in-DRAM TRR implementations. \method{} is based on the new observation that data
retention failures in DRAM enable a side channel that leaks information on how
TRR refreshes potential victim rows. \method{} allows us to (i) understand how
logical DRAM rows are laid out physically in silicon; (ii) study undocumented
on-die TRR mechanisms; and (iii) combine (i) and (ii) to evaluate the RowHammer
security guarantees of modern DRAM chips. We show how \method{} allows us to
craft RowHammer access patterns that successfully circumvent the TRR mechanisms
employed in \numTestedDIMMs{} DRAM modules of the three major DRAM vendors. We
find that the DRAM modules we analyze are vulnerable to RowHammer, having bit
flips in up to \vulnRowsMaxPct{}\% of all DRAM rows.

\section{Motivation and Goal}
\label{crow_sec:motivation}

As the memory demands of applications have been growing, manufacturers have been
scaling the DRAM process technology to keep pace. Unfortunately, while the
density of the DRAM chips has been increasing as a result of scaling, DRAM faces
three critical challenges in meeting application
demands~\cite{mutlu2014research, mutlu2013memory}: (1)~high access latencies and
(2)~high refresh overheads, both of which degrade system performance and energy
efficiency; and (3)~increasing exposure to vulnerabilities, which reduces the
reliability of DRAM.

First, the high DRAM access latency is a challenge \hhmiii{with respect} to
improving system performance and energy efficiency. While DRAM capacity
increased significantly over the last two decades~\cite{jedec2008ddr3,
jedec2012ddr4, son2013reducing, chang2016understanding, lee2013tiered,
lee2015adaptive}, DRAM access latency decreased only
slightly~\cite{lee2013tiered, mutlu2014research, chang2016understanding}. The
high DRAM access latency significantly degrades the performance of many
workloads~\cite{ferdman2012clearing, huang2014moby, gutierrez2011full,
zhu2015microarchitectural, hestness2014comparative}. The performance impact is
particularly large for applications that 1)~have working sets exceeding the
cache capacity of the system, 2)~suffer from high instruction and data cache
miss rates, and 3)~have low memory-level parallelism. While manufacturers offer
latency-optimized DRAM modules~\cite{micron2021rldram, sato1998fast}, these
modules have significantly lower capacity and higher cost compared to commodity
DRAM~\cite{lee2013tiered, chang2016low, kim2012case}. Thus, reducing the high
DRAM access latency \emph{without trading off capacity and cost} in commodity
DRAM remains an important challenge~\cite{ghose2019demystifying, lee2013tiered,
mutlu2013memory}.

Second, the high DRAM refresh overhead is a challenge to improving system
performance and energy consumption. A DRAM cell stores data in a capacitor that
leaks charge over time. To maintain correctness, \emph{every} DRAM cell requires
periodic \emph{refresh} operations that restore the charge level in a cell. As
the DRAM cell size decreases with process technology scaling, newer DRAM devices
contain more DRAM cells than older DRAM devices~\cite{itrs}. As a result, while
DRAM capacity increases, the performance and energy overheads of the refresh
operations scale unfavorably~\cite{chang2014improving, kang2014co,
liu2012raidr}. In modern LPDDR4~\cite{micron-lpddr4} devices, the memory
controller refreshes \emph{every} DRAM cell every \SI{32}{\milli\second}.
Previous studies show that 1)~refresh operations incur large performance
overheads, as DRAM cells \emph{cannot} be accessed when the cells are being
refreshed~\cite{chang2014improving, liu2012raidr, mukundan2013understanding,
nair2014refresh}; and 2)~up to 50\% of the total DRAM energy is consumed by the
refresh operations~\cite{liu2012raidr, chang2014improving}.

Third, the increasing vulnerability of DRAM cells to various failure mechanisms
is an important challenge to maintaining DRAM reliability. As the process
technology shrinks, DRAM cells become smaller and get closer to each other, and
thus become more susceptible to failures~\cite{mandelman2002challenges,
mutlu2013memory, redeker2002investigation, yaney1987meta, konishi1989analysis,
liu2012raidr, liu2013experimental}. A tangible example of such a failure
mechanism in modern DRAM is RowHammer~\cite{kim2014flipping, mutlu2017rowhammer,
mutlu2019rowhammer, mutlu2023fundamentally}. RowHammer causes \emph{disturbance
errors} (i.e., bit flips in vulnerable DRAM cells that are not being accessed)
in DRAM rows physically adjacent to a row that is repeatedly activated many
times.

These three challenges are difficult to solve efficiently by directly modifying
the underlying cell array structure. This is because commodity DRAM implements
an extremely dense \emph{DRAM cell array} that is optimized for \emph{low
area-per-bit}~\cite{lee2013tiered, lee2015adaptive, son2013reducing}. Because of
its density, even a small change in the DRAM cell array structure may incur
non-negligible area overhead~\cite{lee2013tiered, udipi2010rethinking,
zhang2014half, mutlu2013memory}. \textbf{Our goal} in this \hhm{chapter} is to
lower the DRAM access latency, reduce the refresh overhead, and improve DRAM
reliability with \emph{no changes} to the DRAM cell architecture, and with only
\emph{minimal} changes to the DRAM chip.
\section{Copy-Row DRAM}
\label{crow_sec:mechanism}

To efficiently solve the challenges of 1)~high access latency, 2)~high refresh
overhead, and 3)~increasing vulnerability to failure mechanisms in DRAM, without
requiring any changes to the DRAM cell array, we introduce \emph{Copy-Row DRAM}
(\mech). \mech is a new, practical substrate that exploits the existing DRAM
architecture to efficiently duplicate a small number of rows in a subarray at
runtime. \mech is versatile, and enables new mechanisms for improving DRAM
performance, energy efficiency, and reliability, as we show in
\cref{crow_sec:crow_apps}.

\subsection{CROW: A High-Level Overview}
\label{crow_sec:mechanism:overview}

\mech consists of two key components: 1)~a small number of \emph{{\copyrow}s} in
each subarray, which can be used to duplicate or remap the data stored in one of
the remaining rows (called \emph{regular rows}) of the subarray, 2)~a table in
the memory controller, called the \emph{\mech table}, that tracks which rows are
duplicated or remapped.

\mech divides a DRAM subarray into two types of rows: regular rows and
\copyrow{s}. A \copyrow is similar to a regular row in the sense that it has the
same row width and DRAM cell structure. However, {\copyrow}s have their own
small local row decoder within the subarray, separate from the existing local
row decoder for regular rows. This enables the memory controller to activate
{\copyrow}s independently from regular rows (which continue to use the existing
local row decoder). By allowing {\copyrow}s and regular rows to be activated
independently, we enable two DRAM primitives that make use of \emph{multiple-row
activation} (MRA) in \mech.

First, \mech can perform \emph{bulk data movement}, where the DRAM copies an
entire row of data at once from a regular row to a \copyrow. To do so, the
memory controller first activates the regular row, and next activates the
\copyrow, immediately after the local row buffer latches the data of the regular
row. After the second activation, the sense amplifiers restore the data
initially stored only in the regular row to both the regular row and the
\copyrow (similar to the RowClone~\cite{seshadri2013rowclone} mechanism).

Second, \mech can perform \emph{reduced-latency DRAM access}. When a regular row
and a \copyrow contain \emph{the same data}, the memory controller activates
both rows \emph{simultaneously}. This causes two cells on the same bitline to
inject charge into the bitline at a faster total rate than a single cell can.
Thereby, the sense amplifier to operate faster, which leads to lower activation
latency via an effect similar to increasing the amount of charge stored in a
single cell~\cite{choi2015multiple}.

\mech makes use of a table in the memory controller, \mech table, that tracks
which regular rows are duplicated or mapped to \copyrow{s}. A mechanism that
takes advantage of the \mech substrate checks the \mech table and uses the
information in the table to either 1) simultaneously activate duplicate regular
and \copyrow{s} or 2) activate the \copyrow that a regular row is remapped to.
For example, the \mech-cache mechanism (\cref{crow_subsec:in_dram_caching})
updates a \mech table entry with the address of the regular row that has been
copied to the corresponding \copyrow. Prior to issuing an \cmdact command to
activate a target regular row, the memory controller queries the \mech table to
check whether the regular row has a duplicate \copyrow. If so, the memory
controller issues a custom reduced-latency DRAM command to simultaneously
activate both the regular row and the \copyrow, instead of a single activate to
the regular row. Doing so enables faster access to the duplicated data stored in
both rows. Next, we explain \mech in detail.

\subsection{Copy Rows}
\label{crow_subsec:cache_row}

\begin{figure}[h] 
    \centering
    \includegraphics[width=0.6\linewidth]{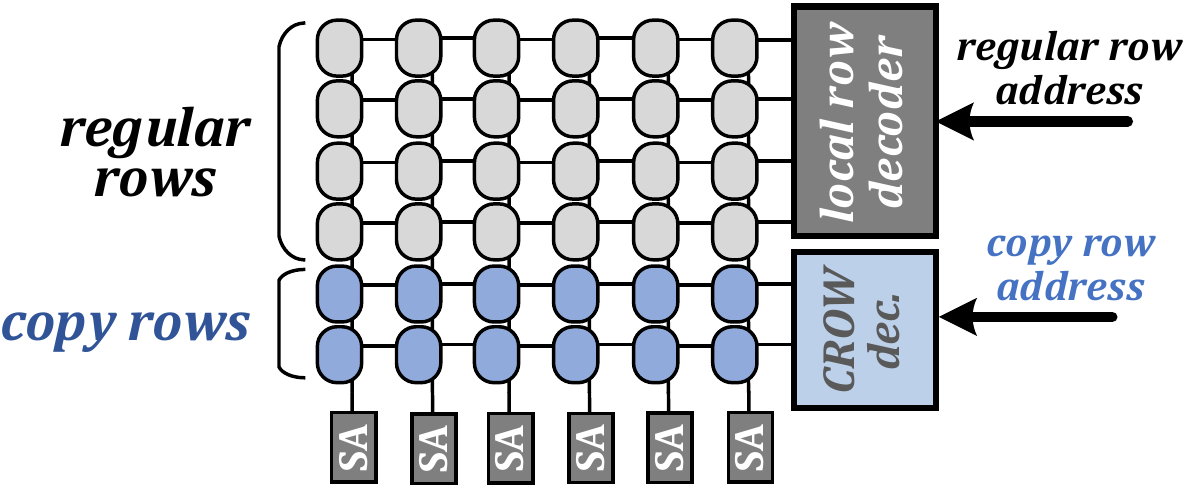}%
    \caption{Regular and copy rows in a subarray in CROW.} 
    \label{crow_fig:crow_subarray}
\end{figure}

\mech logically categorizes the rows in a subarray into two sets, as shown in
Figure~\ref{crow_fig:crow_subarray}: (1)~\emph{regular rows}, which operate the
same as in conventional DRAM; and (2)~\emph{{\copyrow}s}, which can be activated
independently of the regular rows. In our design, we add a small number of
\copyrow{s} to the subarray, and add a second local row decoder (called the
\emph{\mech decoder}) for the \copyrow{s}.\footnote{Alternatively, \mech can use
a small set of the existing rows in a conventional subarray as \copyrow{s}.}
Because the \mech decoder drives a much
smaller number of rows than the existing local row decoder, it has a much
smaller area cost (\cref{crow_subsec:crow_power_area}). The memory controller
provides the regular row and \copyrow addresses to the respective decoders when
issuing \mbox{an \cmdact command}.

\subsection{\mech Table}
\label{crow_subsec:ctable}

\begin{figure}[h] 
    \centering
    \includegraphics[width=0.6\linewidth]{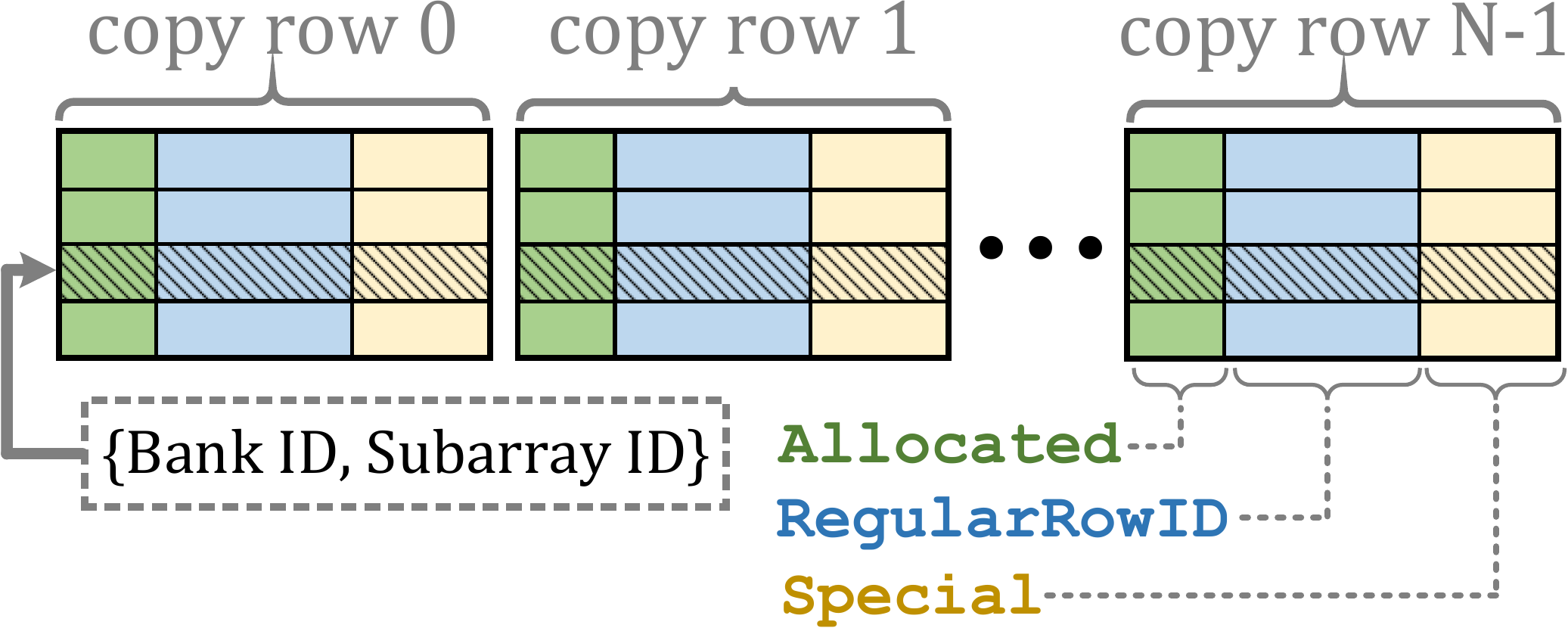}
    \caption{Organization of the \mech table.} 
    \label{crow_fig:crow_table}
\end{figure}

As shown in Figure~\ref{crow_fig:crow_table}, the \mech table in the memory
controller stores information on 1) whether a \copyrow is allocated, and 2)
which regular row is duplicated or remapped to a \copyrow. For example, in our
weak row remapping scheme (\cref{crow_subsec:refresh_reduction_mech}), a
\mech table entry holds the address of the regular row that the \copyrow
replaces. The \mech table stores an entry for each \copyrow in a DRAM channel,
and is \emph{n}-way set associative, where \emph{n} is the number of {\copyrow}s
in each subarray.

The memory controller indexes the \mech table using a combination of bank and
subarray addresses, which are part of the physical address of a memory request.
\hhmiv{However, in current systems, subarray addresses are not exposed to the
memory controller. Therefore, to enable CROW, it is required to expose such
information to the memory controller.} 

The memory controller checks the table for an entry associated with the regular
row that the memory request is to access. Our current design stores three fields
in each \mech table entry:
1) the \texttt{Allocated} field indicates whether the entry is valid or not; 2)
the \texttt{RegularRowID} field stores a pointer to the regular row that the
corresponding \copyrow is associated with\; 3) the \texttt{Special} field stores
additional information specific to the mechanism that is implemented using
\mech.

\section{Row Scout (\rtp{})}
\label{utrr_sec:ret_profiling}

\method{} uses data retention failures as a side channel to determine if and
when a row receives a (TRR-induced or regular) refresh. As such, to know how
long a row can retain its data correctly without being refreshed, \method{}
requires a mechanism for profiling the retention time of a DRAM row. We define
\emph{DRAM row retention time} as the maximum time interval for which all cells
in the row can correctly retain their data without being refreshed.

Unlike existing DRAM retention-time profiling
techniques~\cite{patel2017reaper,liu2013experimental,
liu2012raidr,qureshi2015avatar,das2018vrl}, \method{} does \emph{not} require a
profiler that finds the retention time of \emph{all} rows in a DRAM chip.
Instead, \method{} needs to search for a small set of DRAM rows (i.e., tens of
rows depending on the experiment) that match certain criteria
(\cref{utrr_subsec:profiler_requirements}) as specified by the \method{} user
based on the desired experiment.

\subsection{Row Scout (\rtp{}) Requirements}
\label{utrr_subsec:profiler_requirements}

Depending on the experiment that the \method{} user conducts, \trran{} needs the
data retention time of DRAM rows that match different criteria. We identify the
following general requirements for \rtp{} to enable it to search for DRAM rows
suitable for \trran{}.

\textbf{Rows with uniform retention time.} A TRR mechanism may refresh multiple
victim rows. For instance, during a single-sided RowHammer attack, a TRR
mechanism may refresh the row on either side of the aggressor \emph{or} both at
the same time. To examine whether or not TRR can refresh multiple victim rows at
the same time (i.e., with a single \cmdrefresh{}), \rtp{} must provide multiple
rows (i.e., a \emph{row group}) that have the \emph{same} retention time.

\textbf{Relative positions of profiled rows.} The location of a victim row
depends on the location of the aggressor rows that the \method{} user specifies
for an experiment. For example, for a double-sided RowHammer attack (see
Figure~\ref{bg_fig:rowhammer_access}b), \rtp{} must provide three rows with the same
retention time that are exactly one row apart from each other. \trran{} can then
analyze which of the three victim rows get refreshed by TRR when hammering the
two aggressor rows that are placed between the victim rows. We represent the
relative positions of rows in a row group (i.e., the \emph{row group layout})
using a notation such as \texttt{R-R-R}, where `\texttt{R}' indicates a
retention-profiled row and `\texttt{-}' indicates a distance of one DRAM row.
\rtp{} must find a row group based on the row group layout that the user
specifies.

\textbf{Rows in specific DRAM regions.} TRR may treat rows in different parts of
a DRAM chip differently by operating independently at different granularities.
For example, TRR may operate independently at the granularity of a DRAM bank or
a region of DRAM bank. To identify the granularity at which TRR operates, \rtp{}
must find DRAM rows within a \emph{specific region} of a DRAM chip.

\textbf{Rows with consistent retention time.} 
An \rtp{}-provided row must have a consistent retention time such that \method{}
can accurately infer the occurrence of a TRR-induced refresh operation based on
whether the row contains retention failures after a time period equivalent to
the row's retention time. The main difficulty is a phenomenon known as Variable
Retention Time (VRT)~\cite{kang2014co, khan2014efficacy, liu2013experimental,
mori2005origin, qureshi2015avatar, restle1992dram, yaney1987meta}, which causes
the retention time of certain DRAM cells to change over time. If an
\rtp{}-provided row has an inconsistent retention time that was initially
measured to be $T$, \method{} will not be able to correctly infer the occurrence
of a TRR-induced refresh operation.\footnote{ \method{} fails to correctly infer
a TRR-induced refresh when a row retains its data for significantly longer or
shorter than $T$. If the row retains its data for longer than $T$, \method{}
will \emph{always} infer the occurrence of a TRR-induced refresh operation. If
the row fails too soon (i.e., before $\frac{T}{2}$ or during Step~1 in
\cref{utrr_subsec:trran_overview}), \method{} will always observe retention
failures, since even a TRR-induced refresh will not be able to prevent the bit
flip (in Step~2 in \cref{utrr_subsec:trran_overview}). Consequently, \method{}
will \emph{always} infer that a TRR-induced refresh operation was not issued to
the row.} To \hhmiv{make it likely for} row's retention time \hhmiv{to remain
consistent during the experiment}, \rtp{} validates the retention time of a row
\emph{one thousand times} in order to rule out \hhmiv{rows with inconsistent
retention times} (that are due to VRT).

\textbf{Rows with short retention times.} The time it takes to finish a single
\method{} experiment depends on the retention time of the rows \rtp{} finds.
This is because even retention-weak DRAM rows typically retain their data
correctly for tens or hundreds of
milliseconds~\cite{liu2013experimental,liu2012raidr,patel2017reaper}, whereas other
\trran{} operations (e.g., reading from or writing to a row, hammering a row,
performing refresh) often take much less than a millisecond. Thus, as
Figure~\ref{utrr_fig:general_trr_detection_approach} shows, the duration of a \trran{}
experiment is dominated by retention times (T) of the profiled rows. To reduce
the overall experiment time, it is critical for \rtp{} to identify rows with
short data retention times.

\subsection{Row Scout (\rtp{}) Operation}
\label{subsec:rtp}

We design and implement \rtplong{} (\rtp{}), a DRAM retention time profiler,
such that it satisfies the requirements listed in
\cref{utrr_subsec:profiler_requirements}. We implement \rtp{} using a modified
version of SoftMC~\cite{hassan2017softmc,softmcsource} with DDR4 support
(described in \cref{utrr_subsec:softmc}).

\begin{figure}[!h]
    \includegraphics[width=\linewidth]{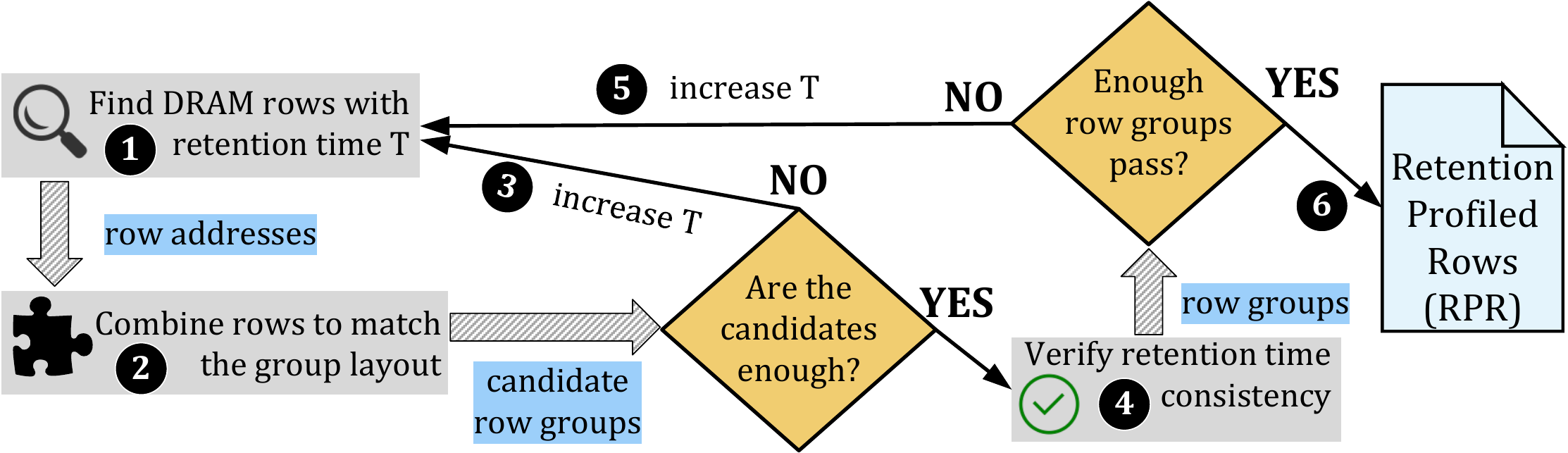}
	\caption{Detailed \rtp{} operation.}
	\label{utrr_fig:retprofiler_operation}
\end{figure}

We illustrate the operation of \rtp{} in
Figure~\ref{utrr_fig:retprofiler_operation}. \circled{1} \rtp{} scans a full
range of DRAM rows within a DRAM bank, as specified in the profiling
configuration (Figure~\ref{utrr_fig:methodology_overview}), and collects the
addresses of rows that experience retention failures if not refreshed over the
time interval $T$.  \rtp{} initially sets $T$ to a small value (e.g.,
\SI{100}{\milli\second}) in order to identify rows with small retention times as
we discuss in the requirements of \rtp{}
(\cref{utrr_subsec:profiler_requirements}). \circled{2} \rtp{} creates candidate
row groups by combining the appropriate row addresses (with retention time $T$)
that match the row group layout specified in the profiling configuration. If the
number of candidate row groups is less than the number of row groups to find
according to the profiling configuration, \circled{3} \rtp{} increases $T$ by a
certain amount (e.g., \SI{50}{\milli\second}) and starts over from~\circled{1}.
Otherwise, \circled{4} \rtp{} tests each row in a candidate row group one
thousand times to \hhmiv{build confidence} that all rows in the candidate row
group have a consistent retention time that is equal to $T$. \hhmiv{Repeating
the test thousand times does \emph{not} fully ensure that the retention time of
the entire candidate row group will remain consistent during the whole
experiment. The retention time of the rows in the group may still change due to
effects such as VRT and result in noise (i.e., observing retention success on a
row after $T$ even though the row is not refreshed) in the results. In our
experiments, we observe that testing the retention time of the candidate row
group thousand times and using only row groups passing the tests largely
eliminate such noise.} If the number of candidate row groups that pass the
retention time consistency test is less than the number of row groups to find
according to the profiling configuration, \circled{5} \rtp{} increases $T$ by a
certain amount (e.g., \SI{50}{\milli\second}) and starts over from~\circled{1}.
Otherwise, \circled{6} \rtp{} provides a list of retention time-profiled rows to
be used by \trran{}.

\section{Analyzing TRR-induced Refresh}
\label{utrr_sec:analyzing_refs}

\trran{} is a configurable and extensible component in \method{} for analyzing
TRR-induced as well as regular refresh operations. We use the \trran{} to
inspect in-DRAM RowHammer mitigations in modules from the three major DRAM
vendors. We implement \trran{} on top of a modified version of
SoftMC~\cite{hassan2017softmc,softmcsource} with DDR4 support. In
\cref{utrr_sec:overview}, we discuss the general operation of \trran{} using
Figure~\ref{utrr_fig:general_trr_detection_approach}. 

\subsection{TRR Analyzer Requirements}
\label{utrr_subsec:trr_analyzer_reqs}

We identify and discuss four key requirements needed to enable reverse
engineering a RowHammer mitigation mechanism. First, to analyze the capability
of TRR in detecting multiple aggressor rows, \trran{} should allow the user to
specify one or more aggressor rows, their corresponding hammer counts, and the
order in which to hammer the aggressor rows.

\begin{req}
    Ability to hammer multiple aggressor rows with individually configurable
    hammer counts in a configurable order.
\end{req}

The user should be able to specify dummy rows\footnote{A dummy row operates
similarly to an aggressor row, but it can be implemented more efficiently in a
SoftMC program since a dummy row does not need to be initialized with specific
data unlike an aggressor row.} that can be hammered to divert the TRR mechanism
to refresh the neighbors of a dummy row instead of victims of an aggressor row.

\begin{req}
Ability to specify dummy rows that are hammered in addition to the aggressor
rows.
\end{req}

To force the TRR mechanism to perform an additional refresh operation when
desired during the experiment, \trran{} should allow flexibly issuing an any
number of \cmdrefresh{} commands at arbitrary times.

\begin{req}
    Ability to flexibly issue \cmdrefresh{} commands.
\end{req}

The TRR mechanism under study may retain its state beyond a single experiment,
potentially causing the TRR mechanism to detect different rows as aggressors
depending on previous experiments. For example, in a \emph{counter-based TRR}
(\cref{utrr_subsec:vendorA_counter_based_TRR}), the TRR mechanism's internal counter
values updated due to a previous experiment might affect the outcome of future
experiments. To isolate an experiment from the past experiments, \trran{} should
reset TRR's internal state to a consistent state after each experiment.

\begin{req}
    Ability to reset TRR mechanism's internal state.
\end{req}

\subsection{TRR Analyzer Operation}
\label{utrr_subsec:trr_operation}

\begin{figure}[!h]
    \centering
    \includegraphics[width=.85\linewidth]{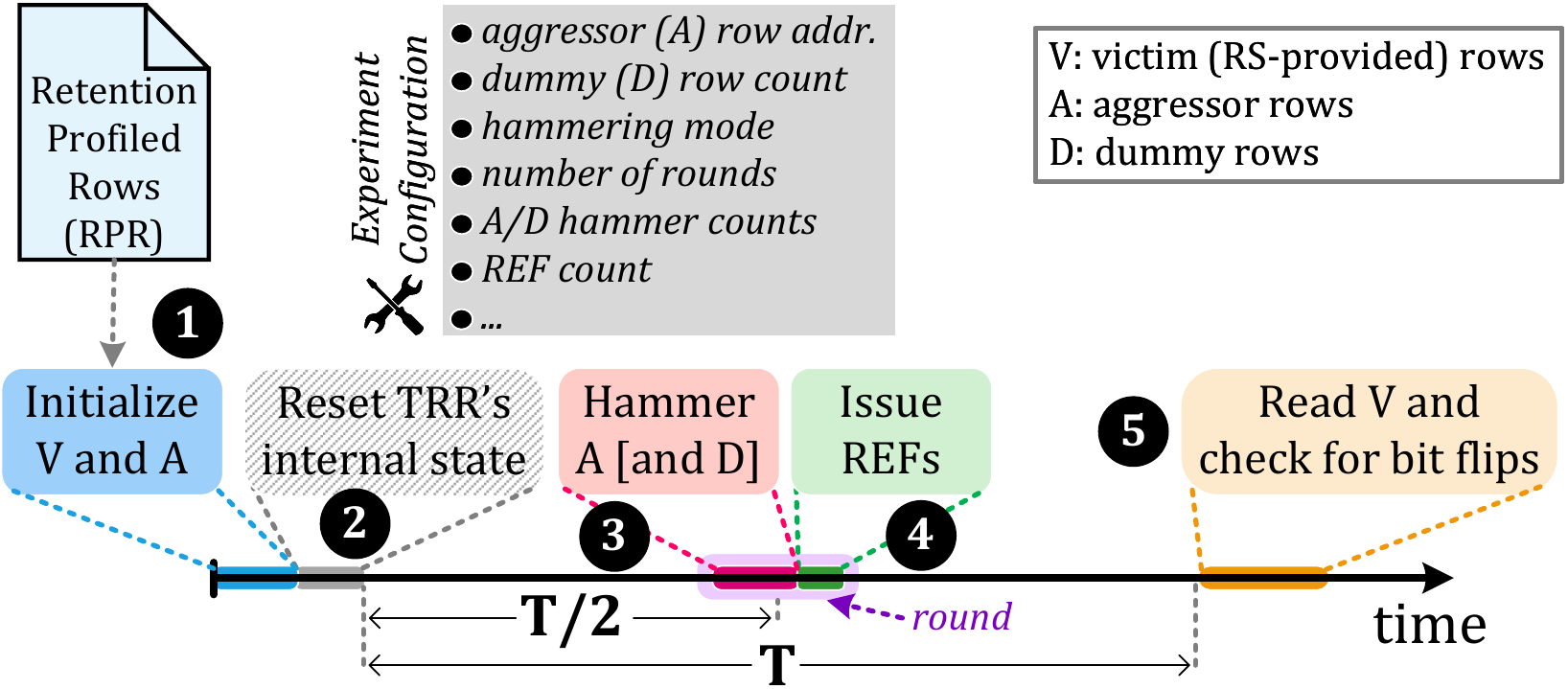}
    \caption{Detailed \trran{} experiment.}
    \label{utrr_fig:trr_analyzer_experiment}
\end{figure}

We explain how \trranlong{} (\trran{}) satisfies all of the requirements
described in \cref{utrr_subsec:trr_analyzer_reqs} to enable detailed experiments
that uncover the implementation details of in-DRAM RowHammer mitigation
mechanisms. Figure~\ref{utrr_fig:trr_analyzer_experiment} illustrates a typical
\trran{} experiment and provides a list of the experiment configuration
parameters.

\textbf{Initializing rows' data.} \trran{} initializes~\circled{1} the
\rtp{}-provided rows by writing into them the same data pattern that that \rtp{}
used for profiling these rows. \trran{} also initializes the aggressor rows,
whose addresses are specified in the experiment configuration, as the RowHammer
vulnerability greatly depends on the data values stored in an aggressor
row~\cite{kim2020revisiting,kim2014flipping}.

\textbf{Resetting internal TRR state.}
To reset the TRR mechanism's internal state~\circled{2} so as to satisfy
Requirement~4, \trran{} performs refresh for multiple refresh periods while
hammering a set of dummy rows. \trran{} issues \cmdrefresh{} commands at the
default refresh rate (i.e., one \cmdrefresh{} every \SI{7.8}{\micro\second}) for
several (e.g., $10$) \SI{64}{\milli\second} refresh periods. During these
refresh operations, the \rtp{}-provided rows do \emph{not} experience bit flips
as they get refreshed by the regular refresh operations. Between two refresh
commands, \trran{} hammers a large number (e.g., $128$) of dummy rows as many
times as the DRAM timing parameters (i.e., \tras{} and
\trp{}~\cite{micronddr4,ddr4operationhynix,kim2012case,
lee2013tiered,lee2015adaptive}) permit. \trran{} automatically selects dummy
rows from the same bank as the \rtp{}-provided and aggressor rows, since a TRR
mechanism may operate independently in each bank. 
To \hhmiv{select a dummy row that} does \emph{not} cause RowHammer bit flips on
the \rtp{}-provided rows \hhmiv{when the dummy row is hammered}, \trran{}
enforces a minimum row distance of $100$ between a selected dummy row and the
\rtp{}-provided rows. \hhmiv{Note that $100$ rows is an arbitrary distance that
we have set in \trran{} for selecting a dummy row, and it does \emph{not}
necessarily ensure that hammering the dummy row will not cause RowHammer bit
flip in the \rtp{}-provided rows. Regardless of how a dummy row is selected,
\trran{} performs a test, where it hammers the selected dummy row to actually
ensure that hammering the dummy row does \emph{not} cause a bit flip in the
\rtp{}-provided rows. If the selected dummy row fails to pass this test (i.e.,
it causes bit flips in a \rtp{}-provided row), \trran{} selects a different
dummy row.} 
We find that the operations we perform
in~\circled{2} make the TRR mechanism clear any internal state that is relevant
to rows activated in past experiments or during~\circled{1}. Resetting TRR's
internal state is an optional step since the user may sometimes want to examine
how TRR operates across multiple experiments.

\textbf{Hammering aggressor and dummy rows.} The experiment configuration
specifies the addresses and individual hammer counts of aggressor rows that
\trran{} accordingly hammers~\circled{3}. Optionally, the configuration
specifies a number of dummy rows that can be hammered to divert the TRR
mechanism away from the actual aggressor rows. \trran{} automatically selects
dummy rows based on the criteria for selecting dummy rows explained above. The
experiment configuration specifies a single hammer count for all dummy rows;
\trran{} hammers each dummy row by that count.

\textbf{Hammering modes.} The order with which DRAM rows get hammered can affect
both the magnitude of the disturbance each hammer causes and how the TRR
mechanism detects an aggressor row. \trran{} supports two \emph{hammering
modes}. \emph{Interleaved} mode hammers each aggressor row one at a time until
all aggressors accumulate their corresponding hammer count. \emph{Cascaded} mode
repeatedly hammers one aggressor row until it accumulates its corresponding
hammer count and then does the same for the other aggressor rows. In our
experiments, we observe that interleaved hammering generally causes more bit
flips (up to four orders of magnitude) compared to cascaded hammering for a
given hammer count. In contrast, we find that cascaded hammering is more
effective at evading the TRR mechanism than interleaved hammering. Therefore, it
is critical to support both hammering modes.

\textbf{Issuing \cmdrefresh{}s.} To cause the TRR mechanism to perform
TRR-induced refresh operations on the victim rows, \trran{} issues a number of
\cmdrefresh{} commands~\circled{3} according to the experiment configuration. As
shown in Figure~\ref{utrr_fig:trr_analyzer_experiment}, \trran{} issues the
\cmdrefresh{} commands \emph{after} hammering the aggressor/dummy rows and
waiting for half of the retention time $T$ of the \rtp{}-provided rows. This is
to 1) allow TRR to potentially detect the aggressor rows hammered in~\circled{3}
and 2) ensure that a victim row refreshed in~\circled{4} does \emph{not}
experience retention failures until its data is read in~\circled{5}.

\textbf{Hammering rounds.} To allow distributing hammers between multiple
\cmdrefresh{} commands, \trran{} performs hammers in multiple \emph{rounds}. A
round consists of 1) hammering the aggressor and dummy rows and 2) issuing
\cmdrefresh{} commands as the last operation of the round. The experiment
configuration specifies the aggressor/dummy row hammer counts and the number of
\cmdrefresh{}s to issue per round.

\subsection{Determining Physical Row Mapping}
\label{utrr_subsec:physical_row_mapping}

DRAM rows that have consecutive \emph{logical} row addresses (from the
perspective of the memory controller) may not be physically adjacent inside a
DRAM chip~\cite{kim2012case,cojocar2020are, jung2016reverse} due to two main
reasons. First, post-manufacturing row repair techniques
(e.g.,~\cite{kang2014co,mandelman2002challenges,nair2013archshield,son2015cidra,horiguchi2011nanoscale,cha2017defect,ddr4,jedec2021ddr5})
may repair a faulty row by remapping the logical row address that points to the
faulty row to a spare row at a different physical location inside a DRAM chip.
Second, a row address decoder in the DRAM chip may not necessarily map
consecutive row addresses to adjacent
wordlines~\cite{kim2012case,cojocar2020are, jung2016reverse}. The row address
decoder can maintain the logical row address order in physical row space but it
may scramble the logical-to-physical row mapping as well depending on the
circuit implementation.

Since a TRR mechanism should refresh rows that are physically adjacent to an
aggressor row, we need to ensure that \trran{} uses physically adjacent rows in
the experiments. For this purpose, we use two methods. First, before we run
\rtp{}, we reverse engineer the logical-to-physical row address mapping of a
DRAM chip by disabling refresh and performing double-sided
RowHammer.\footnote{Other works~\cite{jung2016reverse, lee2017design,
kim2018solar, pessl2016drama, orosa2021deeper} also propose various methods for
reverse engineering DRAM physical layout and their methods can also be used for
our purposes.} We analyze the rows at which RowHammer bit flips appear, so as to
determine the physical adjacency of rows and hence reconstruct the physical row
mapping. If the logical row address order is preserved in the physical space, we
simply observe RowHammer bit flips on logical row addresses $R-1$ and $R+1$ as a
result of hammering $R$. Otherwise, bit flips occur in other logical rows
depending on the logical-to-physical row address mapping of the DRAM chip.
Second, before an experiment, \trran{} verifies that the given aggressor row can
actually successfully hammer the \rtp{}-provided rows by hammering the aggressor
row a large number of times (i.e., $300K$) with refresh disabled. Doing so
ensures that the \rtp{}-provided or aggressor rows are \emph{not} remapped due
to post-manufacturing repair and are still physically adjacent to each other.

\section{Reverse-Engineering TRR}
\label{utrr_sec:insights}

We use \method{}, the components of which we describe in
\cref{utrr_sec:ret_profiling} and \cref{utrr_sec:analyzing_refs}, to gain insights about
TRR implementations by analyzing DDR4 modules from three major DRAM vendors. We
follow a systematic approach in our reverse engineering to gain these insights.
First, we discover which refresh commands perform TRR. Second, we observe how
many rows are concurrently refreshed by TRR. Third, we try to understand the
strategy that the TRR mechanism employs for detecting a DRAM row as an aggressor
row so as to refresh its neighboring victim rows. Unless stated otherwise, we
conduct all experiments at \SI{85}{\celsius} DRAM temperature.

Our approach leads to a number of new insights specific to each DRAM vendor.
Table~\ref{utrr_table:trr_summary} summarizes our key findings regarding the TRR
implementations of the \numTestedDIMMs{} modules we test. We describe the
experiments that lead us to these insights in more detail.

\afterpage{%
    \clearpage
    \begin{landscape}

\begin{table}[!ht]
    
    \centering
    \caption{Summary of our key observations and results on TRR implementations of \numTestedDIMMs{} DDR4 DRAM modules.
    }
    \label{utrr_table:trr_summary}

    \resizebox{\linewidth}{!}{

                \begin{tabular}{ l c c | c c c | c | c c c c c c c c}
                    \toprule

                    \mr{3.5}{\emph{Module}} & \mr{3.5}{\emph{\makecell{Date\\(yy-ww)}}}  & \mr{3.5}{\emph{\makecell{Chip\\Density\\(Gbit)}}} & \multicolumn{3}{c}{\emph{Organization}} & \mr{3.5}{$HC_{first}$$\dag$} & \multicolumn{8}{c}{\emph{Our Key TRR Observations and Results}}\\

                    \cmidrule(lr){4-6}
                    \cmidrule(lr){8-15}


                    \multicolumn{3}{c|}{} & \makecell{\emph{Ranks}} & \emph{Banks} & \emph{Pins} & & \emph{Version} & \emph{\makecell{Aggressor\\Detection}} & \emph{\makecell{Aggressor\\Capacity}} & \emph{\makecell{Per-Bank\\TRR}} &
                        \emph{\makecell{TRR-to-REF\\Ratio}} & \emph{\makecell{Neighbors\\Refreshed}} & \emph{\makecell{\% Vulnerable\\DRAM Rows$\dag$}} & \emph{\makecell{Max. Bit Flips\\per Row per Hammer$\dag$}} \\

                    \midrule

                    \stripe
                    A0     & 19-50 & 8 & 1 & 16 & 8 & $16K$     & $A_{TRR1}$ & Counter-based & 16 & \cmark & $1/9$ & 4 & 73.3\% & 1.16 \\
                    A1-5   & 19-36 & 8 & 1 & 8 & 16 & $13K$-$15K$ & $A_{TRR1}$ & Counter-based & 16 & \cmark & $1/9$ & 4 & 99.2\% - 99.4\% & 2.32 - 4.73 \\
                    \stripe
                    A6-7   & 19-45 & 8 & 1 & 8 & 16 & $13K$-$15K$ & $A_{TRR1}$ & Counter-based & 16 & \cmark & $1/9$ & 4 & 99.3\% - 99.4\% & 2.12 - 3.86 \\
                    A8-9   & 20-07 & 8 & 1 & 16 & 8 & $12K$-$14K$ & $A_{TRR1}$ & Counter-based & 16 & \cmark & $1/9$ & 4 & 74.6\% - 75.0\% & 1.96 - 2.96 \\
                    \stripe
                    A10-12 & 19-51 & 8 & 1 & 16 & 8 & $12K$-$13K$ & $A_{TRR1}$ & Counter-based & 16 & \cmark & $1/9$ & 4 & 74.6\% - 75.0\% & 1.48 - 2.86 \\
                    A13-14 & 20-31 & 8 & 1 & 8 & 16 & $11K$-$14K$ & $A_{TRR2}$ & Counter-based & 16 & \cmark & $1/9$ & 2 & 94.3\% - 98.6\% & 1.53 - 2.78 \\

                    
                    \cmidrule(lr){1-15}
                    \stripe
                    B0     & 18-22 & 4 & 1 & 16 & 8 & $44K$       & $B_{TRR1}$ & Sampling-based & 1 & \xmark & $1/4$ & 2 & 99.9\% & 2.13 \\
                    B1-4   & 20-17 & 4 & 1 & 16 & 8 & $159K$-$192K$ & $B_{TRR1}$ & Sampling-based & 1 & \xmark & $1/4$ & 2 & 23.3\% - 51.2\% & 0.06 - 0.11 \\
                    \stripe
                    B5-6   & 16-48 & 4 & 1 & 16 & 8 & $44K$-$50K$   & $B_{TRR1}$ & Sampling-based & 1 & \xmark & $1/4$ & 2 & 99.9\% & 1.85 - 2.03 \\
                    B7     & 19-06 & 8 & 2 & 16 & 8 & $20K$       & $B_{TRR1}$ & Sampling-based & 1 & \xmark & $1/4$ & 2 & 99.9\% & 31.14 \\
                    \stripe
                    B8     & 18-03 & 4 & 1 & 16 & 8 & $43K$       & $B_{TRR1}$ & Sampling-based & 1 & \xmark & $1/4$ & 2 & 99.9\% & 2.57 \\
                    B9-12  & 19-48 & 8 & 1 & 16 & 8 & $42K$-$65K$   & $B_{TRR2}$ & Sampling-based & 1 & \xmark & $1/9$ & 2 & 36.3\% - 38.9\% & 16.83 - 24.26 \\
                    \stripe
                    B13-14 & 20-08 & 4 & 1 & 16 & 8 & $11K$-$14K$   & $B_{TRR3}$ & Sampling-based & 1 & \cmark & $1/2$ & 4 & 99.9\% & 16.20 - 18.12 \\

                    \cmidrule(lr){1-15}
                    C0-3   & 16-48 & 4 & 1 & 16 & x8 & $137K$-$194K$ & $C_{TRR1}$ & Mix & Unknown & \cmark & $1/17$ & 2 & 1.0\% - 23.2\% & 0.05 - 0.15 \\
                    \stripe
                    C4-6   & 17-12 & 8 & 1 & 16 & x8 & $130K$-$150K$ & $C_{TRR1}$ & Mix & Unknown & \cmark & $1/17$ & 2 & 7.8\% - 12.0\% & 0.06 - 0.08 \\
                    C7-8   & 20-31 & 8 & 1 & 8 & x16 & $40K$-$44K$ & $C_{TRR1}$ & Mix & Unknown & \cmark & $1/17$ & 2 & 39.8\% - 41.8\% & 9.66 - 14.56 \\
                    \stripe
                    C9-11  & 20-31 & 8 & 1 & 8 & x16 & $42K$-$53K$ & $C_{TRR2}$ & Mix & Unknown & \cmark & $1/9$ & 2 & 99.7\% & 9.30 - 32.04 \\
                    C12-14 & 20-46 & 16 & 1 & 8 & x16 & $6K$-$7K$ & $C_{TRR3}$ & Mix & Unknown & \cmark & $1/8$ & 2 & 99.9\% & 4.91 - 12.64 \\
                    
                    \bottomrule
                \end{tabular}
                
                    
    } 
    \begin{footnotesize}
    {\raggedright 
        $\dag$We report the minimum and maximum
        $HC_{first}$, \%~\emph{Vulnerable DRAM Rows}, and    
        \emph{Max. Bit Flips per Row per Hammer} for table rows containing \emph{multiple}
        DRAM modules.

        $HC_{first}$: Minimum activation count per aggressor row in double-sided RowHammer to cause a bit flip. | \emph{Version}: Unique identifier for different TRR implementations we observe across DRAM vendors.
                    
        \emph{Aggressor Detection}: Main method used to detect an aggressor row. | \emph{Aggressor Capacity}: Maximum number of potential aggressor rows TRR can track.
        
        \emph{Per-Bank TRR}: Indicates whether a TRR mechanism operates
        independently in each bank or is shared across banks. | \emph{TRR-to-REF
        Ratio}: Fraction of TRR-capable \cmdrefresh{}s out of all
        \cmdrefresh{}s.
        
        \emph{Neighbors Refreshed}: Number of neighboring victim rows refreshed by a TRR-induced refresh. | \emph{\% Vulnerable DRAM Rows}: Fraction of DRAM rows vulnerable to our custom access patterns.
        
        \emph{Max. Bit Flips per Row per Hammer}: Maximum number of bit flips
    observed in any victim row per each hammer to an aggressor row between two
    \cmdrefresh{}s.\par }
    \end{footnotesize}

\end{table}

\end{landscape}
\clearpage
}

\subsection{Vendor A}
\label{utrr_subsec:vendorA_trr}

Using \method{}, we find that vendor A uses two slightly different TRR
implementations in their modules as we show in Table~\ref{utrr_table:trr_summary}. We
explain how to use \method{} to understand the operation of the $A_{TRR1}$
mechanism but our methodology is also applicable to $A_{TRR2}$. 

\subsubsection{TRR-Capable \cmdrefresh{} Commands}

We first run an experiment to determine whether all of the \cmdrefresh{}
commands issued to DRAM can perform TRR-induced refresh in addition to the
regular refresh operations or only certain \cmdrefresh{} commands are
responsible for TRR-induced refresh.

To uncover TRR-capable \cmdrefresh{} commands, we perform experiments that
follow the general template that we present in
Figure~\ref{utrr_fig:trr_analyzer_experiment}. We use \rtp{} to find $N$ row groups
that match the \texttt{R-R} layout. Among the profiled rows in each row group,
we designate an aggressor row, which we hammer $H$ times. We choose $H$ so that
it does not cause RowHammer bit flips on the profiled rows but at the same time
is large enough to potentially trigger the TRR mechanism to consider the
hammered row as a potential aggressor. To verify that $H$ hammers do not cause
RowHammer bit flips, we simply run a separate experiment where we 1) initialize
the profiled rows, 2) immediately after initialization, we hammer the profiled
rows $H$ times each, without performing any refresh, and 3) read back the
profiled rows and verify that there are no bit flips. 

We issue only one \cmdrefresh{} command to individually analyze each refresh
operation. Without a TRR mechanism, we expect to see retention failures in
\emph{all} of the profiled rows in \emph{almost every} iteration of the
experiment. This is because each row is not refreshed for a long enough period
of time (i.e., for $T$ as in
Figure~\ref{utrr_fig:general_trr_detection_approach}) such that retention
failures occur. Retention failures may not be observed during the \emph{very
few} iterations that a regular refresh operation refreshes one or more of the
profiled rows. Since regular refreshes happen periodically (i.e., a row is
refreshes by a regular refresh at a fixed \cmdrefresh{} command interval
(\cref{utrr_subsec:regular_refresh})), \method{} easily determines when a row is
refreshed by a regular refresh. When we observe a profiled row with no bit flips
when regular refresh is \emph{not} expected, we attribute that to a TRR-induced
refresh operation.

When we run the experiment in Figure~\ref{utrr_fig:trr_analyzer_experiment} with
$N \geq 16$ and $H=5K$, we find an interesting pattern where we see a row group
with no bit flips \emph{only} in every $9^{th}$ iteration of the experiment,
i.e., for every $9^{th}$ \cmdrefresh{} command issued consecutively. This shows
that, for this particular TRR design, not all \cmdrefresh{} commands perform a
TRR-induced refresh but only every $9^{th}$ of them have this capability.

\begin{obsA}
    Every $9^{th}$ \cmdrefresh{} command performs a TRR-induced refresh.
\end{obsA}

We also find with this experiment that, when TRR detects an aggressor row, it
simultaneously refreshes \emph{both} victim rows on each side of the detected
aggressor row with a single \cmdrefresh{}. To check if the refreshes are limited
to these two rows, we repeat the experiment using three profiled rows on each
side of the row that we hammer (i.e., we use row group layout \texttt{RRR-RRR}).
We observe that the TRR mechanism refreshes \emph{four} of the victim rows
closest to the detected aggressor row, i.e., two victims on each side of the
aggressor. This is likely done to protect against the probability that RowHammer
bit flips can occur in victim rows that are two rows apart from the aggressor
rows, as demonstrated by prior works~\cite{yaglikci2021blockhammer,
kim2020revisiting,kim2014flipping, yauglikcci2021security}.

\begin{obsA}
    TRR refreshes four rows that are physically closest to the detected
    aggressor row. When row address $A$ is detected as an aggressor, TRR
    refreshes rows $A\mp1$ and $A\mp2$.
\end{obsA}

We next perform a slightly different experiment to understand in what sequence
TRR detects the hammered rows as aggressor rows in consecutive TRR-capable
\cmdrefresh{} commands. We use two \texttt{R-R} row groups and hammer the
aggressor rows $H_0$ and $H_1$ times, where $H_0 << H_1$ (e.g., typical values
we use are $H_0=50$ and $H_1=5K$). This experiment uncovers that there are two
different types of TRR-induced refresh operations that alternate on every
$9^{th}$ \cmdrefresh{}. These two TRR-induced refresh operations differ in how
they detect an aggressor row to refresh its neighbors. The first type ($TREF_a$)
always detects the row that has accumulated the most hammers since the time TRR
previously detected the same row (e.g., the row that we hammer $H_1$ times in
this experiment). This suggests that this particular TRR mechanism might use a
counter table to keep track of activation counts of the accessed DRAM rows. The
second type ($TREF_b$) detects the same row periodically every $16^{th}$
instance of $TREF_b$. We anticipate that $TREF_b$ uses a pointer that refers to
an entry in the counter table that has $16$ entries. $TREF_b$ refreshes the
neighbor rows of the row address associated with the table entry that the
pointer refers to. After performing a TRR-induced refresh, $TREF_b$ increments
the pointer to refer to the next entry in the table. This is our hypothesis as
to why $TREF_b$ repeatedly detects the same row once every 16 instances of
$TREF_b$, and detects other activated rows that are in the counter table during
other instances of $TREF_b$. In \cref{utrr_subsec:vendorA_counter_based_TRR}, we
uncover the exact reason why we see the neighbors of the same row refreshed
every $16^{th}$ $TREF_b$ operation.

\begin{obsA}
    The TRR mechanism performs two types of TRR-induced refresh operations
    ($TREF_a$ and $TREF_b$) that both use a 16-entry counter table to
    detect aggressor rows.\\
    $TREF_a$: Detects the row that corresponds to the table entry with the
    highest counter value.\\
    $TREF_b$: Traverses the counter table by detecting a row that corresponds to one table entry at each of its instances.
    \label{utrr_obsA:trr_types}
\end{obsA}

\subsubsection{Counter-Based TRR}
\label{utrr_subsec:vendorA_counter_based_TRR}

Observation~\ref{utrr_obsA:trr_types} indicates that the TRR mechanism is capable of
determining which single row is activated (i.e., hammered) more than the others.
This suggests that the TRR mechanism implements a set of counters it associates
with the accessed rows and increments the corresponding counter upon a DRAM row
activation. We perform a set of experiments using the \method{} methodology to
understand more about this counter-based TRR implementation we hypothesize
about.

To find the maximum number of rows that the TRR mechanism can keep track of, we
perform an experiment where we use $N$ \texttt{R-R} row groups and hammer the
rows between the profiled rows $H$ times each in cascaded hammering mode
(\cref{utrr_subsec:trr_operation}). We use $H=1K$ and vary $N$ in the range $1
\leq N \leq 32$. When we repeatedly run the experiment, we observe that all
profiled rows are eventually refreshed by $TREF_a$ or $TREF_b$ when $1 \leq N
\leq 16$. However, with $N \geq 16$, we start observing profiled rows that are
never refreshed (except when they are refreshed due to regular refresh
operations as we discuss in \cref{utrr_subsec:regular_refresh}). Thus, we infer
that this particular TRR mechanism has a counter table capacity for 16 different
row addresses. We also observe that the $TREF_b$ operations detect rows by
repeatedly iterating over the counter table entries, such that a $TREF_b$
detects a row associated with one entry and the next $TREF_b$ detects the row
associated with the next entry. We find this to be the reason for why every
$16^{th}$ $TREF_b$ detects the same aggressor row when the same set of aggressor
rows are repeatedly hammered. Using row groups from different banks, we uncover
that the TRR mechanism keeps track of $16$ different rows in \emph{each} bank,
suggesting that each bank implements a separate counter table.

\begin{obsA}
    The TRR mechanism counts how many times DRAM rows are activated using a
    per-bank counter table, which can keep track of activation counts to $16$
    different row addresses.
\end{obsA}

We next try to find how TRR decides which row to evict from the counter table
when a new row is to be inserted. We perform an experiment where we check if the
TRR mechanism evicts the entry with the smallest counter value from the counter
table. In the experiment, we use 17 \texttt{R-R} row groups and hammer the row
between the two retention-profiled rows in each group. We hammer the aggressors
in the following order. First, we hammer one of the aggressors $H_0$ times.
Next, we hammer the remaining 16 aggressors $H_1$ times, where $H_0 < H_1$
(e.g., $H_0 = 50$ and $H_1 = 100$). Even after running the experiment for
thousands of iterations, we observe that the TRR mechanism never identifies the
row that is hammered $H_0$ times as an aggressor row. This indicates that the
counter table entry with the smallest counter value is evicted from the table
upon inserting a new row address (i.e., the last row that we hammer $H_1$ times
in this case) into the table. 

\begin{obsA}
    When inserting a new row into the counter table, TRR evicts
    the row with the smallest counter value.
\end{obsA}

We have already observed that a DRAM row activation increments the corresponding
counter in the table. However, we do not yet know whether or not a TRR-induced
refresh operation updates the corresponding counter value of the detected
aggressor row (e.g., resets the counter to 0). To check if a TRR-induced refresh
updates the corresponding counter, we conduct another experiment with two
\texttt{R-R} row groups, where we hammer the two aggressors $H_0$ and $H_1$
times. When we run the experiment multiple times with $H_0 < H_1$, we notice
that $TREF_a$ detects an aggressor row based on how many hammers the aggressor
row accumulated since the last time it is detected by $TREF_a$ or $TREF_b$. For
example, with $H_0=2K$ and $H_1=3K$, the corresponding counters accumulate $36K$
and $54K$ hammers, respectively, assuming the $18^{th}$ \cmdrefresh{} performs
$TREF_a$.\footnote{Since $TREF_a$ and $TREF_b$ happen every $9^{th}$
\cmdrefresh{} in an interleaved manner, $TREF_a$ happens every $18^{th}$
\cmdrefresh{}.} Thus, $TREF_a$ detects the aggressor row that is hammered $54K$
times and resets the corresponding counter. Until the subsequent $TREF_a$
operation, the two counters reach $72K$ and $54K$ hammers, respectively, and
$TREF_a$ detects the first aggressor row as its counter value is higher than
that of the second aggressor row since the latter counter was reset earlier.
This experiment shows that a TRR-induced refresh operation resets the counter
that corresponds to the aggressor row detected to refresh the neighbors of.

\begin{obsA}
    When TRR detects an aggressor row, TRR resets the counter
     corresponding to the detected row to zero.
\end{obsA}

We next question whether, once inserted, a row address remains indefinitely in
the counter table or TRR periodically clears out the counter table. To answer
this question, we run \emph{only once} an experiment with one \texttt{R-R} row
group and hammer the row between the profiled rows several times to insert the
aggressor rows into the counter table. Then, we repeat the experiment many times
\emph{without} hammering the aggressor row. After running these experiments, we
observe that the aggressor row is detected by a $TREF_a$ operation only once.
This is expected since we do not access the row except in the first experiment,
and once reset by the first $TREF_a$, the corresponding counter value remains
reset and never becomes a target for $TREF_a$ again. However, we observe that
every $16^{th}$ $TREF_b$ detects the same aggressor row and refreshes its
neighbors. We keep observing the same even after repeating the experiment $32K$
times (i.e., issuing $32K$ \cmdrefresh{} commands that equal the number of
refreshes issued within four \SI{64}{\milli\second} nominal refresh periods).
This shows that the aggressor row remains in the counter table and keeps getting
periodically detected by $TREF_b$ operations. The TRR mechanism does \emph{not}
seem to periodically clear the counter table, for example, based on time or the
number of issued \cmdrefresh{} commands.

\begin{obsA}
	After an entry corresponding to a row is inserted into the counter table,
	the entry remains in the table indefinitely until it is evicted due to
	insertion of a different row.
\end{obsA}

\subsubsection{Analyzing Regular Refresh}
\label{utrr_subsec:regular_refresh}

To refresh every DRAM cell at the default \SI{64}{\milli\second} period, the
memory controller issues a \cmdrefresh{} command once every
\SI{7.8}{\micro\second} according to the DDR4
specification~\cite{ddr4,micronddr4, ddr4operationhynix}. In total, the memory
controller issues $\approx8K$ (\SI{64}{\milli\second}$/$\SI{7.8}{\micro\second})
\cmdrefresh{} commands every \SI{64}{\milli\second}. Therefore, it is expected
that $\approx8K$ \cmdrefresh{} commands refresh each row in the DRAM chip once
to prevent a row from leaking charge for more than \SI{64}{\milli\second}. In
our experiments, we observe that the DRAM chips of vendor A internally refresh
more rows with each \cmdrefresh{} such that a row receives a regular refresh
once every $3758$ (instead of  $\approx8K$) \cmdrefresh{} commands. Thus, the
DRAM chip internally refreshes its rows with a period even smaller than
\SI{32}{\milli\second} instead of the specified \SI{64}{\milli\second}. We
suspect this could be an additional measure vendor A takes 1) to protect against
RowHammer~\cite{kim2014flipping} or 2) in response to the decreasing retention
time as DRAM technology node size becomes
smaller~\cite{liu2012raidr,chang2014improving,mutlu2013memory,mutlu2014research}.

\begin{obsA}
    Periodic DRAM refresh leads to internally refreshing the DRAM chip with a
    period smaller than half of the specified \SI{64}{\milli\second} refresh
    period.
\end{obsA}

In \cref{utrr_sec:case_studies}, we exploit the insights we present in this section
to craft a new DRAM access pattern that effectively circumvents the protection
of the TRR mechanism. This new custom access pattern induces a significantly
higher number of RowHammer bit flips than the state-of-the-art access patterns
presented in~\cite{frigo2020trrespass}.

\subsection{Vendor B}
\label{utrr_subsec:vendorB_trr}

Using \method{}, we find that vendor B uses three slightly different TRR
implementations in their modules, as Table~\ref{utrr_table:trr_summary} shows. We
explain how to use \method{} to understand the operation of the $B_{TRR1}$
mechanism. Our methodology is also applicable to $B_{TRR2}$ and $B_{TRR3}$.

\subsubsection{TRR-Capable \cmdrefresh{} Commands}

Similar to vendor A, we again start with uncovering which \cmdrefresh{} commands
can perform TRR-induced refresh. When we repeatedly run the experiment with one
or more row groups (i.e., $N \geq 16$) while hammering each aggressor row $5K$
times in each iteration of the experiment, we observe that not all \cmdrefresh{}
commands perform TRR-induced refresh. Instead, we find that, in $B_{TRR1}$ only
every $4^{th}$ \cmdrefresh{} command is used for TRR-induced refresh. Similar
experiments on modules that implement $B_{TRR2}$ and $B_{TRR3}$ uncover that
every $9^{th}$ and $2^{nd}$ \cmdrefresh{} command, respectively, is used for
TRR-induced refresh.

\begin{obsB}
    Every $4^{th}$, $9^{th}$, and $2^{nd}$ \cmdrefresh{} command performs a
    TRR-induced refresh in the three TRR mechanisms of vendor B.
\end{obsB}

From the same experiment, we also observe that a TRR-induced refresh operation
refreshes only the two neighboring rows that are immediately adjacent to the
hammered row as opposed to vendor A's TRR implementation, which refreshes the
four physically closest rows to the hammered row.

\begin{obsB}
    The TRR mechanism refreshes the two rows physically closest to the detected
    aggressor row. When row address $A$ is detected as an aggressor, TRR
    refreshes rows $A\mp1$.
\end{obsB}

\subsubsection{Sampling-Based TRR}

We perform experiments to show how the TRR mechanism detects the potential
aggressor rows. When we perform the experiments that we use for vendor A's
modules (described in \cref{utrr_subsec:vendorA_counter_based_TRR}), we do
\emph{not} observe obvious patterns in the rows detected by TRR so as to
indicate a counter-based TRR implementation. Instead, we observe that the
aggressor row that is last hammered before a \cmdrefresh{} command is more
likely to be detected. In particular, when we hammer two aggressor rows $H_0=5K$
and $H_1=3K$ times, respectively, we find that the $4^{th}$ \cmdrefresh{}
\emph{always} refreshes the neighbors of the second aggressor row, which we
hammer $2K$ times less than the first aggressor row. We perform experiments with
different $H_0$ and $H_1$ values and find that, when we hammer the second row at
least $2K$ times and issue a \cmdrefresh{}, the TRR mechanism consistently
refreshes the neighbors of the second row on every $4^{th}$ \cmdrefresh{}.
However, as we reduce $H_1$, the first aggressor row gets detected by TRR with
an increasing probability. 

With further analysis, we determine that $B_{TRR1}$ operates by sampling the row
addresses provided along with \cmdact{} commands. This sampling of \cmdact{}
commands happens with a certain probability such that $2K$ consecutive
activations to a particular row consistently causes the row to be detected for
TRR-induced refresh. We did not analyze this aspect of TRR further; we suspect
(based on some experiments) that the sampling does not happen truly randomly but
is likely based on pseudo-random sampling of an incoming \cmdact{}.

\begin{obsB}
    TRR probabilistically detects aggressor rows by sampling row addresses of
    \cmdact{} commands.
\end{obsB}

To determine how many rows the TRR mechanism can sample and refresh at the same
time, we repeat the previous experiment with the same $H_0=5K$ and $H_1=3K$
hammer counts but by issuing $M$ \cmdrefresh{} commands, instead of just one,
after performing the hammers. Even when we use a large $M$ (e.g., to 100) such
that it contains multiple TRR-capable \cmdrefresh{} commands (e.g., $25$), we
\emph{never} see the neighbors of the first aggressor row to be refreshed but
\emph{always} the neighbors of the \emph{second} aggressor row. This suggests
that, a newly-sampled row overwrites the previously-sampled one. Therefore, we
conclude that the TRR mechanism has a capacity to sample \emph{only} one row
address. Further, we find that this sampling capacity is shared across \emph{all
banks} in a DRAM chip that implements $B_{TRR1}$ and $B_{TRR2}$. When TRR
samples row $R_1$ from DRAM bank $B_1$, it overwrites a previously-sampled row
$R_2$ from bank $B_0$ even though $R_2$'s neighbors may not have been refreshed
yet.

\begin{obsB}
    The TRR mechanism has a sampling capacity of only one row that is shared
    across all banks in a DRAM chip (except for $B_{TRR3}$).
\end{obsB}

Our experiments also uncover that a previously-sampled row address is \emph{not}
cleared when the TRR mechanism performs a TRR-induced refresh on the neighbors
of this aggressor row. Instead, when a new TRR-enabled \cmdrefresh{} is issued,
TRR refreshes (again) the neighbors of the same row.

\begin{obsB}
    A TRR-induced refresh does not clear the sampled row, and therefore the same
    row keeps getting detected until TRR samples another aggressor row.
\end{obsB}

\subsection{Vendor C}
\label{utrr_subsec:vendorC_trr}

Using \method{}, we find that vendor C uses three slightly different TRR
implementations in their modules, as Table~\ref{utrr_table:trr_summary} shows.
For brevity, we omit the details of the experiments as they are largely similar
to the experiments for the modules of vendors A and B
(\cref{utrr_subsec:vendorA_trr} and \cref{utrr_subsec:vendorB_trr}). Instead, we
only describe our key observations.

We start with running experiments to find which \cmdrefresh{} commands are
TRR-capable. Different from the modules of vendors A and B, we find that vendor
C's modules implement a TRR mechanism that \emph{can} perform a TRR-induced
refresh during the execution of \emph{any} \cmdrefresh{} command. The TRR
mechanism performs a TRR-induced refresh once every $17$ consecutive
\cmdrefresh{} commands during a likely RowHammer attack. When likely not under
an attack (i.e., when a small number of row activations happen), TRR can defer a
TRR-induced refresh to any of the subsequent \cmdrefresh{} commands until it
detects an aggressor row. We do not observe TRR-induced refresh more frequently
than once in every $17$ \cmdrefresh{} commands. For $C_{TRR2}$ and $C_{TRR3}$,
we find that every $9^{th}$ and $8^{th}$ \cmdrefresh{} command, respectively,
performs a TRR-induced refresh.

\begin{obsC}
    Every $17^{th}$, $9^{th}$, and $8^{th}$ \cmdrefresh{} command normally
    performs a TRR-induced refresh in the three TRR mechanisms of vendor C. A
    TRR-induced refresh can be deferred to a later \cmdrefresh{} if no potential
    aggressor row is detected.
\end{obsC}

To uncover the logic behind how a potential aggressor row is detected, we run
experiments similar to those we use for the modules from vendors A and B. We
find that vendor C's TRR mechanism detects aggressor rows only from the set of
rows targeted by the first $2K$ \cmdact{} commands (per bank) following a
TRR-induced refresh operation. We also find that 1) TRR probabilistically
detects one of the rows activated within the first $2K$ \cmdact{} commands and
2) the rows that are activated earlier have a higher chance to be targeted by
the subsequent TRR-induced refresh operation. Discovering that TRR detects
aggressor rows based on only the first $2K$ \cmdact{} commands helped us to
craft an effective access pattern (\cref{utrr_subsec:access_patterns}); thus we
did not further analyze vendor C modules to uncover the maximum number of
potential aggressor rows TRR tracks.

\begin{obsC}
    TRR detects an aggressor row only among the first $2K$
    \cmdact{}\footnote{Except for the modules that implement $C_{TRR3}$
    (Table~\ref{utrr_table:trr_summary}). $C_{TRR3}$ detects an aggressor row only
    among the first $1K$ activations to each bank.} (to each bank) following a
    TRR-induced refresh. Rows activated earlier are more likely to be detected
    by TRR.
\end{obsC}

Our experiments uncover a unique DRAM row organization in modules $C0$-$8$. It
appears that two consecutively addressed rows (i.e., physical row addresses $R$
and $R+1$ where $R$ is an even row address) are isolated in pairs such that
hammering one row (e.g., $R$) can induce RowHammer bit flips \emph{only} in its
\emph{pair row} (e.g., $R+1$), and not in any other row in the bank. As
expected, we also observe that TRR issues refresh operations \emph{only} to the
pair row of each aggressor row that it identifies.

\begin{obsC}
	Given any two rows, $R$ and $R+1$, where $R$ is an even number, TRR
	refreshes \emph{only} one of the rows (e.g., $R$) upon detecting the other
	(e.g., $R+1$) as an aggressor row.
\end{obsC}
\section{Bypassing TRR Using \method{} Observations}
\label{utrr_sec:case_studies}

\method{} uncovers critical characteristics of the TRR mechanisms different DRAM
vendors implement in their chips. We leverage those characteristics to craft
custom DRAM access patterns that hammer an aggressor row such that TRR cannot
refresh the aggressor row's neighbors (i.e., victim rows) in a timely manner.
Our results show that these new custom access patterns greatly increase
RowHammer bit flips on the \numTestedDIMMs{} DDR4 modules we test.

\subsection{Custom RowHammer Access Patterns}
\label{utrr_subsec:access_patterns}

\textbf{Vendor A.} Using \method{}, we find that vendor A's modules implement a
counter-based TRR ($A_{TRRx}$\footnote{We refer to all versions of TRR
mechanisms that vendor A's modules implement (i.e., $A_{TRR1}$ and $A_{TRR2}$)
as $A_{TRRx}$. We use a similar terminology for other vendors.}), the details of
which are in \cref{utrr_subsec:vendorA_trr}. Since $A_{TRRx}$ evicts the entry
with the lowest counter value when inserting a new entry to the table, a custom
RowHammer access pattern that takes advantage of our \method{} analysis should
first hammer two aggressor rows in a double-sided manner and then evict the two
aggressor rows from the table by hammering other rows (i.e., \emph{dummy rows})
within the same bank during the remaining time until the memory controller
issues a \cmdrefresh{} command.\footnote{The memory controller issues a
\cmdrefresh{} once every \SI{7.8}{\micro\second} when using the default
\SI{64}{\milli\second} refresh period. This allows at most 149 hammers to a
single DRAM bank assuming typical activation (\SI{35}{\nano\second}), precharge
(\SI{15}{\nano\second}), and refresh (\SI{350}{\nano\second})
latencies~\cite{micronddr4,liu2013experimental,chang2014improving}.}

We show how we can hammer two aggressor rows ($A_0$ and $A_1$) in a double-sided
manner without allowing $A_{TRRx}$ to refresh their victim rows. First, the
attacker should synchronize the memory accesses with the periodic \cmdrefresh{}
commands\footnote{A recent work~\cite{deridder2021smash} shows how to detect
when a memory controller issues a periodic \cmdrefresh{} from an unprivileged
process and from a web browser using JavaScript.} in order to hammer $A_0$ and
$A_1$ \emph{right after} the memory controller issues a \cmdrefresh{}. After
hammering the two aggressor rows, the attacker should then use the remaining
time until the next \cmdrefresh{} to hammer dummy rows in order to steer
$A_{TRRx}$ to identify one of the dummy rows (and \emph{not} rows $A_0$ and
$A_1$) as potential aggressors and refresh the dummy rows' neighboring victim
rows. The particular access pattern that leads to the largest number of bit
flips is hammering $A_0$ and $A_1$ 24 times each, followed by hammering 16 dummy
rows 6 times each. We discover the access pattern that maximizes the bit flip
count by sweeping the number of hammers to $A_0$ and $A_1$ and adjusting the
number of hammers to the 16 dummy rows based on the time that remains until the
next \cmdrefresh{} after hammering the aggressors.

\textbf{Vendor B.} $B_{TRRx}$ operates by probabilistically sampling a single
row address from all \cmdact{} (\cref{utrr_subsec:vendorB_trr}) commands issued
(across all banks for $B_{TRR1}$ and $B_{TRR2}$) to DRAM. Rows neighboring the
sampled row are refreshed during a TRR-induced refresh operation that happens
once in every $4$, $9$, and $2$ \cmdrefresh{} commands for $B_{TRR1}$,
$B_{TRR2}$, and $B_{TRR3}$, respectively. To maximize the probability of
$B_{TRRx}$ detecting a dummy row instead of the aggressor row, our custom access
pattern maximizes the number of hammers to dummy rows after hammering the
aggressor rows and before every TRR-induced refresh operation. Our custom access
pattern first hammers rows $A_0$ and $A_1$ immediately following a TRR-induced
refresh. Then, it simultaneously hammers a single dummy row in each of four
banks\footnote{We do not hammer a dummy row in more than four different banks
due to the Four-Activation-Window (\tfaw{}~\cite{micronddr4,
ddr4,ddr4operationhynix}) DRAM timing constraint.} to perform a large number of
dummy row activations within the limited time until the next TRR-induced
refresh.\footnote{For $B_{TRR3}$, which separately samples \cmdact{} commands to
each bank, we hammer a dummy row from the aggressor row's bank.} We find that
$220$ hammers per aggressor row (leaving 156 hammers for each dummy row in the
four banks) within a window of four consecutive \cmdrefresh{} commands causes
RowHammer bit flips even in the least RowHammer-vulnerable module of the
\numTestedDIMMsFromB{} vendor B modules that we use in our experiments.

\textbf{Vendor C.} $C_{TRR1}$, $C_{TRR2}$, and $C_{TRR3}$ have the ability to
perform a TRR-induced refresh once in every $17$, $9$, and $8$ \cmdrefresh{}
commands, respectively, and they can defer the TRR-induced refresh to a later
\cmdrefresh{} until a potential aggressor row is detected
(\cref{utrr_subsec:vendorC_trr}). $C_{TRRx}$ does not keep track of more than
$2K$ \cmdact{} commands that follow a TRR-induced refresh operation and rows
activated earlier in the set of $2K$ \cmdact{} commands are more likely to be
detected. Therefore, we craft a custom RowHammer access pattern that follows a
TRR-induced refresh operation with a large number (e.g., $2K$) of dummy row
activations and then hammers the aggressor rows $A_0$ and $A_1$ until the next
TRR-induced refresh operation. To properly execute this access pattern, it is
critical to synchronize the dummy and aggressor row hammers with TRR-enabled
\cmdrefresh{} commands.

\subsection{Effect on RowHammer Bit Flip Count}
\label{utrr_subsec:overall_bit_flips}

We implement and evaluate the different custom DRAM access patterns that are
used to circumvent $A_{TRRx}$, $B_{TRRx}$, and $C_{TRRx}$ on our FPGA-based
SoftMC platform~\cite{hassan2017softmc} (\cref{utrr_subsec:softmc}). The SoftMC
program executes each custom access pattern for a fixed interval of time
(determined by each chip's TRR-induced refresh frequency), while also issuing
\cmdrefresh{} commands once every \SI{7.8}{\micro\second} to comply with the
vendor-specified default refresh rate.

\begin{figure}[!ht]
    \centering
    \hspace{-0.4em}
    \begin{subfigure}[t]{.3\linewidth} 
        \centering
        \includegraphics[height=1.93in]{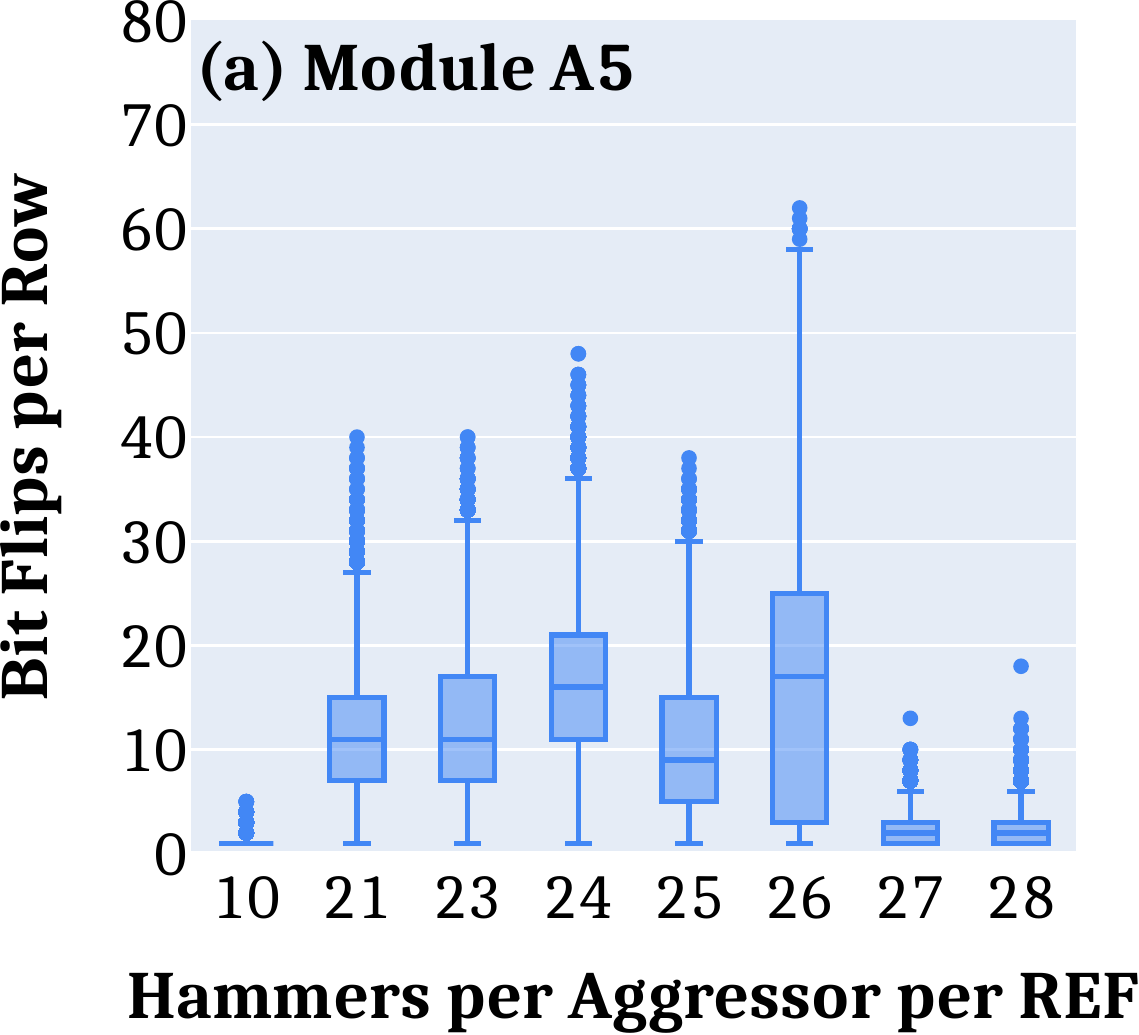}
    \end{subfigure}
    \quad
    \hspace{0.4em}
    \begin{subfigure}[t]{.28\linewidth}
        \centering
        \includegraphics[height=1.93in]{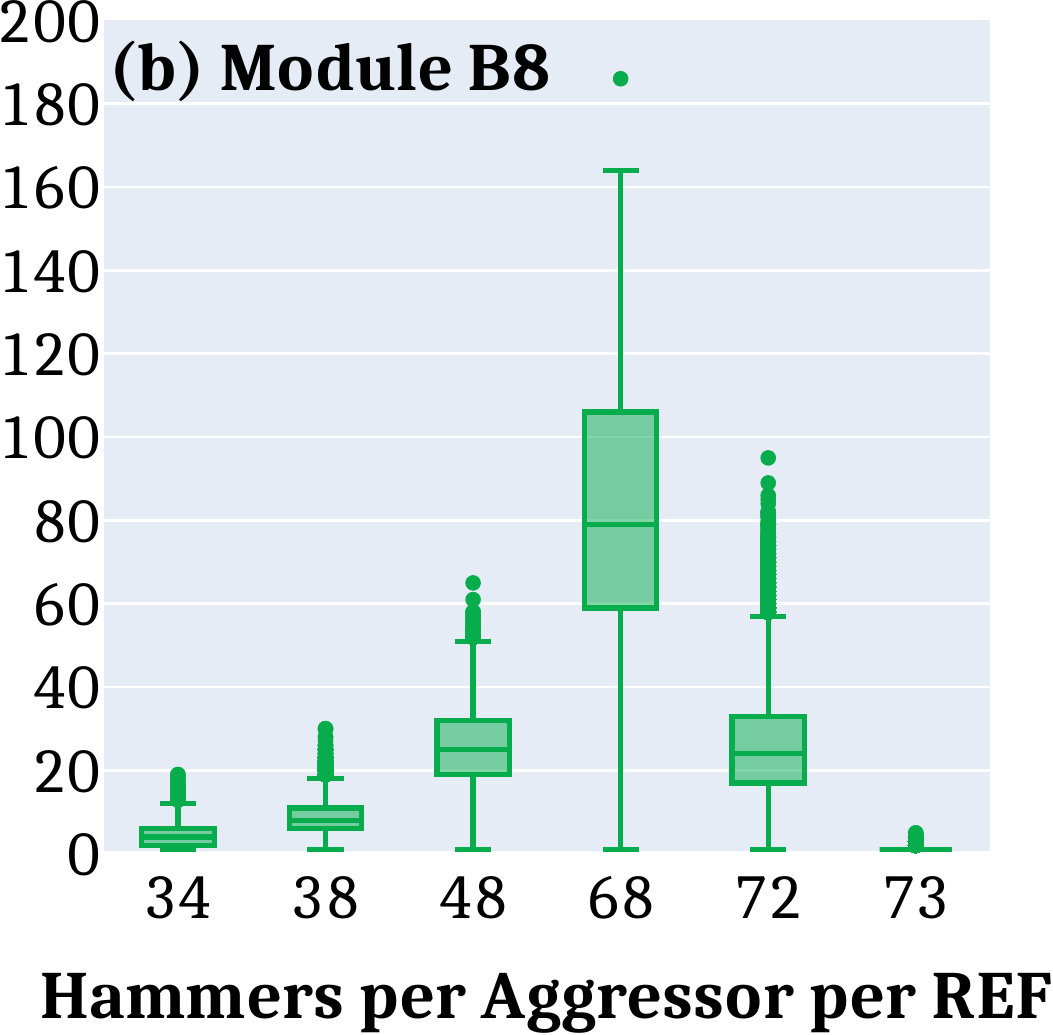}
    \end{subfigure}
    \quad
    \begin{subfigure}[t]{.3\linewidth}
        \centering
        \includegraphics[height=1.91in]{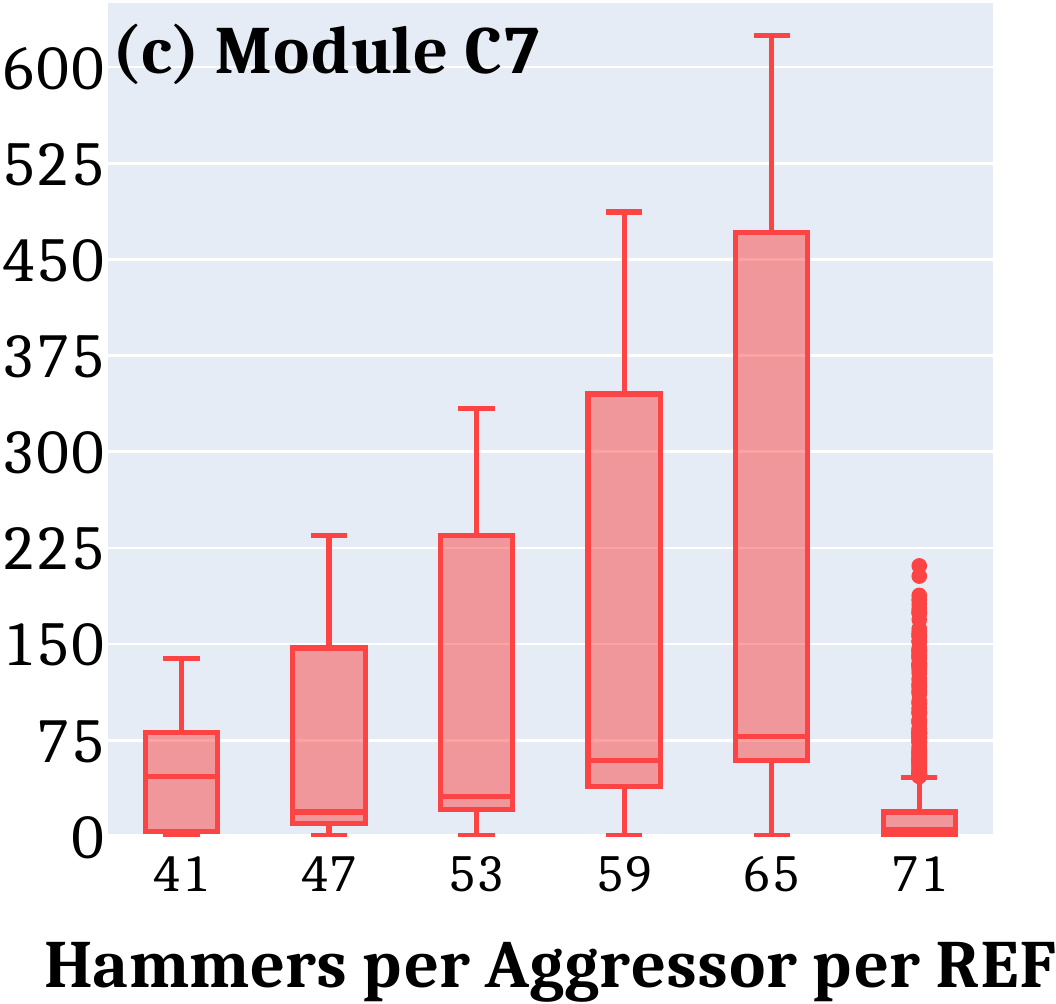}
    \end{subfigure}
    \caption{Distribution of bit flips per DRAM row for different aggressor
    hammer counts in three representative modules.}
    \label{utrr_fig:bitflips_per_bank}
\end{figure}

Figure~\ref{utrr_fig:bitflips_per_bank} shows the distribution of number of bit
flips per row as box-and-whisker plots\footnote{The lower and upper bounds of
the box represent the first quartile (i.e., the median of the first half of a
sorted dataset) and the second quartile (i.e., the median of the second half of
a sorted dataset), respectively. The median line is within the box. The size of
the box represents the inter-quartile range (IQR). The whiskers are placed at
$1.5*IQR$ on both sides of the box. The outliers are represented with dots.} in
modules $A5$, $B8$, and $C7$\footnote{We analyze $A5$, $B8$, and $C7$ as they
are the modules that experience the most RowHammer bit flips and implement
$A_{TRR1}$, $B_{TRR1}$, and $C_{TRR1}$, respectively.} when sweeping the number
of hammers issued to aggressor rows in each custom access pattern. The x-axis is
normalized so that it shows the average number of hammers performed between two
\cmdrefresh{}s to a single aggressor row.\footnote{The x-axis shows the number
of hammers per aggressor row per \cmdrefresh{} to enable easy comparison of the
effectiveness of different RowHammer patterns across different modules. Our
actual experiments perform the aggressor and dummy row hammers as required by
each custom RowHammer access pattern described in
\cref{utrr_subsec:access_patterns}.} We show different hammers per aggressor per
\cmdrefresh{} for each module as the number of hammers we can fit between two
\cmdrefresh{}s depend on the custom RowHammer patterns we craft
(\cref{utrr_subsec:access_patterns}). Each access pattern uses a fixed number of
dummy rows as described in \cref{utrr_subsec:access_patterns}. We perform the
maximum number of hammers that fit between two \cmdrefresh{}s. Therefore, a
lower number of aggressor row hammers translates to a higher number of dummy row
hammers.

\textbf{Vendor A.} We observe the highest bit flip count (i.e., up to 62 bit
flips in a row) when using $26$ hammers per aggressor row, in which case each of
the $16$ dummy rows are hammered $6$ times. The number of bit flips decreases as
we hammer the aggressors more than $26$ times. This is because the aggressors
become less likely to be evicted from the counter table as more activations to
them increment the corresponding counters to higher values. In contrast, the
aggressor rows become more likely to be evicted from the counter table when they
are hammered fewer than $26$ times each. However, we still observe a smaller bit
flip count with fewer than $26$ hammers per aggressor because fewer hammers are
insufficient for many victim rows to experience RowHammer bit flips.

\textbf{Vendor B.} The number of bit flips gradually increases with the number
of hammers per aggressor row. This increases to a point where too many aggressor
row activations leave an insufficient time to perform enough dummy row
activations to ensure that a dummy row is sampled to replace an aggressor row
for the subsequent TRR-induced refresh. According to our experiments, at least
$12$ total dummy row activations simultaneously performed in four banks (leaving
enough time to perform $73$ hammers per aggressor) are needed to induce
RowHammer bit flips. We observe the maximum number of bit flips with $68$
hammers per aggressor row.

\textbf{Vendor C.} We observe bit flips appear when a dummy row is initially
hammered a large number of times to make the subsequent aggressor row
activations less likely to be tracked by $C_{TRRx}$. The access pattern causes
bit flips when performing at least $252$ dummy hammers (right after a
TRR-enabled \cmdrefresh{}) prior to continuously hammering the aggressor rows
until the next TRR-enabled \cmdrefresh{}. This leaves time to perform $71$
hammers per aggressor row per \cmdrefresh{} on average. We observe the maximum
number of bit flips with $65$ hammers per aggressor row.

\subsection{Effect on Individual Rows}
\label{utrr_subsec:row_vulnerability}

To mount a successful system-level RowHammer attack, it is critical to force the
operating system to place sensitive data in vulnerable rows. To make this task
easier, it is important to induce RowHammer bit flips in as many rows as
possible. Ideally, all rows should be vulnerable to RowHammer from an attacker's
perspective.

\begin{figure*}[tp]
    \centering
    \includegraphics[width=\linewidth]{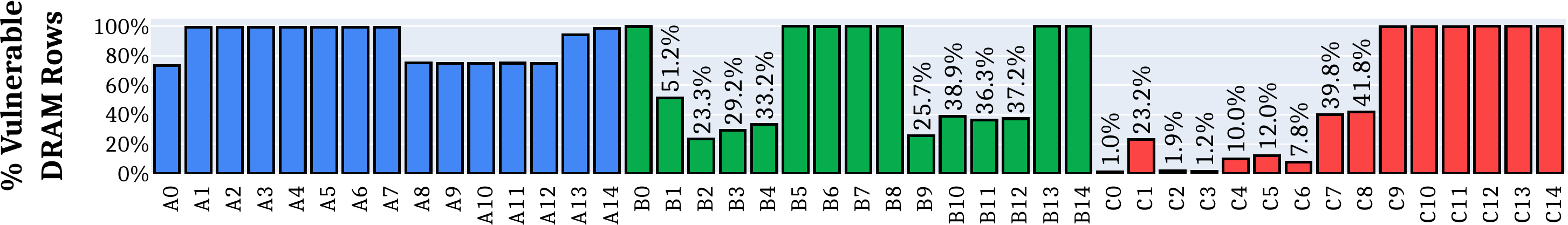}
    \caption{Percentage of rows that experience at least one RowHammer bit flip
    using our custom RowHammer access patterns.}
    \label{utrr_fig:vulnerable_rows}
\end{figure*}

Figure~\ref{utrr_fig:vulnerable_rows} shows the percentage of vulnerable DRAM
rows, i.e., rows that experience at least one RowHammer bit flip
with our custom RowHammer access patterns
(\cref{utrr_subsec:access_patterns}), as a fraction of all rows in a
bank of the tested \numTestedDIMMs{} modules. We report data for a single
bank\footnote{We test a single bank to reduce the experiment time. To
ensure that the results are similar across different banks, we tested multiple
banks from several modules.} from each module. For each DRAM module, we use a
different number of hammers per aggressor that results in the highest percentage
of vulnerable rows in the corresponding module (see
\cref{utrr_subsec:overall_bit_flips}).\footnote{When using the
conventional single- and double-sided RowHammer, we do \emph{not} observe RowHammer bit
flips in any of the \numTestedDIMMs{} DDR4 modules (as expected from our
understanding of the TRR implementations and from~\cite{frigo2020trrespass}).}

In many (i.e., $8$, $7$, and $6$, respectively) modules from vendor A, B, and C
we see bit flips in more than \vulnRowsMaxPct\% of the rows. This shows that the
custom access patterns we use are effective at circumventing the $A_{TRRx}$,
$B_{TRRx}$, and $C_{TRRx}$ implementations. The other modules from vendors A and
B have a smaller yet still a very significant (i.e., $>$23\% in all cases)
fraction of vulnerable rows. We believe that modules $A0$, $A8$-$12$ are
slightly more resistant to our access pattern than the other modules of vendor A
due to having more banks (i.e., $16$ vs. $8$) and smaller banks (i.e., $32K$ vs.
$64K$ rows per bank). $B1$-$4$ have stronger rows that can endure more hammers
than the other modules of vendor B (as shown in $HC_{first}$ column of
\Cref{utrr_table:trr_summary}), and therefore they have a lower fraction of
vulnerable rows. $B9$-$12$ implement a different TRR version ($B_{TRR2}$), for
which our custom access patterns are not as effective.
 
Modules from vendor C that implement $C_{TRR1}$ (i.e., $C0$-$8$) are less
vulnerable to our access patterns than the other modules of the same vendor. We
believe this is due to two main reasons. First, these modules use a unique row
organization that pairs every two consecutive DRAM rows, as we explain in
\cref{utrr_subsec:vendorC_trr}. We only observe bit flips when hammering two
aggressor rows that have odd-numbered addresses but not when the two aggressor
have even-numbered addresses. This essentially halves the number of victim rows
where our access patterns can cause bit flips. Second, $C0$-$6$ have stronger
rows that are less vulnerable to RowHammer than the other vendor C modules (as \Cref{utrr_table:trr_summary} shows), and therefore $C0$-$6$ have even lower
fraction of vulnerable rows than $C7$-$8$.

Overall, even though our custom RowHammer access patterns cause bit flips in
\numTestedDIMMs{} DRAM modules, we could not explore the entire space of both
TRR implementations and custom RowHammer patterns. Therefore, we believe future
work can lead to even better RowHammer patterns via more exhaustive analysis and
testing.

\subsection{Bypassing System-Level ECC Using \method{}}
\label{utrr_subsec:ecc_vs_rowhammer}

Although we clearly show that the custom access patterns we craft induce
RowHammer bit flips in a very large fraction of DRAM rows
(\cref{utrr_subsec:row_vulnerability}), a system that uses Error Correction
Codes (ECC)~\cite{dell1997white, meza2015revisiting, schroeder2009dram,
gong2017dram, kang2014co, micron2017whitepaper, nair2016xed,
patel2019understanding, costello2004ecc, hamming1950error, sridharan2012study,
sridharan2015memory} can potentially protect against RowHammer bit flips if
those bit flips are distributed such that no ECC codeword contains more bit
flips than ECC can correct. 

\begin{figure}[!ht]
    \centering
    \includegraphics[width=.9\linewidth]{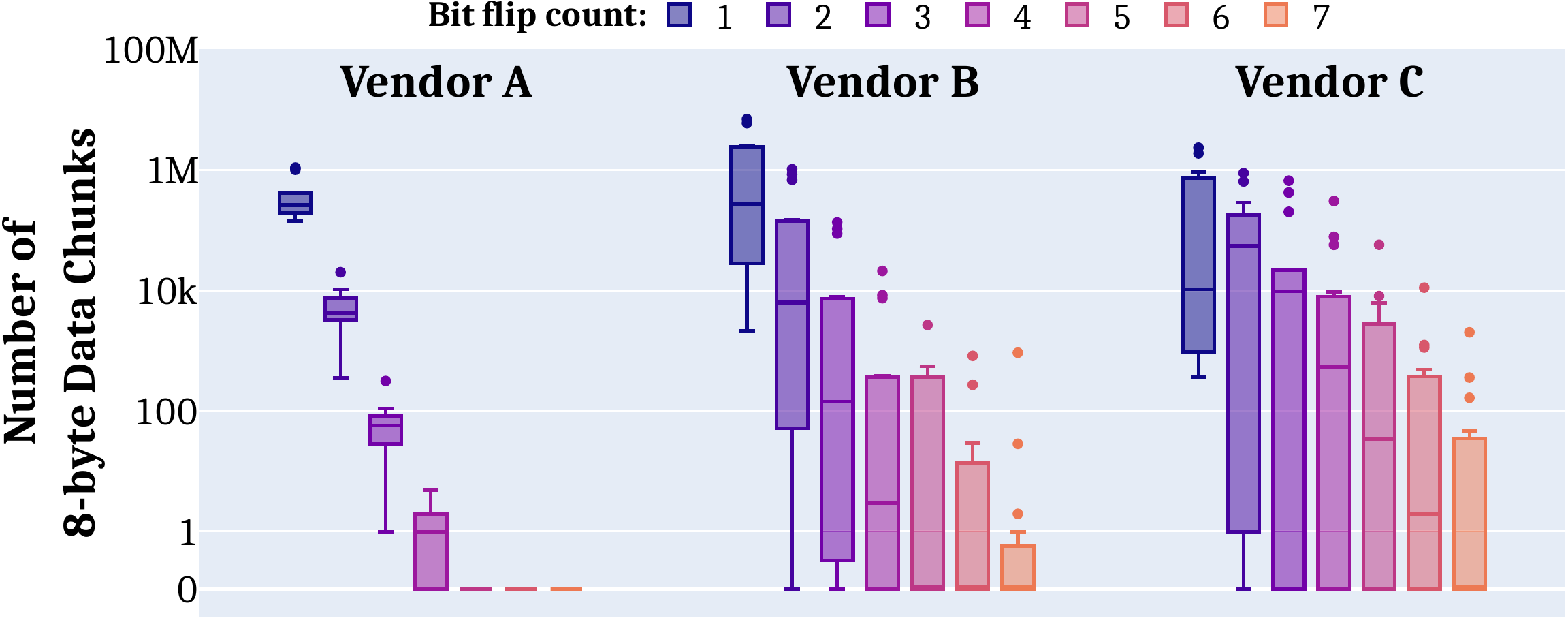}
    \caption{Distribution of 8-byte data chunks (log scale) with different
    RowHammer bit flip counts in a single DRAM bank from each of the
    \numTestedDIMMs{} tested DDR4 modules.}
    \label{utrr_fig:data_chunks_with_bitflips}
\end{figure}

\Cref{utrr_fig:data_chunks_with_bitflips} shows the distribution of
RowHammer bit flips that our custom access patterns induce across
8-byte data chunks as box-and-whisker plots\footnotemark[14] for \emph{all}
\numTestedDIMMs{} DRAM modules we test across three vendors. We use
8-byte data chunks as DRAM ECC typically uses 8-byte or larger
datawords~\cite{micron2017whitepaper, oh2014a, kwak2017a, kwon2017an, im2016im,
son2015cidra, cha2017defect, jeong2020pair}.

The majority of the 8-byte data chunks are those that have only a single
RowHammer bit flip (i.e., up to $6.9$ million 8-byte data chunks with a single
bit flip in one bank of module $B13$), which can be corrected using typical
SECDED ECC~\cite{micron2017whitepaper, oh2014a, kwak2017a, kwon2017an, im2016im,
son2015cidra, cha2017defect, jeong2020pair}. However, our RowHammer access
patterns can cause \hhmiii{$3$ or more} (up to $7$) bit flips in many single
datawords, which the SECDED ECC \emph{cannot} correct or detect, in all three
vendor's modules.

Chipkill~\cite{dell1997white, nair2016xed, amd2009bkdg} is a symbol-based code
conventionally designed to correct errors in one symbol (i.e., one DRAM chip
failure) and detect errors in two symbols (i.e., two DRAM chip failures).
Because our access patterns cause more than two bit flips in \emph{arbitrary}
locations (i.e., different DRAM chips), and thus in arbitrary symbols within an
8-byte data chunk, Chipkill does \emph{not} provide guaranteed protection.

\hhmiii{Still, it is challenging for an attacker to compromise a system protected with ECC due to two reasons. First, identifying a codeword that experiences uncorrectable bit flips after a RowHammer attack is difficult without first causing detectable bit flips, using which the system may recognize the RowHammer attack. Second, to cause a bit flip in a specific bit location in a dataword, the attacker must know the parity check matrix to determine which bit locations to flip on the codeword stored in DRAM.} \hhmiv{However, we believe that determined and powerful attackers can find ways to compromise the system with some effort, and certainly it can be relatively easy to cause data corruption even if the system is not easily compromised.}

\hhmiii{To provide guaranteed protection against our RowHammer patterns using
ECC,} Reed-Solomon codes~\cite{reed1960polynomial} can be designed to
\hhmiii{offer} stronger correction/detection capability at the cost of
additional parity check symbols~\cite{costello2004ecc, huffman2003fundamentals}.
To detect (and correct half of) the maximum number of bit flips (i.e., $7$) that
our access patterns can cause in an 8-byte data chunk, a Reed-Solomon code would
incur a large overhead by requiring at least $7$ parity-check
symbols~\cite{reed1960polynomial}.

We conclude that 1) conventional DRAM ECC \hhmiii{does \emph{not} provide
guaranteed protection} against our new custom RowHammer patterns and 2) an ECC
scheme that \hhmiii{guarantees protection} against our custom patterns requires
a large number of parity check symbols, i.e., large overheads.

\section{Related Work}
\label{crow_sec:related_work}

To our knowledge, \hhm{we are} first to propose a flexible and low-cost DRAM
substrate that enables \emph{multiple} mechanisms for improving DRAM
performance, energy efficiency, and reliability. We briefly discuss closely
related prior work that propose in-DRAM caching, latency reduction, refresh
reduction, and RowHammer protection mechanisms.

\textbf{In-DRAM Caching Mechanisms.} 
In \cref{crow_subsec:comparison_to_salp_tldram}, we qualitatively and
quantitatively compare \mech to the two most-closely related prior works,
TL-DRAM~\cite{lee2013tiered} and SALP~\cite{kim2012case}, which also propose
in-DRAM caching. We show that \mech-cache is lower-cost to implement and
consumes less DRAM energy compared to the two mechanisms. Since TL-DRAM and SALP
exploit \emph{different} observations to enable low-latency regions in DRAM,
\mech-cache can be implemented in combination with both TL-DRAM and SALP to
further improve system performance.

Chang et al.~\cite{chang2016low} propose LISA to enable efficient data movement
between two subarrays. They also propose \emph{LISA-VILLA}, which changes the
bank architecture to include small (but fast) subarrays~\cite{lu2015improving}
to enable dynamic in-DRAM caching. \hhm{FIGARO~\cite{wang2020figaro} proposes a
DRAM architecture that enables fine-grained data relocation within a DRAM bank.}
\hhm{LISA and FIGARO 1)~require} more changes to the DRAM chip compared to
\mech-cache and 2)~\hhm{are} \hhmiv{complementary} to \mech-cache as a \copyrow
can be implemented in the subarrays of \hhm{both designs}.

Hidaka et al.~\cite{hidaka1990cache} propose to add SRAM memory cells inside a
DRAM chip to utilize as a cache. Implementing SRAM-based memory in DRAM incurs
very high area overhead (e.g., 38.8\% of the DRAM chip area for
\SI{64}{\kibi\byte} SRAM cache as shown in~\cite{kim2012case, lee2013tiered}).
Prior works~\cite{rixner2004memory, herrero2012thread, gulur2012multiple} that
propose implementing multiple row buffers (that can be used as cache) also
suffer from high area overhead. \mech-cache can be used in combination with all
these prior mechanisms to reduce the latency of fetching data from a DRAM row to
the in-DRAM cache.

Multiple Clone Row (MCR-DRAM)~\cite{choi2015multiple} is based on an idea that
is similar to simultaneously activating multiple duplicate rows. The key idea is
to dynamically configure a DRAM bank into different access-latency regions,
where a region stores data in a single row or duplicates data into multiple
rows. MCR-DRAM does \emph{not} propose a hardware-managed in-DRAM cache as it
requires support from the operating system (or the application) to manage 1) the
size of each region and 2) data movement between different regions. However,
both operations are difficult to perform for the operating system due to memory
fragmentation, which complicates dynamic resizing of a region and determining
which data would benefit from a low-latency region.
\hhm{CLR-DRAM~\cite{luo2020clr} is a new DRAM architecture that enables
dynamically coupling two adjacent DRAM cells to operate as a single cell for low
latency access. Similar to MCR-DRAM, CLR-DRAM requires software support for
determining which data to place in low-latency regions.} \hhm{Compared to
MCR-DRAM and CLR-DRAM}, \mech is more practical to implement, as it is
hardware managed and thus completely transparent to the software.

\textbf{Mitigating Refresh Overhead.}
Many prior works tackle the refresh overhead problem in DRAM.
RAIDR~\cite{liu2012raidr} proposes to bin DRAM rows based on their retention
times and perform refresh operations at a different rate on each bin. Riho et
al.~\cite{riho2014partial} use a technique based on simultaneous multiple-row
activation. However, their mechanism is specifically designed for optimizing
only refresh operations, and thus it is not as flexible as the \mech{}
substrate, which provides \copyrow{s} that enable multiple orthogonal
mechanisms. Other prior works~\cite{nair2014refresh, qureshi2015avatar,
baek2014refresh, bhati2013coordinated, cui2014dtail, emma2008rethinking,
ghosh2007smart, isen2009eskimo, jung2015omitting, kim2000dynamic,
luo2014characterizing, kim2003block, liu2012flikker, mukundan2013understanding,
nair2013case, patel2005energy, stuecheli2010elastic, chang2014improving,
khan2014efficacy, khan2016parbor, khan2017detecting, venkatesan2006retention,
patel2017reaper,khan2016case} propose various techniques to optimize refresh operations. These
works are specifically designed for \emph{only} mitigating the refresh overhead.
In contrast, \mech provides a versatile substrate that can simultaneously reduce
the refresh overhead, reduce latency, and improve reliability.

\textbf{Reducing DRAM Latency.}
ChargeCache~\cite{hassan2016chargecache} reduces the average DRAM latency based
on the observation that recently-precharged rows, which are fully restored, can
be accessed faster compared to rows that have leaked some charge and are about
to be refreshed soon. Note that ChargeCache enables low-latency access when DRAM
rows are repeatedly activated in very short intervals, e.g.,
\SI{1}{\milli\second}. In contrast, a \copyrow in \mech-cache enables
low-latency access to a regular row for an indefinite amount of time (until the
row is evicted from \mech table). Thus, \mech-cache captures higher amount of
in-DRAM locality. \mech-cache is also orthogonal to ChargeCache and the two
techniques can be implemented together.

Recent studies~\cite{lee2015adaptive, chandrasekar2014exploiting, kim2019d,
kim2018dram, chang2016understanding} propose mechanisms to reduce the DRAM
access latency by reducing the margins present in timing parameters when
operating under appropriate conditions (e.g., low temperature). Other
works~\cite{gulur2012multiple, son2013reducing, zhang2014half,
seshadri2017ambit, chang2014improving, chang2017understanding, lee2017design,
kim2018solar, lee2015adaptive} propose different methods to reduce DRAM latency.
These works are orthogonal to \mech, and they can be implemented along with our
mechanism to further reduce DRAM latency.

\section{Summary}
\label{crow_sec:conclusion}

We propose \mech, a low-cost substrate that partitions each DRAM subarray into
two regions (regular rows and {\copyrow}s) and enables independent control over
the rows in each region. We leverage \mech to design two new mechanisms,
\mech-cache and \mech-ref, that improve DRAM performance and energy-efficiency.
\mech-cache uses a {\copyrow} to duplicate a regular row and simultaneously
activates a regular row together with its duplicated \copyrow to reduce the DRAM
activation latency (by 38\% in our experiments). \mech-ref remaps retention-weak
regular rows to strong {\copyrow}s, thereby reducing the DRAM refresh rate.
{\mech}'s flexibility allows us to simultaneously employ both \mech-cache and
\mech-ref to provide 20.0\% speedup and 22.3\% DRAM energy savings over
conventional DRAM. We conclude that \mech is a flexible DRAM substrate that
enables a wide range of mechanisms to improve performance, energy efficiency,
and reliability. \hhms{We freely open source CROW~\cite{crow_spice_github} to
enable future work in this area.} We hope future work exploits \mech to devise
more use cases that can take advantage of its low-cost and versatile substrate.

\chapter[Self-Managing DRAM]{A Case for Self-Managing DRAM Chips: Enabling Efficient and Autonomous in-DRAM Maintenance Operations}
\label{chap:smd}

\newcommand{\hht}[1]{{#1}\xspace}
\newcommand{\hhf}[1]{{#1}\xspace}
\newcommand{\hext}[1]{{{#1}}\xspace}
\newcommand{\hha}[1]{{{#1}}\xspace}
\newcommand{\agy}[1]{{#1}\xspace}
\newcommand{\agycomment}[1]{{}}

\newcommand{\mechunformatted}{SMD\xspace}
\newcommand{\mech}{\texttt{\mechunformatted{}}\xspace}
\newcommand{\mechlong}{Self-Managing DRAM\xspace}

\newcommand{\fr}{\texttt{SMD-FR}\xspace}
\newcommand{\vr}{\texttt{SMD-VR}\xspace}
\newcommand{\prp}{\texttt{SMD-PRP}\xspace}
\newcommand{\prpplus}{\texttt{SMD-PRP+}\xspace}
\newcommand{\drp}{\texttt{SMD-DRP}\xspace}
\newcommand{\sms}{\texttt{SMD-MS}\xspace}

\newcommand{\cmdrsq}{\texttt{{RSQ}}\xspace}
\newcommand{\ARI}{\texttt{{ARI}}\xspace}
\newcommand{\actnack}{\texttt{ACT\_NACK}\xspace}
\newcommand{\tnackplain}{T_{\actnack{}}}
\newcommand{\tnack}{$\tnackplain$\xspace}


\hhmv{The memory controller is in charge of managing DRAM maintenance operations
(e.g., refresh, RowHammer protection, memory scrubbing) in current DRAM chips.
Implementing new maintenance operations often necessitate modifications in the
DRAM interface, memory controller, and potentially other system components. Such
modifications are only possible with a new DRAM standard, which takes a long
time to develop, leading to slow progress in DRAM systems.}

In this \hhm{chapter}, our goal is to 1) ease the process of enabling new DRAM
maintenance operations and 2) enable more efficient in-DRAM maintenance
operations. Our idea is to set the memory controller free from managing DRAM
maintenance. To this end, we propose \mechlong{} (\mech{}), a new low-cost DRAM
architecture that enables implementing new in-DRAM maintenance mechanisms (or
modifying old ones) with no further changes in the DRAM interface, memory
controller, or other system components. We use \mech{} to implement new in-DRAM
maintenance mechanisms for three use cases: 1) periodic refresh, 2) RowHammer
protection, and 3) memory scrubbing. Our evaluations show that, \mech{}-based
maintenance operations significantly improve the system performance and energy
efficiency while providing higher reliability compared to conventional DDR4
DRAM. A combination of \mech{}-based maintenance mechanisms that perform
refresh, RowHammer protection, and memory scrubbing achieve 7.6\% speedup and
5.2\% lower DRAM energy on average for four-core workloads. \hhmiv{We open
source our Self-Managing DRAM framework~\cite{smdsource}.}

\section{Motivation and Goal}

Ensuring reliable and efficient DRAM operation becomes an even more critical
challenge as a DRAM cell becomes smaller and the distance between adjacent cells
shrink~\cite{awasthi2012efficient,cha2017defect,hong2010memory,kang2014co,kim2020revisiting,kim2014flipping,
lee2016technology, mandelman2002challenges, mutlu2013memory,
mutlu2019rowhammer,nair2013archshield, park2015technology,qureshi2015avatar}. A
modern DRAM chip typically requires three types of maintenance operations for
reliable and secure operation: 1) periodic refresh, 2) RowHammer protection, and
3) memory scrubbing.\footnote{Memory scrubbing is not always required in
consumer systems but often used by cloud service
providers~\cite{jacob2010memory,
mukherjee2004cache,schroeder2009dram,meza2015revisiting, saleh1990reliability,
siddiqua2017lifetime, micron2019whitepaper, gong2018duo}.}

\textbf{DRAM Refresh.} Due to the short retention time of DRAM cells, all DRAM
cells need to be periodically refreshed to ensure reliable operation. In
existing DRAM chips, a cell is typically refreshed every 64 or
\SI{32}{\milli\second} below
\SI{85}{\celsius}~\cite{micronddr4,jedec2012ddr4,jedec2014lpddr4} and every
\SI{16}{\milli\second} above \SI{85}{\celsius}~\cite{jedec2020lpddr5,jedec2020lpddr5}. A
DRAM chip consists of billions of cells, and therefore refreshing such a large
number of cells incurs large performance and energy
overheads~\cite{liu2012raidr, liu2013experimental, qureshi2015avatar}. These
overheads are expected to increase as DRAM chips become
denser~\cite{chang2014improving,kang2014co,liu2012raidr}. Many ideas have been
proposed to mitigate the DRAM refresh overhead~\cite{chang2014improving,
liu2012raidr, qureshi2015avatar,nair2014refresh, baek2014refresh,
bhati2013coordinated, cui2014dtail, emma2008rethinking, ghosh2007smart,
isen2009eskimo, jung2015omitting, kim2000dynamic, luo2014characterizing,
kim2003block, liu2012flikker, mukundan2013understanding, nair2013case,
patel2005energy, stuecheli2010elastic, khan2014efficacy, khan2016parbor,
khan2017detecting, venkatesan2006retention, patel2017reaper, riho2014partial,
hassan2019crow, kim2020charge, nguyen2018nonblocking, kwon2021reducing}.

\textbf{RowHammer Protection.} 
Small and densely-packed DRAM cells suffer from the RowHammer
vulnerability~\cite{kim2014flipping}, which leads to disturbance failures that
stem from electro-magnetic interference created by repeatedly activating (i.e.,
hammering) a DRAM row (i.e., aggressor row) many
times~\cite{yang2019trap,gautam2019row,jiang2021quantifying,park2016experiments,
park2016statistical, ryu2017overcoming, walker2021dram,yang2016suppression}.
Such rapid row activations lead to bit flips in DRAM rows that are physically
nearby an aggressor row. Many works propose mechanisms to protect DRAM chips
against RowHammer~\cite{apple2015about,kim2014flipping,
aweke2016anvil,lee2019twice,park2020graphene,seyedzadeh2017cbt,son2017making,taouil2021lightroad,yauglikcci2021security,you2019mrloc,
yaglikci2021blockhammer,greenfield2016throttling,
hassan2019crow,saileshwar2022randomized, konoth2018zebram,
brasser2016can,van2018guardion, marazzi2022protrr, kim2022mithril, wi2023shadow,
woo2023scalable, marazzi2023rega}.

\textbf{Memory Scrubbing.}
System designers typically use Error Correcting Codes
(ECC)~\cite{hamming1950error,dell1997white} to improve system reliability. DRAM
vendors recently started to equip DRAM chips with on-die
ECC~\cite{jedec2021ddr5, patel2019understanding, patel2020bit, patel2021harp,
nair2016xed}. Detection and correction capability of an ECC scheme is limited.
For example, \emph{Single-Error-Correction Double-Error-Detection (SECDED)} ECC
detects up to two bit errors and corrects one bit error in a codeword. To
prevent independent single-bit errors from accumulating multi-bit errors, memory
scrubbing~\cite{jacob2010memory,
mukherjee2004cache,schroeder2009dram,meza2015revisiting, saleh1990reliability,
siddiqua2017lifetime, micron2019whitepaper, gong2018duo} is commonly employed: the
entire memory is scanned to correct single-bit errors.

We consider periodic refresh, RowHammer protection, and memory scrubbing as
\emph{DRAM maintenance operations} needed for reliable and secure operation of
modern and future DRAM chips. An ideal maintenance operation should efficiently
improve DRAM reliability and security with minimal system performance and energy
consumption overheads.

As DRAM technology node scales to smaller sizes, the reliability and security of
DRAM chips worsen. As a result, new DRAM chip generations necessitate making
existing maintenance operations more intensive (e.g., lowering the refresh
period~\cite{jedec2021ddr5, jedec2020lpddr5,apple2015about}) and introducing new
types of maintenance operations (e.g., targeted
refresh~\cite{hassan2021uncovering,frigo2020trrespass,jattke2022blacksmith} and
DDR5 RFM~\cite{jedec2021ddr5} as RowHammer defenses). As the intensity and
number of different maintenance operations increase, it becomes increasingly
more challenging to keep the overhead of the maintenance operations low.

Unfortunately, modifying the DRAM maintenance operations is difficult due to the
current rigid DRAM interface that places the Memory Controller (MC) completely
in charge of DRAM control. \hhmv{Implementing new maintenance operations (or
changing existing ones) often necessitate modifications in the DRAM interface,
MC, and potentially other system components. Such modifications are only
possible with a new DRAM standard, which takes a long time to develop, leading
to slow progress in DRAM systems. For example, there was a five-year gap between
the DDR3~\cite{jedec2008ddr3} and DDR4~\cite{jedec2012ddr4} standards, and
eight-year gap between the DDR4~\cite{jedec2012ddr4} and
DDR5~\cite{jedec2021ddr5} standards. In addition to waiting for a new standard
to come out, DRAM vendors also need to push their proposal regarding the
maintenance operations through the JEDEC committee so the new standard includes
the desired changes to enable new maintenance operations. Thus, we believe a
flexible DRAM interface that allows DRAM vendors to quickly implement custom
in-DRAM maintenance mechanisms would help to develop more reliable and secure
DRAM chips.}


\textbf{Our goal} is to 1) ease the process of implementing new
DRAM maintenance operations and 2) enable more efficient in-DRAM maintenance
operations. We believe that enabling autonomous in-DRAM maintenance operations
would encourage innovation, reduce time-to-market, and ease adoption of novel
mechanisms that improve DRAM efficiency, reliability, and security. DRAM
architects will quickly deploy their new in-DRAM solutions to tackle reliability
and security problems of DRAM without demanding further changes in the DRAM
interface, MC, or other system components.

\section{Motivation}
\label{sec:drawbacks_current_maintenance}

\hh{In current DRAM chips, the \hht{Memory Controller (MC)} is in charge of
managing DRAM maintenance operations such as periodic refresh, RowHammer
protection, and memory scrubbing. When DRAM vendors make modifications in a DRAM
maintenance mechanism, the changes often need to be reflected to the MC design
\hht{as well as the DRAM interface}, which makes such modifications \hht{very}
difficult. As a result, implementing new or modifying existing maintenance
operations \hht{requires} multi-year effort by multiple parties that are part of
the JEDEC committee. A prime example to support our argument is the \hht{most
recent} DDR5 standard~\cite{jedec2021ddr5}\hht{, which took almost a decade to
develop after the initial release of DDR4}. DDR5 introduces changes to \hht{key
issues we study in this paper:} DRAM refresh, RowHammer protection, and memory
scrubbing. We discuss these changes \hht{as motivating examples to show the
shortcomings of the status quo in DRAM}.}

\subsection{DRAM Refresh}

DDR5 introduces \emph{Same Bank Refresh (SBR)}, \hh{which refreshes one bank in
each bank group at a time instead of simultaneously refreshing all banks as in
DDR4~\cite{ddr4,micronddr4}. \emph{SBR} improves \hh{bank} availability as the
MC can access the non-refreshing banks while certain banks are being refreshed}.
\hh{DDR5 implements \emph{SBR} by introducing a new \texttt{REFsb}
command~\cite{jedec2021ddr5}. \bgyellow{Implementing \texttt{REFsb}}
necessitates changes in the DRAM interface \emph{and} MC.}

\subsection{RowHammer Protection}

\hh{In DDR4, DRAM vendors implement in-DRAM RowHammer protection mechanisms by
performing Targeted Row Refresh (TRR) operations within the slack time available
when performing regular refresh. However, prior works have shown that in-DRAM
TRR is vulnerable to certain memory access
patterns~\cite{hassan2021uncovering,frigo2020trrespass,jattke2022blacksmith}.
DDR5 specifies a \emph{Refresh Management (RFM)} mechanism that an MC implements
to aid \hht{the RowHammer protection implemented in DRAM chips}. As part of RFM,
the MC \hht{uses} a set of counters to keep track of row activations to each
DRAM bank. When a counter reaches a specified threshold value, the MC issues the
new \texttt{RFM} command to DRAM \hht{chips}. \hht{A DRAM chip then internally
performs an undocumented operation with the time allowed for \texttt{RFM} to
mitigate the RowHammer effect.} Since the \hht{first RowHammer
work~\cite{kim2014flipping}} in 2014, it took about 7 years to introduce DDR5
RFM, which requires significant changes in the DRAM interface and the MC design.
Still, RFM is likely not a definitive solution for RowHammer as \hht{it does not
outline a specific RowHammer defense with security proof. Rather, RFM provides
additional time to a DRAM chip for internally mitigating the RowHammer effect.}}

\subsection{Memory Scrubbing}

DDR5 adds support for on-die ECC and in-DRAM scrubbing. \hh{A DDR5 chip
internally performs ECC encoding and decoding when the chip \hht{is accessed}.
To perform DRAM scrubbing, the MC must periodically issue the new
\emph{scrub command} (for manual scrubbing) or \emph{all-bank refresh command}
(for automatic scrubbing). Similar to \emph{Same Bank Refresh} and \emph{RFM},
enabling in-DRAM scrubbing \hht{necessitated} changes in the DRAM interface and
in the MC design.}

\section{Self-Managing DRAM}
\label{sec:sm_dram}

We introduce the \emph{Self-Managing DRAM} (\mech{}) architecture.

\subsection{Overview of \mechunformatted{}}
\label{sec:overview}


\mech{} has a flexible interface that enables efficient implementation of
multiple DRAM maintenance operations within the DRAM chip. \bgyellow{Our key insight
is to prevent the MC from accessing a DRAM row that is under maintenance by
prohibiting the MC from activating the row until the maintenance is complete.}
\textbf{The key idea} of \mech{} is to provide a DRAM chip with the ability to
reject an \cmdact{} command via a \hht{single-bit}
\emph{negative-acknowledgment} (\actnack{}) signal. An \actnack{} informs the MC
that the DRAM row it tried to activate is under maintenance\hh{, and thus
temporarily unavailable}. Leveraging the ability to reject an \cmdact{}, a
maintenance operation can be implemented \emph{completely within} a DRAM chip.

\begin{figure}[!h]
    \centering
    \includegraphics[width=.9\linewidth]{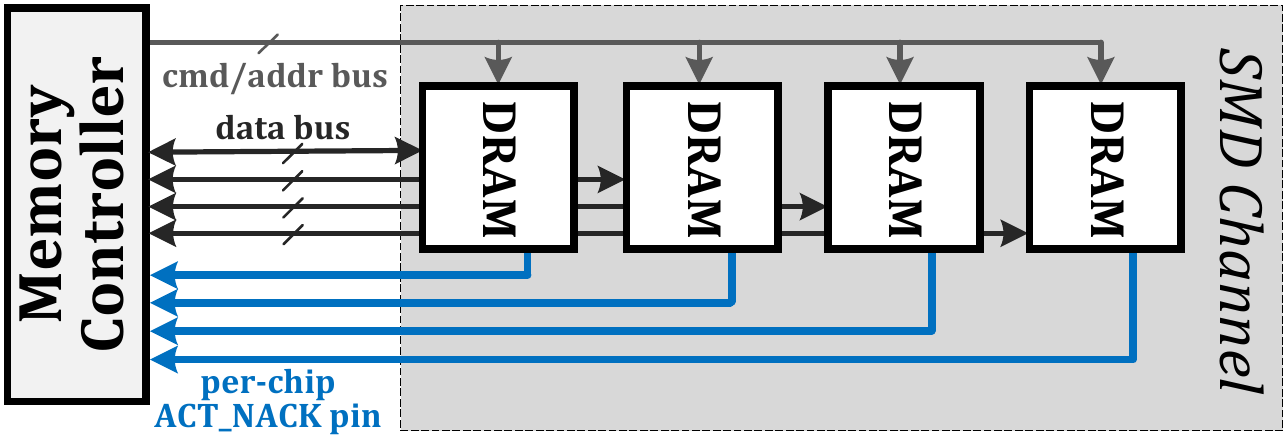}
    \caption{A DRAM channel \hht{with four} \mech{} chips.}
    \label{fig:smd_interface}
\end{figure}

As Fig.~\ref{fig:smd_interface} shows, \mech{} preserves the general DRAM
interface\footnote{We consider \hht{the widely-available} DDR4 as a baseline
DRAM and explain the changes required to implement \mech{} on top of DDR4.
However, other DRAM interfaces (e.g., DDR5, LPDDR5, GDDR6, HBM2) can also be
modified in a similar way to support \mech{}.} and only adds a single
uni-directional pin to the physical interface of a DRAM chip. The extra pin is
used for transmitting \hht{the} \actnack{} signal from the DRAM chip to the MC.
\bgyellow{
Upon receiving an \actnack{}, the MC notices that the previous row activation is
rejected, and thus it need to be re-issued at later time. While introducing
changes in the MC to handle \actnack{}, \mech{} also simplifies MC design and
operation as the MC no longer implements control logic to periodically issue
DRAM maintenance commands. For example, the MC does \emph{not} 1) prepare a bank
for refresh by precharging it, 2) implement timing parameters relevant to
refresh, and 3) issue \cmdrefresh{} commands. The MC still maintains the bank
state information (e.g., whether and which row is open in a bank) and respects
the DRAM timing parameters associated with commands for performing access (e.g.,
\cmdact{}, \cmdpre{}, \cmdread{}, \cmdwrite{}).}

\begin{figure}[!h]
    \centering
    \includegraphics[width=\linewidth]{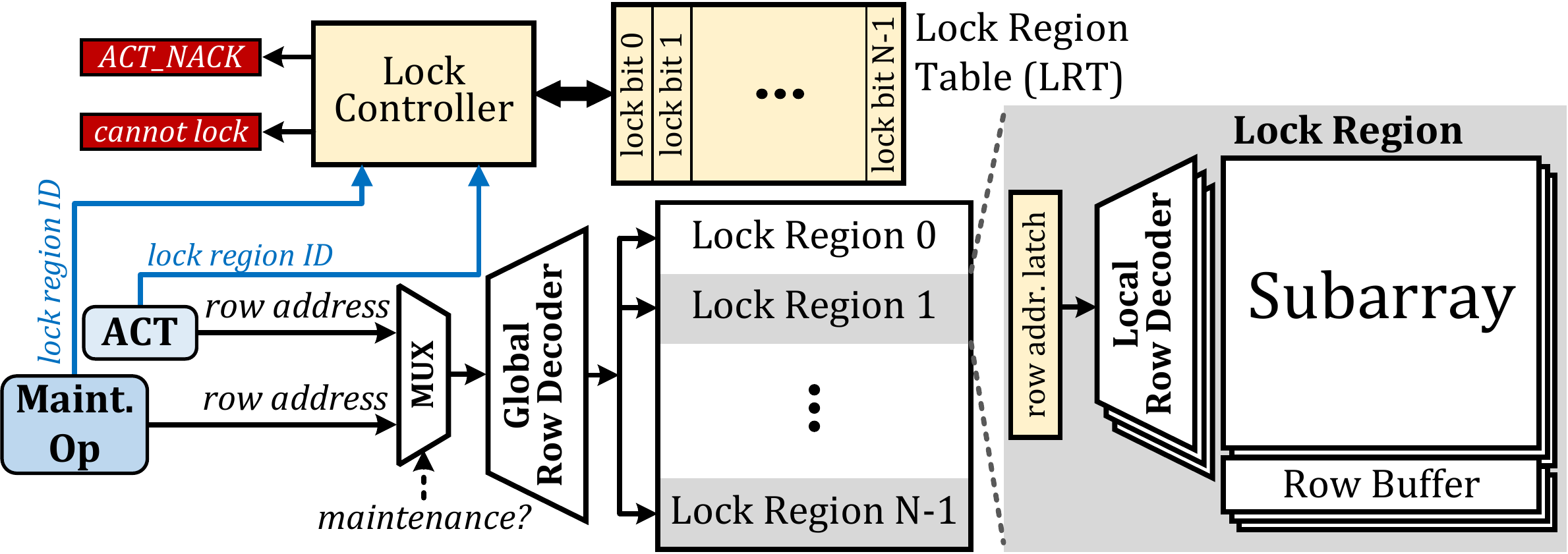}
    \caption{\mech{} bank organization \hht{in DRAM chip}.}
    \label{fig:smd_bank_organization}
\end{figure}

Fig.~\ref{fig:smd_bank_organization} shows the organization of a bank in an
\mech{} chip. \mech{} divides the rows in a DRAM bank into multiple \emph{lock
regions}. \hht{\mech{} provides the number and the size of the lock regions to
the MC via DRAM \emph{mode registers}~\cite{ddr4}.} For each lock
region, \mech{} stores an entry in a \emph{Lock Region Table (LRT)} to indicate
whether or not a lock region is \hht{reserved} for performing a maintenance
operation. 

A maintenance operation and an \cmdact{} can be performed concurrently \hht{in
the same bank} on different lock regions. To enable this, \mech{} implements a
\emph{row address latch} in order to drive two local row decoders with two
independent row addresses.\footnote{Having a row address latch per lock region
enables two or more maintenance operations to concurrently happen on different
lock regions within a bank. However, to keep our design simple, we restrict a
maintenance operation to happen on one lock region while the MC
can access another lock region. \hht{Our design can be easily extended to
increase concurrency of maintenance operations across multiple lock regions.}}
\hh{When a maintenance operation and an \cmdact{} arrive \hht{at} the same lock
region at the same time}, the multiplexer shown in
Fig.~\ref{fig:smd_bank_organization} prioritizes the maintenance operation and
the \mech{} chip rejects the \cmdact{}. 

For a maintenance operation to take place, the \emph{Lock Controller} first sets
the \emph{lock bit} in the LRT entry that corresponds to the lock region \hh{on
which the maintenance operation is to be performed}. When the MC attempts to
open a row in a locked region, the Lock Controller generates an \actnack{}. 

\subsection{Handling an ACT\_NACK Signal}
\label{subsec:rejecting_acts}

\begin{figure}[!h]
    \centering
    \includegraphics[width=\linewidth]{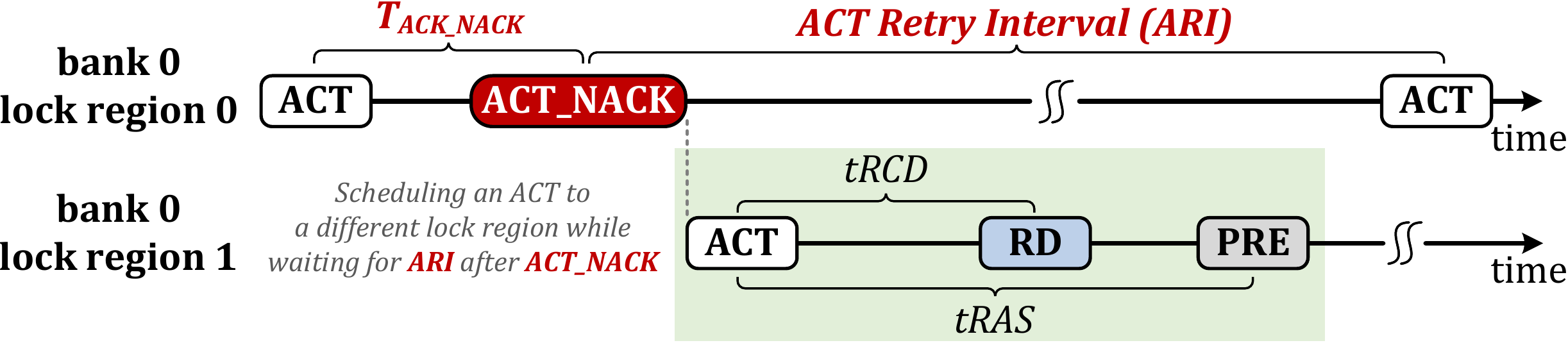}
    \caption{Handling \actnack{} \hht{in MC}.}
    \label{fig:smd_act_nack_timeline}
\end{figure}

Fig.~\ref{fig:smd_act_nack_timeline} depicts a timeline that shows how the
MC handles an \actnack{} signal. \hh{Upon receiving} an
\actnack{}, the MC waits for \emph{ACT Retry Interval} (\ARI)
\hht{time} \hh{to re-issue} the same \cmdact{}. \hh{It} keeps re-issuing the
\cmdact{} command once every \ARI{} until the DRAM chip accepts the \cmdact{}.
While waiting for \ARI{}, the MC can attempt activating a row
\hh{from a different lock region or bank}, to overlap the \ARI{} latency with a
useful operation.

\subsubsection{Setting \hht{the} \emph{ACT Retry Interval} (\ARI)} 

Setting \ARI{} to a very low or high value relative to the expected duration of
the maintenance operations can have a negative impact on system performance and
energy efficiency. When too low, the MC frequently re-issues an
\cmdact{}, and the DRAM chip rejects many of these \cmdact{} commands. The
rejected \cmdact{} commands waste energy and potentially delay other row
activations that the MC could have issued to serve other memory
requests. When \ARI{} is too high, the MC may delay re-issuing a
rejected \cmdact{} for too long, \hht{and} the \cmdact{} gets delayed
significantly even after the completion of the maintenance. We empirically find
that $\ARI=\SI{\valARI}{\nano\second}$ is a favorable configuration for the six
maintenance mechanisms that we evaluate in this work
(\cref{sec:maintenance_mechanisms}).

\subsubsection{\tnack{} Latency}


An \mech{} chip sends an \actnack{} (in case an \cmdact{} command is to be
rejected) \tnack{} DRAM command bus cycles after receiving the \cmdact{}. Thus,
the MC \hht{knows} that an \cmdact{} is successful when it does
\emph{not} receive an \actnack{} in \tnack{} cycles after issuing the \cmdact{}.
%
%
\hht{It is desirable for \tnack{} to be as low as possible so that the MC can be quickly notified when an \cmdact{} fails, and it can attempt
to activate a different row in the same bank while waiting for \ARI{} to retry
the rejected \cmdact{}.} 
Further, if \tnack{} is larger than \trcd{}, the latency of \cmdread{} and
\cmdwrite{} commands increases (by $\tnackplain{}-\trcd{}$) as the MC should not issue a \cmdread{} or \cmdwrite{} without ensuring that
the row is successfully opened.

The \tnack{} latency has three components: 1) the propagation delay (from the
MC to the DRAM chip) of the \cmdact{}, 2) the latency of
determining whether or not the row to be activated belongs to a locked region,
and 3) the propagation delay (from the DRAM chip to the MC) of
the \tnack{} signal. 

We estimate the overall \tnack{} latency based on the latency of an \cmdread{}
in a conventional DRAM. An \cmdread{} has a latency breakdown that resembles the
latency breakdown of a \tnack{}. An \cmdread{} command 1) propagates from the MC
to the DRAM chip, 2) accesses data in a portion of the row buffer in the
corresponding bank, and 3) sends the data back to the MC. In the DDR4
standard~\cite{ddr4}, the latency between issuing an \cmdread{} and the first
data beat appearing on the data bus is defined as \tcl{} \hht{(typically
22~cycles for DDR4-3200)}. The latency components 1) and 3) of \cmdread{} are
similar to those of \tnack. Thus, the main difference between \tnack{} and
\tcl{} arises from the second component. According to our evaluation, the
latency of accessing the Lock Region Table (LRT) is \SI{0.053}{\nano\second}
(\cref{sec:hw_overhead}). Given the relatively low complexity of the LRT
compared to the datapath that is involved during an \cmdread{}, we believe the
\hht{second component of the latency breakdown and thus the} overall \tnack{}
latency can be designed to be much smaller than \tcl{}. \hht{We assume \tnack{}
$= 5$ cycles unless stated otherwise. In our evaluations, we find that small
\tnack{} latencies (e.g., $\leq \tcl{}$) have negligible effect on system
performance mainly because the number of rejected \cmdact{}s constitute a small
portion of all \cmdact{}s.}


\subsubsection{\actnack{} Divergence Across DRAM Chips}

\mech{} maintenance operations \hht{take place} independently in each DRAM chip.
Therefore, when a DRAM rank, which can be composed of multiple DRAM chips
operating in lock step\footnote{DDRx systems typically consist of DRAM ranks
built with multiple chips (e.g., eight chips with 8-bit data bus in each chip),
whereas LPDDRx systems typically consist of DRAM ranks built with a single DRAM
chip.}, receives an \cmdact{} command, some of the chips may \hh{send an
\actnack{} while others do not.}
\hh{Normally, \actnack{} divergence would not happen for maintenance mechanisms
that perform the exact same operation at the exact same time in all DRAM chips.
However, a mechanism can also operate differently in each DRAM chip, e.g.,
the Variable Refresh (\vr{}) mechanism} (\cref{subsubsec:variable_refresh}).
As a result \hh{of this divergence}, the row becomes partially open in chips
that do \emph{not} send \actnack{}. 

\hh{As a solution}, the MC can take one of the following two approaches. First,
\hht{it} can issue a \cmdpre{} command to close the partially open row. Upon
receiving the \cmdpre{}, chips that \hh{sent an \actnack{}} and do not have an
open row can simply ignore the \cmdpre{}. The advantage of this approach is
that, after the \cmdpre{}, the MC can attempt to open another row from the same
bank. However, as a downside, precharging the partially activated row would
waste energy. As a second approach, the MC can wait for \ARI{} and re-issue the
same \cmdact{} command to attempt activating the row in the chips that
previously sent an \actnack{} signal. This approach does not waste row
activation energy but it prevents the MC from attempting to activate another row
in the same bank while waiting for \ARI{}. \hh{We empirically find that the
second approach performs slightly better \hht{(by 0.04\% on average)} mainly
because} \hht{the benefits of the first approach are limited as} a partially
activated row cannot be immediately precharged \hht{due to} \tras{}, which
constitutes a large part of \ARI{}\footnote{By default, we set
$\ARI=\SI{\valARI}{\nano\second}$ while \tras{} is about
\SI{35}{\nano\second}~\cite{micronddr4}.}). \hh{Our analysis shows that
under the worse-case \actnack{} divergence (i.e., when each of the 16 chips in a
rank perform refresh on a different region at different time), \vr{} performs
15.4\% worse than \vr{} with no \actnack{} divergence, which still provides
7.8\% average speedup over conventional DDR4 refresh for memory-intensive
four-core workloads.}

\subsection{Region Locking Mechanism}
\label{subsec:locking_mechanism}


\hht{\mech{} locks rows at a \emph{lock region} granularity.}

\subsubsection{Lock Region Size}
\label{subsubsec:lock_region_size}

A \emph{lock region} consists of a fixed number of \hht{consecutively-addressed}
DRAM rows. \hht{To simplify the design,} we set the lock region size \hht{such
that a lock region spans} one or multiple subarrays. This is because a
maintenance operation \hht{uses} the local row buffers in a subarray, and thus
having a subarray shared by multiple lock regions will cause a conflict when
accessing a row in a different lock region that maps to the subarray under
maintenance. We design and evaluate \mech{} assuming \hht{a default} lock region
size of $16$ 512-row subarrays. However, a lock region can be designed to
\hht{span fewer than $16$ subarrays} or \hht{it can} span the entire bank. 



\begin{yellowb}
Modern DRAM chips typically use the density-optimized \emph{open-bitline}
architecture~\cite{keeth2007dram,jacob2010memory}, which places sense amplifiers
both on top and bottom sides of a subarray and adjacent subarrays share sense
amplifiers. With the open-bitline architecture, the MC should not be allowed to
access a row in a subarray that is adjacent to one under maintenance. To achieve this, the Lock Controller simply sends an \actnack{} when the MC attempts to activate a row in a subarray adjacent to a subarray in a locked region. Consequently, when SMD locks a region that spans $16$ 512-row subarrays, it prevents the MC from accessing $18$ subarrays in a bank with $128K$ rows ($256$ subarrays). In the \emph{folded-bitline} architecture~\cite{keeth2007dram, jacob2010memory}, adjacent subarrays do not share sense amplifiers with each other. Therefore, SMD prevents access only to the subarrays in a locked region. We evaluate SMD using the density-optimized open-bitline architecture.
\end{yellowb}

\subsubsection{Locking a Region}
\label{subsubsec:region_locking}

A maintenance operation can be performed \hht{only} on \hht{DRAM rows in a
locked region}. Therefore, a maintenance mechanism must lock the region that
includes the rows that should undergo a maintenance. A maintenance mechanism can
\emph{only} lock a region that is \emph{not} \hht{already} locked. This is to
prevent different maintenance mechanisms from interfering with each other. 
Once having locked a region, \hht{the region remains locked} until the
completion of the maintenance. Afterward, the maintenance mechanism releases the
lock to allow 1) the MC \hht{to} access the region and 2) other
maintenance mechanisms \hht{to} lock the same region. 

\subsubsection{Ensuring Timely Maintenance Operations}
\label{subsubsec:ensuring_timely_maintenance}

A maintenance mechanism \emph{cannot} lock a region that has an active DRAM row.
To lock the region, the maintenance mechanism must wait for the MC to precharge
the row. \hht{The MC may keep a DRAM row active for too long, which may delay a
maintenance operation to a point that affects DRAM reliability (e.g., delaying a
refresh operation too long as such a retention failure occurs). To prevent this,
we rely on the existing time limit for a row to remain open, i.e., \tras{} has
an upper limit of 9x~\trefi{}~\cite{ddr4,micronddr4,ddr4operationhynix}).} 

\begin{yellowb}
\subsection{Impact to Request Scheduling}
\label{subsec:impact_to_scheduling}

\mech{} partitions DRAM into small regions to perform maintenance operations.
This allows \mech{} to pause accesses to small DRAM regions while the rest of
the DRAM remains available to access. In contrast, existing DRAM chips perform
maintenance operations at a larger granularity. For example, with existing DRAM
chips, MCs perform refresh at bank or rank granularity, pausing accesses to an
entire bank/rank. Although the MC can postpone a refresh operation for a small
time interval (e.g., 8x \trefi{}), it does so just based on information (e.g.,
queue occupancy) available to the MC, but without the knowledge of future
requests. As a result, processing units are stalled with existing DRAM chips
when requests enter the MC after a maintenance operation starts. \mech{}
significantly improves the overall availability of DRAM by pausing accesses to
small regions and improves performance as we show in
\cref{subsec:single_core_perf} and \cref{subsec:multi_core_perf} (e.g., for
memory intensive four-core workloads, a system with \mech{}-based refresh
achieves 8.7\% average speedup compared to a system with conventional DDR4).
Therefore, \mech{} improves the overall availability of DRAM, and thus reduces
the performance overhead of DRAM maintenance operations. 

\end{yellowb}
\section{SMD Maintenance Mechanisms}
\label{sec:maintenance_mechanisms}

\hht{We propose \mech{}-based maintenance mechanisms for three use cases.}
However, \mech{} is not limited to these three use cases and it can be used to
support more operations in DRAM. 

\subsection{Use Case 1: DRAM Refresh}
\label{subsec:smd_refresh}

In conventional DRAM, the MC periodically issues \cmdrefresh{} commands to
initiate a DRAM refresh operation. This approach is inefficient due to
\hht{three main} reasons. First, transmitting $8192$ \cmdrefresh{} commands over
the DRAM command bus within the refresh period \hht{(e.g., 64, 32, or even
\SI{16}{\milli\second} depending on the refresh rate)} consumes energy and
increases the command bus utilization. Due to transmitting a \cmdrefresh{} over
the command bus, the MC may delay a command to another rank in the same channel,
which increases DRAM access latency~\cite{chang2014improving}. Second, an entire
bank becomes inaccessible while being refreshed although a \cmdrefresh{} command
refreshes only a few rows in the bank. \hht{This incurs up to 50\% loss in
memory throughput~\cite{liu2012raidr}.} \hht{Third, new mechanisms that reduce refresh
overhead (e.g., by skipping refreshes on retention-strong rows) are difficult to
implement. Even a simple optimization of enabling per-bank refresh in DDR4
requires changes to the DRAM interface and the MC.}

Leveraging \mech{}, we design two maintenance mechanisms for DRAM refresh:
\emph{Fixed-Rate Refresh} and \emph{Variable Refresh}. 

\subsubsection{Fixed-Rate Refresh (\fr{})}
\label{subsubsec:fixed_refresh}

\fr{} refreshes DRAM rows in a fixed time interval (i.e., \trefi{}), similar to
conventional DRAM refresh. \hht{It enables refresh-access parallelization
without any burden on the MC beyond the changes required for implementing the
\mech{} interface. As with any maintenance operation in \mech{}, the MC can
activate a row (in any bank) in regions that are not locked while \fr{}
refreshes rows in a locked region.} \fr{} refreshes $RG$ (Refresh Granularity)
number of rows from a lock region and switches to refreshing another region to
1) limit the time for which a single lock region remains under maintenance and
2) prevent memory requests from waiting too long for the region to become
available for access. 

\begin{figure}[!h]
    \centering
    \includegraphics[width=.95\linewidth]{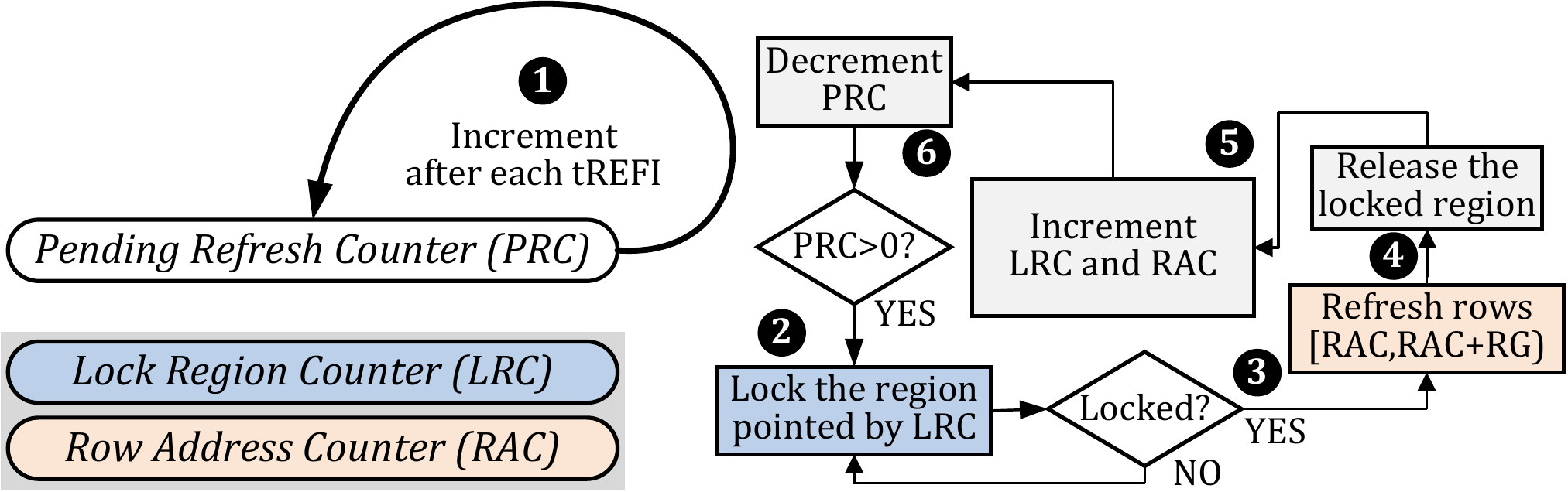}
    \caption{Fixed-Rate Refresh (\fr{}) Operation.}
    \label{fig:smd_fixed_refresh}
\end{figure}

\fr{} operates independently in each DRAM bank and \fr{} uses three counters for
managing the refresh operations in a bank: \emph{pending refresh counter},
\emph{lock region counter}, and \emph{row address counter}.
Fig.~\ref{fig:smd_fixed_refresh} illustrates \fr{} operation. The \emph{pending
refresh counter} is \hht{used to slightly delay refresh operations within the
slack\footnote{DDR4~\cite{ddr4,micronddr4,ddr4operationhynix} allows the MC to postpone issuing up to $8$ \cmdrefresh{} commands in order to
serve \hht{pending} memory requests first.} given in the DRAM standard until the
region to be refreshed can be locked.} The \emph{pending refresh counter} is
initially zero and \fr{} increments it by one at the end of a \trefi{}
interval~\circled{1}. \fr{} allows up to $8$ refresh operations to be
accumulated in the \emph{pending refresh counter}. As we explain in
\cref{subsubsec:ensuring_timely_maintenance}, because the MC can keep a row open
for a limited time, the \emph{pending refresh counter} never exceeds the value
$8$, and this is how \mech{} ensures that the refresh operations do \emph{not}
\hht{lag} behind more than the window of $8$ refresh operations.

The \emph{lock region} and \emph{row address} counters indicate the next row to
refresh. When \emph{pending refresh counter} is greater than zero, \fr{}
attempts to lock the region indicated by the \emph{lock region counter} every
clock cycle until it successfully locks the region~\circled{2}. \hht{In} some
cycles, \fr{} may fail to lock the region either because the lock region
contains an active row or the region is locked by another maintenance mechanism.
When \fr{} successfully locks the region, it initiates a refresh operation that
refreshes $RG$ number of \hht{sequentially-addressed} rows in the lock region
starting from the row indicated by the \emph{row address counter}~\circled{3}.
\hht{The refresh operation is carried out similarly as in conventional DRAM,
essentially by activating and precharging a row to restore the charge in its
DRAM cells. Making $RG$ larger increases the latency of a single refresh
operation, which prolongs the time that a lock region remains unavailable for
access. In contrast, small $RG$ causes switching from one lock region to another
very often, increasing interference with accesses. We empirically find $RG = 16$
to be a favorable design point.}

After the refresh operation completes, \fr{} releases the locked
region~\circled{4} and  increments \emph{only} the \emph{lock region counter}.
\hht{When} the \emph{lock region counter} \hht{rolls back to zero}, \fr{} also
increments by one the \emph{row address counter}~\circled{5}. Finally, \fr{}
decrements the \emph{target refresh counter}~\circled{6}.




\subsubsection{Variable Refresh (\vr{})}
\label{subsubsec:variable_refresh}

In conventional DRAM, all DRAM rows are uniformly refreshed with the same
refresh period. However, the \hht{actual data retention times of different rows}
in the same DRAM chip \hht{greatly vary} mainly due to manufacturing
\hht{process variation} and design-induced
variation~\cite{liu2012raidr,qureshi2015avatar,liu2013experimental,patel2017reaper,lee2017design,nair2014refresh}.
In fact, only hundreds of ``weak'' \hht{rows} across an entire
\SI{32}{\giga\bit} DRAM chip require to be refreshed at the default refresh
rate, and \hht{a vast} majority of the \hht{rows} can correctly operate when the
refresh period is doubled or quadrupled~\cite{liu2012raidr}. Eliminating unnecessary
refreshes to DRAM rows that do not contain weak cells can significantly mitigate
the performance and energy consumption overhead of DRAM refresh~\cite{liu2012raidr}. 

We \hht{develop} \emph{Variable Refresh} (\vr{}), a mechanism \hht{that}
refreshes different rows at different refresh rates depending on the retention
time characteristics of the weakest cell in each row. \hht{\mech{} enables
implementing \vr{} with no further changes in the DRAM interface and the MC.}
Our \vr{} design demonstrates the versatility of \mech{} in supporting different
DRAM refresh mechanisms. 

The high-level idea of \vr{} is to group DRAM rows into multiple retention time
bins and refresh a DRAM row based on the retention time bin that it belongs to.
To achieve low design complexity, \vr{} uses only two bins: 1)
\emph{retention-weak} rows that have retention time less than $RT_{weak\_row}$
and 2) rows that have retention time more than $RT_{weak\_row}$. \hht{Inspired
by RAIDR~\cite{liu2012raidr}, \vr{} stores the addresses of retention-weak rows
using a per-bank Bloom Filter~\cite{bloom1970space}, which is a space-efficient
probabilistic data structure for representing set membership. We assume
retention-weak rows are already inserted into the Bloom Filters by the DRAM
vendors during post-manufacturing tests.\footnote{\hht{Alternatively, \mech{}
can be used to develop a maintenance mechanism that performs retention
profiling. We believe \mech{} enables such profiling in a seamless way. Due to
limited space, we leave the development and analysis of it to future work.}}}

\begin{figure}[!h]
    \centering
    \includegraphics[width=.95\linewidth]{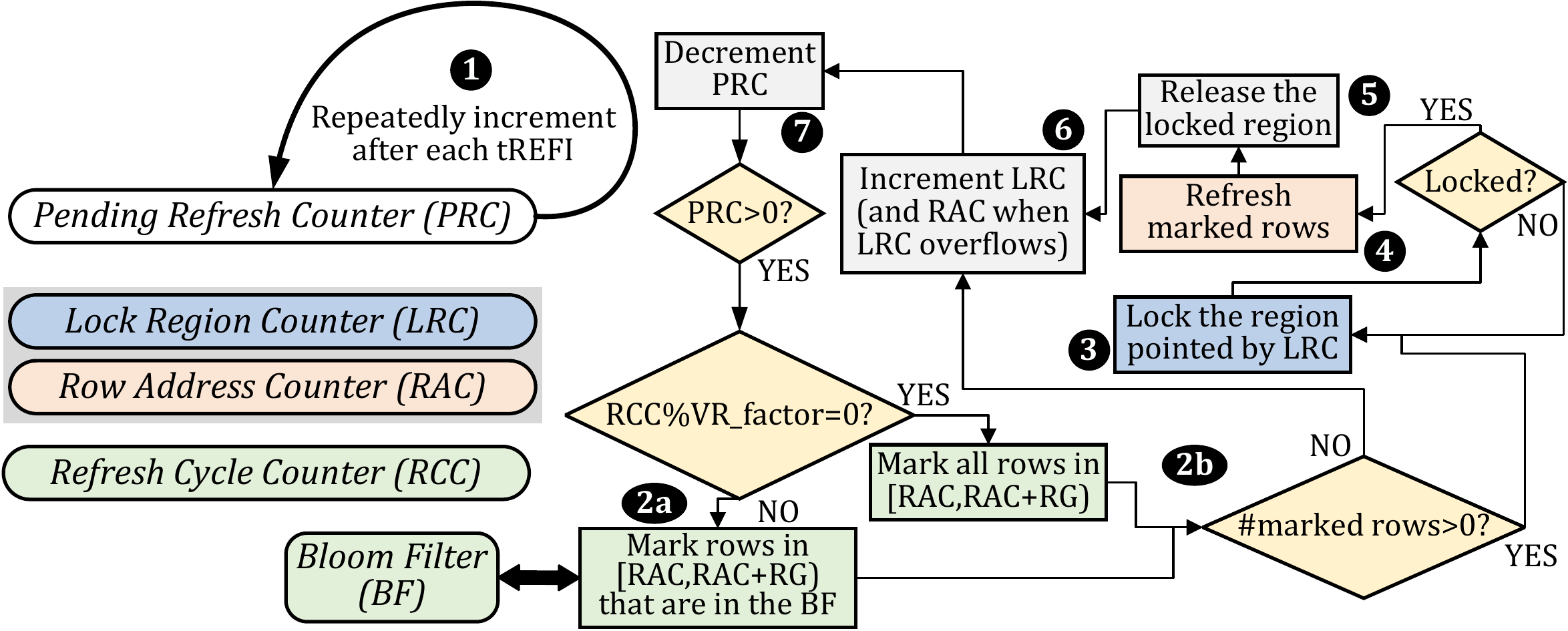}
    \caption{Variable Refresh (\vr{}) Operation.}
    \label{fig:smd_variable_refresh}
\end{figure}

The operation of \vr{} resembles the operation of \fr{} with the \hht{key}
difference that \vr{} \hht{sometimes} skips refreshes to a row that is not in
the Bloom Filter, i.e., a row with high retention time.
Fig.~\ref{fig:smd_variable_refresh} illustrates how \vr{} operates. \vr{} uses
the same three counters \hht{as} \fr{}. 
\vr{} \hht{also} uses a \emph{refresh cycle counter}, which is \hht{used to
indicate the refresh period when all rows (including retention-strong rows) must
be refreshed.} The \emph{refresh cycle counter} is initially zero and gets
incremented at the end of every refresh period, i.e., when the entire DRAM is
refreshed. 

\vr{} increments the \emph{pending refresh counter} by one at the end of a
\trefi{} interval~\circled{1}. When the \emph{pending refresh counter} is
greater than zero, \vr{} determines whether or not the $RG$ number of rows,
starting from the address indicated by the \emph{lock region} and \emph{row
address} counters, are retention-weak rows by testing their row addresses using
the bank's Bloom Filter~\circled{2a}. \vr{} refreshes the rows that are present
in the Bloom Filter every time when it is their turn to be refreshed, as
indicated by the \emph{lock region} and \emph{row address} counters. In
contrast, \vr{} refreshes the rows that are \emph{not} present in the Bloom
Filter \emph{only} when the \emph{refresh cycle counter} has a value that is
multiple of $VR\_factor$~\circled{2b}, which is specified by
Equation~\ref{eq:vr_factor}.

\begin{equation}
    VR\_factor = \frac{RT_{weak\_row}}{RefreshPeriod}
    \label{eq:vr_factor}
\end{equation}

\noindent Unless stated otherwise, we assume $RT_{weak\_row} =
\SI{128}{\milli\second}$ and $RefreshPeriod=\SI{32}{\milli\second}$. Therefore,
\vr{} refreshes the DRAM rows that are not in the Bloom Filter once every four
consecutive refresh \hht{periods} as these rows can retain their data correctly
for at least four refresh periods.

After determining which DRAM rows need refresh, \vr{} \hht{operates in a way
similar to \fr{}.}


\subsection{Use Case 2: RowHammer Protection}
\label{subsec:smd_rowhammer_protection}

RowHammer is a DRAM reliability issue mainly caused due to the small DRAM cell
size and the short cell-to-cell distance in modern DRAM chips. Repeatedly
activating and precharging (i.e., hammering) a \hht{(aggressor)} DRAM row
creates a disturbance effect, which accelerates charge leakage and causes bit
errors in the cells of a nearby \hht{(victim)} DRAM
row~\cite{yang2019trap,gautam2019row,jiang2021quantifying,park2016experiments,
park2016statistical, ryu2017overcoming, walker2021dram,yang2016suppression}. 

The three major DRAM manufacturers equip their \hht{existing} DRAM chips with
in-DRAM RowHammer protection mechanisms, generally referred to as Target Row
Refresh (TRR)~\cite{hassan2021uncovering,frigo2020trrespass,
jattke2022blacksmith}. At a high level, TRR protects against RowHammer by
detecting an aggressor row and refreshing \hht{its} victim rows. Because a
conventional DRAM chip \emph{cannot} initiate a refresh operation by itself, TRR
refreshes victim rows by taking advantage of the slack time available in the
\hht{refresh} latency (i.e., \trfc{}), originally used to perform \emph{only}
periodic DRAM
refresh~\cite{hassan2021uncovering,frigo2020trrespass,jattke2022blacksmith}.
This approach has two major \hht{shortcomings} that limit the protection
capability of TRR~\cite{hassan2021uncovering}. First, the limited slack time in
\trfc{} puts a hard constraint on the number of \hht{TRR-induced} refreshes.
\hht{This restriction results in insufficient protection to victim rows,
especially when DRAM becomes more RowHammer vulnerable with DRAM technology
scaling.} Second, \hht{TRR-induced refresh takes} place only at a certain rate
(i.e., at most on every \cmdrefresh{}
). Recent
works~\cite{hassan2021uncovering,frigo2020trrespass,jattke2022blacksmith,deridder2021smash}
demonstrate a variety of \hht{new} RowHammer access patterns that circumvent the
TRR protection in chips of all \hht{three} major DRAM vendors, proving that the
existing DRAM interface is not well suited to enable strong RowHammer protection
\hht{especially as RowHammer becomes a bigger problem with DRAM technology
scaling}.

\hht{We use \mech{} to develop two maintenance mechanisms that overcome the
limitations of the existing TRR mechanisms by initiating victim row refresh
within the DRAM chip.}

\subsubsection{\hht{Probabilistic} RowHammer Protection}
\label{subsubsec:srp_para}

\begin{figure}[!h]
    \centering
    \includegraphics[width=\linewidth]{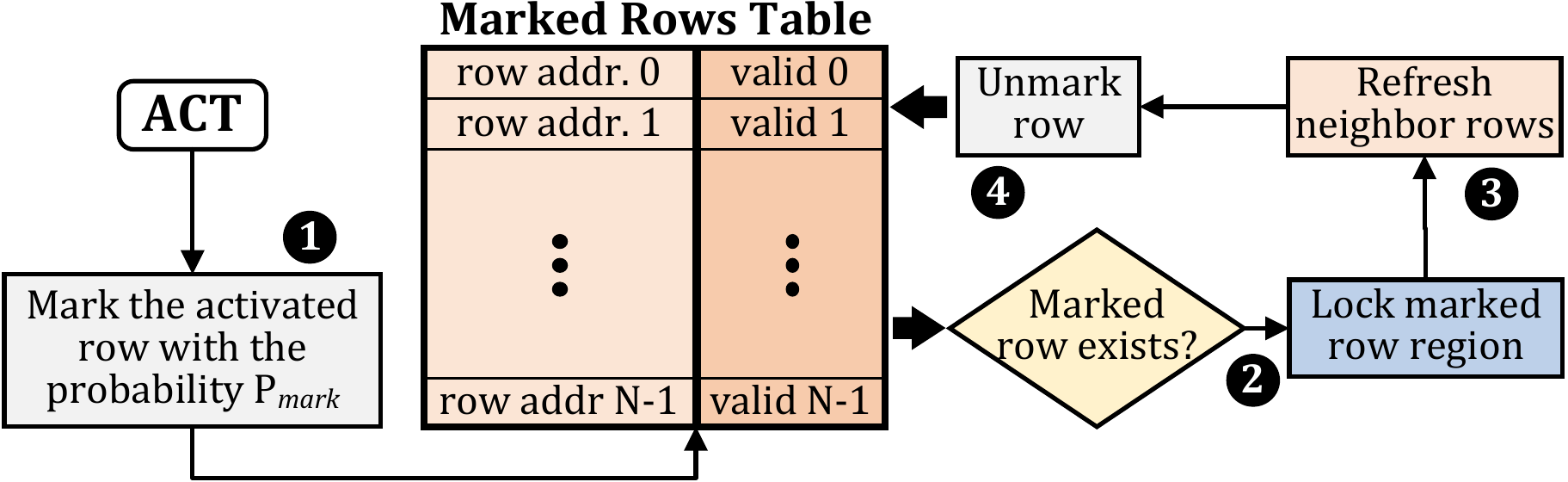}
    \caption{Probabilistic RowHammer Protection (\prp{}) Operation.}
    \label{fig:smd_para}
\end{figure}

Inspired by PARA~\cite{kim2014flipping}, we implement an in-DRAM maintenance
mechanism called \hht{Probabilistic RowHammer Protection (\prp{})}. \hht{The
high level idea is to refresh the nearby rows of an activated row with a small
probability. PARA is proposed as a mechanisms in the MC, which makes it
difficult to adopt since victim rows are not always known to the MC. \mech{}
enables us to overcome this issue by implementing the PARA-inspired \prp{}
mechanism completely within the DRAM chip. In addition, \prp{} avoids explicit
\cmdact{} and \cmdpre{} commands to be sent over the DRAM bus.}
Fig.~\ref{fig:smd_para} illustrates the operation of \prp{}. 

On a DRAM row activation, \prp{} marks the activated row as an aggressor with
\hht{a small} probability of $P_{mark}$~\circled{1}. \hht{It} marks \hht{an
aggressor} row using a \hht{per-bank} \emph{Marked Rows Table (MRT)}, that
contains an entry for each lock region in the bank. \hht{An} entry consists of
the \hht{marked row address} and a \emph{valid} bit. The bit length of the
address depend on the size of a lock region. For example, an MRT entry has a
13-bit address field when the lock region size is $8192$ rows. When MRT contains
a marked row, \prp{} locks the corresponding region~\circled{2} and refreshes
the neighbor rows of the marked row~\circled{3}. \hht{This step can easily
accommodate blast radius~\cite{yaglikci2021blockhammer} and any address
scrambling.} Once the neighbor rows are refreshed, \prp{} \hht{unlocks the}
region and unmarks the row in MRT~\circled{4}.

\subsubsection{\prp{} with Aggressor Row Detection}
\label{subsubsec:srp_blockhammer}

\hht{\prp{} refreshes victim rows with a small probability on every row
activation, even does so for a row that has been activated only a few times.
This \hht{results} in unnecessary victim refreshes, especially for high
$P_{mark}$ values that strengthen the RowHammer protection as DRAM cells become
more vulnerable to RowHammer with technology scaling.}

\hht{We propose \prpplus{}, which detects potential aggressor rows and
probabilistically refreshes only their victim rows.} The key idea of \prpplus{}
is to track frequently-activated rows in a DRAM bank and refresh the neighbor
rows of these rows using the region locking mechanism of \mech{}. 

\begin{figure}[!h]
    \centering
    \includegraphics[width=\linewidth]{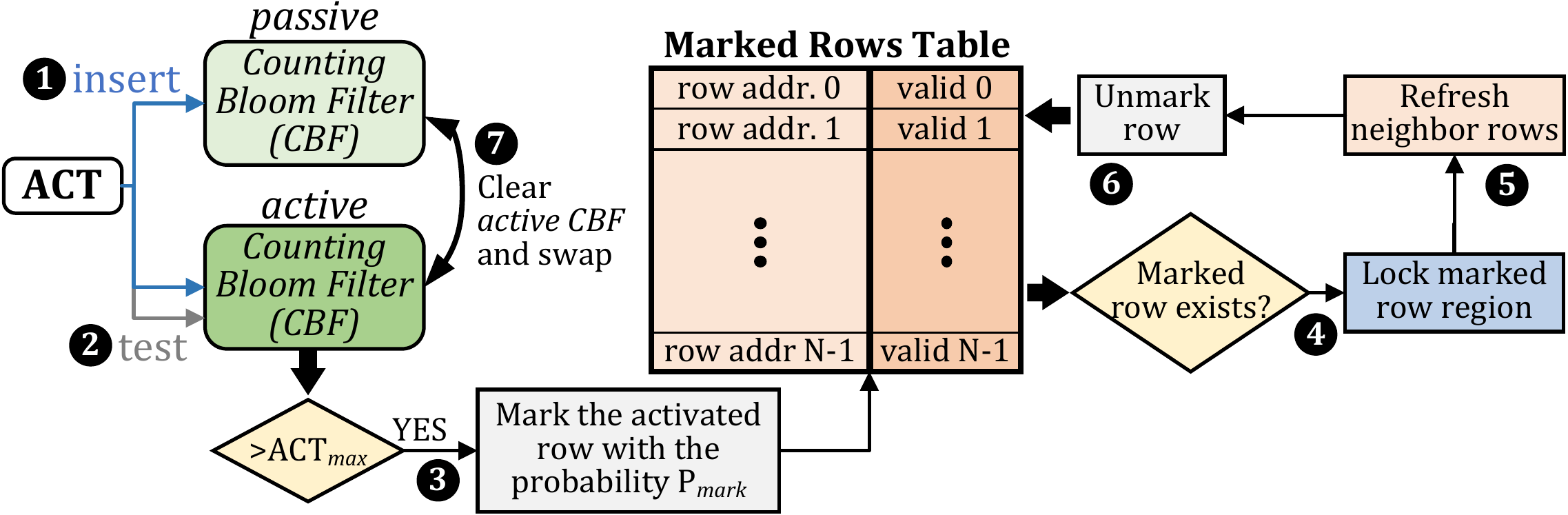}
    \caption{\prpplus{} Operation.}
    \label{fig:smd_blockhammer}
\end{figure}

\prpplus{} tracks frequently-activated rows, within a rolling time window of
length $L_{RTW}$, using two Counting Bloom Filters (CBF)~\cite{fan1998summary}
that operate in time-interleaved manner. CBF is a Bloom Filter variant that
\hht{represents the upperbound for} the number of times an element is inserted
into the CBF. \hht{We refer the reader to prior work for background on
CBFs~\cite{yaglikci2021blockhammer,pontarelli2016improving}.}
Depicted in Fig.~\ref{fig:smd_blockhammer}, \prpplus{} operation is based on 1)
detecting a row that has been activated more than $ACT_{max}$ times within the
most recent $L_{RTW}$ interval and 2) refreshing the neighbor rows of this row.
\hht{$ACT_{max}$ must be set according to the minimum hammer count needed to
cause a RowHammer bit flip in a given DRAM chip. A recent work shows that $4.8K$
activations can cause bit flips in LPDDR4 chips~\cite{kim2020revisiting}. We
conservatively set $ACT_{max}$ to $1K$. $L_{RTW}$ must be equal to or larger
than the refresh period in order to track all activations that happen until rows
get refreshed by regular refresh operations. We set $L_{RTW}$ equal to the
refresh period.}

\hht{Initially, one of the CBFs is in \emph{active} mode while the other is in
\emph{passive} mode.} When activating a row, \prpplus{} inserts the address of
the activated row to \hht{both} CBFs~\circled{1}. \hht{Then}, \prpplus{} tests
the active CBF to check if the accumulated insertion count exceeds the
$ACT_{max}$ threshold~\circled{2}. If \hht{so}, \prpplus{} marks the row in the
\emph{Marked Rows Table} to refresh its neighbor rows with \hht{the}
probability of $P_{mark}$~\circled{3}. \prpplus{} does not always mark the row
because it is impossible to reset \emph{only} the counters that correspond to
the marked row address in the CBFs. \hht{If always marked, a} subsequent
activation of the same row \hht{would again cause} the row to be marked,
\hht{leading to unnecessary neighbor row refresh} until all CBF counters are
reset\hht{, which happens} at the end of the $\frac{L_{RTW}}{2}$ interval.
\hht{After a row is marked, steps~\circled{4}-~\circled{6} are same as
steps~\circled{2}-~\circled{4} in \prp{}. Finally, at the end of an
$\frac{L_{RTW}}{2}$ interval,}
\prpplus{} clears the active CBF and swaps the two CBFs to continuously track
row activations within the most recent $L_{RTW}$ window~\circled{7}.

\begin{figure}[!h]
    \begin{yellowb}
    \centering
    \includegraphics[width=\linewidth]{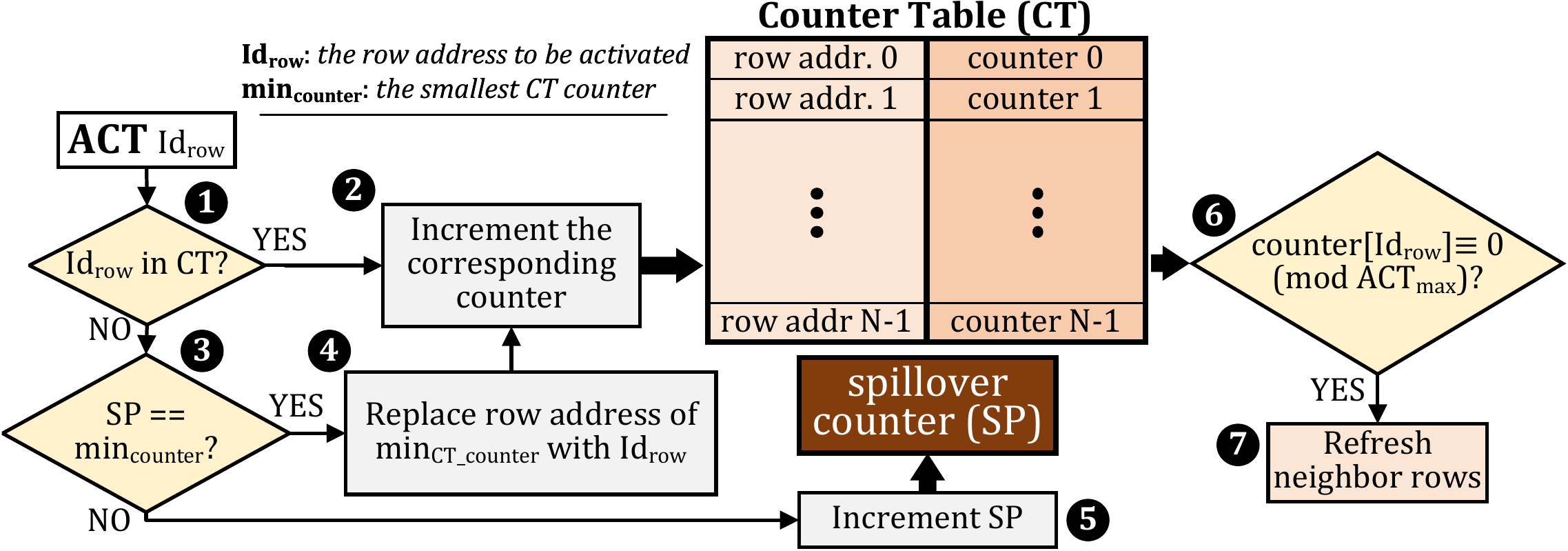}
    \end{yellowb}
    \caption{\drp{} Operation.}
    \label{fig:smd_graphene}
\end{figure}

\begin{yellowb_break}

\subsubsection{Deterministic RowHammer Protection}
\label{subsubsec:smd_graphene}

\prp{} and \prpplus{} are probabilistic mechanisms that provide
statistical security against RowHammer attacks with $P_{mark}$ being the
security parameter. On an extremely security-critical system, a deterministic
RowHammer protection mechanisms can guarantees mitigation of RowHammer bit
flips at all times.

We use \mech{} to implement \drp{}, a deterministic RowHammer protection
mechanism based on the Graphene~\cite{park2020graphene} mechanism, which uses
the Misra-Gries algorithm~\cite{misra1982finding} for detecting frequent
elements in a data stream to keep track of frequently activated DRAM rows.
Graphene is provably secure RowHammer protection mechanism with zero chance to
miss an aggressor row activated more than a predetermined threshold. Different
from Graphene, we implement \drp{} completely within a DRAM chip by taking
advantage of \mech{}, whereas Graphene requires the memory controller to issue
neighbor row refresh operations when necessary.

The key idea of \drp{} is to maintain a per-bank \emph{Counter Table (CT)} to
track the $N$ most-frequently activated DRAM rows within a certain time interval
(e.g., refresh period of \trefw{}). Fig.~\ref{fig:smd_graphene} illustrates the
operation of \drp{}.

\end{yellowb_break}

\begin{yellowb_break}

When \drp{} receives an \cmdact{}, it check if the activated row address
($Id_{row}$) is already in CT~\circled{1}. If so, \drp{} increments the
corresponding counter in CT by one~\circled{2}. If $Id_{row}$ is not in CT,
\drp{} finds the smallest counter value ($min_{counter}$) in CT and compares it
to the value of the \emph{spillover counter (SP)}~\circled{3}, which is
initially zero. If SP is equal to $min_{counter}$, \drp{} replaces the row
address corresponding to $min_{counter}$ in CT with $Id_{row}$~\circled{4} and
increments the corresponding CT counter by one~\circled{2}. If SP is smaller
than $min_{counter}$, \drp{} increments SP by one~\circled{5}. When a CT counter
is incremented in step~\circled{2}, \drp{} checks if the counter value is a
multiple of $ACT_{max}$~\circled{6}, which is the maximum number of times a row
can be activated without refreshing its neighbors. If so, \drp{} refreshes the
neighbors of $Id_{row}$~\circled{7}. To prevent the counters from overflowing,
\drp{} resets the CT counters and SP on every \trefw{} interval. 

To ensure that no row is activated more than $ACT_{max}$ without refreshing its
neighbor rows, the number of CT counters ($N$) must be configured according to the
following formula:

\begin{equation}
    N > \frac{ACT_{\trefw}}{ACT_{max}} - 1
\end{equation}

\noindent where $ACT_{\trefw}$ is the maximum number of activations that the
memory controller can perform within a \trefw{} interval in a single bank.

We refer the reader to~\cite{park2020graphene} for more details on the Graphene
mechanism and its security proof. The operation of \drp{} is similar to the
operation of Graphene with the difference of \drp{} being implemented completely
within a DRAM chip, which does not affect the underlying operation, and thus the
security proof of Graphene applies to \drp{}.

\end{yellowb_break}

\subsection{Use Case 3: Memory Scrubbing}
\label{subsec:smd_ecc_scrubbing}

To mitigate the increasing bit error rates that are mainly caused by the
continued DRAM technology scaling, DRAM vendors equip their DRAM chips with
on-die Error Correction \hht{Codes}
(ECC)~\cite{patel2019understanding,patel2020bit,nair2016xed,kang2014co}. 
%
On-die ECC is designed to correct a single bit error assuming that a failure
mechanism is unlikely to incur more than one bit error in a
codeword~\cite{patel2020bit}. However, even \hht{when} the assumption always
holds, a failure mechanism can gradually incur two or more bit \hht{errors} over
time. A widely-used technique for preventing the accumulation of an
uncorrectable number of bit errors is \emph{memory scrubbing}. Memory scrubbing
describes the process of periodically scanning the memory for bit errors in
order to correct \hht{them before more errors occur}.


\hht{We propose \mech{}-based Memory Scrubbing (\sms{}), which is an in-DRAM
maintenance mechanism that periodically performs scrubbing on DRAM chips with
on-die ECC. Using \mech{}, we implement \sms{} without any further changes to the
DRAM interface and the MC. Compared to conventional MC-based scrubbing, \sms{}
eliminates moving data to the MC by scrubbing within the DRAM chip, and thus
reduces the performance and energy overheads of memory scrubbing.}

\sms{} operation resembles the operation of \fr{}. Similar to \fr{}, \sms{}
maintains \emph{pending scrub counter}, \emph{lock region counter}, and
\emph{row address counter}. \sms{} \hht{increments} the pending scrub counter at
fixed intervals of $tScrub$. When the pending scrub counter is greater than
zero, \sms{} attempts to lock the region indicated by the \emph{lock region
counter}. After locking the region, \sms{} performs scrubbing operation on the
row indicated by the \emph{row address counter}. The scrubbing operation takes
more time than a refresh operation as performing scrubbing on a row consists of
three time consuming steps for each codeword in the row: 1) reading the
codeword, 2) performing ECC decoding and checking for bit errors, and 3)
encoding and writing back the new codeword into the row only when the decoded
codeword contains a bit error. Refreshing a row takes $tRAS+tRP \approx 50ns$,
whereas scrubbing a row takes $tRCD+128*tBL+tRP \approx \SI{350}{\nano\second}$
when no bit errors are detected.\footnote{\hht{We assume ECC decoding/encoding
hardware is fully pipelined. In case of a bit error, writing a corrected
codeword incurs $4*tBL = \SI{2.5}{\nano\second}$ additional latency.}}
Therefore, \sms{} keeps a region locked for much longer than \fr{}. When the
scrubbing operation is complete, \sms{} releases the lock region and increments
the \emph{lock region} and \emph{row address} counters, and decrements the
\emph{pending scrub counter} as in \fr{}.

\section{Experimental Methodology}
\label{sec:methodology}


We extend Ramulator~\cite{kim2015ramulator,ramulatorgithub} to implement
\hht{and evaluate} the six \mech{} maintenance mechanisms \hht{(\fr{}, \vr{},
\prp{}, \prpplus{}, \drp{}, and \sms{})} that we describe in
\cref{sec:maintenance_mechanisms}. We use DRAMPower~\cite{drampowergithub,
chandrasekar2011improved} to evaluate DRAM energy consumption. We use Ramulator
in CPU-trace driven mode \hht{executing} traces of representative sections of
\hht{our workloads collected with} a custom Pintool~\cite{luk2005pin}. \hht{We}
warm-up the caches by fast-forwarding 100 million instructions. We simulate each
\hht{representative trace} for 500 million instructions (for \hht{multi}-core
simulations, we simulate until each core executes at least 500 million
instructions).

We use the system configuration provided in Table~\ref{table:system_config} in
our evaluations. Although our evaluation is based on DDR4 DRAM, the
modifications required to enable \mech{} can be adopted in other DRAM standards,
as we explain in \cref{sec:hw_overhead}.

\begin{table}[h!] \caption{Simulated system
    configuration.}
    \vspace{-6pt}
    \centering \renewcommand{\arraystretch}{1.4}
    \resizebox{\linewidth}{!}{
    \begin{scriptsize}
    \begin{tabular}{m{2.2cm} m{4.8cm}}
        \hline
        \textbf{Processor} & \SI{4}{\giga\hertz} \& 4-wide issue CPU core, 1-4
        cores, 8~MSHRs/core, 128-entry instruction window\\
        \hline
        \textbf{Last-Level Cache} & \SI{64}{\byte} cache-line, 8-way associative, \SI{4}{\mebi\byte}/core \\
        \hline
    \textbf{Memory Controller} & \makecell[l]{64-entry read/write request
        queue,\\ FR-FCFS-Cap~\cite{mutlu2007stall}} \\
        \hline
        \textbf{DRAM} & DDR4-3200~\cite{ddr4}, \SI{32}{\milli\second} refresh period,
            4~channels, 2~ranks, 4/4~bank groups/banks, 128K-row bank, 512-row
            subarray, \SI{8}{\kibi\byte} row size \\
        \hline
    \end{tabular}
    \end{scriptsize}
    } 
\label{table:system_config}
\end{table}

\textbf{Workloads.} We evaluate 44 single-core applications from four benchmark
suites: SPEC CPU2006~\cite{spec2006}, TPC~\cite{tpc}, STREAM~\cite{stream}, and
MediaBench~\cite{fritts2005mediabench}. We classify the workloads in three
groups based on their memory intensity, which we measure using
\hht{misses-per-kilo-instructions} (MPKI) in the last-level cache (LLC). The
low\hht{, medium, and high} memory intensity \hht{groups consist} of workloads
with $MPKI < 1$, $1 \leq MPKI \leq 10$, and $MPKI \geq 10$. We randomly combine
single-core workloads to create multi-programmed workloads. \hht{Each
multi-programmed workload group}, \texttt{4c-low}, \texttt{4c-medium}, and
\texttt{4c-high}, contains $20$ four-core workloads with \hht{low, medium, and
high} memory intensity.

\textbf{Metrics.} We use Instructions Per Cycle (IPC) to evaluate the
performance of single-core workloads. For multi-core workloads, we evaluate the
system throughput using the weighted speedup
metric~\cite{eyerman2008system,snavely2000symbiotic, michaud2012demystifying}.
\section{Evaluation}
\label{sec:evaluation}

We evaluate the performance and energy efficiency of \mech{}-based maintenance
mechanisms.

\subsection{Single-core Performance}
\label{subsec:single_core_perf}

Fig.~\ref{fig:single_core_perf} shows the speedup of single-core workloads when
using the following maintenance mechanisms: 1) \fr{}, 2) \vr{}, 3) \prp{}, 4)
\prpplus{}, 5) \drp{}, 6) \sms{}, and 7) the combination of \vr{}, \prp{}, and
\sms{} \hht{(\mech{}-Combined)}.\footnote{\hht{We individually evaluate \prp{},
\prpplus{}, \drp{}, and \sms{} using \fr{} as a refresh mechanism since \fr{} is
very simple and easy to implement.}} \bgyellow{In \mech{}-Combined, we prefer
\texttt{SMD-PRP} over \texttt{SMD-PRP+} and \texttt{SMD-DRP} to keep the DRAM
chip area overhead minimal.} \hht{Based on~\cite{liu2012raidr}, for \vr{}, we
conservatively assume 0.1\% of rows in each bank need to be refreshed every
\SI{32}{\milli\second} while the rest retain their data correctly for
\SI{128}{\milli\second} and more. \vr{} uses a $8K$-bit Bloom Filter with $6$
hash functions. \prp{} and \prpplus{} refresh the victims of an activated row
with a high probability, i.e., $P_{mark} = 1\%$. \sms{} operates with an
aggressive 5-minute scrubbing period.}\footnote{\hht{We analyze \prp{},
\prpplus{}, \drp{}, and \sms{} at various configurations and find that their
overheads are relatively low even at more aggressive settings.}}
We calculate the speedup compared to DDR4 DRAM with \hht{a nominal}
\SI{32}{\milli\second} refresh period. To demonstrate the maximum speedup
achievable by eliminating the entire DRAM refresh overhead,
Fig.~\ref{fig:single_core_perf} also \hht{includes} the performance of
hypothetical DRAM that does not require refresh (\emph{NoRefresh}). The figure
shows $22$ workloads that have at least 10 Last-Level Cache \emph{Misses per
Kilo Instruction (MPKI)}. \hht{\emph{GMEAN-22} provides the geometric mean of
the $22$ workloads and \emph{GMEAN-ALL} of all $44$ single-core workloads that
we evaluate.}

\ifisthesis
    \afterpage{%
        \clearpage
        \begin{landscape}
            \begin{figure*}[!t]
                \centering
                \includegraphics[width=\linewidth]{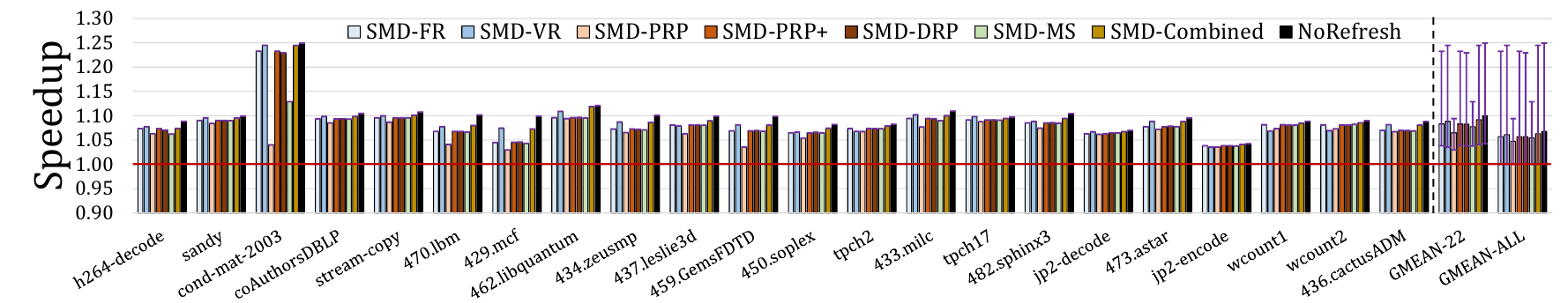}
                \caption{Single-core speedup \hht{over a DDR4 system}}
                \label{fig:single_core_perf}
            \end{figure*}
        \end{landscape}
        \clearpage
    }
\fi

We make five key observations. First, \fr{} provides 5.7\% average speedup (up
to 23.2\%) over conventional DDR4 refresh (84.1\% of the speedup that
\emph{NoRefresh} provides). Although \fr{} and DDR4 have similar average latency
to refresh a single DRAM row, \fr{} outperforms DDR4 refresh due to two main
reasons: 1) enabling \hht{concurrent access} and \hht{refresh of} two different
\emph{lock regions} within a DRAM bank and 2) \hht{completely} eliminating the
\cmdrefresh{} commands on the DRAM commands bus to reduce the command bus
utilization. \hht{\fr{} requires very low area overhead and no information with
respect to DRAM cell characteristics, yet provides significant speedup.} 

Second, \vr{} \hht{provides} 6.1\% average speedup (up to 24.5\%), achieving
89.9\% of the speedup that \emph{NoRefresh} provides. \vr{} provides \hht{higher
speedup} due to eliminating unnecessary refresh operations to
\hht{retention-strong rows. Compared to \fr{}, \vr{}'s performance benefits are
limited because \fr{} significantly mitigates the refresh overhead by
overlapping refresh operations with accesses to other lock regions, leaving
small opportunity for further improvement by skipping refreshes to
retention-strong rows.}

Third, both \fr{} and \vr{} \hht{perform} close to the hypothetical
``NoRefresh'' DRAM. On average, \fr{}/\vr{} \hht{provide} 83.8\%/86.8\% of the
speedup that ``NoRefresh'' \hht{provides} compared to the DDR4 baseline.

Fourth, although the \hht{overheads} of the \prp{}, \drp{}, and \sms{}
mechanisms partially negate the \hht{performance benefits of} \fr{}, \hht{they
still} achieve average \hht{(maximum)} speedup of 4.7\%/5.7\%/5.4\%
\hht{(9.4\%/23.2\%/12.9)} over the DDR4 baseline\hht{, when integrated on top of
\fr{}}.

Fifth, \mech{}-Combined improves performance by 6.3\% (up to 24.5\%),
\hht{showing that \mech{} enables robust RowHammer protection and frequent
(5-minute) memory scrubbing while still improving system performance}.

\subsection{Multi-core Performance}
\label{subsec:multi_core_perf}

\begin{figure}[!h]
    \centering
    \includegraphics[width=\linewidth]{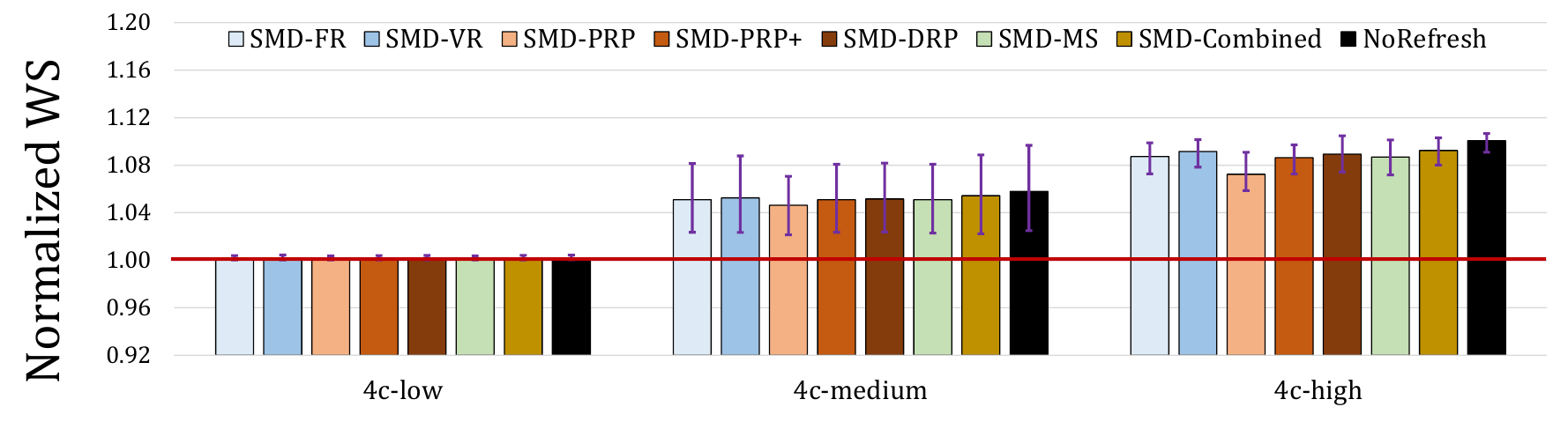}
    \caption{Four-core weighted speedup}
    \label{fig:multi_core_perf}
\end{figure}

Fig.~\ref{fig:multi_core_perf} shows weighted speedup (normalized to the
weighted speedup of the DDR4 baseline) for \hht{60 four-core workloads (20 per
memory intensity level)}. \hht{The bars (error lines) represent the average
(minimum and maximum) weighted speedup across the 20 workloads in the
corresponding group.} We make three major observations.

First, we observe the largest \hht{average} speedup in \hht{memory-intensive}
four-core workloads, i.e., \texttt{4c-high}. \fr{}/\vr{} improve \hht{average}
speedup by 8.7\%/9.2\% \hht{(9.9\%/10.2 maximum)}, which corresponds to
86.8\%/91.2\% of the speedup that \hht{hypothetical} ``NoRefresh'' achieves over
DDR4.

Second, combining \vr{}+\drp{}+\sms{} provides 9.1\% average \hht{(10.1\%
maximum)} improvement in weighted speedup, which indicates that \drp{} and \sms{}
incur \hht{only small} performance overhead on top of \vr{}.

Third, \hht{the performance benefits become smaller as the memory intensity of
the workloads decrease. Specifically, \texttt{4c-medium} achieves lower speedup
than \texttt{4c-high}, and \texttt{4c-low} achieves lower speedup than
\texttt{4c-medium}.}

\hht{\textbf{Performance Summary.}} The new DRAM refresh mechanisms \fr{} and
\vr{} significantly improve system performance, achieving \hht{speedup} close to
that provided by the hypothetical ``NoRefresh'' DRAM. \hht{\mech{} enables
efficient and robust RowHammer protection (\prp{},\prpplus{}, and \drp{}) and
memory scrubbing (\sms{}) that still provide significant speedup when integrated
with \fr{}. Overall, \mech{} enables DRAM that is both faster and more robust
than DDR4.}

\subsection{Energy Consumption}
\label{subsec:energy_results}

\begin{figure}[!h]
    \centering
    \includegraphics[width=\linewidth]{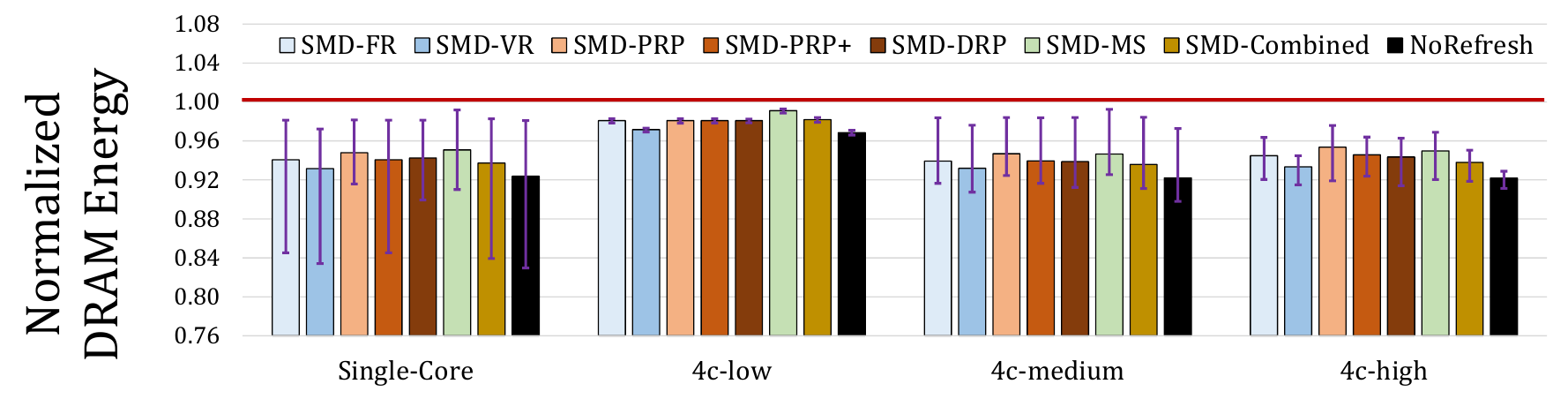}
    \caption{DRAM energy consumption}
    \label{fig:dram_energy}
\end{figure}

Fig.~\ref{fig:dram_energy} shows the average DRAM energy consumption
normalized to the DDR4 baseline for single- and four-core workloads. We make
three major observations.

First, \fr{} reduces DRAM energy for both single- and four-core workloads
\hht{(by 5.9\% and 5.5\% on average) over DDR4} mainly due to reduction in
execution time.

Second, \vr{} reduces DRAM energy by 6.8\%/6.7\% on average for single-core and
\texttt{4c-high} workloads by \hht{especially} eliminating \hht{some}
unnecessary refresh operations. The maximum DRAM energy reduction that can be
achieved by completely eliminating DRAM refresh operation is 7.8\% as indicated
by ``NoRefresh'' DRAM.

Third, \hht{\mech{}-Combined reduce DRAM energy by 6.1\% for \hht{both}
single-core and \texttt{4c-high} workloads. This shows that the energy overhead
of \drp{} and \sms{} are small.}

The \hht{new \mech{}-based refresh} mechanisms eliminate \hht{most} of the
energy overhead of DRAM refresh. \mech{}-based RowHammer protection and memory
scrubbing \hht{improve DRAM reliability and security with} low DRAM energy cost.


\subsection{Sensitivity Studies}
\label{subsec:sensitivity_studies}

We analyze the performance \hext{and DRAM energy} effects of different \mech{}
configuration parameters.

\subsubsection{DRAM Refresh Period}
\label{subsec:ref_period_sensitivity}

\begin{figure}[!h]
    \begin{subfigure}{\linewidth}
        \begin{yellowb}
            \centering
            \includegraphics[width=\linewidth]{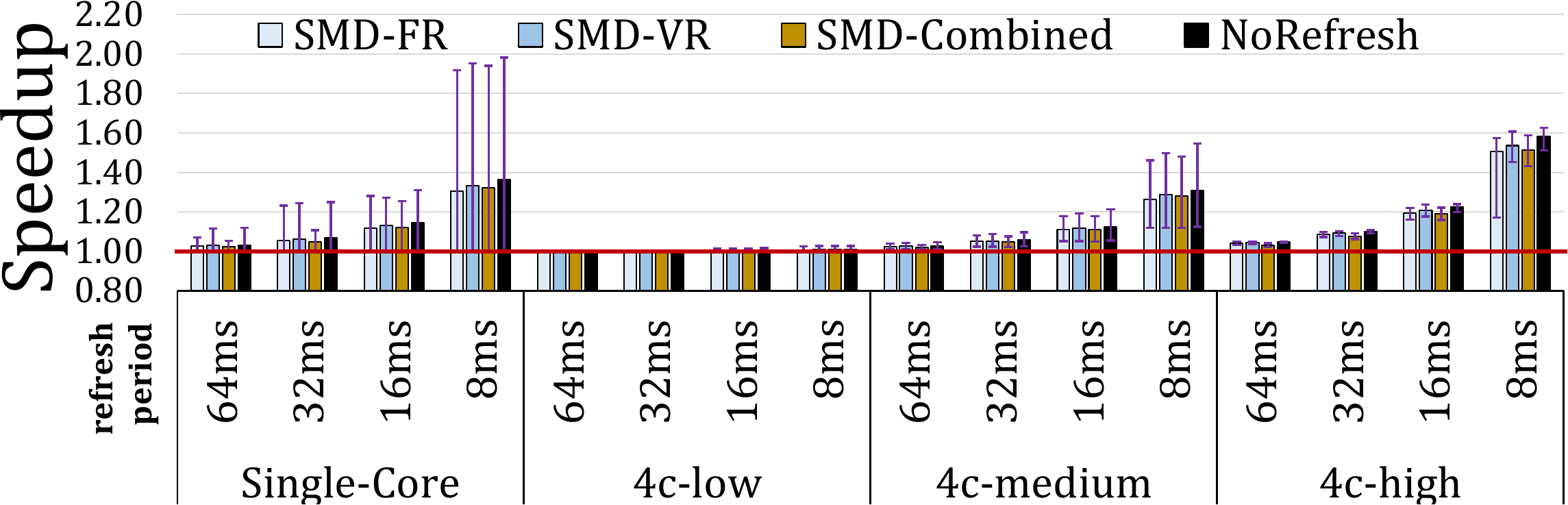}
        \end{yellowb}
    \end{subfigure}
    \begin{subfigure}{\linewidth}
        \begin{yellowb}
            \centering
            \includegraphics[width=\linewidth]{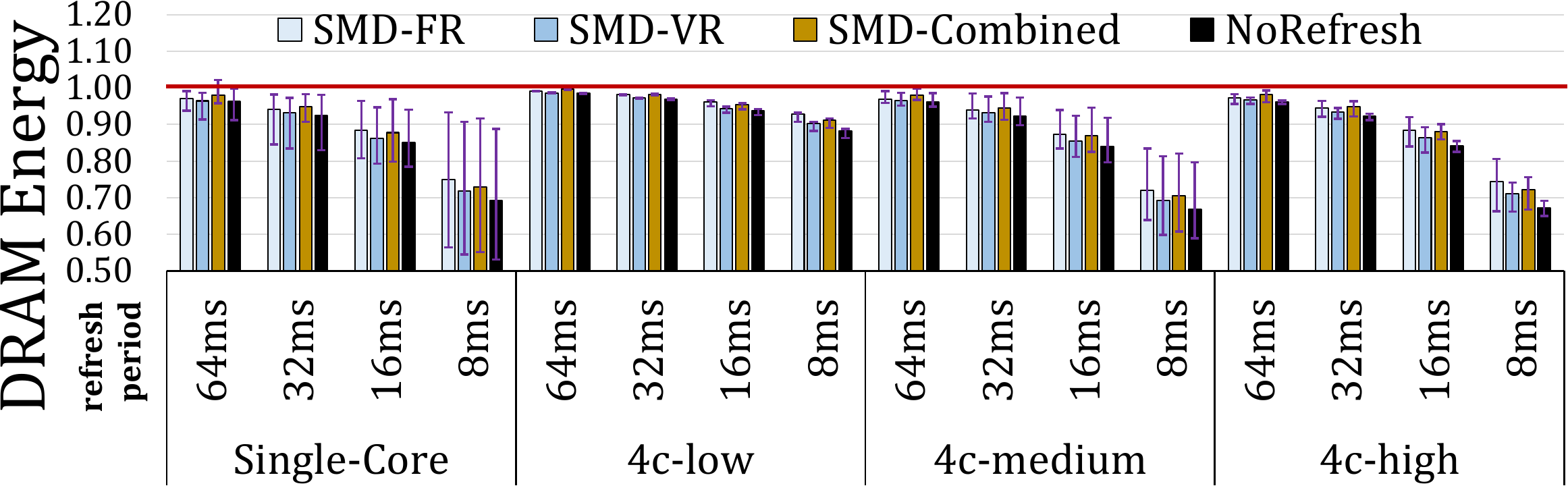}
        \end{yellowb}
    \end{subfigure}

    \caption{Sensitivity to Refresh Period}
    \label{fig:refresh_period_sweep}
\end{figure}

Fig.~\ref{fig:refresh_period_sweep} plots the speedup \hext{and energy
reduction} that \fr{}, \hht{\vr{}, and \mech{}-combined achieve} over DDR4 for
different refresh periods across single- and four-core workloads. We make
\hht{three} key observations. First, the performance \hhf{and energy} benefits
of \fr{}/\vr{} increase as the refresh period reduces, achieving
50.5\%/\hht{53.6\%} speedup \hhf{and 25.6\%/29.0\% DRAM energy reduction on
average} at \SI{8}{\milli\second} refresh period across \texttt{4c-high}
workloads. Thus, \hht{\mech{}-based refresh mechanisms} will become even more
\hht{beneficial} to employ in future DRAM chips that \hht{are expected to}
require more frequent
refresh~\cite{liu2012raidr,kang2014co,liu2013experimental}. \hht{Second,
\mech{}-combined provides 51.5\% speedup \hhf{and 27.9\% DRAM energy reduction
on average} across \texttt{4c-high} at \SI{8}{\milli\second} refresh period,
showing that the overheads of \mech{}-based RowHammer protection and memory
scrubbing mechanisms remain low even at high refresh rates. Third, all
\mech{}-based maintenance mechanism eliminate most of the performance overhead
of refresh across all refresh periods (e.g., \mech{}-combined reaches 95.6\% of
the hypothetical \emph{NoRefresh} DRAM at \SI{8}{\milli\second}).}

\hht{We conclude that \mech{}-based maintenance mechanisms improve DRAM
reliability and security while greatly reducing the overhead of DDR4 refresh.
Their performance benefits are expected to increase for future DRAM chips that
are likely to require high refresh rates.}



\begin{figure}[!h]
    \begin{subfigure}{\linewidth}    
        \begin{yellowb}
            \centering
            \includegraphics[width=\linewidth]{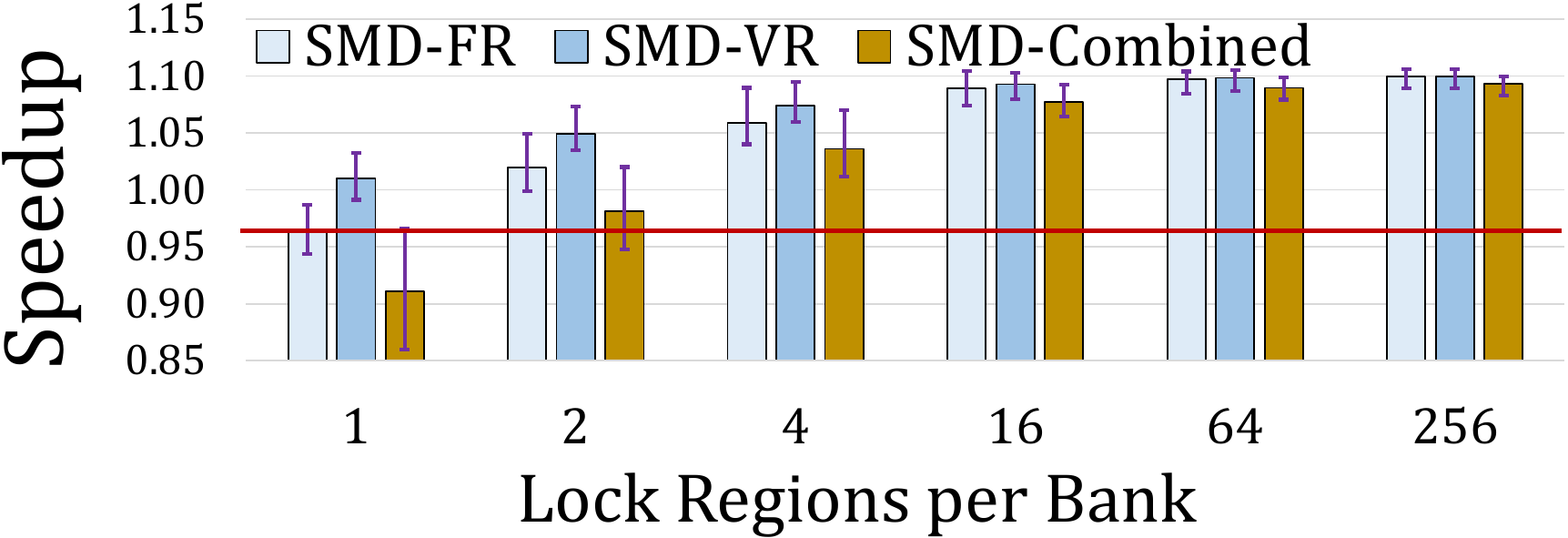}
        \end{yellowb}
    \end{subfigure}
    \begin{subfigure}{\linewidth}    
        \begin{yellowb}
            \centering
            \includegraphics[width=\linewidth]{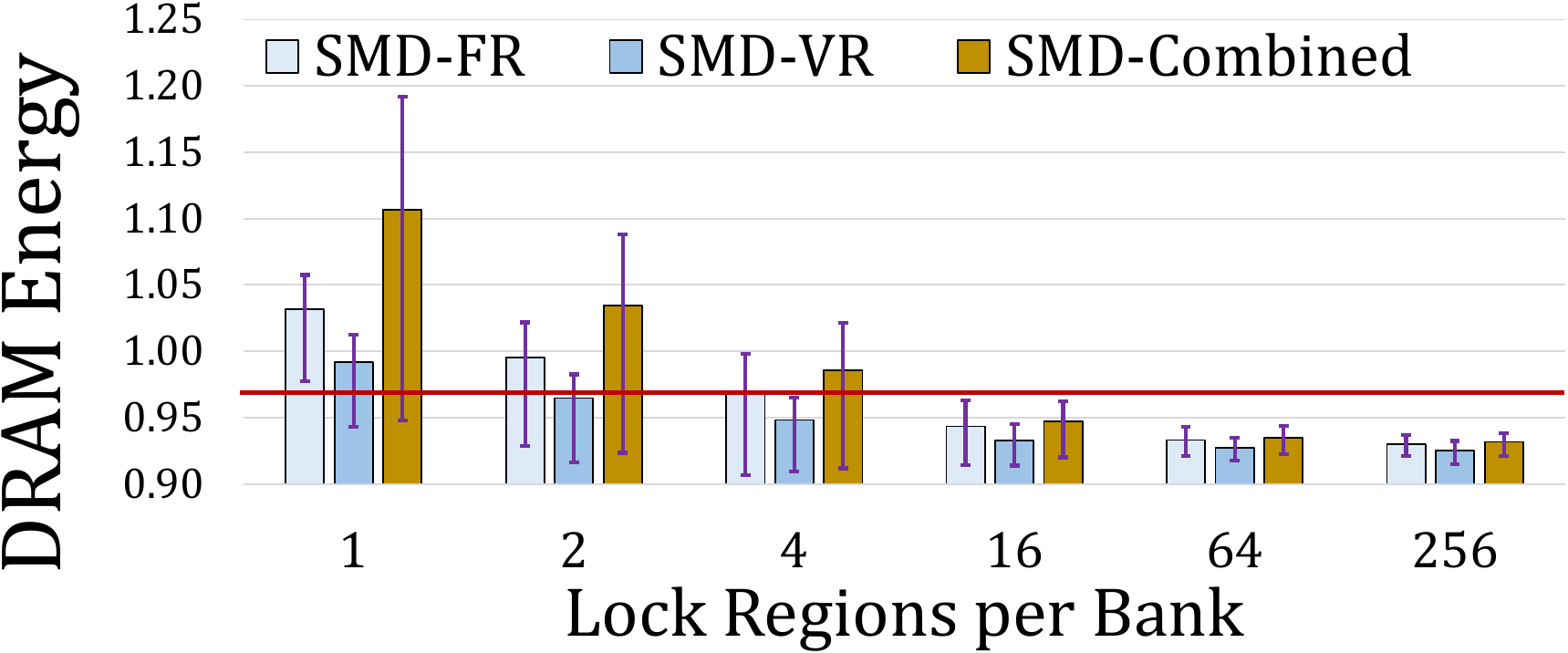}
        \end{yellowb}
    \end{subfigure}

    \caption{Sensitivity to Lock Region Size.}
    \label{fig:region_size_sweep}
\end{figure}

\begin{yellowb_break}
\subsubsection{Lock Region Size}
\label{subsec:region_size_sensitivity}

The size of a \mech{} lock region affects the performance and energy of
\mech-based maintenance mechanisms. Generally, it is desirable to have many
small-sized lock regions to enable locking smaller portions of the DRAM chip for
maintenance and allow accesses to non-locked regions. However, the DRAM chip
area overhead of \mech{} increases as the number of lock regions in a bank
increase. We analyze the performance and energy impact of different lock region
size configurations.

We use the following equation to calculate the number of lock regions per bank ($Num_{LRs/bank}$) by dividing the number of rows in a bank ($Num_{rows/bank}$) to the number of rows in a lock region ($Num_{rows/LR}$).

\vspace{-6pt}
\begin{equation}
    Num_{LRs/bank} = \frac{Num_{rows/bank}}{Num_{rows/LR}}
\end{equation}

Figure~\ref{fig:region_size_sweep} shows the speedup and DRAM energy savings
that \fr{}, \vr{}, and \mech{}-Combined achieve for different number of lock
regions per bank across \texttt{4c-high} workloads. We make two key
observations. First, with a single lock region per bank, \fr{}/\mech{}-Combined
on average perform 3.7\%/8.9\% worse than the DDR4 baseline, which leads to
3.2\%/10.7\% higher average DRAM energy consumption. This is because a
maintenance operation locks the entire bank, and thus the MC waits for \ARI{}
upon receiving an \actnack{} to issue another \cmdact{} to the same bank.
Although \vr{} suffers from the same issue, it still outperforms the baseline by
eliminating unnecessary refreshes to rows with high retention times.

\end{yellowb_break}

\begin{yellowb_break}

Second, when the lock regions are smaller, and thus a bank has four or more lock
regions, all mechanisms outperform the baseline for all \texttt{4c-high}
workloads and save DRAM energy. This shows that even a small number of lock
regions are sufficient to overlap the maintenance operation latency with
accesses to non-locked regions.

Third, the performance and energy benefits of increasing the number of lock
regions gradually diminish. We find that $16$ 8192-row lock regions incur low
area overhead and achieve a significant portion of the speedup that smaller lock
regions provide (e.g., \mech{}-Combined achieves 7.7\%/9.3\% speedup with 16/256
lock regions). Therefore, by default, we use $16$ lock regions per bank in our
evaluations. 

We conclude that \mech{} significantly reduces the overhead of DRAM maintenance
operations even with large lock regions.

\end{yellowb_break}

\begin{figure}[!h]
    \begin{subfigure}{\linewidth}
        \begin{yellowb}
            \centering
            \includegraphics[width=\linewidth]{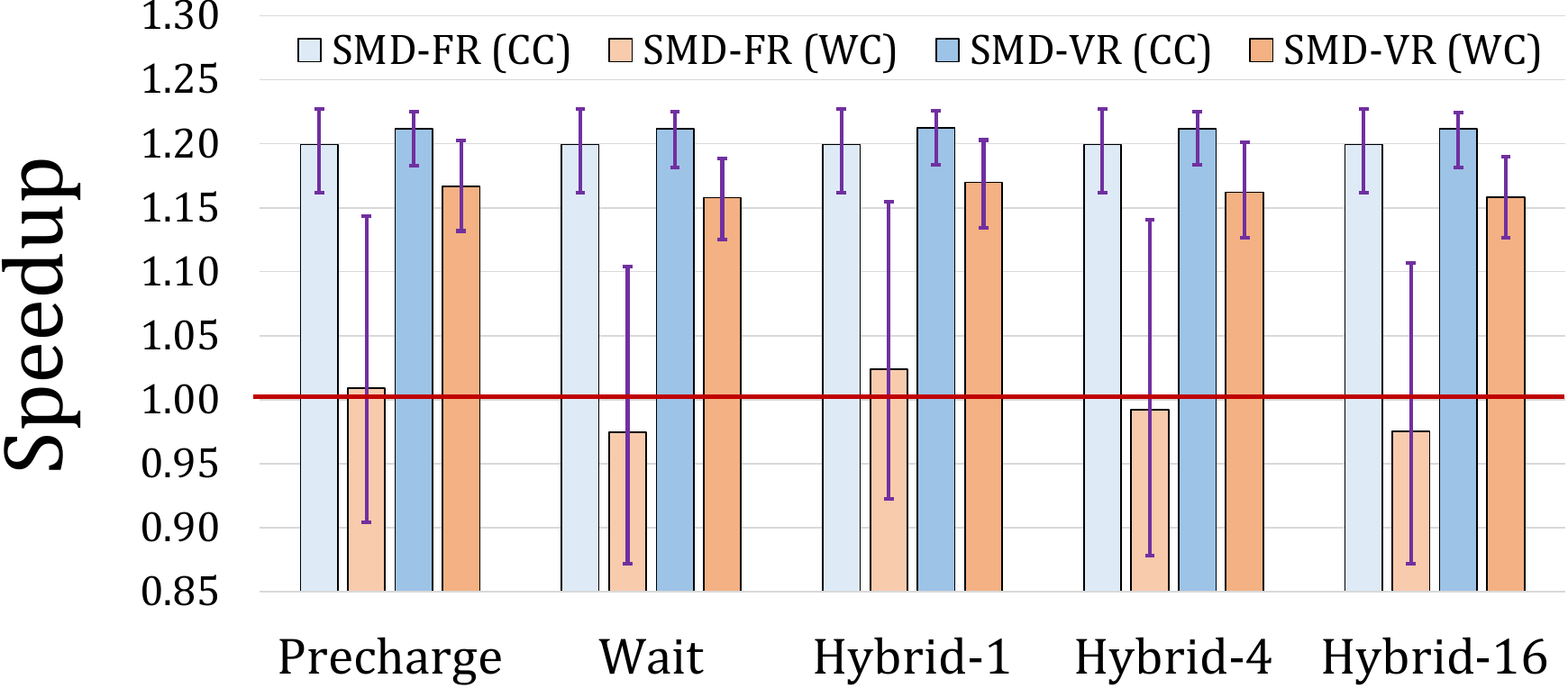}
        \end{yellowb}
    \end{subfigure}

    \begin{subfigure}{\linewidth}
        \begin{yellowb}
            \centering
            \includegraphics[width=\linewidth]{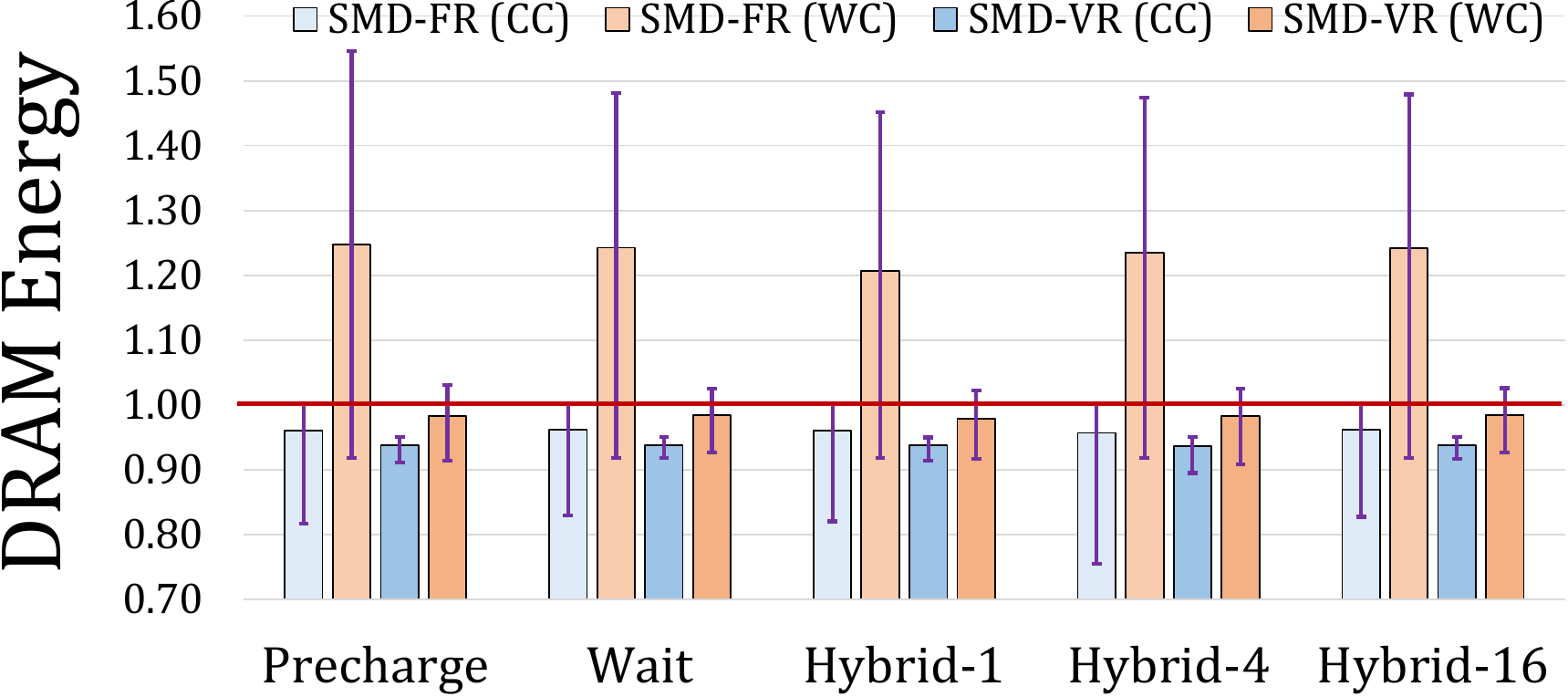}
        \end{yellowb}
    \end{subfigure}

    \caption{Comparison of different policies for handling refresh divergence across DRAM chips.}
    \label{fig:worst_case_ref_distrib}
\end{figure}

\begin{yellowb_break}
\subsubsection{\actnack{} Divergence Across Chips}
\label{subsec:ack_nack_divergence}


In Section~\ref{subsec:rejecting_acts}, we explain divergence in \mech{}
maintenance operations can happen when different DRAM chips in the same rank
perform maintenance operation at different times. Such a divergence leads to a
partial row activation when the activated row is in a locked region in some DRAM
chips but not in others. To handle partial row activations, we develop three
policies.

\noindent\textbf{Precharge.} With the \emph{Precharge} policy, the MC
issues a \cmdpre{} command to close the partially activated row when some DRAM
chips send \actnack{} but others do not. After closing the partially activated
row, the MC can attempt to activate a row from a different lock
region in the same bank.

\noindent\textbf{Wait.} With the \emph{Wait} policy, the MC issues
multiple \cmdact{} commands until a partially activated row becomes fully
activated. When some chips send \actnack{} for a particular \cmdact{} but others
do not, the MC waits for \ARI{} and issues a new \cmdact{} to
attempt activating the same row in DRAM chips that previously sent \actnack{}.

\noindent\textbf{Hybrid.} We also design a \emph{Hybrid} policy, where the memory
controller uses the \emph{Precharge} policy to close a partially activated row
if the request queue contains $N$ or more requests that need to access rows in
different lock regions in the same bank. If the requests queue has less than $N$
requests to different lock regions, the MC uses the \emph{Wait}
policy to retry activating the rest of the partially activated row.

In Fig.~\ref{fig:worst_case_ref_distrib}, we compare the performance \hext{and
energy savings} of \fr{} and \vr{} when using the three \actnack{} divergence
handling policies \hext{across} \texttt{4c-high} workloads. \hext{The plots show
results for the common-case (CC) and worst-case (WC) scenarios with regard to
when maintenance operations happen across different \mech{} chips in the same
rank. In the common-case scenario, the DRAM chips generally refresh the same row
at the same time due to sharing the same DRAM architecture design. However,
refresh operations in some of the DRAM chips may still diverge during operation
depending on the refresh mechanism that is in use. For example, \vr{} refreshes
retention-weak rows, whose locations may differ across the DRAM chips, at a
higher rate compared to other rows, resulting in divergence in refresh
operations across the DRAM chips in a rank. In the worst-case scenario, we
deliberately configure the DRAM chips to refresh different rows at different
times. For this, we 1) delay the first refresh operation in $chip_i$ by $i
\times l{ref}$, where $0 \leq i < Num{chips/rank}$ and $l_{ref}$ is the latency
of a single refresh operation, and 2) set the \emph{Lock Region Counter (LRC)}
of $chip_i$ to $i$.}

\end{yellowb_break}

\begin{figure}[!h]
    \begin{subfigure}{\linewidth}
        \begin{yellowb}
            \centering
            \includegraphics[width=.9\linewidth]{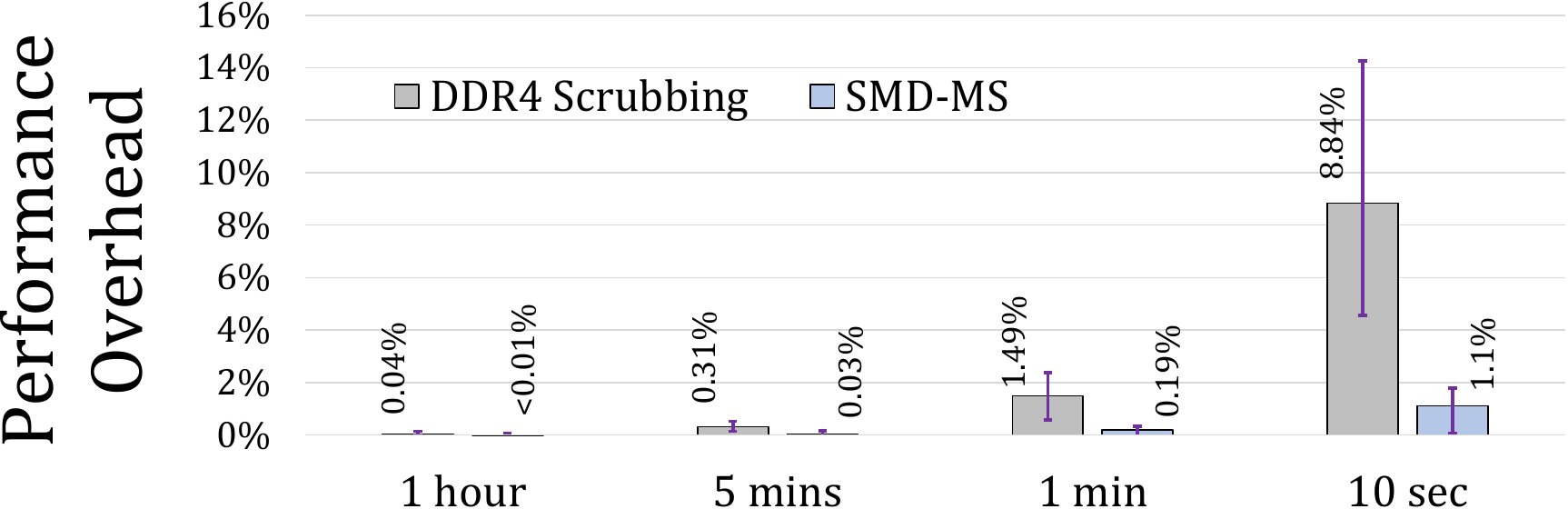}
        \end{yellowb}
    \end{subfigure}
    \begin{subfigure}{\linewidth}
        \begin{yellowb}
            \centering
            \includegraphics[width=.9\linewidth]{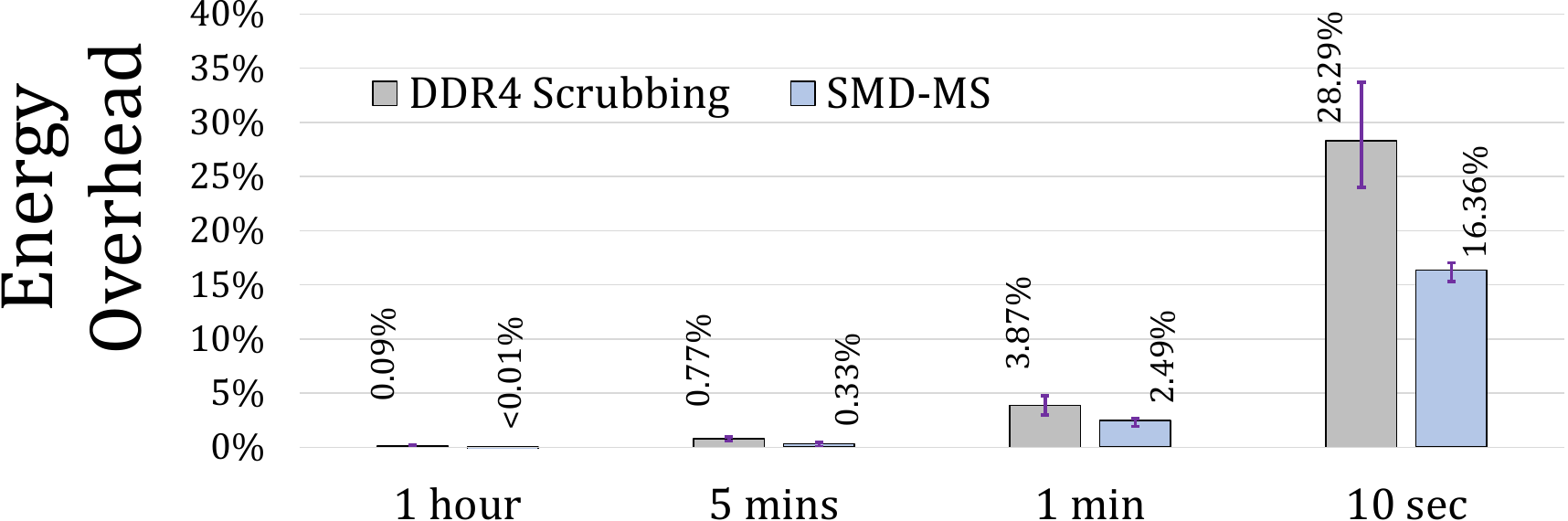}
        \end{yellowb}
    \end{subfigure}
    \caption{DDR4 Scrubbing vs. \sms{}.}
    \label{fig:scrubbing_comparison}
\end{figure}

\begin{yellowb_break}

We make three key observations. First, the performance \hext{and DRAM energy
consumption} only slightly varies across different \hext{divergence handling}
policies \hext{in both common-case and worst-case scenarios}. This is because,
after a partial row activation happens, a different row that is not in a locked
region in all off of the DRAM chips often does not exist in the memory request
queue. As a result, the \emph{Precharge} and \emph{Hybrid} policies perform
similarly to the \emph{Wait} policy. 

Second, \fr{} performs worse \hext{than the DDR4 baseline} in the worst-case
refresh distribution scenario (i.e., average slowdown of 2.6\% with the
\emph{Wait} policy). Certain individual workloads experience even higher
slowdown (up to 12.8\% \hext{with the \emph{Wait} policy}). The reason is that,
when $N$ different DRAM chips lock a region at different times, the total
duration during which the lock region is unavailable becomes $N$ times the
duration when all chips simultaneously refresh the same lock region. This
significantly increases the performance overhead of refresh operations. 

Third, \vr{} does not suffer \hext{much} from the divergence problem \hext{and
outperforms the DDR4 baseline even in the worst-case scenario. This is} because
\vr{} mitigates the DRAM refresh overhead by significantly reducing the number
of total refresh operations by exploiting retention-time variation across DRAM
rows. \hext{Thus, the benefits of eliminating many unnecessary refresh
operations surpass the overhead of \actnack{} divergence.}

We conclude that 1) although \fr{} suffers from noticeable slowdown for the
worst-case scenario, which should not occur in a well-designed system, it still
provides comparable performance to conventional DDR4 and 2) \hext{\vr{}
outperforms the baseline and saves DRAM energy even in the worst-case scenario}.

\subsubsection{Comparison to Conventional DRAM Scrubbing}
\label{subsec:conventional_dram_scrubbing}

Fig.~\ref{fig:scrubbing_comparison} compares the performance \hext{and energy
overheads} of \hext{conventional DDR4 scrubbing} to \sms{} across
\texttt{4c-high} workloads. In the figure, \emph{DDR4 Scrubbing}
\hext{represents the performance and energy overhead of conventional scrubbing
compared to DDR4 without memory scrubbing. Similarly, \sms{} represents the
performance and DRAM energy overhead of \mech{}-based scrubbing compared to
\fr{}}. 

\end{yellowb_break}

\begin{yellowb_break}

\hext{We make two observations. First,} both DDR4 scrubbing and \sms{} have
negligible performance overhead for scrubbing periods \hext{of 5 minutes and
larger} because scrubbing operations are infrequent at such periods.
\hext{Second,} DDR4 scrubbing causes up to 1.49\%/8.84\% average slowdown for 1
minute/10 second scrubbing period (up to 2.37\%/14.26\%), while \sms{} causes
only up to 0.34\%/1.78\% slowdown. DDR4 scrubbing has high overhead at low
scrubbing periods because moving data from DRAM to the MC to perform scrubbing
is inefficient compared to performing scrubbing within DRAM using \sms{}.
Scrubbing at high rates may become necessary for future DRAM chips as their
reliability characteristics continuously worsen. Additionally, mechanisms that
improve DRAM performance at the cost of reduced
reliability~\cite{qureshi2015avatar,sharifi2017online} can use frequent DRAM
scrubbing to achieve \hext{the} desired DRAM reliability level.

We conclude that SMS performs memory scrubbing more efficiently than
conventional MC based scrubbing and it enables scrubbing at high rates with
\hext{small performance and energy overheads}.

\end{yellowb_break}

\subsubsection{Comparison to PARA in Memory Controller}
\label{subsec:para_comparison}

\begin{figure}[!h]
    \begin{subfigure}{.48\linewidth}
        \begin{yellowb}
            \centering
            \includegraphics[width=\linewidth]{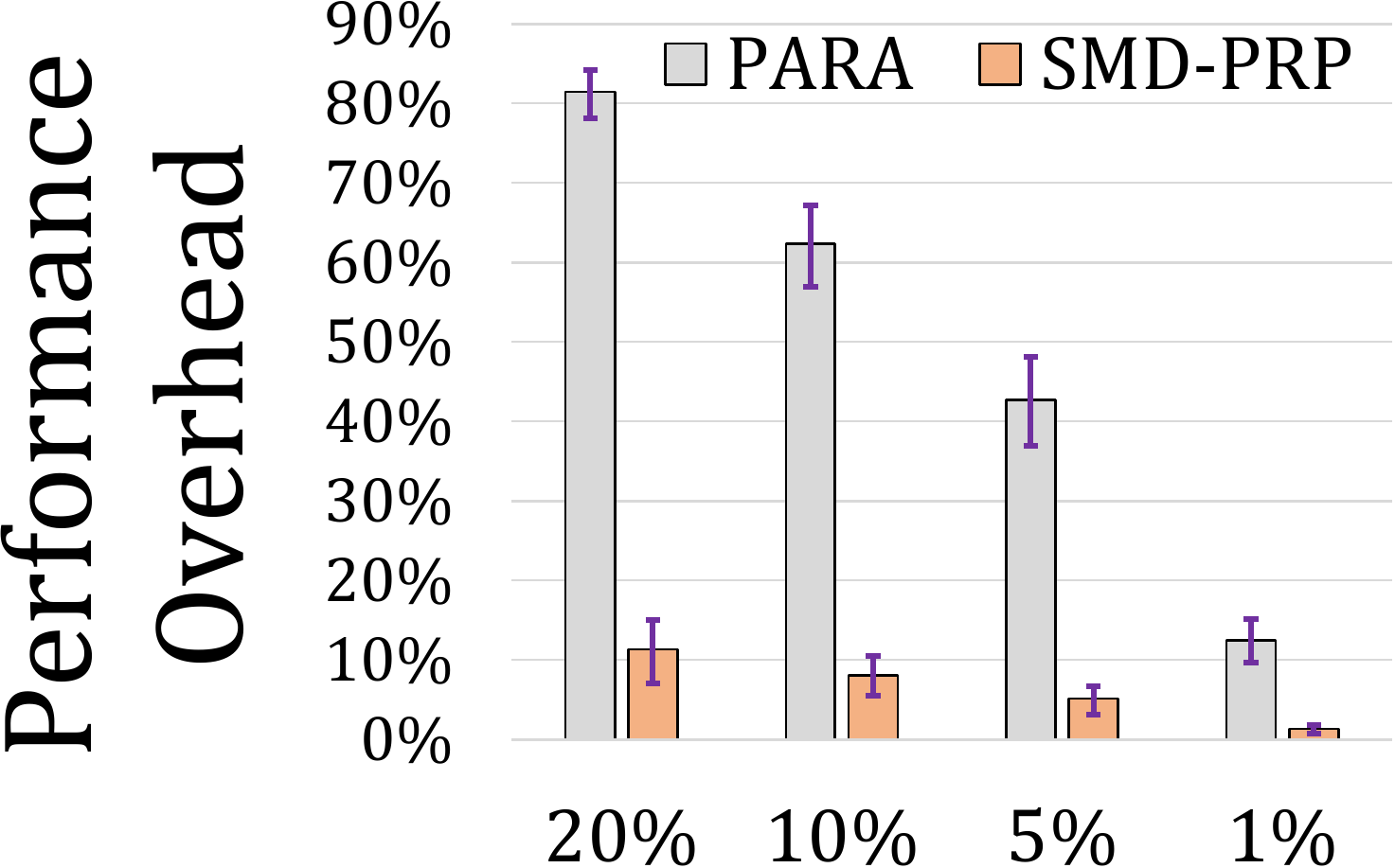}
        \end{yellowb}
    \end{subfigure}\hfill
    \begin{subfigure}{.48\linewidth}
        \begin{yellowb}
            \centering
            \includegraphics[width=\linewidth]{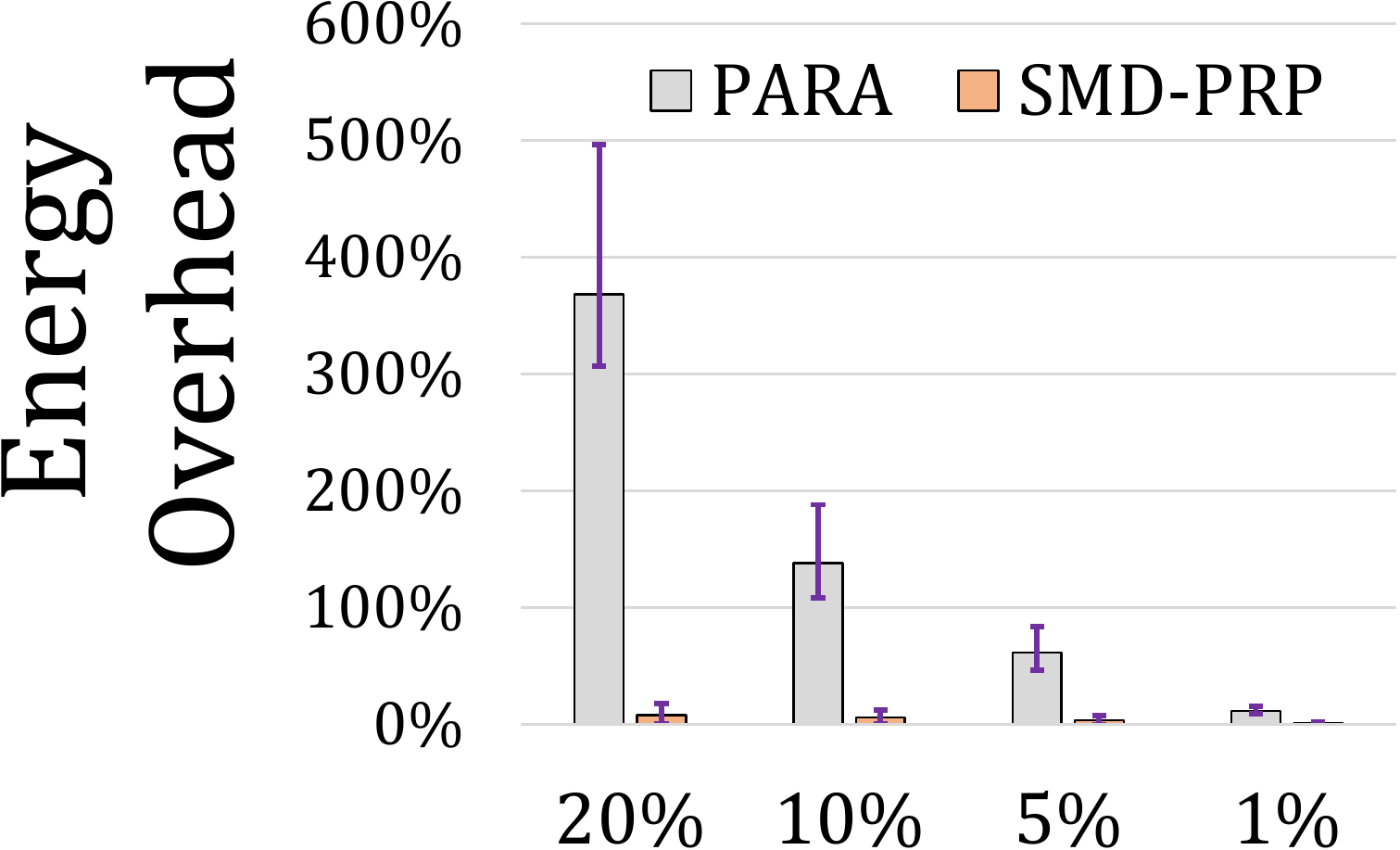}
        \end{yellowb}
    \end{subfigure}
    \caption{PARA vs. \prp{}.}
    \label{fig:para_comparison}
\end{figure}

\begin{yellowb_break}

\hext{Fig.~\ref{fig:para_comparison} compares the performance and energy
overheads of PARA implemented in the MC (as proposed by Kim et
al.~\cite{kim2014flipping}) for DDR4 and \prp{} for different neighbor row
activation probabilities (i.e., $P_{mark}$) across \texttt{4c-high} workloads.
\emph{PARA} represents the performance and energy overheads with respect to
conventional DDR4 with no RowHammer protection. Similarly, \prp{} represents the
performance and energy overheads with respect to a \mech{} chip, which uses
\fr{} for periodic refresh, with no RowHammer protection.}

\end{yellowb_break}

\begin{yellowb_break}

\hext{We make two observations. First, the performance and energy consumption of
MC-based PARA scales poorly with $P_{mark}$. At the default $P_{mark}$ of 1\%,
PARA incurs 12.4\%/11.5\% average performance/DRAM energy overhead. For higher
$P_{mark}$, the overheads of PARA increase dramatically to 81.4\%/368.1\% at
$P_{mark}$ of 20\%. Second, \prp{} is significantly more efficient than PARA. At
the default $P_{mark}$ of 1\%, \prp{} incurs only 1.4\%/0.9\% performance/DRAM
energy overheads and at $P_{mark}$ of 20\% the overheads become only
11.3\%/8.0\%. \prp{} is more efficient than PARA mainly due to enabling access
to non-locked regions in a bank while \prp{} performs neighbor row refreshes on
the locked region.

We conclude that \prp{} is a highly-efficient RowHammer protection that incurs
small performance and DRAM energy overheads even with high neighbor row refresh
probability, which is critical to protect future DRAM chips that may have
extremely high RowHammer vulnerability.}

\end{yellowb_break}

\begin{figure}[!h]
    \begin{yellowb}
        \centering
        \includegraphics[width=\linewidth]{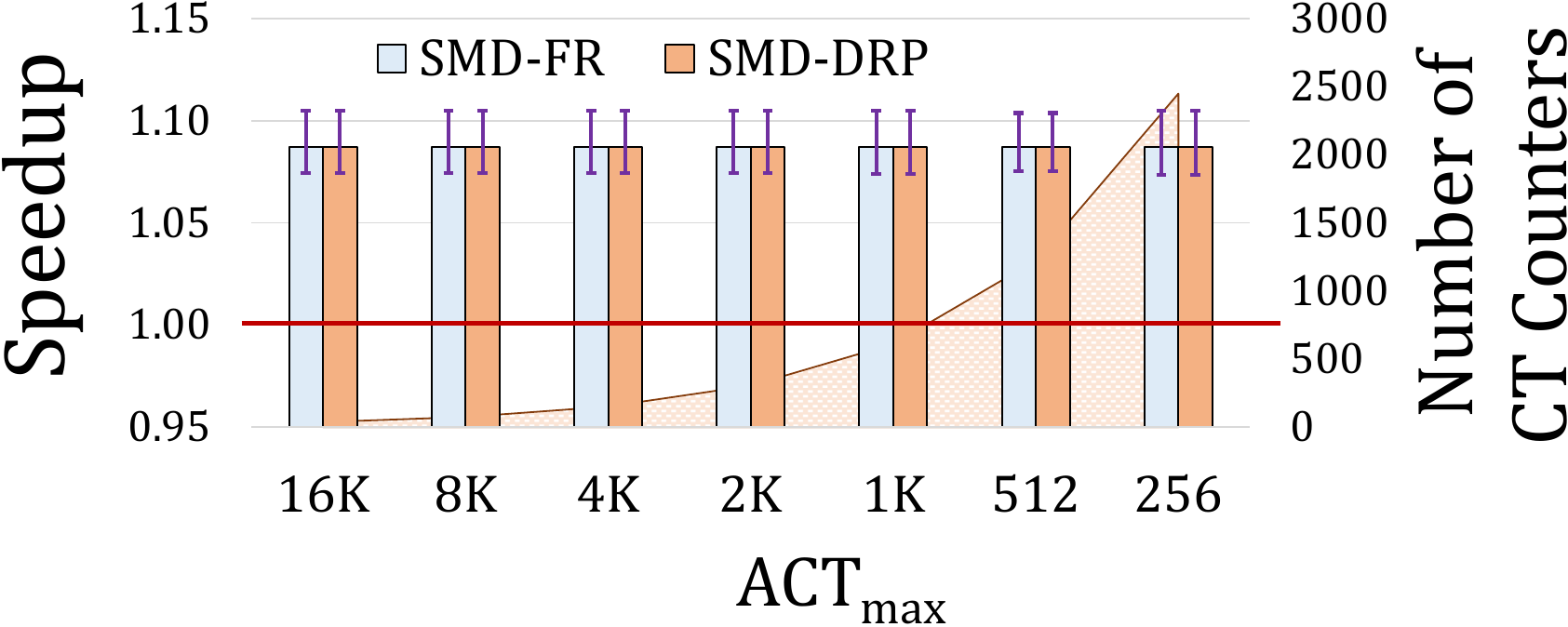}
    \end{yellowb}
    
    \caption{\drp{}'s sensitivity to $ACT_{max}$.}
    \label{fig:graphene_act_threshold_sweep}
\end{figure}

\begin{yellowb_break}

\subsubsection{\drp{} Maximum Activation Threshold}
\label{subsec:graphene_act_threshold_sweep}

We analyze the \drp{}'s sensitivity to the maximum row activation threshold
($ACT_{max}$). Fig.~\ref{fig:graphene_act_threshold_sweep} shows the average
speedup that \fr{} and \drp{} achieve for different $ACT_{max}$ values across
\texttt{4c-high} workloads compared to the DDR4 baseline. When evaluating
\drp{}, we use \fr{} as a DRAM refresh mechanism.

\end{yellowb_break}

\begin{yellowb_break}

We observe that \drp{} incurs negligible performance overhead on top of \fr{}
even for extremely small $ACT_{max}$ values. This is because \drp{} generates
very few neighbor row refreshes as \texttt{4c-high} is a set of benign workloads
that do not repeatedly activate a single row many times.

Although the performance overhead of \drp{} is negligible, the number of Counter Table (CT) entries required is significantly large for small $ACT_{max}$ values.
For $ACT_{max} = 16K$, \drp{} requires $38$ counters per bank, and the number of counters required increase linearly as $ACT_{max}$ reduces, reaching $2449$ counters at the lowest $ACT_{max} = 256$ that we evaluate.

\end{yellowb_break}

\begin{figure}[!h]
    \begin{yellowb}
        \centering
        \includegraphics[width=\linewidth]{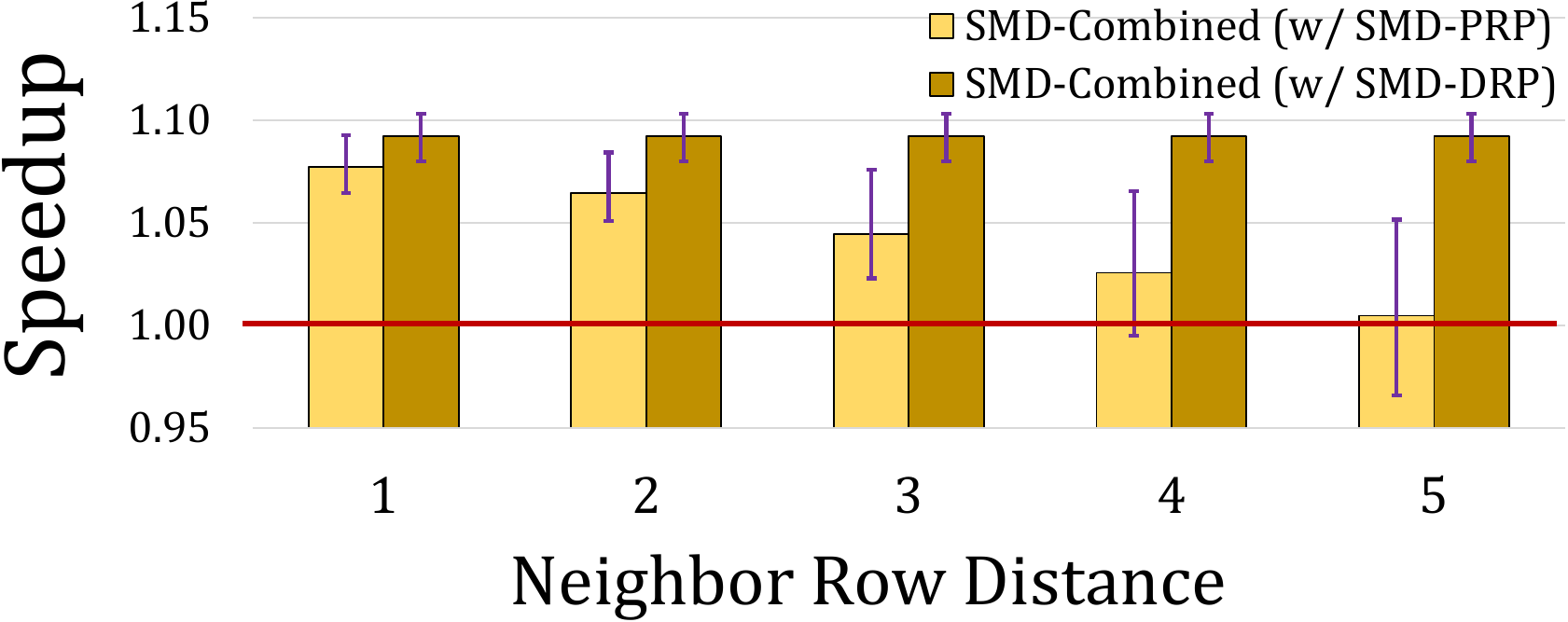}
    \end{yellowb}
    
    \caption{Sensitivity to the number of neighbor rows affected by RowHammer.}
    \label{fig:blast_radius_sweep}
\end{figure}

\begin{yellowb_break}

\subsubsection{Number of Vulnerable Neighbor Rows}
\label{subsec:blast_radius_sweep}
    
We analyze the performance overheads of \prp{} and \drp{} when refreshing a
different number of neighbor rows upon detecting a potentially aggressor row.
Kim et al.~\cite{kim2020revisiting} show that, in some DRAM chips, an aggressor
row can cause bit flips also in rows that are at a greater distance than the two
victim rows surrounding the aggressor row. Thus, it may be desirable to configure a RowHammer protection mechanism to refresh more neighbor rows than the two rows that are immediately adjacent to the aggressor row. 

Fig.~\ref{fig:blast_radius_sweep} shows the average speedup that
\mech{}-Combined (separately with \prp{} and \drp{}) achieves for different
number of neighbor rows refreshed across \texttt{4c-high} workloads compared to
the DDR4 baseline. The \emph{Neighbor Row Distance} values on the x-axis
represent the number of rows refreshes on each of the two sides of an aggressor
row (e.g., for neighbor row distance of $2$, \prp{} and \drp{} refresh four
victim rows in total).

\end{yellowb_break}

\begin{yellowb_break}

We make two key observations from the figure. First, \prp{} incurs large performance overheads as the neighbor row distance increases. This is because, with $P_{mark} = 1\%$, \prp{} perform neighbor row refresh approximately for every 100th \cmdact{} command, and the latency of this refresh operation increases with the increase in the number of victim rows. Second, the performance overhead of \drp{} is negligible even when the neighbor row distance is five. This is because, \drp{} detects aggressor rows with a higher precision than \prp{} using area-expensive counters. As the \texttt{4c-high} workloads do not repeatedly activate any single row, \drp{} counters rarely exceed the maximum activation threshold, and thus trigger neighbor row refresh only a few times.

\end{yellowb_break}
\subsection{Hardware Overhead}
\label{sec:hw_overhead}

\hht{An \mech{} chip introduces an extra physical pin to transmit \actnack{}
signals from a DRAM chip to a Memory Controller (MC).} \bgyellow{For systems
that use rank-based DRAM chip organizations (e.g., DDR4 DIMM), the total number
of extra pins depends on the number of chips per rank. A typical ECC DIMM with
18 x4 DDR4 chips requires 18 \actnack{} pins per channel. The following two
approaches can be used to reduce the \actnack{} pin count for rank-based
organizations. First, the RCD chip on Registered DIMMs can act as a proxy
between the DRAM chips and the MC by sending the MC a single \actnack{} when any
of the per-chip \actnack{} signals are asserted. A similar approach is used with
the per-chip \texttt{alert\_n} signals in existing DDR4 modules. This approach
requires a single \actnack{} pin per channel. Second, \mech{} can use the
\texttt{alert\_n} pin to transmit an \actnack{}. \texttt{alert\_n} is currently
used to inform the MC that the DRAM chip detected a CRC or parity check failure
on the issued command, and thus the MC must issue the command again. The
\texttt{alert\_n} signal can simply be asserted not only on CRC or parity check
failure, but also when the MC attempts accessing a row in a locked region. This
approach does not require additional physical pins. Several DRAM standards
(e.g., LPDDR, HBM, and GDDR) are typically organized as a single chip per
channel. For such memories, \mech{} introduces only one \actnack{} pin per
channel.}

\hht{We individually discuss
the other changes required on the existing DRAM chip and the MC circuitry.}

\subsubsection{DRAM Chip Modifications}

%
We use CACTI~\cite{muralimanohar2009cacti} to evaluate the hardware overhead of
the changes that \mech{} introduces over a conventional DRAM bank
\hht{(highlighted in Fig.~\ref{fig:smd_bank_organization})} assuming
\SI{22}{\nano\meter} technology. \hht{Lock Region Table (LRT)} is a small table
that stores a single bit for each lock region to indicate whether or not the
lock region is under maintenance. In our evaluation, we assume that a bank is
divided into $16$ lock regions. Therefore, LRT consists of only $16$ bits, which
are indexed using a 4-bit lock region address. According to our evaluation, an
LRT incurs \emph{only} \SI{32}{\micro\meter\squared} area overhead per bank. The
area overhead of all LRTs in a DRAM chip is \emph{only} 0.001\% of a
\SI{45.5}{\milli\meter\squared} DRAM chip. The access time of an LRT is
\SI{0.053}{\nano\second}, which is \emph{only} 0.4\% of typical row activation
latency (\trcd{}) of \SI{13.5}{\nano\second}.

\mech{} adds a \hht{per-region} RA-latch to enable accessing one lock region
while another is under maintenance. An RA-latch stores a pre-decoded row
address, which is provided by the global row address decoder, and drives the
local row decoders where the row address is fully decoded. According to our
evaluation, \hht{all} RA-latches incur \hht{a} total area overhead of 1.63\% of
a \SI{45.5}{\milli\meter\squared} DRAM chip. An RA-latch has only
\SI{0.028}{\nano\second} latency, which is negligible compared to \trcd{}.

Besides these changes that are the core of the \mech{} substrate, a
particular maintenance mechanism may incur additional area overhead. 
\begin{yellowb}
We evaluate the DRAM chip area overhead of the maintenance mechanisms presented
in \cref{sec:maintenance_mechanisms}. The simple refresh mechanism, \fr{},
requires only \SI{77.1}{\micro\meter\squared} additional area in a DRAM chip.
\vr{} maintains bloom filters that constitute a large portion of the required
\SI{31663}{\micro\meter\squared}. The DRAM chip area overhead of each refresh
mechanism is less than 0.1\% of a typical \SI{45.5}{\milli\meter\squared} DRAM
chip. \prp{} requires \SI{863.6}{\micro\meter\squared} area, which is a
similarly small portion of a typical DRAM chip. \drp{} requires a large
\emph{Counter Table} with 1224 counters per bank for the RowHammer threshold
value $ACT_{max}=512$ that we use in our performance evaluation. Across a DRAM
chip, \drp{} requires \SI{3.2}{\milli\meter\squared} area, which is 7.0\% of a
typical DRAM chip size. The control logic of \sms{} is similar to the control
logic of \fr{} and it requires only \SI{77.1}{\micro\meter\squared} additional
area excluding the area of the ECC engine, which is already implemented by DRAM
chips that support in-DRAM ECC.

\end{yellowb}

\subsubsection{Memory Controller Modifications}

We slightly modify the MC's request scheduling mechanism to retry a rejected
\cmdact{} command as we explain in \cref{subsec:rejecting_acts}. \hht{Upon
receiving an \actnack{}, the MC marks the corresponding bank as precharged.}
\hht{An existing} MC already implements control circuitry to pick an appropriate
request from the request queue and issue the necessary DRAM command based on the
DRAM bank state (e.g., \cmdact{} to a precharged bank or \cmdread{} \hht{if the
corresponding row is already open}) by respecting the DRAM timing parameters.
The \emph{ACT Retry Interval (ARI)} is simply a new timing parameter that
specifies the minimum time interval for issuing an \cmdact{} to a lock region
after receiving an \actnack{} \hht{from} the same region. Therefore, \mech{} can
be implemented in existing MCs with \hht{only} slight modifications by
leveraging the existing request scheduler \hht{(e.g.,
FRFCFS~\cite{mutlu2007stall})}.

No further changes are required in the MC to support different maintenance
mechanisms enabled by \mech{}. Thus, \mech{} enables the DRAM vendors to update
existing maintenance mechanisms and implement new ones \hht{without any further
changes to} MCs that support \mech{}.

\section{Related Work}
\label{sec:related_work}

To our knowledge, this is the first paper to \bgyellow{enable DRAM chips that
autonomously and efficiently perform various maintenance operations with} simple
changes to existing DRAM interfaces.
\hht{Our new \mechlong{} (\mech{}) design} enables \bgyellow{new} maintenance
operations completely within DRAM without requiring further changes in the DRAM
interface, Memory Controller (MC) or \hht{other system components}. 
\bgyellow{No prior work proposes setting the MC free from managing DRAM maintenance
 operations nor studies the system-level performance and energy impact of
 autonomous maintenance mechanisms.}
\hht{We briefly discuss relevant prior works.}

\textbf{Changing the DRAM Interface.}
%
\hht{Several prior works~\cite{udipi2011combining, ham2013disintegrated,
fang2011memory, cooper2012buffer} propose using high-speed serial links and
packed-based protocols in DRAM chips.}
\mech{} differs from prior works in two key aspects. First, none of \hht{these}
works describe how to implement maintenance mechanisms completely within DRAM.
\hht{We propose six new \mech{}-based maintenance mechanisms
(\cref{sec:maintenance_mechanisms}).}
Second, prior works significantly overhaul the existing DRAM interface, which
makes their proposals more difficult to adopt compared to \mech{}, which adds
only a single \actnack{} signal to the existing DRAM interface \hht{and requires
slight modifications in the MC}.

\textbf{Mitigating DRAM Refresh Overhead.}
\hht{Many} previous works~\cite{chang2014improving, liu2012raidr,
qureshi2015avatar, nair2014refresh, baek2014refresh, bhati2013coordinated,
cui2014dtail, emma2008rethinking, ghosh2007smart, isen2009eskimo,
jung2015omitting, kim2000dynamic, luo2014characterizing, kim2003block,
liu2012flikker, mukundan2013understanding, nair2013case, patel2005energy,
stuecheli2010elastic, khan2014efficacy, khan2016parbor, khan2017detecting,
venkatesan2006retention, patel2017reaper, riho2014partial, hassan2019crow,
kim2020charge, nguyen2018nonblocking, kwon2021reducing} propose techniques to
reduce DRAM refresh overhead. 
%
%
\mech{} not only enables maintenance mechanisms for reducing DRAM refresh
overhead but also other efficient maintenance mechanisms for improving DRAM
reliability and security, described in \cref{sec:maintenance_mechanisms}.

\textbf{RowHammer Protection.}
Many prior works~\cite{lee2019twice, son2017making,yaglikci2021blockhammer,
park2020graphene, you2019mrloc, seyedzadeh2018cbt, aweke2016anvil,
cojocar2019exploiting, kim2014flipping, frigo2020trrespass, herath2015these,
van2018guardion, brasser2016can, konoth2018zebram, kim2014flipping,
yaglikci2021blockhammer, marazzi2023rega, woo2023scalable, wi2023shadow} propose
techniques for RowHammer protection. \hht{Although we propose \prp{} and
\prpplus{}, \mech{} can be used to implement existing or new RowHammer
protection mechanisms.}

\textbf{Memory Scrubbing.}
Although prior works report that the overhead of memory scrubbing is small as
low scrubbing rate (e.g., scrubbing period of 24
hours~\cite{siddiqua2017lifetime, micron2019whitepaper, jedec2021ddr5}, 45 minutes
per 1~GB scrubbing rate~\cite{schroeder2009dram}, scrubbing only when
idle~\cite{meza2015revisiting}) \hht{is generally} sufficient, the cost of
scrubbing \hht{can} dramatically increase for future DRAM chips due to
increasing DRAM bit error rate and increasing DRAM chip density. \sms{} enables
efficient scrubbing \hht{by eliminating off-chip data transfers}.



\section{Summary}

\label{sec:conclusion}

\hht{To set the memory controller free from managing DRAM maintenance
operations, \mechlong{} (\mech{}) introduces minimal changes to the existing
DRAM chips and memory controllers.} \mech{} enables in-DRAM maintenance
operations with no further changes to the DRAM interface, memory controller, or
other system components. Using \mech{}, we implement efficient maintenance
mechanisms for DRAM refresh, RowHammer protection, and memory scrubbing. \hhf{We
show that these mechanisms altogether enable a higher performance, more energy
efficient and at the same time more reliable and secure DRAM system.} \hhms{To
foster research and development in this direction, we open source our
Self-Managing DRAM framework~\cite{smdsource} and early-release our
Self-Managing DRAM work on arXiv.org~\cite{hassan2022self}.}
We believe \hhf{and hope that} \mech{} will enable practical adoption of
innovative ideas in DRAM design.  
\chapter[Copy-Row DRAM]{CROW: A Low-Cost Substrate for Improving DRAM Performance, Energy Efficiency, and Reliability}
\label{chap:crow}

\renewcommand{\mech}{{CROW}\xspace}
\newcommand{\copyrow}{{copy row}\xspace}

\newcommand{\cmdactc}{\texttt{{ACT-c}}\xspace}
\newcommand{\cmdactt}{\texttt{{ACT-t}}\xspace}

DRAM has been the dominant technology for architecting main memory for decades.
Recent trends in multi-core system design and large-dataset applications have
amplified the role of DRAM as a critical system bottleneck.  We propose Copy-Row
DRAM (\mech), a flexible substrate that enables new mechanisms for improving
DRAM performance, energy efficiency, and reliability. We use the \mech substrate
to implement 1)~a low-cost in-DRAM caching mechanism that lowers DRAM activation
latency to frequently-accessed rows by 38\% and 2)~a mechanism that avoids the
use of short-retention-time rows to mitigate the performance and energy overhead
of DRAM refresh operations. {\mech}'s flexibility allows the implementation of
both mechanisms at the same time. Our evaluations show that the two mechanisms
synergistically improve system performance by 20.0\% and reduce DRAM energy by
22.3\% for memory-intensive four-core workloads, while incurring 0.48\% extra
area overhead in the DRAM chip and \SI{11.3}{\kibi\byte} storage overhead in the
memory controller, and consuming 1.6\% of DRAM storage capacity, for one
particular implementation. \hhmiv{To aid future research and development, we
open source CROW~\cite{crow_spice_github}.}

\section{Applications of \mech}
\label{crow_sec:crow_apps}

\mech is a versatile substrate that enables multiple mechanisms for improving
DRAM performance, energy efficiency, and reliability. We propose three such new
mechanisms based on \mech: 1) \mech-cache, a mechanism that reduces DRAM access
latency by exploiting \mech's ability to simultaneously activate a regular row
and a \copyrow that store the same data; 2) \mech-ref, a mechanism that reduces
DRAM refresh overhead by exploiting \mech's ability to remap weak rows with low
retention times to strong {\copyrow}s. This mechanism relies on retention time
profiling~\cite{patel2017reaper, qureshi2015avatar, liu2013experimental} to
determine weak and strong rows in DRAM; 3) a mechanism that protects against the
RowHammer~\cite{kim2014flipping} vulnerability by identifying rows that are
vulnerable to RowHammer-induced errors and remapping the vulnerable rows to
\copyrow{s}.

\subsection{In-DRAM Caching (\mech-cache)}
\label{crow_subsec:in_dram_caching}

Our in-DRAM caching mechanism, \mech-cache, exploits both of our new
multiple-row activation primitives in \mech
(\cref{crow_sec:mechanism:overview}). The key idea of \mech-cache is to use a
\copyrow{} as a duplicate of a recently-accessed regular row within the same
subarray, and to activate \emph{both} the regular row and the \copyrow. By
simultaneously activating both rows, \mech-cache reduces activation latency and
starts servicing read and write requests sooner.

\subsubsection{Copying a Regular Row to a Copy Row}
\label{crow_subsubsec:row_copy}

Given $N$ {\copyrow}s, we would like to duplicate the $N$ regular rows that will
be most frequently activated in the near future to maximize the benefits of
\mech-cache. However, determining the $N$ most frequently-activated rows is a
difficult problem, as applications often access main memory with irregular
access patterns. Therefore, we follow a similar approach used by processor
caches, where the most-recently-used rows are cached instead of the most
frequently-accessed rows. Depending on how many {\copyrow}s are available,
\mech-cache maintains copies of the $N$ \emph{most-recently-activated} regular
rows in each subarray. When the memory controller needs to activate a regular
row that is not already duplicated, the memory controller copies the regular row
to an available \copyrow{}, using multiple-row activation. While this is a
simple scheme, we achieve a significant hit rate on the \mech table with a small
number of \copyrow{s} (see \cref{crow_subsec:crow_perf}).

To efficiently copy a regular row to a \copyrow within a subarray, we introduce
a new DRAM command, which we call \emph{Activate-and-copy} (\cmdactc). The
\cmdactc command performs a three-step row copy operation in DRAM, using a
technique similar to RowClone~\cite{seshadri2013rowclone}. First, the memory
controller issues \cmdactc to DRAM and sends the addresses of 1) the
\emph{source} regular row and 2) the \emph{destination} \copyrow{}, over
multiple command bus cycles. Second, upon receiving an \cmdactc command, DRAM
enables the wordline of only the regular row. The process of reading and
latching the row's data into the sense amplifiers completes as usual
(see~\cref{sec:dram_operation}). Third, as soon as the data is latched in the
sense amplifiers, DRAM enables the wordline of the \copyrow. This causes the
sense amplifiers to restore charge to both the regular row \emph{and} the
\copyrow. On completion of the restoration phase, both the regular row and the
{\copyrow} contain the same data. 

The \cmdactc operation slightly increases the restoration time compared to
regular \cmdact. This is because, after the wordline of the \copyrow is enabled,
the capacitance that the sense amplifier drives the bitline against increases,
since two cell capacitors are connected to the same bitline. In our
circuit-level SPICE model, for the \cmdactc command, we find that the
activate-to-precharge latency (\tras) increases by 18\%. However, this
activation latency overhead has a very limited impact on overall system
performance because \mech table typically achieves a high hit rate
(\cref{crow_subsec:crow_perf}), which means that the latency of duplicating the
row is amortized by the reduced activation latency of future accesses to the
row.

\subsubsection{Reducing Activation Latency}

\mech-cache introduces a second new DRAM command, \emph{Activate-two}
(\cmdactt), to simultaneously activate a regular row and its duplicate \copyrow.
If the \mech table (see \cref{crow_subsec:ctable}) contains an entry
indicating that a \copyrow is a duplicate of the regular row to be activated,
the memory controller issues \cmdactt to perform MRA on both rows, achieving
low-latency activation. In \cref{crow_sec:spice}, we perform detailed
circuit-level SPICE simulations to analyze the activation latency reduction with
\cmdactt.

Note that the modifications needed for \cmdactc and \cmdactt in the row decoder
logic are nearly identical, as they both perform multiple-row activation. The
difference is that rather than activating the \copyrow \emph{after} the regular
row as with \cmdactc, \cmdactt activates both of the rows concurrently.

\subsubsection{Reducing Restoration Latency}
\label{crow_subsec:reducing_tras}

In addition to decreasing the amount of time required for charge sharing during
activation (see \cref{sec:dram_operation}), the increased capacitance on the
bitline due to the activation of multiple cells enables the reduction of \tras
by using \emph{partial restoration}~\cite{zhang2016restore, wang2018reducing}.
The main idea of the partial restoration technique is to terminate row
activation earlier than normal, i.e., issue \cmdprech by relaxing (i.e.,
reducing) \tras, on a multiple-row activation such that the cells are not fully
restored. While this degrades the retention time of each cell individually, the
partially restored cells contain just enough charge to maintain data integrity
until the next refresh operation~\cite{riho2014partial} (we verify this using
SPICE simulations in \cref{crow_sec:spice}). We combine partial cell restoration
(i.e., reduction in \tras) with the reduction in \trcd in order to provide an
even greater speedup than possible by simply decreasing \trcd alone (see
\cref{crow_sec:results}).

Taking the concept one step further, we can also apply partial restoration to
write accesses, decreasing the amount of time required for the \cmdwrite
command. Since the write restoration latency (\twr) is analogous to cell
restoration during activation (\cref{sec:dram_operation}), we can terminate the
write process early enough such that the written cells are only partially
charged.

\subsubsection{Evicting an Entry from the \mech Table}
\label{crow_subsec:replacing_row}

As described in \cref{crow_subsec:reducing_tras}, \mech-cache relaxes \tras for
the \cmdactt command to further improve the DRAM access latency. Relaxing \tras
may terminate the restoration operation early, which puts the precharged regular
and {\copyrow}s into a \emph{partially-restored} state.

Note that there are two cases where relaxing \tras does not necessarily lead to
partial restoration. First, the memory controller can exploit an open regular
row + copy row pair to serve multiple memory requests, if available in the
request queue, from different portions of the row. The time needed to serve
these requests can delay precharging the associated regular and copy rows until
they are fully restored. Second, if there are no other memory requests to
different rows in the same bank, the currently-opened regular row and \copyrow
can be left open, providing enough time for full charge restoration.

\mech-cache maintains the restoration state of the paired regular and
{\copyrow}s by utilizing a single bit of the \texttt{Special} field of the
\mech table, which in the context of \mech-cache we refer to as the
\texttt{isFullyRestored} field. \mech-cache sets \texttt{isFullyRestored} to
\texttt{false} only if 1)~\cmdactt was used to activate the currently open row,
and 2)~the memory controller issues a \cmdprech command to close the rows
\emph{before} the default \tras is satisfied. In contrast, \mech-cache sets
\texttt{isFullyRestored} to \texttt{true} if the memory controller issues the
next \cmdprech after the default \tras (i.e., a time interval sufficient for
fully restoring the open regular row and \copyrow). Note that existing memory
controllers already maintain timing information for conventional command
scheduling, e.g., when the last \cmdact to each bank was issued. By taking
advantage of the existing timing information, \mech-cache does not incur
significant area overhead for managing the \texttt{isFullyRestored} field.

Partially restoring the duplicate rows helps to significantly reduce restoration
latency (see \cref{crow_sec:spice}). However, partial restoration introduces a
new challenge to the design of \mech-cache, because a partially-restored row can
only be correctly accessed using \cmdactt, which activates a regular row along
with its duplicate \copyrow. Duplicating a new regular row ($RR_{new}$) to a
\copyrow that is already a duplicate of another regular row ($RR_{old}$), i.e.,
evicting $RR_{old}$ from the \mech table, causes $RR_{old}$ to be activated as a
single row during a future access. If $RR_{old}$ is partially restored prior to
eviction from the \mech table, performing a single-row activation on $RR_{old}$
in the future may lead to data corruption. Thus, the memory controller needs to
guarantee that a partially-restored regular row is never evicted from the
\mech table.

To prevent data corruption due to the eviction of a partially-restored row from
the \mech table, if the \emph{isFullyRestored} field is \emph{false}, the memory
controller first issues an \cmdactt and fully restores $RR_{old}$ by conforming
to the \emph{default} \tras. Since this operation sets \emph{isFullyRestored} to
\emph{true}, $RR_{old}$ can now safely be evicted from the \mech table to be
replaced with $RR_{new}$. The disadvantage of this approach is the overhead of
issuing an additional \cmdactt followed by a \cmdprech to perform full
restoration. However, in \cref{crow_subsec:crow_perf}, we show that this
overhead has a negligible performance impact since \mech table has a high hit
rate.

\subsubsection{Implementing the New DRAM Commands}
\label{crow_subsec:overhead_of_the_new_commands}

To implement \mech-cache, we introduce the \cmdactc and \cmdactt DRAM commands,
both of which activate a regular row \emph{and} a \copyrow. The wordline of a
regular row is enabled in the same way as in conventional DRAM. To enable the
wordline of a \copyrow, we extend the existing local row decoder logic such that
it can drive the wordline of a {\copyrow} independently of the regular row. As
we show in \cref{crow_subsec:crow_power_area}, for the default configuration
of the \mech substrate with eight {\copyrow}s, our modifications increase the
decoder area by 4.76\%, leading to 0.48\% area overhead in the entire DRAM chip.
The additional eight {\copyrow}s per subarray require only 1.6\% of the DRAM
storage capacity.

Unlike \cmdact, which specifies a single row address, \cmdactc and \cmdactt need
to specify the addresses of both a regular and a \copyrow. \cmdactc and \cmdactt
need only a small number of \copyrow address bits in addition to the regular row
address bits since the corresponding \copyrow is in the same subarray as the
activated regular row. For \mech-8, where each subarray has eight {\copyrow}s,
we need only three additional bits to encode the \copyrow address. We can either
add three wires to the address bus (likely undesirable) or send the address over
multiple cycles as done in current LPDDR4 chips for the \cmdact, \cmdread, and
\cmdwrite commands~\cite{jedec2014lpddr4}.\footnote{In our evaluations, we assume
an additional cycle on the command/address bus to send the \copyrow address to
DRAM. This additional cycle does not always impact the activation latency, as
the memory controller can issue the \cmdactc and \cmdactt commands one cycle
earlier if the command/address bus is idle. Also, DRAM does not immediately need
the address of the \copyrow for \cmdactc.}

\subsection{Reducing Refresh Overhead (\mech-ref)}
\label{crow_subsec:refresh_reduction_mech}

To reduce the DRAM refresh overhead, we take advantage of the \mech substrate to
develop \mech-ref, a software-transparent weak row remapping scheme. \mech-ref
extends the refresh interval of the \emph{entire} DRAM chip beyond the
worst-case value defined in DRAM standards (\SI{64}{\milli\second} for DDR3 and
DDR4, \SI{32}{\milli\second} for LPDDR4) by avoiding the use of the very small
set of weak rows in a given DRAM chip, i.e., rows that would fail when the
refresh interval is extended (due to the existence of at least one weak cell in
the row). \mech-ref consists of three key components. First, \mech-ref uses
retention time profiling at system boot or during runtime to identify the weak
rows. Second, \mech-ref utilizes the strong {\copyrow}s provided by \mech in
each subarray to store the data that would have originally been stored in weak
regular rows. Third, \mech-ref uses the \mech table to maintain the remapping
information, i.e., which strong \copyrow replaces which weak regular row.

\subsubsection{Identifying Weak Rows}
\label{crow_subsec:identify_weak_rows}

To identify the weak rows in each subarray, we rely on retention time profiling
methodologies proposed by prior work~\cite{patel2017reaper, qureshi2015avatar,
liu2013experimental, khan2014efficacy, khan2016parbor, khan2016case,
kim2014flipping, restle1992dram, yaney1987meta}. A retention time profiler tests
the DRAM device with various data patterns and a wide range of operating
temperatures to cover all DRAM cells that fail at a chosen target retention
time. Prior works find that very few DRAM cells fail when the refresh interval
is extended by 2x-4x. For example, Liu et al.~\cite{liu2013experimental} show
that \emph{only} $\sim$1000 cells in a \SI{32}{\gibi\byte} DRAM module fail when
the refresh interval is extended to \SI{256}{\milli\second}. Assuming the
experimentally-demonstrated uniform random distribution of these weak cells in a
DRAM chip~\cite{patel2017reaper, qureshi2015avatar, liu2013experimental,
liu2012raidr, baek2014refresh}, we calculate that the \emph{bit error rate
(BER)} is $~4\cdot10^{-9}$ when operating at a \SI{256}{\milli\second} refresh
interval. 

Based on this observation, we can determine how effective \mech-ref is likely to
be for a given number of \copyrow{s}. First, using the BER, we can calculate
$P_{\text{\em weak\_row}}$, the probability that a row contains at least one
weak cell, as follows:
\begin{equation}\label{crow_eq:p_weak_row}
P_{\text{\em weak\_row}} = 1 - (1 - BER)^{N_{\text{\em cells\_per\_row}}}
\end{equation}
\noindent where $N_{\text{\em cells\_per\_row}}$ is the number of DRAM cells in a row.
Second, we can use $P_{\text{\em row}}$ to determine $P_{\text{\em subarray}-n}$, which is
the probability that a subarray with $N_{\text{\em rows}}$ rows
contains more than $n$~weak rows:
\begin{equation}\label{crow_eq:p_subarray_n}
P_{\text{\em subarray}-n} = 1 - \sum\limits_{k=0}^n \binom {N_{\text{\em rows}}}k P_{\text{\em row}}^k(1 - P_{\text{\em row}})^{N_{\text{\em rows}}-k}
\end{equation}

Using these equations, for a DRAM chip with 8~banks, 128~subarrays per bank,
512~rows per subarray, and \SI{8}{\kibi\byte} per row, the probability of any
subarray having more than 1/2/4/8 weak rows is $0.99/3.1 \times 10^{-1}/3.3
\times 10^{-4}/3.3 \times 10^{-11}$. We conclude that the probability of having
a subarray with a large number of weak rows is extremely low, and thus \mech-ref
is highly effective even when the \mech substrate provides only 8~\copyrow{s}
per subarray. In the unlikely case where the DRAM has a subarray with more rows
with weak cells than the number of available {\copyrow}s, \mech-ref falls back
to the default refresh interval, which does not provide performance and energy
efficiency benefits but ensures correct DRAM operation.\footnote{Alternatively,
\mech-ref can be combined with a heterogeneous refresh-rate scheme, e.g.,
RAIDR~\cite{liu2012raidr} or AVATAR~\cite{qureshi2015avatar}.}

\subsubsection{Operation of the Remapping Scheme}
\label{crow_subsec:remapping_operation}

In each subarray, \mech-ref remaps weak regular rows to \emph{strong}
{\copyrow}s in the same subarray.\footnote{We do \emph{not} use a weak \copyrow
to replace a weak regular row, as a weak \copyrow would also not maintain data
correctly for an extended refresh interval. The retention time profiler also
identifies the weak \copyrow{s}.} \mech-ref tracks the remapping using the
\mech table. When a weak regular row is remapped to a strong \copyrow, \mech-ref
stores the row address of the regular row into the \texttt{RegularRowID} field
of the \mech table entry that corresponds to the \copyrow. When the memory
controller needs to activate a row, it checks the \mech table to see if any of
the \texttt{RegularRowID} fields contains the address of the row to be
activated. If one of the \texttt{RegularRowID} fields matches, the memory
controller issues an activate command to the \copyrow that replaces the regular
row, instead of to the regular row. On a \mech table miss, which indicates that
the row to be activated contains only strong cells, the memory controller
activates only the original regular row. This row remapping operates
transparently from the software, as the regular-to-copy-row remapping is
\emph{not} exposed to the software.

\subsubsection{Support for Dynamic Remapping}

\mech-ref can dynamically change the remapping of weak rows if new weak rows are
detected at runtime. This functionality is important as DRAM cells are known to
be susceptible to a failure mechanism known as \emph{variable retention time}
(VRT)~\cite{qureshi2015avatar, patel2017reaper, liu2013experimental,
khan2016case, khan2016parbor, restle1992dram, yaney1987meta, mori2005origin}. As
a VRT cell can nondeterministically transition between high and low retention
states, new VRT cells need to be continuously identified by periodically
profiling DRAM, as proposed by prior work~\cite{qureshi2015avatar,
patel2017reaper, khan2014efficacy}. To remap a newly-identified weak regular row
at runtime, the memory controller simply allocates an unused strong \copyrow
from the subarray, and issues an \cmdactc to copy the regular row's data into
the \copyrow. Once this is done, the memory controller accesses the strong
\copyrow instead of the weak regular row, as we explain in
\cref{crow_subsec:remapping_operation}.

\subsection{Mitigating RowHammer}
\label{crow_subsec:crow_hammer}

We propose a third mechanism enabled by the \mech substrate that protects
against the RowHammer vulnerability~\cite{kim2014flipping, mutlu2017rowhammer,
mutlu2019rowhammer}. Due to the small DRAM cell size and short distance between
DRAM cells, electromagnetic coupling effects that result from rapidly activating
and precharging (i.e., hammering) a DRAM row cause bit flips on the
physically-adjacent (i.e., victim) rows~\cite{kim2014flipping,
mutlu2017rowhammer, mutlu2019rowhammer}. This effect is known as
\emph{RowHammer}, and has been demonstrated on real DRAM
chips~\cite{kim2014flipping}. Aside from the decreased reliability of DRAM due
to RowHammer, prior works (e.g., \cite{seaborn2015exploiting, van2018guardion,
veen2016drammer, gruss2016rowhammer, cojocar2019exploiting, qiao2016new,
kwong2020rambleed, pessl2016drama, bhattacharya2016curious, jang2017sgx,
zhang2020pthammer, cojocar2020are, weissman2020jackhammer, ji2019pinpoint,
gruss2018another, bosman2016dedup, frigo2020trrespass, deridder2021smash,
frigo2018grand, razavi2016flip, xiao2016one, tatar2018defeating,
lipp2018nethammer, yao2020deephammer, hong2019terminal}) exploit RowHammer to
perform attacks that gain privileged access to the system. Therefore, it is
important to mitigate the effects of RowHammer in commodity DRAM.

We propose a \mech-based mechanism that mitigates RowHammer by remapping the
victim rows adjacent to the hammered row to \copyrow{s}. By doing so, the
mechanism prevents the attacker from flipping bits on the data originally
allocated in the victim rows. Several prior works~\cite{ghasempour2015armor,
lee2019twice, seyedzadeh2017mitigating, kim2015architectural} propose techniques
for detecting access patterns that rapidly activate and precharge a specific
row. Typically, these works implement a counter-based structure that stores the
number of \cmdact commands issued to each row.  Our RowHammer mitigation
mechanism can use a similar technique to detect a RowHammer attack. When the
memory controller detects a DRAM row that is being hammered, it issues two
\cmdactc commands to copy the victim rows that are adjacent to the hammered row
to two of the available \copyrow{s} in the same subarray. Similar to the
\mech-ref mechanism (\cref{crow_subsec:refresh_reduction_mech}), our RowHammer
mitigation mechanism tracks a remapped victim row using the
\texttt{RegularRowID} field in each \mech table entry, and looks up the
\mech table every time a row is activated to determine if a victim row has been
remapped to a \copyrow.

\mech enables a simple yet effective mechanism for mitigating the RowHammer
vulnerability that can reuse much of the logic of \mech-ref. We leave the
evaluation of our RowHammer mitigation mechanism to future work.

\section{Circuit Simulations}
\label{crow_sec:spice}

We perform detailed circuit-level SPICE simulations to find the latency of 1)
simultaneously activating two DRAM rows that store the same data (\cmdactt) and
2) copying an entire regular row to a \copyrow inside a subarray (\cmdactc).
Table~\ref{crow_table:timings_summary} shows a summary of change in the \trcd,
\tras, and \twr timing parameters that we use for the two new commands, based on
the SPICE simulations.  We discuss in detail how the timings are derived for
\cmdactt in \cref{crow_sec:spice:actt}, and for \cmdactc in
\cref{crow_sec:spice:actc}.

\begin{table}[h!]
    \caption{Timing parameters for new DRAM commands.}
    \centering 
    \resizebox{.9\linewidth}{!}{
    \begin{threeparttable}
        \begin{tabular}{cc||c|c|c}
            \multicolumn{2}{c||}{\textbf{DRAM Command}} & \textbf{\trcd} & \textbf{\tras} &
            \textbf{\twr} \\
            \hline
            \multirow{2}{*}{\cmdactt} & \emph{activating fully-restored rows} & -38\% &
            -7\%\tnote{a}\phantom{x}
            (-33\%\tnote{b}\phantom{x}) &
            +14\%\tnote{a}\phantom{x} (-13\%\tnote{b}\phantom{x}) \\
            \cline{2-5}
            & \emph{activating partially-restored rows} & -21\% &
            -7\%\tnote{a}\phantom{x}
            (-25\%\tnote{b}\phantom{x}) &
            +14\%\tnote{a}\phantom{x} (-13\%\tnote{b}\phantom{x}) \\
            \hline
            \multicolumn{2}{c||}{\cmdactc} & 0\% & +18\%\tnote{a}\phantom{x}
            (-7\%\tnote{b}\phantom{x}) & +14\%\tnote{a}\phantom{x}
            (-13\%\tnote{b}\phantom{x}) \\
        \end{tabular}
        \begin{tablenotes}
            \begin{multicols}{2}
            \item[a] \begin{small} When fully restoring the charge. \end{small}
            \item[b] \begin{small} When terminating charge restoration early. \end{small} 
            \end{multicols}
        \end{tablenotes}
    \end{threeparttable}
    } 
\label{crow_table:timings_summary}
\end{table}

In our simulations, we model the entire cell array of a modern DRAM chip using
\SI{22}{\nano\meter} DRAM technology parameters, which we obtain by scaling the
reference \SI{55}{\nano\meter} technology parameters~\cite{rambus} based on the
ITRS roadmap~\cite{vogelsang2010understanding, itrs}. We use
\SI{22}{\nano\meter} PTM low-power transistor models~\cite{zhaoptm, ptmweb} to
implement the access transistors and the sense amplifiers. In our SPICE
simulations, we run $10^4$ iterations of Monte-Carlo simulations with a 5\%
margin on every parameter of each circuit component, to account for
manufacturing process variation. Across all iterations, we observe that the
\cmdactt and \cmdactc commands operate correctly. We report the latency of the
\cmdactt and \cmdactc commands based on the Monte-Carlo simulation iteration
that has the highest access latency for each of these commands. We release our
SPICE model as an open-source tool~\cite{crow_spice_github}.

\subsection{Simultaneous Row Activation Latency}
\label{crow_sec:spice:actt}

\begin{figure}[!ht] \centering
    \begin{subfigure}[t]{0.49\linewidth}
        \centering
        \includegraphics[width=.98\linewidth]{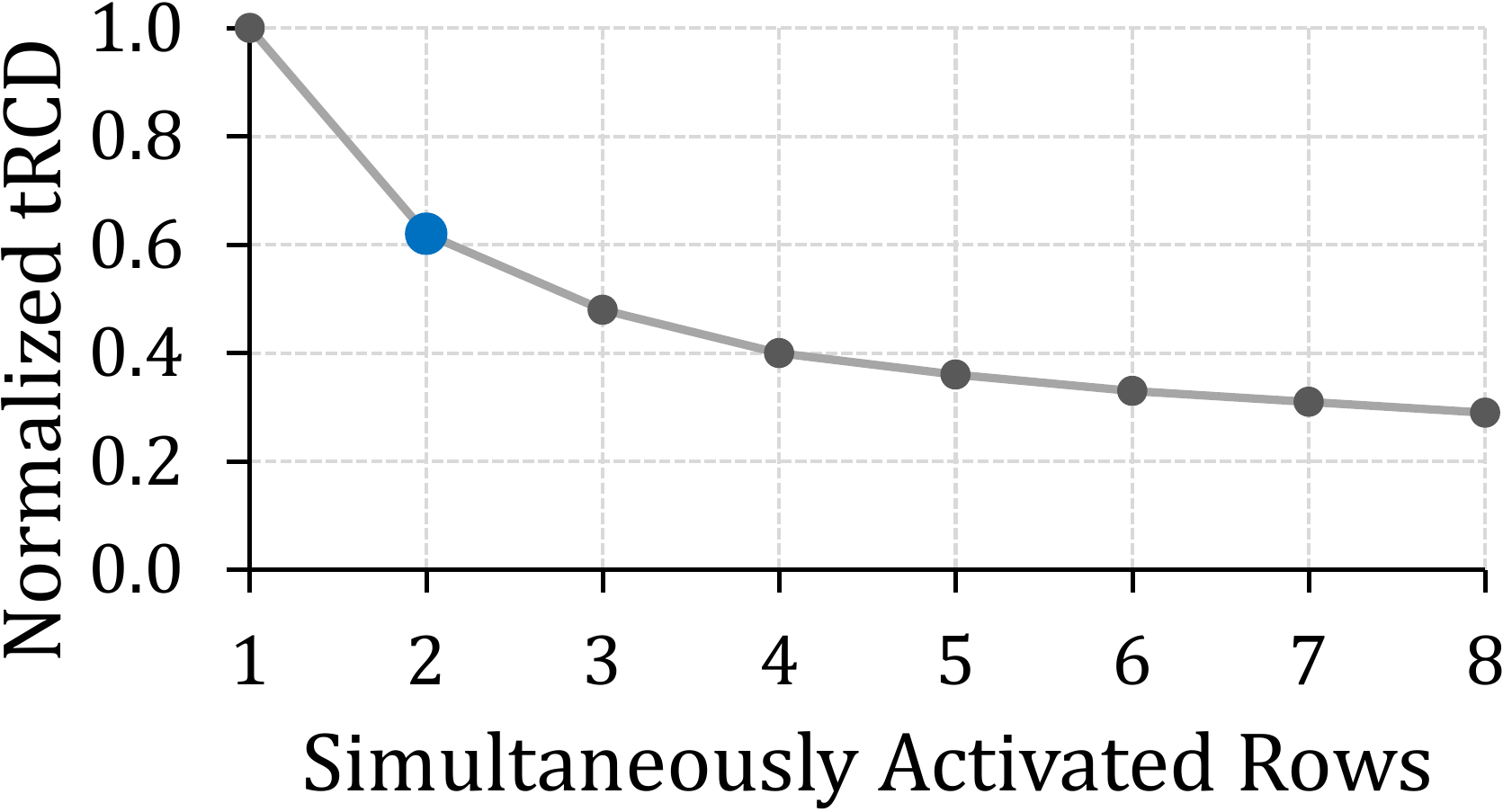}%
        \caption{\trcd~(\SI{18}{\nano\second})}
        \label{crow_fig:trcd_vs_num_rows}
    \end{subfigure}
    \begin{subfigure}[t]{0.49\linewidth}
        \centering
        \includegraphics[width=\linewidth]{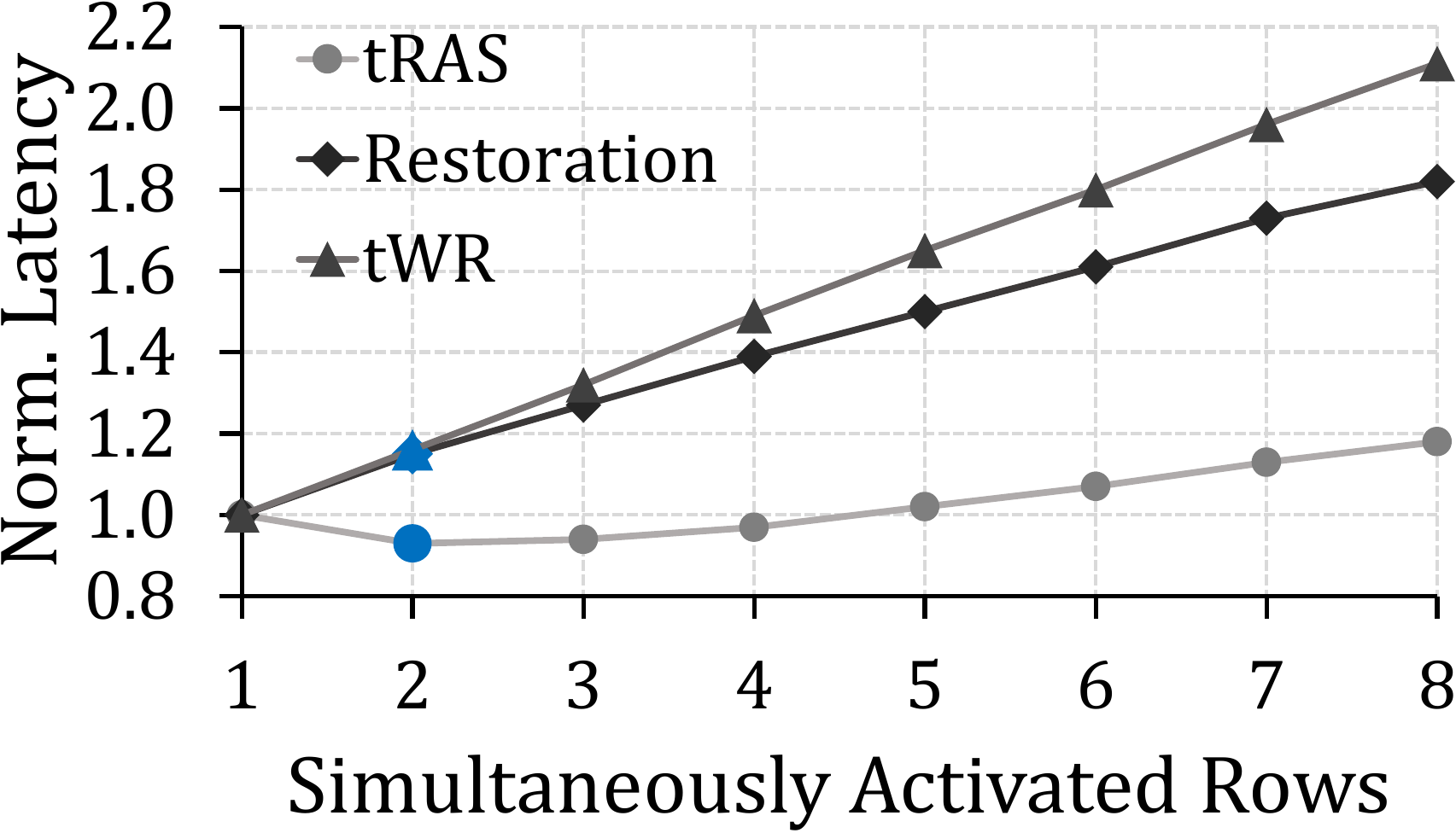}%
        \caption{\tras{} (\SI{42}{\nano\second}), \twr{}
        (\SI{18}{\nano\second}), and restoration
        (\SI{24}{\nano\second})}
        \label{crow_fig:tras_vs_num_rows}
    \end{subfigure}
    \caption{Change in various DRAM latencies with different number of
    simultaneously-activated rows, normalized to baseline DRAM timing parameters
    (absolute baseline values in parentheses).}
\end{figure}

Simultaneously activating multiple rows that store the same data accelerates the
\emph{charge-sharing} process, as the increased amount of charge driven on each
bitline perturbs the bitline faster compared to single-row activation.
Figure~\ref{crow_fig:trcd_vs_num_rows} plots the reduction in activation latency
(\trcd) for a varying number of simultaneously-activated rows. As seen in the
figure, we observe a \trcd reduction of 38\% when simultaneously activating
\emph{two} rows. \trcd reduces further when we increase the number of
simultaneously-activated rows, but the latency reduction per additional
activated row becomes smaller. We empirically find that simultaneously
activating only two rows rather than more rows achieves a large portion of the
maximum possible \trcd reduction potential (i.e., as we approach an infinite
number of rows being activated simultaneously) with low area and power overhead
(see \cref{crow_subsec:crow_power_area}).

\textbf{Change in Restoration Latency.} Although two-row activation reduces
\trcd, \emph{fully} restoring the capacitors of two cells takes more time
compared to fully restoring a single cell. Therefore, two-row activation can
potentially increase \tras and \twr, which could offset the benefits of the
\trcd reduction. Figure~\ref{crow_fig:tras_vs_num_rows} shows the change in
\tras, restoration time, and \twr for a varying number of
simultaneously-activated rows. Although restoration time always increases with
the number of rows, we see a slight decrease in \tras when the number of rows is
small. This is because the reduction in \trcd, which is part of \tras{} (as we
explain in \cref{sec:dram_operation}), is larger than the increase in
restoration time. However, for five or more simultaneously-activated rows, the
overhead in restoration time exceeds the benefits of \trcd reduction, and, as a
result, \tras increases. \twr always increases with the number of
simultaneously-activated rows, because writing to a DRAM cell is similar to
restoring a cell (\cref{sec:dram_operation}).

\textbf{Terminating Restoration Early.} We use \mech to enable in-DRAM caching
by duplicating recently-accessed rows and using two-row activation to reduce
\trcd significantly for future activation of such rows. However, to make the
in-DRAM caching mechanism more effective, we aim to reduce \tras further, as it
is a critical timing parameter that affects how quickly we can switch between
rows in the event of a row buffer conflict. We make \emph{three} observations
based on two-row activation, which leads us to a trade-off between reducing
\trcd and reducing \tras. First, when two DRAM rows are used to store the same
data, the data is correctly retained for a longer period of time compared to
when the data is stored in a single row. This is due to the increased aggregate
capacitance that storing each bit of data using two cells provides. Second,
since data is correctly retained for a longer period of time when stored in two
rows, we can terminate the restoration operation early to reduce \tras and still
achieve the target retention time, e.g., \SI{64}{\milli\second}. Third, as the
amount of charge stored in a cell capacitor decreases, the activation latency
increases~\cite{hassan2016chargecache, lee2015adaptive}. Therefore, terminating
restoration early reduces \tras at the expense of a slight increase in \trcd{}
(due to less charge).

\begin{figure}[h] 
    \centering
    \includegraphics[width=0.67\linewidth]{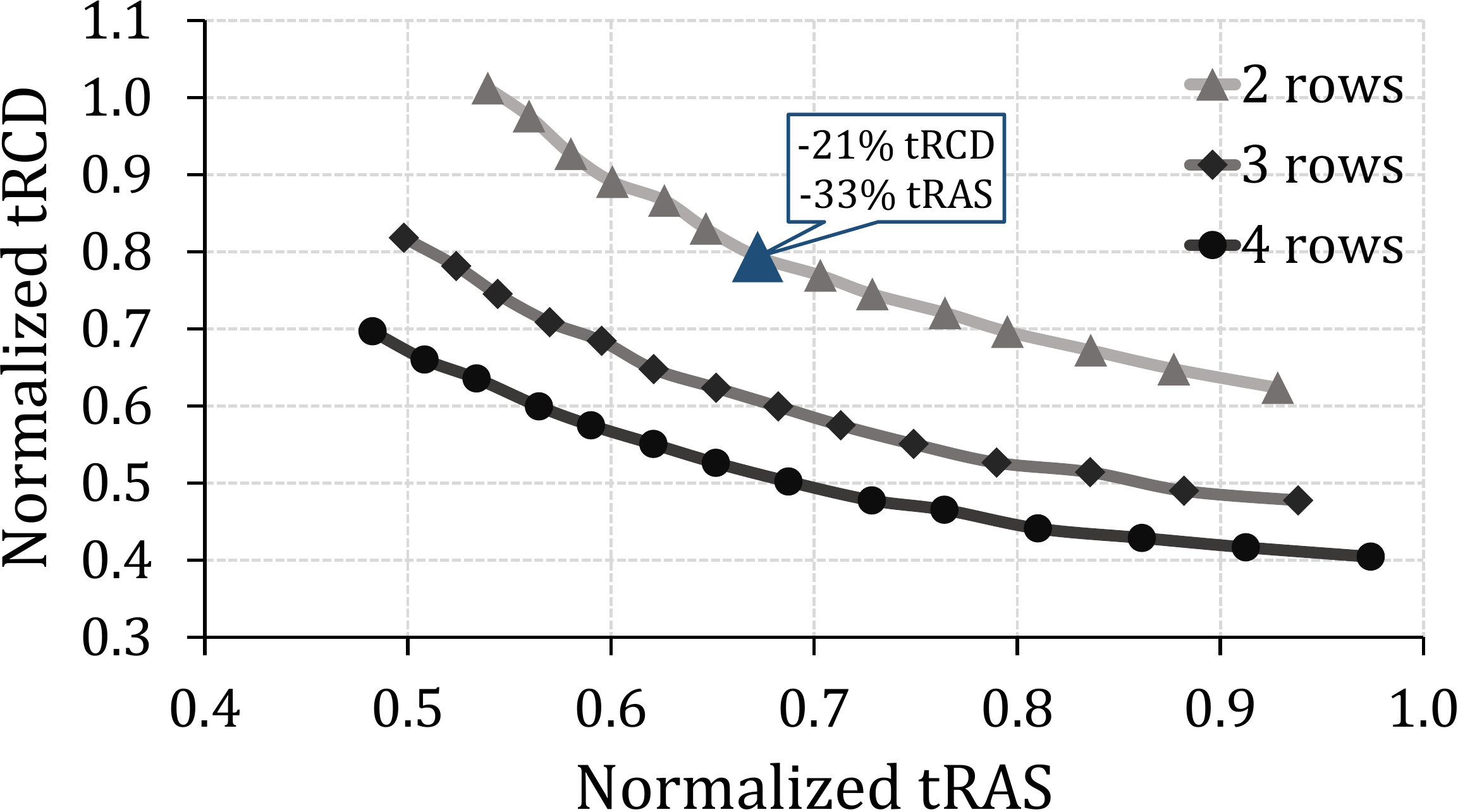}%
    \caption{Normalized \trcd latency as a function of normalized \tras latency for
    different number of simultaneously activated DRAM rows.}
    \label{crow_fig:trcd_tras_tradeoff2}
\end{figure}

We explore the trade-off space between reducing \trcd and reducing \tras for a
varying number of simultaneously-activated rows using our SPICE model. In
Figure~\ref{crow_fig:trcd_tras_tradeoff2}, we show the different \trcd and \tras
latencies that can be used with multiple-row activation (MRA) while still
ensuring data correctness. For two-row activation, we empirically find that a
21\% reduction in \trcd and a 33\% reduction in \tras provides the best
performance on average for the workloads that we evaluate
(\cref{crow_subsec:crow_perf}).

We also eliminate the \twr overhead of MRA by terminating the restoration step
early when writing to simultaneously-activated DRAM rows. Since reducing \twr
increases the \trcd latency for the next activation of the row, there exists a
trade-off between reducing \trcd and reducing \twr, similar to the trade-off
between reducing \trcd and reducing \tras. In our evaluations, we find that to
achieve a 21\% reduction in \trcd, we can reduce \twr by 13\% (see
Table~\ref{crow_table:timings_summary}) by terminating the restoration step early
during a write to simultaneously-activated DRAM rows.

\subsection{MRA-Based Row Copy Latency}
\label{crow_sec:spice:actc}

As we explain in \cref{crow_subsubsec:row_copy}, \mech-cache uses the new
\cmdactc command to efficiently copy a regular row to a \copyrow in the same
subarray by activating the \copyrow slightly after activating the regular row.
This gives the sense amplifiers sufficient time to correctly read the data
stored in the regular row. Using our SPICE model, we find that \cmdactc does not
affect \trcd because the \copyrow is activated after satisfying \trcd. This is
because, to start restoring the data of the regular row to both the regular row
and the \copyrow, the local row buffer first needs to correctly latch the data
of the regular row. In contrast to \trcd, the \cmdactc command increases \tras
by 18\% (reduces \tras by 7\% when terminating restoration early), as restoring
two DRAM rows requires more time than restoring a single row. We model this
additional time in all of our evaluations.
\section{Hardware Overhead}
\label{crow_sec:hw_overhead}

Implementing the \mech substrate requires only a small number of changes in the
memory controller and the DRAM die. 

\subsection{Memory Controller Overhead}
\label{crow_subsec:crow_table_overhead}

\mech introduces the \mech table (\cref{crow_subsec:ctable}), which
incurs modest storage overhead in the memory controller. The storage requirement
for each entry ($Storage_{\text{\em entry}}$) in \mech table can be calculated
in terms of bits using the following equation:
\begin{equation}\label{crow_eq:storage_entry}
Storage_{\text{\em entry}} = \lceil log_2(RR) \rceil + Bits_\text{\em Special} + Bits_\text{\em Allocated}
\end{equation}
where $RR$ is the number of regular rows in each subarray (we take the log of
$RR$ to represent the number of bits needed to store the \texttt{RegularRowID}
field), $Bits_\text{\em Special}$ is the size of the \texttt{Special} field in
bits, and $Bits_\text{\em Allocated}$ is set to 1 to indicate the single bit
used to track whether the entry is currently valid. Note that the
\texttt{RegularRowID} field does not have to store the entire row address.
Instead, it is sufficient to store a pointer to the position of the row in the
subarray. For an example subarray with 512 regular rows, an index range of
0--511 is sufficient to address all regular rows in same subarray. We assume one
bit for the \texttt{Special} field, which we need to distinguish between the
\mech-cache and \mech-ref mechanisms (\cref{crow_sec:crow_apps}).

We calculate $Storage_\text{\em \mech table}$, the storage requirement for 
the entire \mech table, as:
\begin{equation}\label{crow_eq:storage_crow_table}
Storage_\text{\em \mech table} = Storage_\text{\em entry}*CR*SA
\end{equation}
where \emph{CR} is the number of {\copyrow}s in each subarray, and
\emph{SA} is the total number of subarrays in the DRAM. We calculate the
storage requirement of the \mech table for a single-channel memory system
with 512 regular rows per subarray, 1024 subarrays (8 banks; 128
subarrays per bank), and 8 {\copyrow}s per subarray to be
\SI{11.3}{\kibi\byte}. Thus, the processor die overhead is small.

The \mech table storage overhead is proportional to the memory size. Although
the overhead is small, we can further optimize the \mech table implementation to
reduce its storage overhead for large memories. One such optimization shares one
set of \mech table entries across multiple subarrays. While this limits the
number of \copyrow{s} that can be utilized simultaneously, \mech can still
capture the majority of the benefits that would be provided if the \mech
substrate had a separate \mech table for each subarray. From our evaluations,
when sharing each \mech table entry between 4 subarrays (i.e., approximately a
factor of 4 reduction in \mech table storage requirements), we observe that the
average speedup that \mech-cache provides for single-core workloads reduces from
7.1\% to only 6.1\%.

We evaluate the access time of a \mech table with the configuration given in
Table~\ref{crow_table:system_config} using CACTI~\cite{muralimanohar2009cacti}, and
find that the access time is only \SI{0.14}{\nano\second}. We do \emph{not}
expect the table access time to have any impact on the overall cycle time of the
memory controller.

\subsection{DRAM Die Area Overhead}
\label{crow_subsec:crow_power_area}

To activate multiple rows in the same subarray, we modify the row decoder to
drive multiple wordlines at the same time. These modifications incur a small
area overhead and cause MRA to consume more power compared to a single-row
activation. 

\begin{figure}[h] \centering
    \begin{subfigure}[b]{0.49\linewidth}
        \centering
        \includegraphics[width=\linewidth]{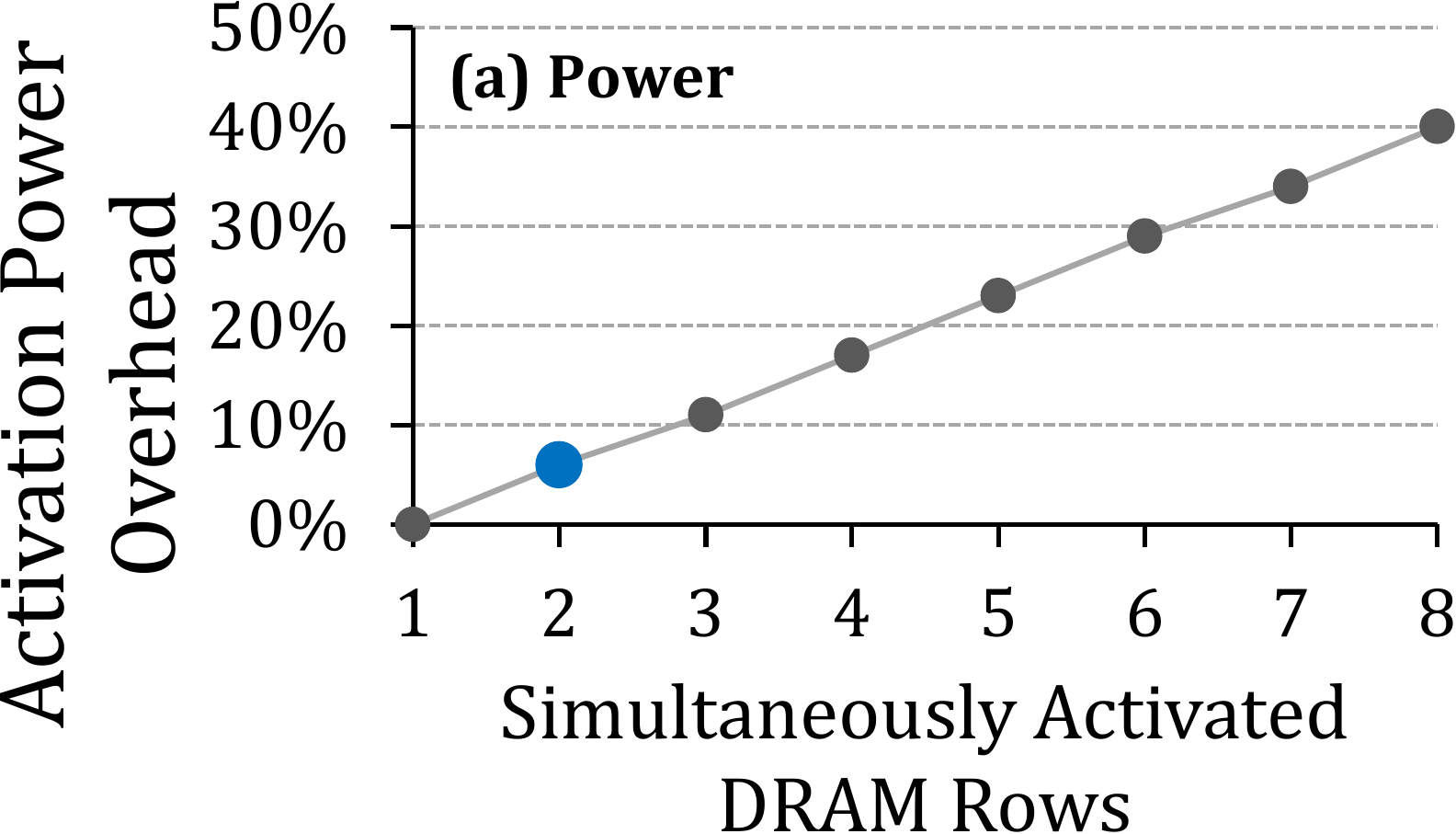}%
    \end{subfigure}
    \begin{subfigure}[b]{0.49\linewidth}
        \centering
        \includegraphics[width=\linewidth]{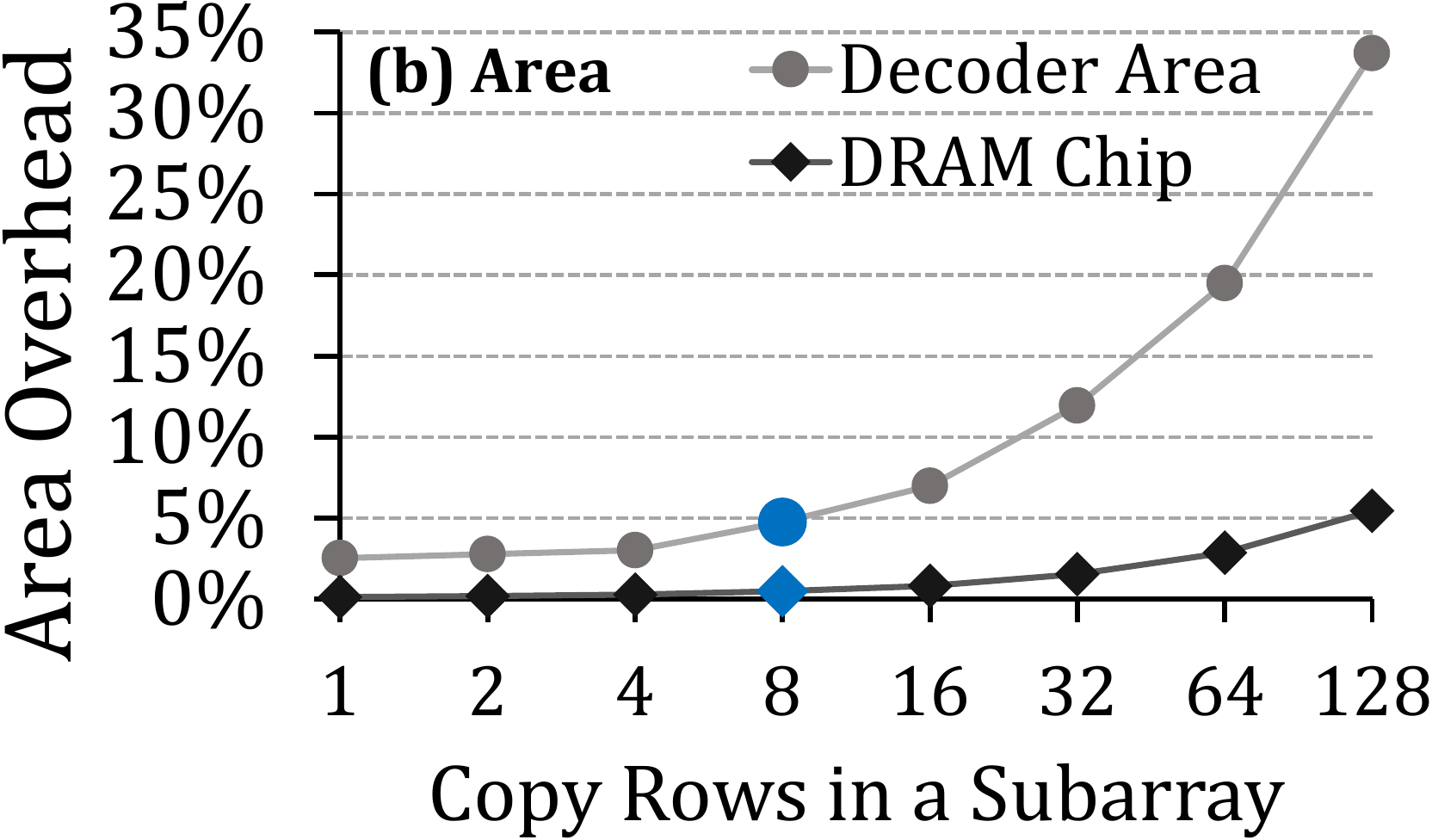}%
    \end{subfigure}
    \caption{Power consumption and area overhead of MRA.}
    \label{crow_fig:crow_power_area}
\end{figure}

In Figure~\ref{crow_fig:crow_power_area} (left), we show the activation power
overhead for simultaneously activating up to nine DRAM rows in the same
subarray. The \cmdactc and \cmdactt commands simultaneously activate two rows,
and consume only 5.8\% additional power compared to a conventional \cmdact{}
command that activates only a single row. The slight increase in power
consumption is mainly due to the small \copyrow decoder that \mech{} introduces
to drive the {\copyrow}s.

Figure~\ref{crow_fig:crow_power_area} (right) plots the area overhead of a \copyrow
decoder that enables one of the \copyrow{s} in the subarray independently from
the existing regular row decoder. The figure shows the \copyrow decoder area
overhead as we vary the number of \copyrow{s} in a subarray. For \mech-8, which
has eight \copyrow{s} per subarray, the decoder area increases by 4.8\%, which
corresponds to only a 0.48\% area overhead for the entire DRAM chip. The area
overhead of \mech-8 is small because the decoder required for eight \copyrow{}s
is as small as \SI{9.6}{\micro\meter\squared}, while our evaluations show that
the local row decoder for 512~regular rows occupies
\SI{200.9}{\micro\meter\squared}.

\section{Methodology}
\label{crow_sec:methodology}

We use Ramulator~\cite{kim2015ramulator, ramulatorgithub}, a cycle-accurate DRAM
simulator, to evaluate the performance of the two mechanisms that we propose
based on the \mech substrate. We run Ramulator in CPU-trace-driven mode and
collect application traces using a custom Pintool~\cite{luk2005pin}. The traces
include virtual addresses that are accessed by the application during the trace
collection run. In Ramulator, we perform virtual-to-physical address translation
by randomly allocating a \SI{4}{\kibi\byte} physical frame for each access to a
new virtual page, which emulates the behavior of a steady-state
system~\cite{park2013regularities}.

Table~\ref{crow_table:system_config} provides the system configuration we evaluate.
Unless stated otherwise, we implement \mech with eight {\copyrow}s per subarray.
We analyze the performance of \mech-cache and \mech-ref on single- and
multi-core system configurations using typical LPDDR4 timing parameters, as
shown in the table. Although our evaluation is based on LPDDR4 DRAM, which is
predominantly used in various low-power systems~\cite{ghose2019demystifying}, the
mechanisms that we propose can also reduce access latency in other DRAM-based
memories, including 3D-stacked DRAM~\cite{lee2016simultaneous}.

\begin{table}[h!] \caption{Simulated system configuration.}
    \centering 
    \renewcommand{\arraystretch}{1.6} 
    \begin{tabular}{m{3.9cm} m{10.5cm}}
        \hline
        \textbf{Processor} & 1-4 cores, \SI{4}{\giga\hertz} clock frequency, 4-wide issue,
        8~MSHRs per core, 128-entry instruction window\\ 
        \hline 
        \textbf{Last-Level Cache} & \SI{64}{\byte} cache-line, 8-way associative, \SI{8}{\mebi\byte} capacity \\ 
        \hline 
	\textbf{Memory Controller} & 64-entry read/write
        request queue, FR-FCFS-Cap\footnotemark~scheduling
        policy~\cite{mutlu2007stall}, timeout-based row
        buffer policy\footnotemark \\
        \hline
        \textbf{DRAM} & LPDDR4~\cite{jedec2014lpddr4}, \SI{1600}{\mega\hertz} bus
            frequency, 4 channels, 1 rank,
        8 banks/rank, 64K rows/bank, 512 rows/subarray, \SI{8}{\kibi\byte} row buffer size,
        \trcd/\tras/\twr~29 (18)/67 (42)/29 (18) cycles (ns) \\ 
        \hline
    \end{tabular}
\label{crow_table:system_config} 
\end{table} 
\addtocounter{footnote}{-2}
\stepcounter{footnote}\footnotetext{We use a variation of the
FR-FCFS~\cite{rixner2000memory, zuravleff1997controller} policy, called
FR-FCFS-Cap~\cite{mutlu2007stall}, which improves fairness by enforcing an upper
limit for the number of read/write requests that a row can service once
activated. This policy performs better, on average, than the conventional
FR-FCFS policy~\cite{rixner2000memory, zuravleff1997controller}, as also shown
in~\cite{mutlu2007stall, kim2010thread, subramanian2014blacklisting,
subramanian2016bliss}.}
        
\stepcounter{footnote}\footnotetext{The timeout-based row buffer management
policy closes an open row after $75~ns$ if there are no requests in the memory
controller queues to that row.}

\textbf{Workloads.} We evaluate 44 single-core applications from four benchmark
suites: SPEC CPU2006~\cite{spec2006}, TPC~\cite{tpc}, STREAM~\cite{stream}, and
MediaBench~\cite{fritts2005mediabench}. In addition, we evaluate two synthetic
applications~\cite{mutlu2007memory} (excluded from our average performance
calculations): 1)~\emph{random}, which accesses memory locations at random and
has very limited row-level locality; and 2)~\emph{streaming}, which has high
row-level locality because it accesses contiguous locations in DRAM such that
the time interval between two consecutive memory requests is long enough for the
memory controller to precharge the recently-open row.

We classify the applications into three groups based on the
misses-per-kilo-instruction (MPKI) in the last-level cache. We obtain the MPKI
of each application by analyzing SimPoint~\cite{simpoint} traces (200M
instructions) of each application's representative phases using the single-core
configuration. The three groups are:
\begin{list}{$\bullet$} {
        \setlength{\itemsep}{-1pt}
        \setlength{\parsep}{2pt}
        \setlength{\topsep}{0pt}
        \setlength{\partopsep}{0pt}
        \setlength{\leftmargin}{2.0em}
        \setlength{\labelwidth}{1em}
        \setlength{\labelsep}{0.5em}
    }
    \item L (low memory intensity): $MPKI < 1$
    \item M (medium memory intensity): $1 \leq MPKI < 10$
    \item H (high memory intensity): $MPKI \geq 10$
\end{list}

We create eight multi-programmed workload groups for the four-core
configuration, where each group consists of 20 multi-programmed workloads. Each
group has a mix of workloads of different memory intensity classes. For example,
\emph{LLHH} indicates a group of 20 four-core multi-programmed workloads, where
each workload consists of two randomly-selected single-core applications with
low memory intensity~(L) and two randomly-selected single-core applications with
high memory intensity~(H). In total, we evaluate 160 multi-programmed workloads.
We simulate the multi-core workloads until each core executes at least 200
million instructions. For all configurations, we initially warm up the caches by
fast-forwarding 100 million instructions. 

\textbf{Metrics.} We measure the speedup for single-core applications using the instructions per
cycle (IPC) metric. For multi-core evaluations, we use the weighted speedup
metric~\cite{snavely2000symbiotic}, which prior work shows is a good
measure of job throughput~\cite{eyerman2008system}.

We use CACTI~\cite{muralimanohar2009cacti} to evaluate the DRAM area and power
overhead of our two mechanisms (i.e., \mech-cache and \mech-ref) and prior works
(i.e., TL-DRAM~\cite{lee2013tiered} and SALP~\cite{kim2012case}) that we
compare against. We perform a detailed evaluation of the latency impact of MRA
using our circuit-level SPICE model~\cite{crow_spice_github}. We use
DRAMPower~\cite{chandrasekar2012drampower} to estimate DRAM energy consumption
of our workloads.

\section{Evaluation}
\label{crow_sec:results}

We evaluate the performance, energy efficiency, and area overhead of the
\mech-cache and \mech-ref mechanisms.

\subsection{\mech-cache}
\label{crow_sec:results:cache}

We evaluate \mech-cache with different numbers of {\copyrow}s per subarray in
the \mech substrate. We denote each configuration of \mech in the form of
\mech-$C_r$, where $C_r$ specifies the number of {\copyrow}s (e.g., we use
\mech-8 to refer to \mech with eight {\copyrow}s). To show the potential of
\mech-cache, we also evaluate a hypothetical configuration with a 100\%
\mech table hit rate, referred to as \emph{Ideal \mech-cache}.

\subsubsection{Single-core Performance.}
\label{crow_subsec:crow_perf}

Figure~\ref{crow_fig:crow_sc_perf} (top) shows the speedup of \mech-cache over the
baseline and the \mech table hit rate (bottom) for single-core applications. We
also list the last-level cache MPKI of each application to indicate memory
intensity. We make four observations from the figure.

\afterpage{%
    \clearpage
    \begin{landscape}

\begin{figure}[tp] \centering
    \begin{subfigure}{\linewidth}
        \begin{subfigure}[c]{0.84\linewidth}
            \centering
            \includegraphics[width=\linewidth]{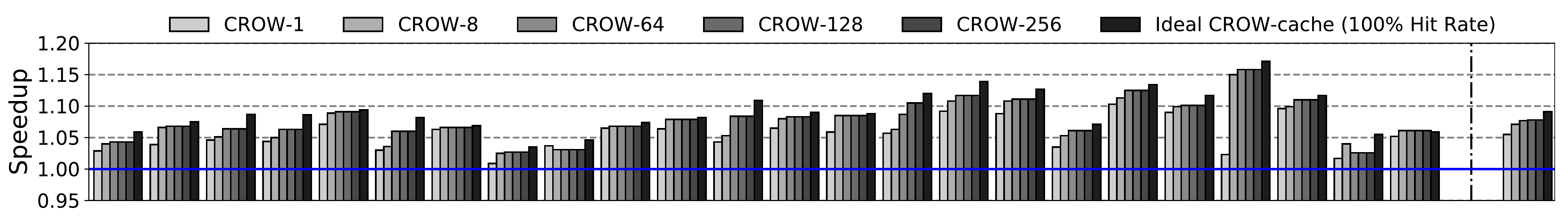}%
            \label{fig:speedup_sc}
            \end{subfigure}
        \begin{subfigure}[c]{0.145\linewidth}
            \centering
            \vspace{1.8mm}
            \includegraphics[width=\linewidth]{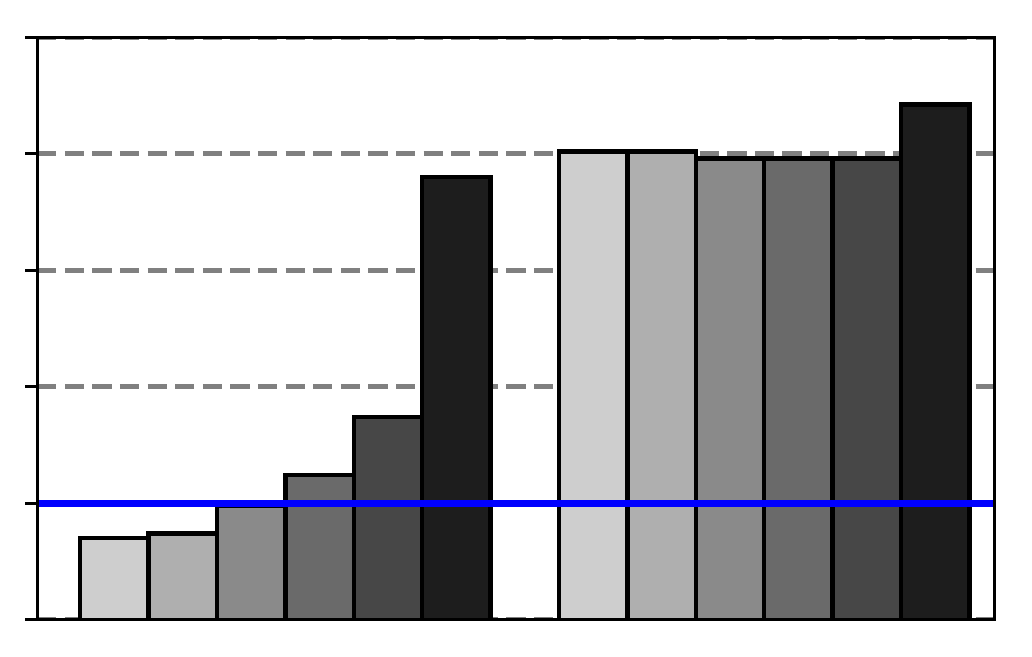}%
            \label{crow_fig:speedup_sc_synth}
        \end{subfigure}
    \end{subfigure}
    \begin{subfigure}{\linewidth}
        \begin{subfigure}[c]{0.84\linewidth}
            \centering
            \includegraphics[width=\linewidth]{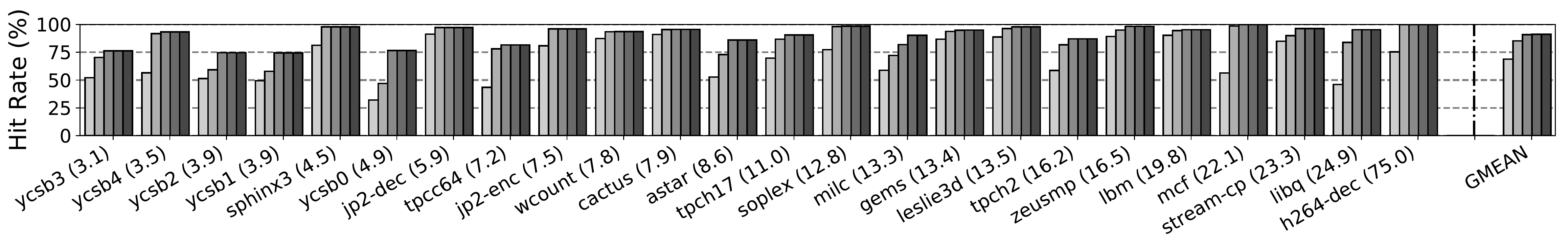}%
            \label{crow_fig:hit_rates_sc}
        \end{subfigure}
        \begin{subfigure}[c]{0.155\linewidth}
            \centering
            \vspace{1.2mm}
            \includegraphics[width=\linewidth]{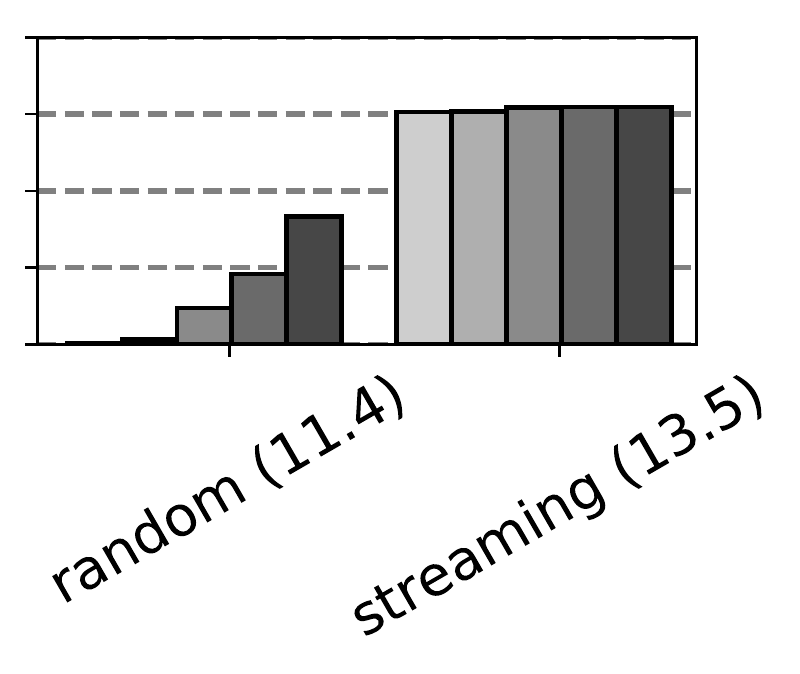}%
            \label{crow_fig:hit_rates_sc_synth}
        \end{subfigure}
    \end{subfigure}

        \caption{Speedup and \mech table hit rate of different configurations
        of \mech-cache for single-core applications. The MPKI of each
        application is listed in parentheses.}
        \label{crow_fig:crow_sc_perf}
\end{figure}
\end{landscape}
\clearpage
}

First, \mech-cache provides significant speedups. On average, \mech-1, \mech-8,
and \mech-256 improve performance by 5.5\%, 7.1\%, and 7.8\%, respectively, over
the baseline.\footnotemark\ In general, the \mech-cache speedup increases with
the application's memory intensity. Some memory-intensive applications (e.g.,
\emph{libq}, \emph{h264-dec}) have lower speedups than less-memory-intensive
applications because they exhibit high row buffer locality, which reduces the
benefits of \mech-cache. Applications with low MPKI ($<3$, not plotted for
brevity) achieve speedups below 5\% due to limited memory activity, but
\emph{no} application experiences slow down.

Second, for most applications, \mech-1, with \emph{only} one \copyrow per
subarray, achieves most of the performance of configurations with more
{\copyrow}s (on average, 60\% of the performance improvement of Ideal
\mech-cache).

Third, the \mech table hit rate is very high for most of the applications. This
is indicative of high in-DRAM locality and translates directly into performance
improvement. On average, the hit rate for \mech-1/\mech-8/\mech-256 is
68.8\%/85.3\%/91.1\%. 

Fourth, the overhead of fully restoring rows evicted from the \mech table is
negligible (not plotted). For \mech-1, which has the highest overhead from these
evictions, restoring evicted rows accounts for only 0.6\% of the total
activation operations.

\footnotetext{We notice that, for some applications (e.g., \emph{jp2-encode}),
using fewer {\copyrow}s slightly outperforms a configuration with more
{\copyrow}s. We find the reason to be the changes in memory request service
order due to the fact that memory controller makes different command scheduling
decisions for different configurations.}

\subsubsection{Multi-Core Performance.}
\label{crow_subsec:crow_perf_mc}

\begin{figure}[t] 
    \centering
    \includegraphics[width=0.99\linewidth]{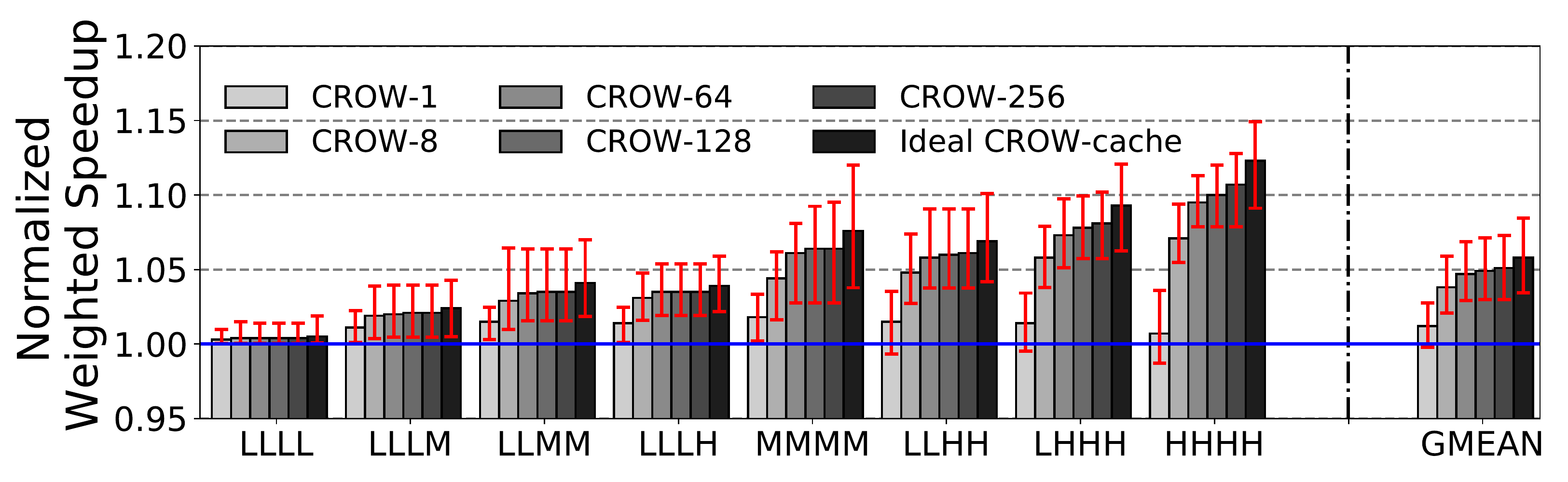}%
    \caption{\mech-cache speedup (160 four-core workloads).}
    \label{crow_fig:four_core_speedup}
\end{figure}

Figure~\ref{crow_fig:four_core_speedup} shows the weighted speedup \mech-cache
achieves with different \mech substrate configurations for four-core workload
groups. The bars show the average speedup for the workloads in the corresponding
workload group, and the vertical lines show the maximum and minimum speedup
among the workloads in the group.

We observe that \emph{all} \mech-cache configurations achieve a higher speedup
as the memory intensity of the workloads increase. On average, \mech-8 provides
7.4\% speedup for the workload group with four high-intensity workloads (i.e.,
\emph{HHHH}), whereas it provides only 0.4\% speedup for \emph{LLLL}. 

In contrast to single-core workloads, \mech-8 provides significantly better
speedup compared to \mech-1 on four-core configurations. This is because
simultaneously-running workloads are more likely to generate requests that
compete for the same subarray. As a result, the \mech table cannot achieve a
high hit rate with a single \copyrow per subarray. In most cases, \mech-8
performs close to Ideal \mech-cache with 100\% hit rate, and requires only 1.6\%
of the DRAM storage capacity for in-DRAM caching.

\subsubsection{DRAM Energy Consumption}
\begin{figure}[h]
    \centering
    \begin{subfigure}[t]{0.23\linewidth}
        \includegraphics[width=\linewidth]{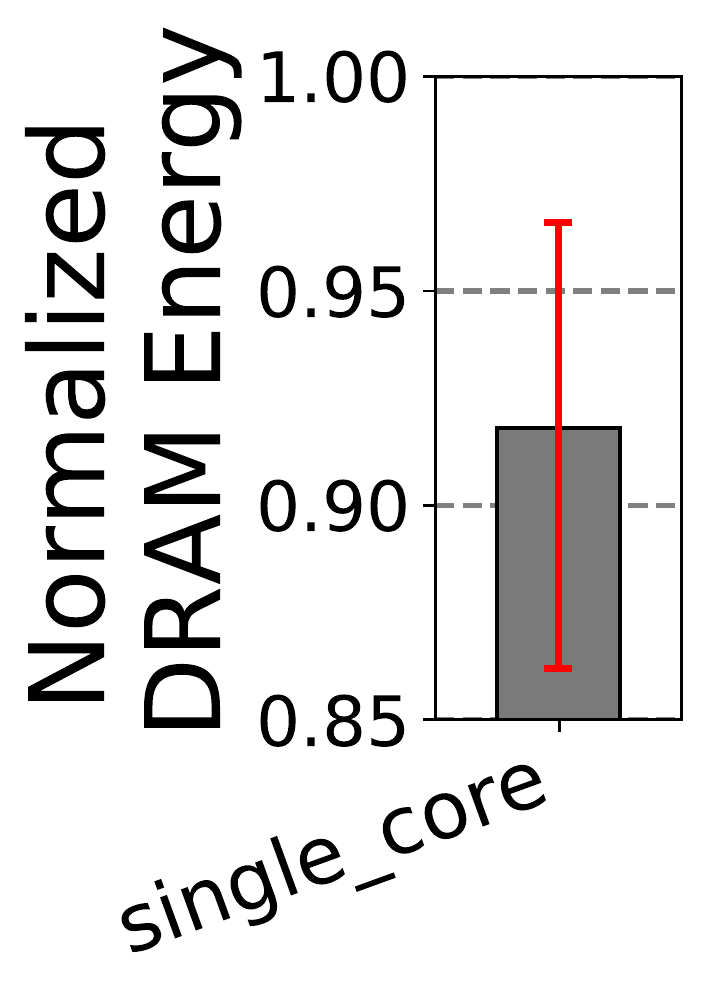}%
    \end{subfigure}
    \begin{subfigure}[t]{0.46\linewidth}
        \includegraphics[width=\linewidth]{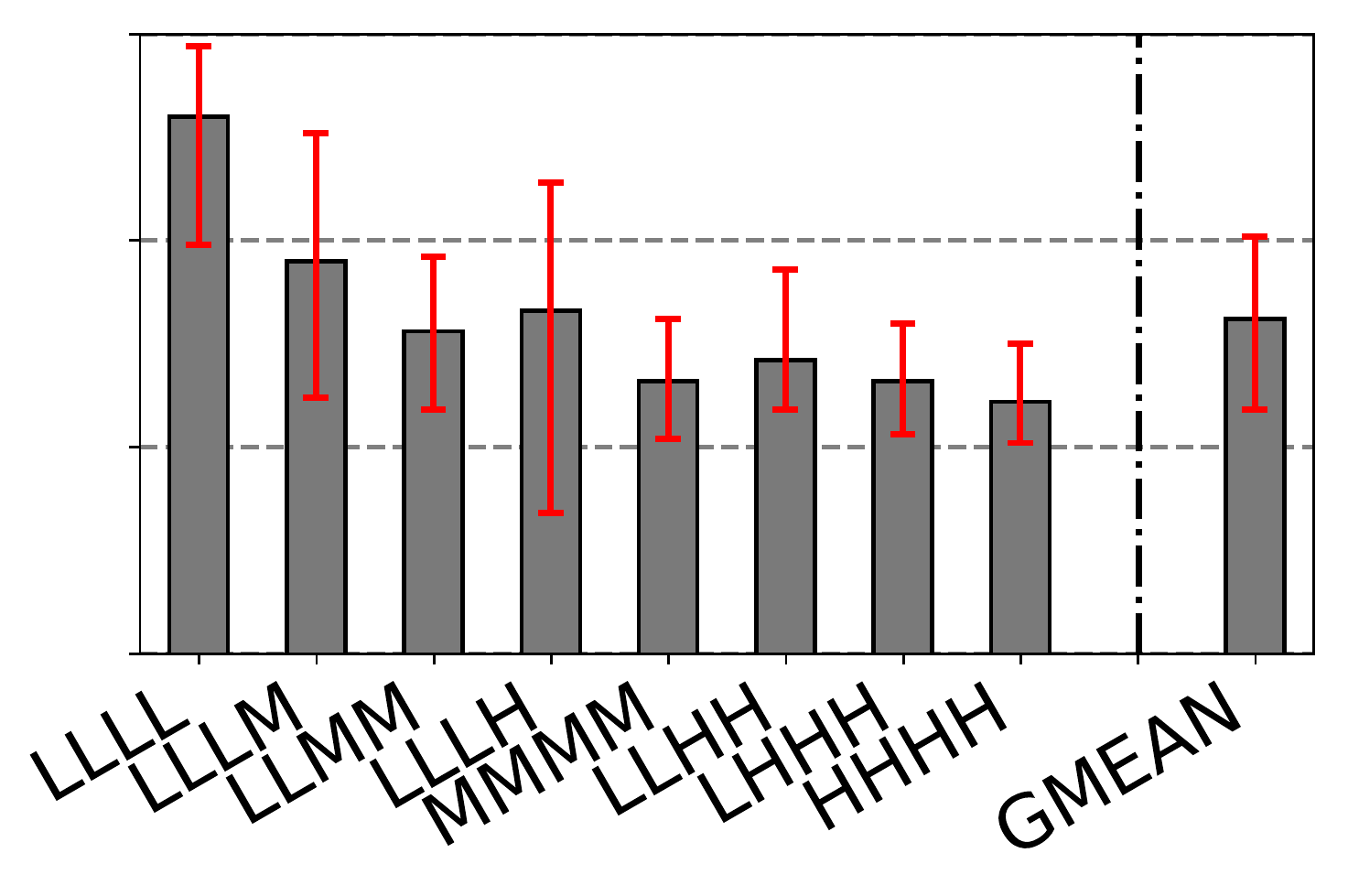}%
    \end{subfigure}
    \caption{DRAM energy consumption with \mech-cache.}
    \label{crow_fig:dpower_crow}
\end{figure}

We evaluate the total DRAM energy consumed during the execution of single-core
and four-core workloads. Figure~\ref{crow_fig:dpower_crow} shows the average DRAM
energy consumption with \mech-cache normalized to the baseline. Although each
\cmdactc and \cmdactt{} command consumes more power than \cmdact, \mech-cache
reduces the total DRAM energy due to the improvement in execution time. On
average, \mech-cache decreases DRAM energy consumption by 8.2\% and 6.9\% on
single- and four-core systems, respectively.

\subsubsection{Comparison to TL-DRAM and SALP}
\label{crow_subsec:comparison_to_salp_tldram}

We compare \mech-cache to two previous works that also enable in-DRAM caching in
different ways: TL-DRAM~\cite{lee2013tiered} and SALP~\cite{kim2012case}. 

TL-DRAM uses isolation transistors on each bitline to split a DRAM subarray into
a \emph{far} and a \emph{near} segment. When the isolation transistors are
turned off, the DRAM rows in the \emph{near segment}, which is composed of a
small number of rows close to the sense amplifiers, can be accessed with low
\trcd and low \tras due to reduced parasitic capacitance and resistance on the
short bitline. In contrast, accessing the far segment requires a slightly higher
\trcd and \tras than in conventional DRAM due to the additional parasitic
capacitance and resistance that the isolation transistor adds to the bitline. We
extend our circuit-level DRAM model to evaluate the reduction in DRAM latencies
for different far and near segment sizes. We use the notation TL-DRAM-$N_r$,
where $N_r$ specifies the number of rows in the near segment. TL-DRAM uses the
rows in the near segment as a cache by copying the most-recently accessed rows
in the far segment to the near segment. Thus, similar to our caching mechanism,
TL-DRAM requires an efficient in-DRAM row copy operation. We reuse the \cmdactc
command that we implement for \mech-cache to perform the row copy in TL-DRAM.

SALP~\cite{kim2012case} modifies the row decoder logic to enable parallelism
among subarrays. As opposed to a conventional DRAM bank where only a single row
can be active at a time, SALP enables the activation of multiple local row
buffers independently from each other to provide fast access to the
most-recently-activated row of each subarray. We evaluate the SALP-MASA mode,
which outperforms the other two modes that Kim et al.~\cite{kim2012case}
propose. We evaluate SALP with a different number of subarrays per bank, which
we indicate as SALP-$N_s$, where $N_s$ stands for the number of subarrays in a
bank. For SALP, we use both the timeout-based and the open-page (denoted as
SALP-$N_s$-O) row buffer management policies. The open-page policy keeps a row
open until a local row buffer conflict. Although this increases the performance
benefit of SALP by preventing a local row buffer from being precharged before it
is reused, SALP with open-page policy consumes more energy since rows remain
active for a longer time. Note that, in SALP, the in-DRAM cache capacity changes
with the number of subarrays in the DRAM chip as each subarray can cache a
single row in its local row buffer. To evaluate SALP with higher cache capacity,
we reduce the number of rows in each subarray by increasing the number of
subarrays in a bank, thus keeping the DRAM capacity constant.

\begin{figure}[!ht]
    \centering
    \begin{subfigure}[t]{0.49\linewidth}
        \centering
        \includegraphics[width=\linewidth]{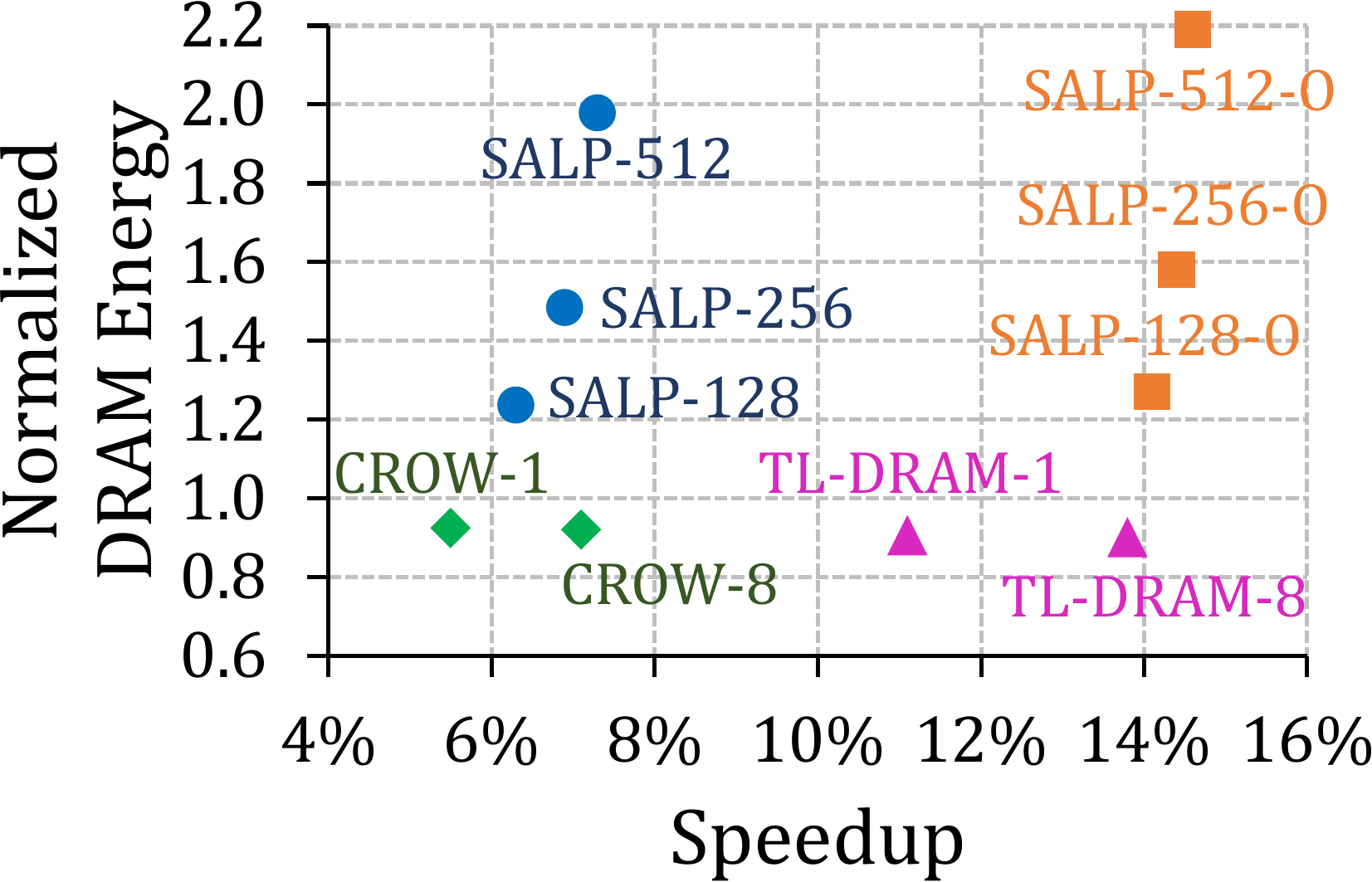}%
        \caption{Energy vs. speedup}
        \label{crow_fig:energy_vs_speedup}
    \end{subfigure}
    \begin{subfigure}[t]{0.49\linewidth}
        \centering
        \includegraphics[width=\linewidth]{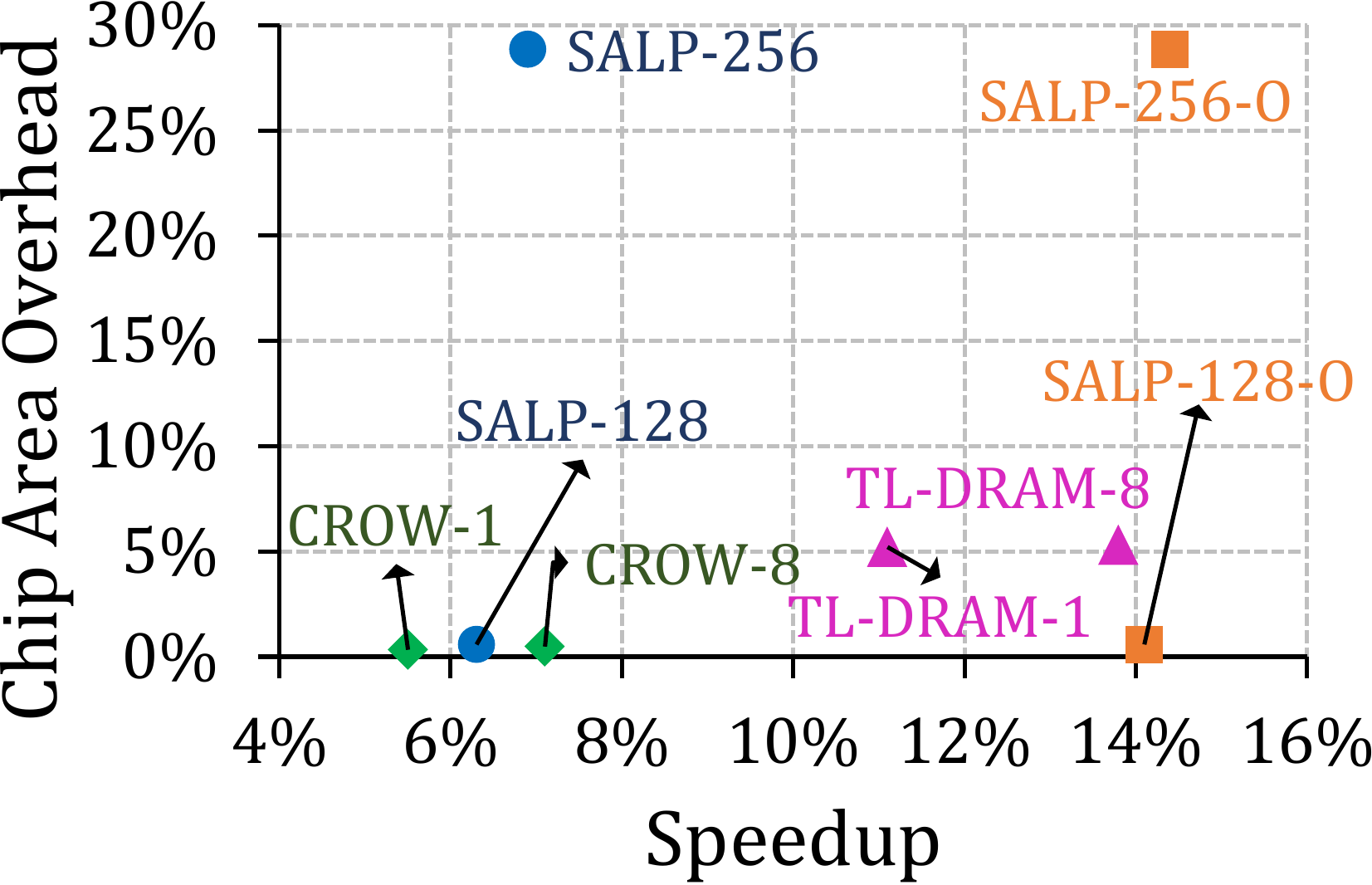}%
        \caption{Chip area vs. speedup}
        \label{crow_fig:area_vs_speedup}
    \end{subfigure}
    \caption{\mech-cache vs. TL-DRAM~\cite{lee2013tiered} \& SALP~\cite{kim2012case}.}
    \label{crow_fig:crow_tldram_salp_comparison}
\end{figure}

Figure~\ref{crow_fig:crow_tldram_salp_comparison} compares the performance, energy
efficiency, and DRAM chip area overhead of different configurations of
\mech-cache, TL-DRAM~\cite{lee2013tiered}, and SALP~\cite{kim2012case} for
single-core workloads. We draw four conclusions from the figure. 

First, all SALP configurations using the open-page policy outperform
\mech-cache. However, SALP increases the DRAM energy consumption significantly
as it frequently keeps multiple local row buffers active, each of which consume
significant static power (as also shown in~\cite{kim2012case}). An idle LPDDR4
chip that has only a single bank with an open row draws 10.9\% more current
($I_{DD}3N$) compared to the current ($I_{DD}2N$) when all banks are in closed
state~\cite{micron-lpddr4}. In SALP, the static power consumption is much higher
than in conventional DRAM, since multiple local row buffers per bank can be
active at the same time, whereas only one local row buffer per bank can be
active in conventional DRAM.

Second, increasing the number of subarrays in a bank increases the in-DRAM cache
capacity and, thus, the performance benefits of SALP. However, with the
open-page policy, SALP-256 has a 58.4\% DRAM energy and 28.9\% area overhead
over the baseline, which is much higher than the 27.3\% DRAM energy and 0.6\%
area overhead of SALP-128. SALP-512 has even larger DRAM energy (119.3\%) and
area (84.5\%) overheads (not plotted). In contrast, \mech-8 \emph{reduces} DRAM
energy (by 8.2\%) at very low (0.48\%) DRAM chip area overhead.

Third, TL-DRAM-8 provides a higher speedup of 13.8\% compared to \mech-8, which
provides 7.1\% speedup (while reserving the same DRAM storage capacity for
caching as TL-DRAM-8). This is because the latency reduction benefit of a very
small TL-DRAM near-segment, which comprises only one or eight rows, is higher
than the latency reduction benefit of \mech's two-row activation. According to
our circuit-level simulations, a TL-DRAM near-segment with eight rows can be
accessed with a 73\% reduction in \trcd and an 80\% reduction in \tras. However,
this comes at the cost of high DRAM chip area overhead, as we explain next. 

Fourth, in TL-DRAM, the addition of an isolation transistor to each bitline
incurs high area overhead. As seen in Figure~\ref{crow_fig:area_vs_speedup},
TL-DRAM-8 incurs 6.9\% DRAM chip area overhead, whereas \mech-8 incurs
\emph{only} 0.48\% DRAM chip area overhead. 

We conclude that \mech enables a more practical and lower cost in-DRAM caching
mechanism than TL-DRAM and SALP.

\subsubsection{\mech-cache and Prefetcher.} 
\begin{figure}[h] 
    \centering
    \includegraphics[width=1.0\linewidth]{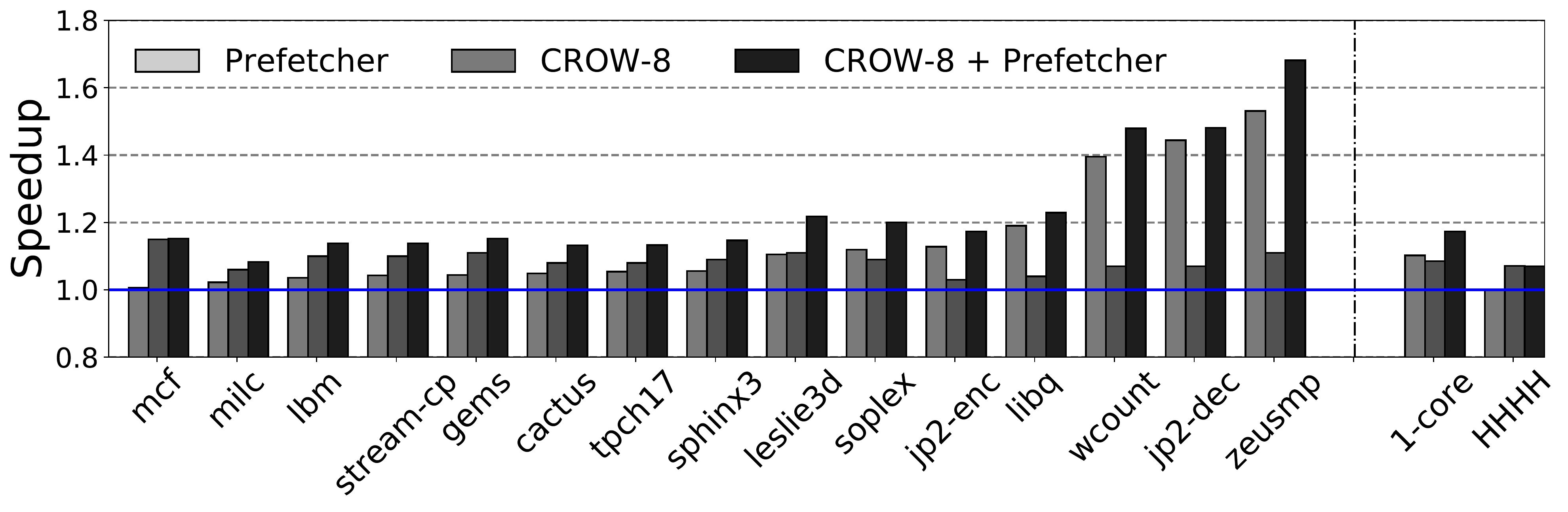}%
    \caption{\mech-cache and prefetching.}
    \label{crow_fig:res_sc_prefetch}
\end{figure}

We evaluate the performance benefits of \mech-cache on a system with a stride
prefetcher, which we implement based on the RPT
prefetcher~\cite{iacobovici2004effective}. In Figure~\ref{crow_fig:res_sc_prefetch},
we show the speedup that the prefetcher, \mech-cache, and the combination of the
prefetcher and \mech-cache achieve over the baseline, which does not implement
prefetching. For brevity, we only show results for a small number of workloads,
which we sampled to be representative of workloads where the prefetcher provides
different levels of effectiveness. We observe that, in most cases, \mech-cache
operates synergistically with the prefetcher, i.e., \mech-cache serves both read
and prefetch requests with low latency, which further improves average system
performance by 5.7\% over the prefetcher across all single-core workloads.

\subsection{\mech-ref}

Our weak row remapping scheme, \mech-ref, extends the refresh interval from
\SI{64}{\milli\second} to \SI{128}{\milli\second} of DRAM chip by eliminating
the small set of rows that have retention time below \SI{128}{\milli\second}.
For our evaluations, we assume three weak rows for each subarray, which is much
more than expected, given data from real DRAM devices~\cite{liu2012raidr,
patel2017reaper, kim2009new}.\footnote{Based on
Equation~\ref{crow_eq:p_subarray_n} in \cref{crow_subsec:identify_weak_rows},
the probability of a having subarray with more than 3 weak rows across the
entire DRAM is $9.3 \times 10^{-3}$.} Extending the refresh interval provides
performance and energy efficiency benefits as fewer refresh operations occur and
they interfere less with application requests. \mech-ref does not incur any
additional overhead other than allocating a few of the {\copyrow}s that are
available with the \mech substrate. The rest of the \copyrow{s} can potentially
be used by other \mech-based mechanisms, e.g., \mech-cache.

\begin{figure}[h] \centering
    \begin{subfigure}[b]{0.7\linewidth}
        \centering
        \includegraphics[width=\linewidth]{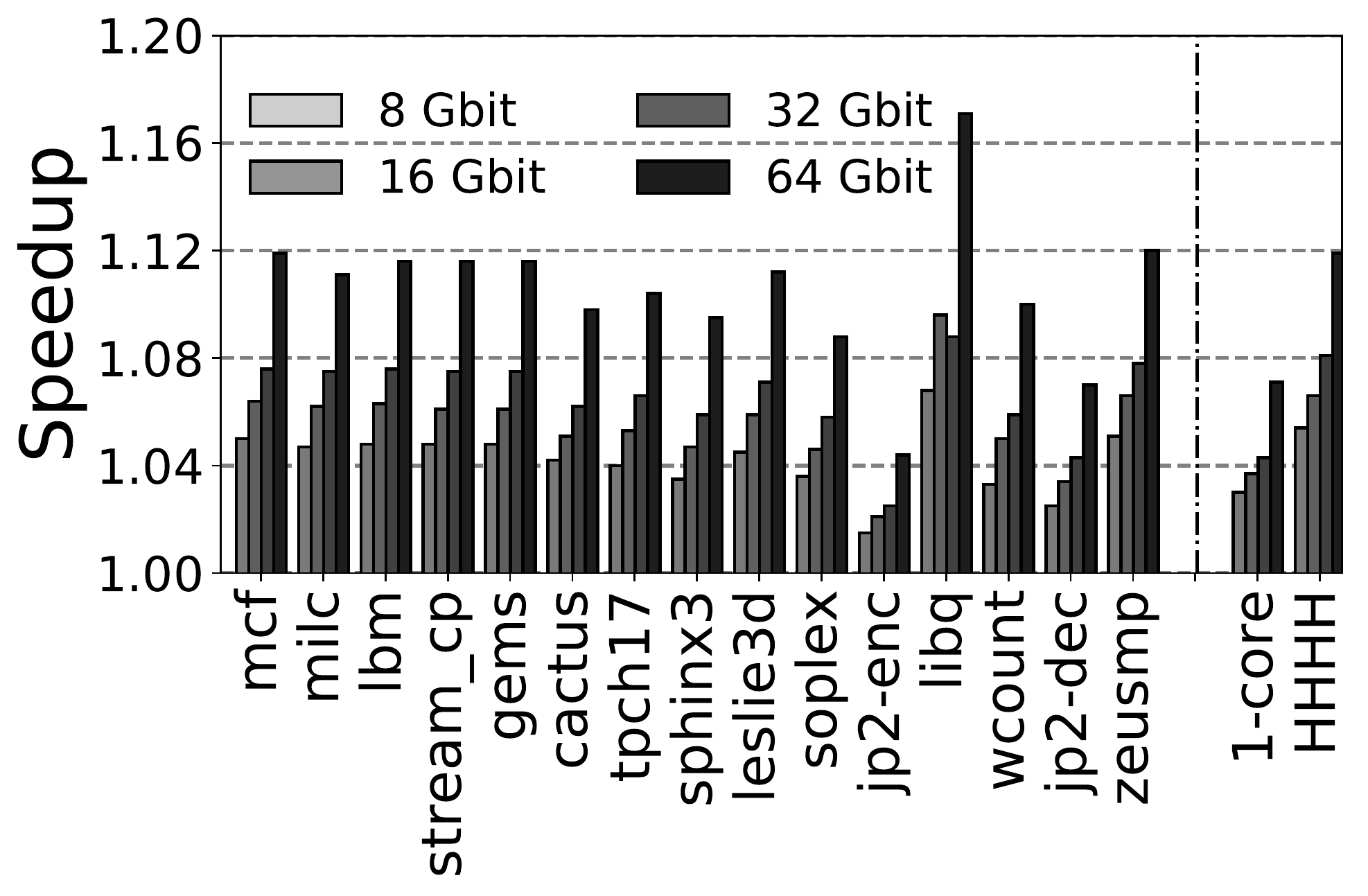}%
        \label{crow_fig:crow_ref_speedup}
    \end{subfigure}
    \begin{subfigure}[b]{0.75\linewidth}
        \centering
        \includegraphics[width=\linewidth]{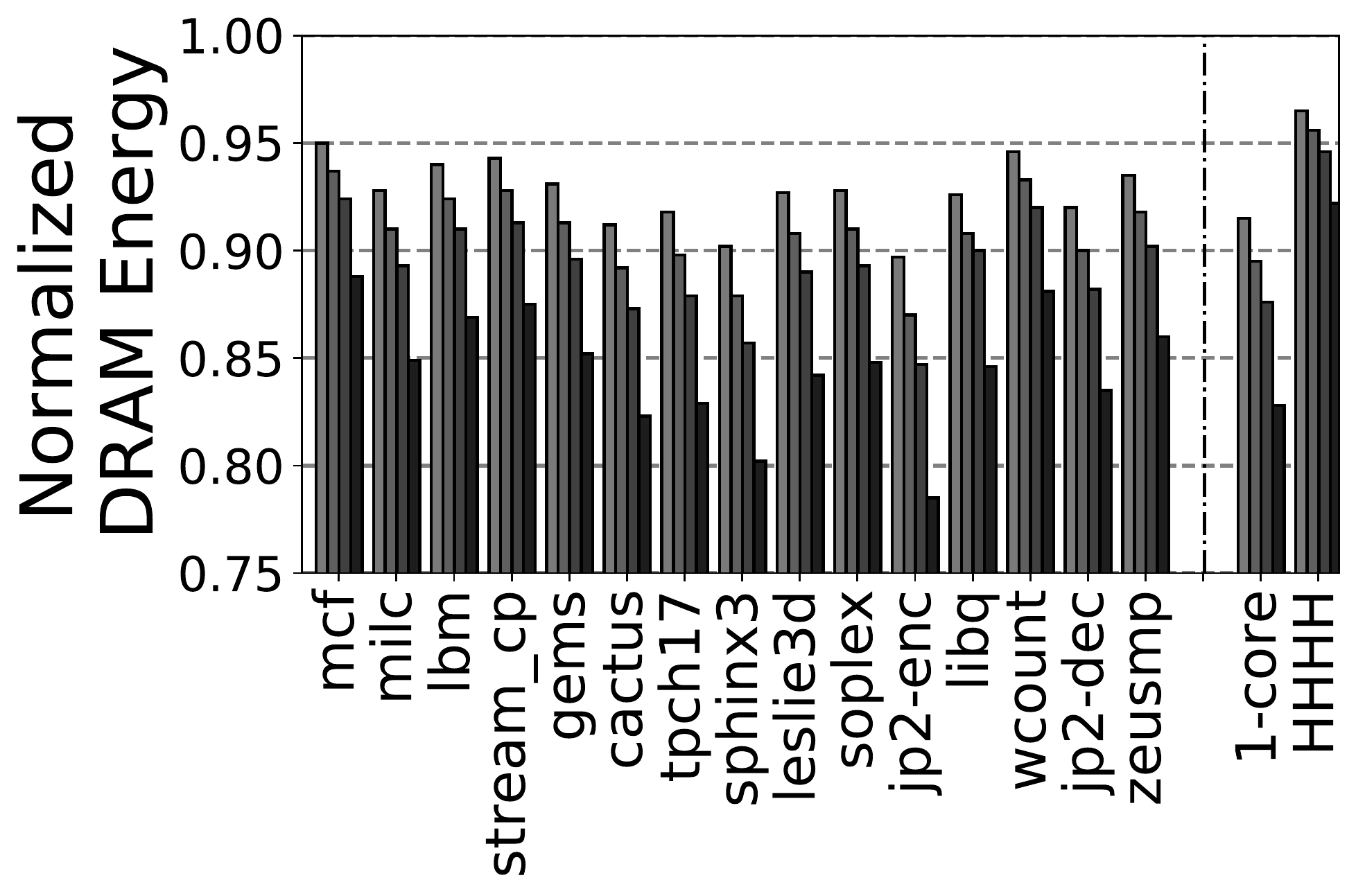}%
        \label{crow_fig:crow_ref_energy}
    \end{subfigure}
    \caption{\mech-ref speedup and DRAM energy.}
    \label{crow_fig:crow_ref_results}
\end{figure}

Figure~\ref{crow_fig:crow_ref_results} shows \mech-ref's average speedup and
normalized DRAM energy consumption for all single-core workloads and memory
intensive four-core workloads (\emph{HHHH}) for four DRAM chip densities
(\SIlist[list-units=single,list-final-separator = {, }, list-pair-separator= {,
}]{8;16;32;64}{\giga\bit}). We observe that, for a futuristic \SI{64}{\giga\bit}
DRAM chip, \mech-ref improves average performance by 7.1\%/11.9\% and reduces
DRAM energy consumption by 17.2\%/7.8\% for single-/multi-core workloads. The
energy benefits are lower for four-core workloads as DRAM accesses contribute a
larger portion of the overall DRAM energy compared to single-core workloads that
inject relatively few requests to the DRAM.

\subsection{Combining \mech-cache and \mech-ref}
\label{crow_subsec:results_combined}

In \cref{crow_sec:results:cache}, we note that not all applications
effectively use all available {\copyrow}s in \mech-cache since \mech-1 provides
speedup close to \mech-cache with more \copyrow{s}. Similarly, for \mech-ref,
previous work shows that there are only a small number (e.g., < 1000 on a
\SI{32}{\gibi\byte} DRAM~\cite{liu2012raidr}) of DRAM rows that must be
refreshed at the lowest refresh intervals, so it is very unlikely to have more
than a few weak rows (or even one) in a subarray. These two observations lead us
to believe that the two mechanisms, \mech-cache and \mech-ref, can be combined
to operate synergistically. We combine the two mechanisms such that \mech-cache
utilizes {\copyrow}s that remain available \emph{after} the weak row remapping
of \mech-ref. Combining the two mechanisms requires only a single additional bit
per \mech table entry to indicate whether a \copyrow is allocated for
\mech-cache or \mech-ref.

Figure~\ref{crow_fig:crow_cache_ref_results} summarizes the performance and energy
efficiency benefits of \mech-cache, \mech-ref, and their combination for an LLC
capacity ranging from \SI{512}{\kibi\byte} to \SI{32}{\mebi\byte} and
\SI{64}{\giga\bit} DRAM chip density. The figure compares the benefits of the
two mechanisms against a hypothetical \mech-based mechanism that has an ideal
\mech-cache with 100\% hit rate and that does not require any DRAM refresh. We
make three observations from the figure.

\afterpage{%
    \clearpage
    \begin{landscape}

\begin{figure}[tp] 
    \centering
    \begin{subfigure}[b]{.9\linewidth}
        \centering
        \includegraphics[width=\linewidth]{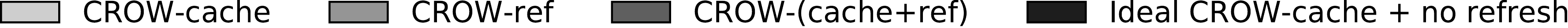}%
    \end{subfigure}

    \begin{subfigure}[b]{\linewidth}
        \begin{subfigure}[b]{0.49\linewidth}
            \centering
            \includegraphics[width=\linewidth]{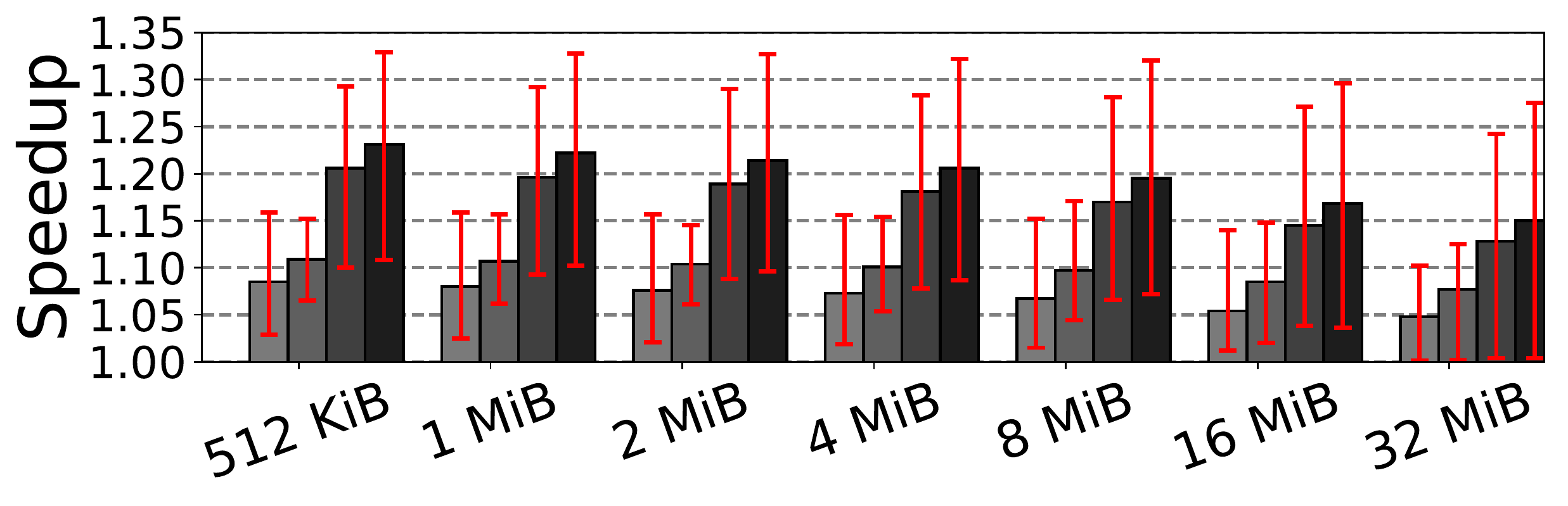}%
            \label{crow_fig:crow_cache_ref_sc_speedup}
        \end{subfigure}
        \begin{subfigure}[b]{0.49\linewidth}
            \centering
            \includegraphics[width=\linewidth]{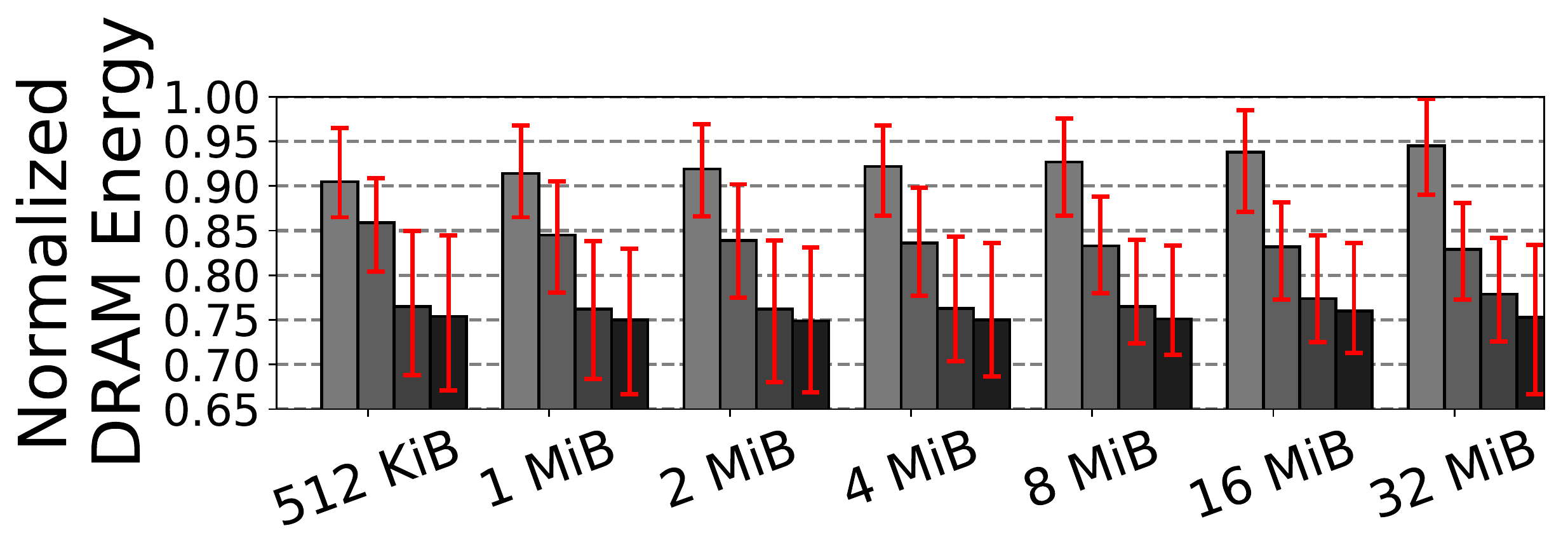}%
            \label{crow_fig:crow_cache_ref_sc_energy}
        \end{subfigure}
        \caption{Single-core workloads}
    \end{subfigure}
   
    \begin{subfigure}[b]{\linewidth}
        \begin{subfigure}[b]{0.49\linewidth}
            \centering
            \includegraphics[width=\linewidth]{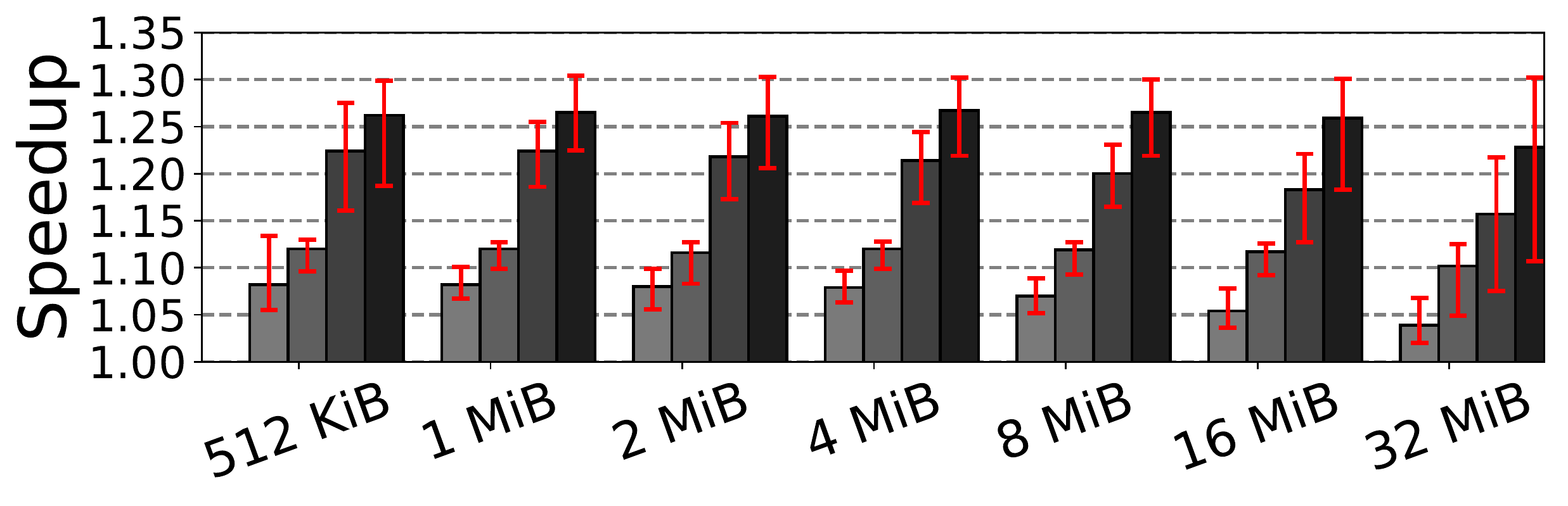}%
            \label{crow_fig:crow_cache_ref_ws_speedup}
        \end{subfigure}
        \begin{subfigure}[b]{0.49\linewidth}
            \centering
            \includegraphics[width=\linewidth]{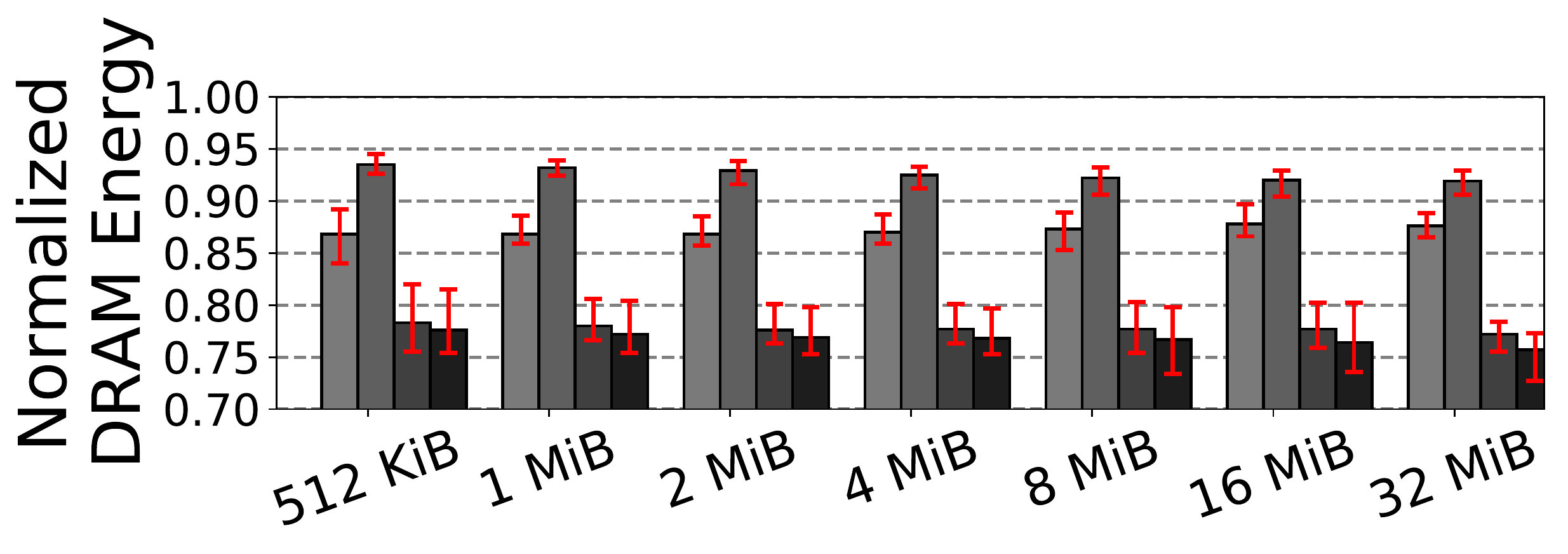}%
            \label{crow_fig:crow_cache_ref_ws_energy}
        \end{subfigure}
        \caption{Four-core workloads}
    \end{subfigure}

    \caption{\mech-(cache+ref) speedup and DRAM energy for different LLC
capacities.}
\label{crow_fig:crow_cache_ref_results}
\end{figure}

\end{landscape}
\clearpage    
}

First, the two mechanisms combined together provide higher performance and
energy efficiency than either mechanism alone. This is because each mechanism
improves a different aspect of DRAM. Using an \SI{8}{\mebi\byte} LLC, the
combination of the two mechanisms provides a 20.0\% performance improvement and
22.3\% DRAM energy reduction for four-core workloads. The performance and energy
benefits of the combined \mech-cache/\mech-ref mechanisms is larger than the the
\emph{sum} of the benefits of each individual mechanism. This is because by
eliminating many refresh operations, \mech-ref decreases the interference of
refresh operations with DRAM access operations. This helps \mech-cache, as a
row that would have been closed by a refresh operation may now stay open, and no
longer needs to be reactivated.  As a result, \mech-cache can use its limited
\copyrow{s} to help accelerate the activation of other rows than the ones that
would have been closed by refresh.

Second, \mech-cache and \mech-ref significantly improve performance
and reduce DRAM energy for \emph{all} LLC capacities. For four-core
workloads, we observe 22.4\%/20.0\%/15.7\% performance improvement and
22.0\%/22.3\%/22.8\% DRAM energy reduction for
\SIlist[list-units=single,list-final-separator={/},
list-separator={/}]{1;8;32}{\mebi\byte} LLC capacity.

Third, the combination of \mech-cache and \mech-ref achieves performance and
DRAM energy improvements close to the ideal mechanism (100\% hit rate
\mech-cache and no refresh). Using an \SI{8}{\mebi\byte} LLC, the combination
of the two mechanisms achieves 71\% of the performance and 99\%\footnote{The
combination of \mech-cache and \mech-ref almost reaches the energy reduction of
the ideal mechanism because the ideal mechanism \emph{always} hits in the
\mech table, and thus \emph{always} uses an \cmdact-t command to activate a row
(which consumes more energy than a regular \cmdact).} of the DRAM energy
improvements of the ideal mechanism.

We conclude that \mech is a flexible substrate that enables multiple different
mechanisms to operate simultaneously, thereby improving both DRAM performance
and energy efficiency.

\chapter{Conclusions and Future Directions}
\label{chap:conc}



\hhm{In this dissertation, we \hhmii{demonstrate that \hhmiv{new understanding
developed based on} rigorous characterization \hhmiv{of real DRAM chips} leads
to new mechanisms for improving DRAM performance, energy consumption,
reliability, and security.} 
%
%
We enable detailed characterization of modern DRAM chips by designing
and prototyping a flexible FPGA-based \hhmii{DRAM testing} infrastructure
\hhmii{(the first of its kind)}, which we use to develop \hhmii{new insights}
and \hhmii{new} mechanisms for improving system performance, efficiency,
reliability, and security.}

First, \hhmii{in~\cref{chap:softmc},} we introduce \emph{SoftMC}, the first
publicly available FPGA-based DRAM testing infrastructure. SoftMC provides a
flexible and easy-to-use software interface that enables testing any combination
of any standard DRAM operations. We demonstrate the capability, flexibility, and
programming ease of SoftMC via two example use cases. 
Our experimental analyses demonstrate the effectiveness of SoftMC as a new tool
to \emph{(i)} perform detailed characterization of various DRAM parameters
(e.g., refresh interval and access latency) as well as the relationships between
them, and \emph{(ii)} test the expected effects of existing or new mechanisms
(e.g., whether or not highly-charged cells can be accessed faster in existing
DRAM chips). We believe and hope that SoftMC, with its flexibility and ease of
use, can enable many other studies, ideas and methodologies in the design of
future memory systems, by making memory control and characterization easily
accessible to a wide range of software and hardware developers. \hhms{In fact,
the improved DDR4 version of SoftMC, called DRAM Bender, is released as an
open-source paper~\cite{olgun2022dram} and tool~\cite{drambendergithub} to
enable such further studies.}

Second, \hhmii{in~\cref{chap:utrr},} we use SoftMC to develop \texttt{U-TRR}, a
\hhmii{new} methodology for uncovering the operation of in-DRAM Target Row
Refresh (TRR) mechanisms implemented in existing DRAM chips as a protection
against \hhmii{the RowHammer vulnerability}. Using \texttt{U-TRR}, we 1) provide
insights into the inner workings of existing proprietary and undocumented TRR
mechanisms and 2) develop custom DRAM access patterns to efficiently circumvent
TRR in \numTestedDIMMs{} DDR4 DRAM modules from three major vendors. We
demonstrate that existing in-DRAM TRR mechanisms are not secure against
RowHammer and can be easily circumvented using the new RowHammer access patterns
that we develop based on \hhmii{new} insights gained via \texttt{U-TRR}
experiments. \hhmii{We believe and hope that \texttt{U-TRR} will inspire future
research and lead to the development of both new RowHammer attacks and secure
RowHammer protection mechanisms.}

Third, \hhmii{in~\cref{chap:smd},} we propose Self-Managing DRAM (\texttt{SMD}),
a new DRAM interface and architecture that \hhm{enables \hhmii{easy adoption of
existing and} new DRAM} maintenance operations \hhm{that improve DRAM
performance, efficiency, and robustness}. To achieve this, \texttt{SMD} sets the
memory controller free from managing DRAM maintenance operations with minimal
changes to the existing DRAM chips and memory controllers. Using \texttt{SMD},
we implement efficient maintenance mechanisms for DRAM refresh, RowHammer
protection, and memory scrubbing. We show that these mechanisms altogether
enable a higher performance, more energy efficient and at the same time more
reliable and secure DRAM system. \hhmii{We believe and hope that \texttt{SMD}
will enable practical \hhmiv{and easier} adoption of innovative ideas in DRAM
design \hhmiv{as well as development of efficient in-DRAM maintenance
mechanisms}.}

Finally, \hhmii{in~\cref{chap:crow},} we propose Copy-Row DRAM (\texttt{CROW}),
a low-cost substrate that enables mechanisms for improving DRAM latency and
energy efficiency. \texttt{CROW} partitions a DRAM subarray into two regions
(regular rows and {\copyrow}s) and enables independent control over the rows in
each region. We use \texttt{CROW} to design two new mechanisms.
\texttt{CROW}-cache reduces DRAM access latency by duplicating a regular row's
data to a \copyrow and simultaneously activating both rows to reduce row
activation latency by 38\%. \texttt{CROW}-ref reduces DRAM refresh rate, and
thus the performance and energy overhead of DRAM refresh overhead, by remapping
retention-weak regular rows to strong {\copyrow}s. We simultaneously employ both
\texttt{CROW}-cache and \texttt{CROW}-ref to provide 20.0\% speedup and 22.3\%
DRAM energy savings over conventional DRAM. We conclude that \texttt{CROW} is a
flexible DRAM substrate that enables a wide range of mechanisms to improve
performance, energy efficiency, and reliability. \hhmii{We believe and hope that
future work will take advantage of CROW to devise more mechanisms that improve
various aspects of DRAM.}

\hhmii{In conclusion, this dissertation shows that understanding DRAM
characteristics and operation via rigorous experiments using real DRAM chips is
critical for developing new mechanisms for improving DRAM performance,
efficiency, reliability, and security. We believe and hope that the DRAM
characterization infrastructure, understanding, and techniques we develop in
this dissertation will encourage more innovation in the design of future memory
systems.}

\section{Future Research Directions}

SoftMC provides a practical infrastructure for studying the reliability and
latency characteristics of real DRAM chips. Our follow-up works, U-TRR, SMD, and
CROW show that \hhmii{new} observations \hhmii{and insights} on DRAM
reliability/latency characteristics and understanding of intrinsic DRAM
operation can be exploited to improve performance, \hhmii{efficiency,}
reliability, and security aspects of DRAM. We believe our works will enable new
research directions. In this section, we discuss \hhmii{several} such potential
future research directions.

\subsection{Deeper DRAM Characterization}

Many works~\cite{koppula2019eden, chang2017understanding, frigo2020trrespass,
gao2019computedram, kim2019d,
khan2017detecting,ghose2018your,chang2016understanding, orosa2021deeper,
kim2020revisiting, olgun2021quac,
orosa2021codic,hassan2021uncovering,yauglikcci2022understanding,
talukder2018exploiting, talukder2019prelatpuf, talukder2018ldpuf,
bepary2022dram, gao2022fracdram, yaglikci2022hira} use SoftMC to perform
experiments using real DRAM chips. Going forward, we expect SoftMC to continue
having a key role in enabling future research on various aspects of DRAM and
potentially other memory technologies. We discuss \hhmii{four} such example
research directions.

First, continuous DRAM technology node scaling leads to increasing reliability
and latency issues in DRAM~\cite{mutlu2013memory, mandelman2002challenges,
kim2005technology, mueller2005challenges, liu2012raidr, kang2014co,
liu2013experimental, khan2016case, khan2016parbor,
khan2014efficacy,khan2017detecting,mutlu2019rowhammer,kim2020revisiting,orosa2021deeper}.
Thus, it is essential to continuously study the characteristics of new DRAM
generations to develop mechanisms for improving DRAM reliability and performance
of future DRAM chips. Our SoftMC infrastructure can help researchers answer
various questions to this end: How are the DRAM cell characteristics,
reliability, and latency changing with different generations of technology
nodes? Do all DRAM operations and cells get affected by scaling at the same
rate? Which DRAM operations are getting worse? What are the \hhmii{future}
trends in the RowHammer vulnerability of DRAM? Do other types of disturbance
errors manifest in new technology node generations?

Second, the existing research on the impact of aging on DRAM reliability and
latency characteristics is very
limited~\cite{meza2015revisiting,schroeder2009dram}. A recent
work~\cite{bepary2022dram} performs experiments with 3 DDR3 modules to study the
effects of accelerated aging on the retention behavior of the DRAM cells. In
their analysis, the authors observe increase in DRAM retention errors after 8
and 16 hours of accelerated aging. Although this work presents valuable
preliminary experimental data, we need to perform thorough experiments across a
much larger set of DRAM chips across different generations. Additionally, we
need to study how aging effects errors rates with reduced DRAM timing
parameters. Overall, the causes, characteristics, and impact of aging have
remained largely unstudied. Using SoftMC, it is possible to devise controller
experiments to analyze and characterize aging over very long time periods.
SoftMC can help researchers answer questions such as: How prevalent are
aging-related failures? What types of usage accelerate aging? Can operation
under high temperature and voltage accelerate aging \hhmii{and if so, to what
degree}? How can we design architectural \hhmii{or system-level} techniques that
can slow down the aging process?

Third, prior works show that the failure rate of DRAM modules in large data
centers is significant, largely affecting the cost and downtime in data
centers~\cite{schroeder2009dram, sridharan2013feng, meza2015revisiting,
luo2014characterizing}. Unfortunately, there is no study that analyzes DRAM
modules that have failed in the field to determine the common causes of failure.
Our SoftMC infrastructure can test faulty DRAM modules and help answer various
research questions: What are the dominant types of DRAM failures at runtime? Are
failures correlated to any location or specific structure in DRAM? Do all chips
from the same generation exhibit the same failure characteristics?

\hhmii{Fourth, memories that can operate at extremely low temperatures (e.g.,
\SI{77}{\kelvin} and below) \hhmiv{are} critical for \hhmiv{the} realization of
cryogenic processors, e.g., quantum
computers~\cite{tannu2017cryogenic,wang2018dram} \hhmiv{and} superconducting
processors~\cite{wang2018dram,ware2017superconducting}. Although recent works
\hhmiv{(e.g., ~\cite{wang2018dram,
kelly2019some,tannu2017cryogenic,lee2021cryoguard,lee2019cryogenic,garzon2021gain})}
analyze the retention and reliability properties of DRAM operating at cryogenic
temperatures, we need to conduct more rigorous experimental studies using real
DRAM chips to fully understand the reliability and latency characteristics of
modern DRAM operating at very low temperatures. SoftMC can help researchers
answer research questions like: How reliably can a DRAM chip operate at very low
temperatures? How do different DRAM designs perform under cryogenic
temperatures? How can we improve DRAM reliability at very low temperatures? How
are different DRAM latencies \hhmiv{affected} by very low temperatures and how
can we improve DRAM access latency \hhmiv{under such operating conditions}?}

\subsection{Extending SoftMC to Support Other DRAM Standards and Memory Technologies}

SoftMC can be used out-of-the-box to test DDR3\hhms{ chips, and the new version
of SoftMC, called DRAM Bender~\cite{olgun2022dram,drambendergithub} can be used
to test DDR4 chips}. However, it can be extended to support other DRAM
interfaces such as different generations of HBM~\cite{gurumurthi2021hbm3},
GDDR~\cite{jedec2016gddr6}, and LPDDR~\cite{jedec2020lpddr5}. Besides enabling
tests on a wide range of DRAM chips, such a design also makes the scope of chips
that SoftMC can test go beyond just DRAM. With the emergence of
\hhmiv{byte-addressable} \hhmii{Non-Volatile Memories (NVM)} (e.g.,
PCM~\cite{raoux2008phare,lee2010phase, lee2009architecting, qureshi2009scalable,
seong2013tri, wong2010phase},
STT-RAM~\cite{kawahara2008spram,kultursay2013evaluating,hamdioui2017test,qureshi2011pay,tavana2017remap,zhang2012memory,everspin2021sttmram,huai2008spin},
ReRAM~\cite{akinaga2010resistive,
wong2012metal,apalkov2013spin,chun2012scaling}) \hhm{in DDR-compatible form
\hhmiv{factors}~\cite{micron2016xpoint,everspin2021sttmram,
akram2021performance,shanbhag2020large}, it is critical to characterize and
analyze such NVM chips in order to \hhmiv{understand and exploit them as well as
improve their design}.} We believe that SoftMC can be seamlessly used to
characterize these chips, and can help enable future mechanisms for NVM.

\subsection{Improving RowHammer Attacks and Defenses}
Prior
works~\cite{yaglikci2021blockhammer,kim2020revisiting,orosa2021deeper,kim2014flipping,cojocar2020are,frigo2020trrespass,park2016statistical,yauglikcci2022understanding,
kim2023ddr5, marazzi2023rega, woo2023scalable, wi2023shadow} show that DRAM
technology node scaling leads to 1) reduction in minimum row activation count
($HC_{first}$) for causing a RowHammer bit flip and 2) increase in RowHammer
blast radius (i.e., the range of neighbor rows on which hammering an aggressor
row causes RowHammer bit flips). To efficiently protect future DRAM chips with
increasing vulnerability to RowHammer, the performance, energy, and area
overheads of a RowHammer mitigation mechanism must be small even for extremely
low $HC_{first}$. 
The SoftMC infrastructure and the U-TRR methodology can be \hhmii{used} to study
the efficiency and security guarantees of future RowHammer protection
mechanisms.

As RowHammer defenses improve, studying the security guarantees of such defenses
implemented in commodity DRAM chips will remain as a critical challenge.
Specifically, there is a need for discovering new access patterns that can
circumvent the RowHammer protection mechanisms and exploring the effects of
various factors (e.g., data pattern dependence, operating temperature and
voltage) on the RowHammer vulnerability level of future DRAM chips will remain
as an important research direction, especially for undocumented RowHammer
protection mechanisms that rely on security-through-obscurity \hhmii{but also
for well-documented mechanisms that are implemented in real hardware}.

In U-TRR, we show that our \hhmiv{carefully crafted} access \hhmiv{patterns} can
cause \hhmii{too many bit flips in a codeword for \hhmiv{simple} ECC to detect
and correct}. However, \hhmiv{on systems equipped with ECC,} mounting a
RowHammer attack \hhmiii{that \hhmiv{could manipulate sensitive data} by
flipping bits in DRAM is challenging}. \hhmiv{This is because an attacker may
unintentionally cause a detectable error while searching for a DRAM location and
access/data pattern that would result in undetectable number of errors.}
When \hhmii{a detectable (but uncorrectable) error occurs}, a system may hang,
restart, or crash, leading to \hhmii{denial-of-service and} disruption of the
RowHammer attack. \hhmiv{Developing a methodology that overcomes this challenge
and efficiently compromises a system despite of the presence of ECC is a
promising research direction.}


\subsection{New DRAM Maintenance Mechanisms}

In~\cref{chap:smd}, we show the limitations of existing DRAM interfaces
\hhmii{impose against} designing efficient DRAM maintenance mechanisms. Using
\hhmii{Self-Managing DRAM (\texttt{SMD})}, we demonstrate how to design
efficient in-DRAM maintenance mechanisms for periodic refresh, RowHammer
protection, and memory scrubbing. As SMD eases the adoption of new \hhmii{DRAM}
maintenance mechanisms, researchers \hhmii{can} further optimize these
maintenance mechanisms or propose new mechanisms for the same purposes \hhmii{or
for other maintenance operations}. We believe SMD can also be used to design
maintenance mechanisms for other purposes. For example, using SMD, one can
implement various profiling-guided mechanisms for improving DRAM performance and
energy efficiency by exploiting design-induced variation across DRAM
cells~\cite{lee2017design,kim2018solar}. A maintenance mechanism can
automatically identify low-latency DRAM rows that can be reliably accesses with
lowered timing parameters. As in other SMD-based maintenance mechanisms, SMD can
prevent the memory controller from accessing a region that is being profiled. By
conveying the profiling results to the memory controller and the operating
system, such low-latency rows can be used to place frequently-accessed data.

\hhmii{\texttt{SMD} can also ease managing the \hhmiv{coordination} between a
near-/in-DRAM processing
engine\hhmiv{~\cite{gao2019computedram,seshadri2017ambit,seshadri2013rowclone,orosa2021codic,
olgun2021pidram,hajinazar2021simdram,
seshadri2015gather,ahn2016scalable,ahn2015pim,fromm1997energy,ghose2019processing,li2016pinatubo,
mutlu2019processing,shin2018mcdram,ferreira2022pluto,gao2022fracdram,deng2018dracc,li2017drisa,
xin2020elp2im,akin2015data,augusta2015jafar,farmahini2015nda,gao2016hrl,gao2017tetris,hsieh2016accelerating,
kim2016neurocube,nai2017graphpim,zhang2014top,zhu2013accelerating,mutlu2021primer,
cho2020chonda, oliveira2022accelerating, boroumand2021google}} and a memory
controller. The ability to reject row activations, which \texttt{SMD} provides,
can help resolve access conflicts between a near-/in-DRAM processing engine and
a memory controller in the same system. To do so, \texttt{SMD} can treat the
near-/in-DRAM processing engine as a maintenance mechanism. The engine can use
\texttt{SMD} to lock a DRAM region that it will operate on. Because \texttt{SMD}
does not allow the memory controller to activate a row in a locked region, only
the near-/in-DRAM processing engine will have access to the locked region until
it completes the processing and releases the region.}

\subsection{Enabling in-DRAM Data Movement}

The CROW substrate \hhmii{we presented in~\cref{chap:crow}} exploits
simultaneous multiple row activation and in-subarray data movement to develop
mechanisms for in-DRAM caching and reducing DRAM refresh overhead. CROW can
enable various other mechanisms, such as the RowHammer mitigation mechanism we
discuss in \cref{crow_subsec:crow_hammer} \hhmii{and reduce the overheads of new
RowHammer mitigation proposals, e.g.,~\cite{saileshwar2022randomized,
woo2023scalable, wi2023shadow}.} We believe even wider range of mechanisms for
improving DRAM performance, energy efficiency, and reliability can be enabled
with \hhmii{either CROW or similar} simple changes in the DRAM architecture.

\section{Final Concluding Remarks}

In this dissertation, we demonstrate the importance of \hhmiv{rigorously}
understanding DRAM characteristics and operation of existing DRAM chips on
developing new mechanisms for improving DRAM performance, energy efficiency,
reliability, and security. 
%
\hhmii{   
We hope that the contributions in this dissertation enable future research in
\hhmiv{many area in future DRAM and memory systems, including} understanding the
characteristics of current and future memory \hhmiv{chips and systems},
improving RowHammer attacks and defenses, adopting efficient DRAM maintenance
mechanisms, exploiting in-DRAM data movement, and many other areas.}

\appendix
\cleardoublepage%
\chapter{Other Works of the Author}

During my doctoral studies, I had the opportunity to contribute on several
different areas with researchers from ETH Z\"urich and other institutions. In
this chapter, I acknowledge these works.

Besides the works presented in this dissertation, I worked on many other DRAM
projects. 
In collaboration with Jeremie S. Kim, we developed 1) D-RaNGE~\cite{kim2019d}, a
mechanism for generating high-throughput true random numbers using DRAM, 2) DRAM
Latency PUF~\cite{kim2018dram}, a new class of fast and reliable DRAM-based
Physical Unclonable Functions (PUFs), and 3) Solar-DRAM~\cite{kim2018solar}, a
mechanism for reducing DRAM access latency by exploiting variation in bitlines.
In collaboration with Minesh Patel, we analyzed modern DRAM chips with on-die
ECC and developed a methodology to extract pre-correction error characteristics
of DRAM chips~\cite{patel2019understanding,eccsimgithub} and a methodology to
determine the full DRAM on-die ECC function~\cite{patel2020bit,beergithub}.
In collaboration with Kevin Chang, we characterized the latency and reliability
characteristics of DRAM chips to understand their intrinsic latency
variation~\cite{chang2016understanding} and operation under reduced
voltage~\cite{chang2017understanding}.

I contributed to several other DRAM projects.
CLR-DRAM~\cite{luo2020clr,clrdramgithub} is a new DRAM architecture that enables
dynamic capacity-latency trade-off by enabling a DRAM row to switch between
max-capacity and high-performance modes. In CAL~\cite{wang2018reducing}, we
proposed a mechanism that reduces DRAM charge restoration time, and thus reduces
DRAM access latency, by partially restoring DRAM cell charge when a DRAM row is
predicted to be soon re-activated. VRL-DRAM~\cite{das2018vrl} proposes a
mechanism that reduces DRAM refresh overhead by partially refreshing a row when
it can correctly retain its data until the next time it gets refreshed. \hhm{In
Refresh Triggered Computation~\cite{jafri2020refresh}, we propose a technique
for eliminating redundant refresh operations by exploiting the regular memory
access patterns of convolutional neural network applications.}
In~\cite{ghose2018your}, we characterize the power consumption of real DRAM
chips and develop an accurate DRAM power model based on the observations we
develop from our characterization study. 
DR-STRaNGe~\cite{bostanci2022dr} is an end-to-end design for enabling DRAM-based
true random number generation mechanisms.
\hhmiv{HIRA~\cite{yaglikci2022hira} reduces the DRAM refresh latency in
off-the-shelf DRAM chips by exploiting back-to-back activation of two rows to
allow refreshing one of the activated rows while accessing the other one. DRAM
Bender~\cite{olgun2022dram,drambendergithub} is a new DRAM characterization
infrastructure that extends SoftMC by enabling support for DDR4 chips and
introducing a new programming model that overcomes the program size limitation
of SoftMC in developing DRAM experiments.} \hhms{in Sectored
DRAM~\cite{olgun2022sectored}, we propose a low-overhead DRAM substrate that
enables fine-grained DRAM row activation and access.}

I contributed to works that enable computation and other functionality within
DRAM chips. Ambit~\cite{seshadri2017ambit} proposes an in-DRAM accelerator for
performing bulk bitwise operations. CODIC~\cite{orosa2021codic} enables
fine-grained control over previously fixed internal DRAM timings and develops a
new DRAM-based PUF and a mechanism for preventing DRAM cold boot attacks.
In~\cite{olgun2021pidram,pidramgithub, olgun2022pidram}, we develop PiDRAM, a flexible
\hhm{open-source} framework that enables system integration of mechanisms that
perform processing using memory.


I also worked on projects that analyze the RowHammer vulnerability of DRAM and
existing defense mechanisms against it. In~\cite{kim2020revisiting}
and~\cite{orosa2021deeper}, we characterize the RowHammer vulnerability of many
DRAM chips across different vendors and technology nodes and study the
sensitivity of RowHammer to factors such as operating temperature and aggressor
row active time. TRRespass~\cite{frigo2020trrespass} shows that many modern DRAM
chips equipped with a Target Row Refresh (TRR) mechanism, which is advertised as
complete solution to the RowHammer problem, are in fact vulnerable to access
patterns that utilize many aggressor rows.
BlockHammer~\cite{yaglikci2021blockhammer,blockhammergithub} is a low-cost and
easy-to-adopt RowHammer mitigation mechanism that efficiently tracks aggressor
rows using Bloom Filters and throttles future activations to the aggressor row
until its neighbors are refreshed. \hhm{In~\cite{yauglikcci2022understanding},
we characterize the effects of reducing DRAM voltage on the RowHammer
vulnerability.}

I also contributed to works outside my main research area. In
CoNDA~\cite{boroumand2019conda} \hhm{(and an earlier
version~\cite{boroumand2016lazypim})}, we propose an efficient cache coherence
mechanism for systems equipped with a near-data accelerator.
\hhm{MetaSys~\cite{vijaykumar2022metasys,metasysgithub} is an open-source
FPGA-based framework for rapidly evaluating new cross-layer techniques.} In
GRIM-Filter~\cite{kim2018grim}, we propose a DNA read mapping seed filtering
algorithm that is optimized to exploit 3D-stacked memory systems to perform seed
filtering in memory. GateKeeper~\cite{alser2017gatekeeper,gatekeepergithub} is a
highly-accurate hardware accelerator for accelerating the pre-alignment step of
short DNA read mapping. Shouji~\cite{alser2019shouji,shoujigithub} is a
highly-parallel and accurate DNA read mapping pre-alignment filter that
significantly reduces the need for expensive dynamic programming algorithms.

\cleardoublepage
\balance
\begin{singlespace}
\bibliographystyle{IEEEtran}
\bibliography{references}

\newcommand{\noopsort}[1]{} \newcommand{\printfirst}[2]{#1}
  \newcommand{\singleletter}[1]{#1} \newcommand{\switchargs}[2]{#2#1}
\begin{thebibliography}{100}
\providecommand{\url}[1]{#1}
\csname url@samestyle\endcsname
\providecommand{\newblock}{\relax}
\providecommand{\bibinfo}[2]{#2}
\providecommand{\BIBentrySTDinterwordspacing}{\spaceskip=0pt\relax}
\providecommand{\BIBentryALTinterwordstretchfactor}{4}
\providecommand{\BIBentryALTinterwordspacing}{\spaceskip=\fontdimen2\font plus
\BIBentryALTinterwordstretchfactor\fontdimen3\font minus
  \fontdimen4\font\relax}
\providecommand{\BIBforeignlanguage}[2]{{%
\expandafter\ifx\csname l@#1\endcsname\relax
\typeout{** WARNING: IEEEtran.bst: No hyphenation pattern has been}%
\typeout{** loaded for the language `#1'. Using the pattern for}%
\typeout{** the default language instead.}%
\else
\language=\csname l@#1\endcsname
\fi
#2}}
\providecommand{\BIBdecl}{\relax}
\BIBdecl

\bibitem{hassan2017softmc}
H.~Hassan, N.~Vijaykumar, S.~Khan, S.~Ghose, K.~Chang, G.~Pekhimenko, D.~Lee,
  O.~Ergin, and O.~Mutlu, ``{SoftMC: A Flexible and Practical Open-Source
  Infrastructure for Enabling Experimental DRAM Studies},'' in \emph{HPCA},
  2017.

\bibitem{lee2013tiered}
D.~Lee, Y.~Kim, V.~Seshadri, J.~Liu, L.~Subramanian, and O.~Mutlu,
  ``{Tiered-Latency DRAM: A Low Latency and Low Cost DRAM Architecture},'' in
  \emph{HPCA}, 2013.

\bibitem{kim2012case}
Y.~Kim, V.~Seshadri, D.~Lee, J.~Liu, and O.~Mutlu, ``{A Case for Exploiting
  Subarray-Level Parallelism (SALP) in DRAM},'' in \emph{ISCA}, 2012.

\bibitem{dennard1968field}
R.~H. Dennard, ``{Field-Effect Transistor Memory},'' 1968, {US} Patent
  3,387,286.

\bibitem{mutlu2013memory}
O.~Mutlu, ``{Memory Scaling: A Systems Architecture Perspective},'' in
  \emph{IMW}, 2013.

\bibitem{mutlu2014research}
O.~Mutlu and L.~Subramanian, ``{Research Problems and Opportunities in Memory
  Systems},'' in \emph{SUPERFRI}, 2014.

\bibitem{jedec2008ddr3}
JEDEC, \emph{{DDR3 SDRAM Specification}}, 2008.

\bibitem{jedec2012ddr4}
JEDEC, \emph{{DDR4 SDRAM Specification}}, 2012.

\bibitem{son2013reducing}
Y.~H. Son, O.~Seongil, Y.~Ro, J.~W. Lee, and J.~H. Ahn, ``{Reducing Memory
  Access Latency with Asymmetric DRAM Bank Organizations},'' in \emph{ISCA},
  2013.

\bibitem{chang2016understanding}
K.~K. Chang, A.~Kashyap, H.~Hassan, S.~Ghose, K.~Hsieh, D.~Lee, T.~Li,
  G.~Pekhimenko, S.~Khan, and O.~Mutlu, ``{Understanding Latency Variation in
  Modern DRAM Chips: Experimental Characterization, Analysis, and
  Optimization},'' in \emph{SIGMETRICS}, 2016.

\bibitem{lee2015adaptive}
D.~Lee, Y.~Kim, G.~Pekhimenko, S.~Khan, V.~Seshadri, K.~Chang, and O.~Mutlu,
  ``{Adaptive-Latency DRAM: Optimizing DRAM Timing for the Common-Case},'' in
  \emph{HPCA}, 2015.

\bibitem{lee2016reducing}
D.~Lee, ``{Reducing DRAM Latency at Low Cost by Exploiting Heterogeneity},''
  Ph.D. dissertation, Carnegie Mellon University, 2016.

\bibitem{patel2022enabling}
M.~Patel, ``{Enabling Effective Error Mitigation in Modern Memory Chips that
  Use On-Die Error-Correcting Codes},'' Ph.D. dissertation, ETH Z\"urich, 2022.

\bibitem{patel2022case}
M.~Patel, T.~Shahroodi, A.~Manglik, A.~G. Yaglikci, A.~Olgun, H.~Luo, and
  O.~Mutlu, ``{A Case for Transparent Reliability in DRAM Systems},''
  \emph{arXiv preprint arXiv:2204.10378}, 2022.

\bibitem{luo2020clr}
H.~Luo, T.~Shahroodi, H.~Hassan, M.~Patel, A.~Giray~Ya{\u{g}}l{\i}k{\c{c}}{\i},
  L.~Orosa, J.~Park, and O.~Mutlu, ``{CLR-DRAM: A Low-Cost DRAM Architecture
  Enabling Dynamic Capacity-Latency Trade-Off},'' in \emph{ISCA}, 2020.

\bibitem{borkar2011future}
S.~Borkar and A.~A. Chien, ``{The Future of Microprocessors},'' in \emph{CACM},
  2011.

\bibitem{hassan2016chargecache}
H.~Hassan, G.~Pekhimenko, N.~Vijaykumar, V.~Seshadri, D.~Lee, O.~Ergin, and
  O.~Mutlu, ``{ChargeCache: Reducing DRAM Latency by Exploiting Row Access
  Locality},'' in \emph{HPCA}, 2016.

\bibitem{ferdman2012clearing}
M.~Ferdman, A.~Adileh, O.~Kocberber, S.~Volos, M.~Alisafaee, D.~Jevdjic,
  C.~Kaynak, A.~D. Popescu, A.~Ailamaki, and B.~Falsafi, ``{Clearing The
  Clouds: A Study Of Emerging Scale-Out Workloads On Modern Hardware},''
  \emph{ASPLOS}, 2012.

\bibitem{huang2014moby}
``Moby: A mobile benchmark suite for architectural simulators.''

\bibitem{gutierrez2011full}
A.~Gutierrez, R.~G. Dreslinski, T.~F. Wenisch, T.~Mudge, A.~Saidi, C.~Emmons,
  and N.~Paver, ``{Full-System Analysis And Characterization Of Interactive
  Smartphone Applications},'' in \emph{IISWC}, 2011.

\bibitem{zhu2015microarchitectural}
Y.~Zhu, D.~Richins, M.~Halpern, and V.~J. Reddi, ``{Microarchitectural
  Implications Of Event-Driven Server-Side Web Applications},'' in
  \emph{MICRO}, 2015.

\bibitem{hestness2014comparative}
J.~Hestness, S.~W. Keckler, and D.~A. Wood, ``{A Comparative Analysis Of
  Microarchitecture Effects On CPU and GPU Memory System Behavior},'' in
  \emph{IISWC}, 2014.

\bibitem{ghose2019demystifying}
S.~Ghose, T.~Li, N.~Hajinazar, D.~S. Cali, and O.~Mutlu, ``{Demystifying
  Complex Workload-DRAM Interactions: An Experimental Study},''
  \emph{SIGMETRICS}, 2019.

\bibitem{mutlu2003runahead}
O.~Mutlu, J.~Stark, C.~Wilkerson, and Y.~N. Patt, ``{Runahead Execution: An
  Alternative to Very Large Instruction Windows for Out-of-Order Processors},''
  in \emph{HPCA}, 2003.

\bibitem{bera2019dspatch}
R.~Bera, A.~V. Nori, O.~Mutlu, and S.~Subramoney, ``{DSPatch: Dual Spatial
  Pattern Prefetcher},'' in \emph{MICRO}, 2019.

\bibitem{boroumand2018google}
A.~Boroumand, S.~Ghose, Y.~Kim, R.~Ausavarungnirun, E.~Shiu, R.~Thakur, D.~Kim,
  A.~Kuusela, A.~Knies, P.~Ranganathan, and O.~Mutlu, ``{Google Workloads for
  Consumer Devices: Mitigating Data Movement Bottlenecks},'' in \emph{ASPLOS},
  2018.

\bibitem{ghose2019processing}
S.~Ghose, A.~Boroumand, J.~S. Kim, J.~G{\'o}mez-Luna, and O.~Mutlu,
  ``{Processing-In-Memory: A Workload-Driven Perspective},'' \emph{IBM Journal
  of Research and Development}, 2019.

\bibitem{kanev2015profiling}
S.~Kanev, J.~P. Darago, K.~Hazelwood, P.~Ranganathan, T.~Moseley, G.-Y. Wei,
  and D.~Brooks, ``{Profiling a Warehouse-Scale Computer},'' in \emph{ISCA},
  2015.

\bibitem{koppula2019eden}
S.~Koppula, L.~Orosa, A.~G. Ya{\u{g}}l{\i}k{\c{c}}{\i}, R.~Azizi, T.~Shahroodi,
  K.~Kanellopoulos, and O.~Mutlu, ``{EDEN: Enabling Energy-Efficient,
  High-Performance Deep Neural Network Inference Using Approximate DRAM},'' in
  \emph{MICRO}, 2019.

\bibitem{liu2019binary}
X.~Liu, D.~Roberts, R.~Ausavarungnirun, O.~Mutlu, and J.~Zhao, ``{Binary Star:
  Coordinated Reliability in Heterogeneous Memory Systems for High Performance
  and Scalability},'' in \emph{MICRO}, 2019.

\bibitem{wilkes2001memory}
M.~V. Wilkes, ``{The Memory Gap and the Future of High Performance Memories},''
  \emph{ACM SIGARCH Computer Architecture News}, 2001.

\bibitem{wulf1995hitting}
W.~A. Wulf and S.~A. McKee, ``{Hitting the Memory Wall: Implications of the
  Obvious},'' \emph{ACM SIGARCH Computer Architecture Mews}, 1995.

\bibitem{oliveira2021damov}
G.~F. Oliveira, J.~G{\'o}mez-Luna, L.~Orosa, S.~Ghose, N.~Vijaykumar,
  I.~Fernandez, M.~Sadrosadati, and O.~Mutlu, ``{DAMOV: A New Methodology and
  Benchmark Suite for Evaluating Data Movement Bottlenecks},'' \emph{IEEE
  Access}, 2021.

\bibitem{boroumand2021google}
A.~Boroumand, S.~Ghose, B.~Akin, R.~Narayanaswami, G.~F. Oliveira, X.~Ma,
  E.~Shiu, and O.~Mutlu, ``{Google Neural Network Models for Edge Devices:
  Analyzing and Mitigating Machine Learning Inference Bottlenecks},'' in
  \emph{PACT}, 2021.

\bibitem{boroumand2022polynesia}
A.~Boroumand, S.~Ghose, G.~F. Oliveira, and O.~Mutlu, ``{Polynesia: Enabling
  High-Performance and Energy-Efficient Hybrid Transactional/Analytical
  Databases with Hardware/Software Co-Design},'' in \emph{ICDE}, 2022.

\bibitem{boroumand2020thesis}
A.~Boroumand, ``{Practical Mechanisms for Reducing Processor-Memory Data
  Movement in Modern Workloads},'' Ph.D. dissertation, Carnegie Mellon
  University, 2020.

\bibitem{micron2021rldram}
\emph{{RLDRAM Memory}}, {Micron Technology}, 2021,
  \url{https://www.micron.com/products/dram/rldram-memory}.

\bibitem{sato1998fast}
Y.~Sato, T.~Suzuki, T.~Aikawa, S.~Fujioka, W.~Fujieda, H.~Kobayashi, H.~Ikeda,
  T.~Nagasawa, A.~Funyu, Y.~Fuji, K.~Kawasaki, M.~Yamazaki, and M.~Taguchi,
  ``{Fast Cycle RAM (FCRAM); A 20-ns Random Row Access, Pipe-Lined Operating
  DRAM},'' in \emph{VLSIC}, 1998.

\bibitem{chang2016low}
K.~K. Chang, P.~J. Nair, D.~Lee, S.~Ghose, M.~K. Qureshi, and O.~Mutlu,
  ``{Low-Cost Inter-Linked Subarrays (LISA): Enabling Fast Inter-Subarray Data
  Movement in DRAM},'' in \emph{HPCA}, 2016.

\bibitem{itrs}
ITRS, ``{International Technology Roadmap for Semiconductors Executive
  Summary},'' 2013, http://www.itrs2.net/2013-itrs.html.

\bibitem{chang2014improving}
K.~K. Chang, D.~Lee, Z.~Chishti, A.~R. Alameldeen, C.~Wilkerson, Y.~Kim, and
  O.~Mutlu, ``{Improving DRAM Performance by Parallelizing Refreshes with
  Accesses},'' in \emph{HPCA}, 2014.

\bibitem{kang2014co}
U.~Kang, H.-s. Yu, C.~Park, H.~Zheng, J.~Halbert, K.~Bains, S.~Jang, and J.~S.
  Choi, ``{Co-Architecting Controllers and DRAM to Enhance DRAM Process
  Scaling},'' in \emph{The Memory Forum}, 2014.

\bibitem{liu2012raidr}
J.~Liu, B.~Jaiyen, R.~Veras, and O.~Mutlu, ``{RAIDR: Retention-Aware
  Intelligent DRAM Refresh},'' in \emph{ISCA}, 2012.

\bibitem{jedec2021ddr5}
JEDEC, ``{DDR5 SDRAM Specification - JESD79-5A},'' 2021.

\bibitem{mukundan2013understanding}
J.~Mukundan, H.~Hunter, K.-h. Kim, J.~Stuecheli, and J.~F. Mart{\'\i}nez,
  ``{Understanding and Mitigating Refresh Overheads in High-Density DDR4 DRAM
  Systems},'' in \emph{ISCA}, 2013.

\bibitem{nair2014refresh}
P.~J. Nair, C.-C. Chou, and M.~K. Qureshi, ``{Refresh Pausing in DRAM Memory
  Systems},'' in \emph{TACO}, 2014.

\bibitem{jafri2020refresh}
S.~M. Jafri, H.~Hassan, A.~Hemani, and O.~Mutlu, ``{Refresh Triggered
  Computation: Improving the Energy Efficiency of Convolutional Neural Network
  Accelerators},'' \emph{TACO}, 2020.

\bibitem{das2018vrl}
A.~Das, H.~Hassan, and O.~Mutlu, ``{VRL-DRAM: Improving DRAM Performance Via
  Variable Refresh Latency},'' in \emph{DAC}, 2018.

\bibitem{qureshi2015avatar}
M.~K. Qureshi, D.-H. Kim, S.~Khan, P.~J. Nair, and O.~Mutlu, ``{AVATAR: A
  Variable-Retention-Time (VRT) Aware Refresh for DRAM Systems},'' in
  \emph{DSN}, 2015.

\bibitem{baek2014refresh}
S.~Baek, S.~Cho, and R.~Melhem, ``{Refresh Now and Then},'' in \emph{TC}, 2014.

\bibitem{bhati2013coordinated}
I.~Bhati, Z.~Chishti, and B.~Jacob, ``{Coordinated Refresh: Energy Efficient
  Techniques for DRAM Refresh Scheduling},'' in \emph{ISPLED}, 2013.

\bibitem{bhati2015flexible}
I.~Bhati, Z.~Chishti, S.-L. Lu, and B.~Jacob, ``{Flexible Auto-Refresh:
  Enabling Scalable and Energy-Efficient DRAM Refresh Reductions},'' in
  \emph{ISCA}, 2015.

\bibitem{cui2014dtail}
Z.~Cui, S.~A. McKee, Z.~Zha, Y.~Bao, and M.~Chen, ``{DTail: A Flexible Approach
  to DRAM Refresh Management},'' in \emph{SC}, 2014.

\bibitem{emma2008rethinking}
P.~G. Emma, W.~R. Reohr, and M.~Meterelliyoz, ``{Rethinking Refresh: Increasing
  Availability and Reducing Power in DRAM for Cache Applications},'' in
  \emph{MICRO}, 2008.

\bibitem{ghosh2007smart}
M.~Ghosh and H.-H.~S. Lee, ``{Smart Refresh: An Enhanced Memory Controller
  Design for Reducing Energy in Conventional and 3D Die-Stacked DRAMs},'' in
  \emph{MICRO}, 2007.

\bibitem{kim2020charge}
S.~Kim, W.~Kwak, C.~Kim, D.~Baek, and J.~Huh, ``{Charge-Aware DRAM Refresh
  Reduction with Value Transformation},'' in \emph{HPCA}, 2020.

\bibitem{liu2012flikker}
S.~Liu, K.~Pattabiraman, T.~Moscibroda, and B.~G. Zorn, ``{Flikker: Saving DRAM
  Refresh-Power Through Critical Data Partitioning},'' in \emph{ASPLOS}, 2012.

\bibitem{nair2013case}
P.~Nair, C.-C. Chou, and M.~K. Qureshi, ``{A Case for Refresh Pausing in DRAM
  Memory Systems},'' in \emph{HPCA}, 2013.

\bibitem{stuecheli2010elastic}
J.~Stuecheli, D.~Kaseridis, H.~C. Hunter, and L.~K. John, ``{Elastic Refresh:
  Techniques to Mitigate Refresh Penalties in High Density Memory},'' in
  \emph{MICRO}, 2010.

\bibitem{zhang2014cream}
T.~Zhang, M.~Poremba, C.~Xu, G.~Sun, and Y.~Xie, ``{CREAM: A
  Concurrent-Refresh-Aware DRAM Memory Architecture},'' in \emph{HPCA}, 2014.

\bibitem{mandelman2002challenges}
J.~A. Mandelman, R.~H. Dennard, G.~B. Bronner, J.~K. DeBrosse, R.~Divakaruni,
  Y.~Li, and C.~J. Radens, ``{Challenges and Future Directions for the Scaling
  of Dynamic Random-Access Memory (DRAM)},'' in \emph{IBM JRD}, 2002.

\bibitem{redeker2002investigation}
M.~Redeker, B.~F. Cockburn, and D.~G. Elliott, ``{An Investigation Into
  Crosstalk Noise in DRAM Structures},'' in \emph{MTDT}, 2002.

\bibitem{yaney1987meta}
D.~S. Yaney, C.-Y. Lu, R.~A. Kohler, M.~J. Kelly, and J.~T. Nelson, ``{A
  Meta-Stable Leakage Phenomenon in DRAM Charge Storage-Variable Hold Time},''
  in \emph{IEDM}, 1987.

\bibitem{konishi1989analysis}
Y.~Konishi, M.~Kumanoya, H.~Yamasaki, K.~Dosaka, and T.~Yoshihara, ``{Analysis
  of Coupling Noise Between Adjacent Bit Lines in Megabit DRAMs},''
  \emph{JSSC}, 1989.

\bibitem{liu2013experimental}
J.~Liu, B.~Jaiyen, Y.~Kim, C.~Wilkerson, and O.~Mutlu, ``{An Experimental Study
  of Data Retention Behavior in Modern DRAM Devices: Implications for Retention
  Time Profiling Mechanisms},'' in \emph{ISCA}, 2013.

\bibitem{khan2014efficacy}
S.~Khan, D.~Lee, Y.~Kim, A.~R. Alameldeen, C.~Wilkerson, and O.~Mutlu, ``{The
  Efficacy of Error Mitigation Techniques for DRAM Retention Failures: A
  Comparative Experimental Study},'' in \emph{SIGMETRICS}, 2014.

\bibitem{khan2016case}
S.~Khan, C.~Wilkerson, D.~Lee, A.~R. Alameldeen, and O.~Mutlu, ``{A Case for
  Memory Content-Based Detection and Mitigation of Data-Dependent Failures in
  DRAM},'' in \emph{IEEE CAL}, 2016.

\bibitem{khan2016parbor}
S.~Khan, D.~Lee, and O.~Mutlu, ``{PARBOR: An Efficient System-Level Technique
  to Detect Data-Dependent Failures in DRAM},'' in \emph{DSN}, 2016.

\bibitem{khan2017detecting}
S.~Khan, C.~Wilkerson, Z.~Wang, A.~R. Alameldeen, D.~Lee, and O.~Mutlu,
  ``{Detecting and Mitigating Data-Dependent DRAM Failures by Exploiting
  Current Memory Content},'' in \emph{MICRO}, 2017.

\bibitem{kim2020revisiting}
J.~S. Kim, M.~Patel, A.~G. Ya{\u{g}}l{\i}k{\c{c}}{\i}, H.~Hassan, R.~Azizi,
  L.~Orosa, and O.~Mutlu, ``{Revisiting RowHammer: An Experimental Analysis of
  Modern Devices and Mitigation Techniques},'' in \emph{ISCA}, 2020.

\bibitem{orosa2021deeper}
L.~Orosa, A.~G. Ya{\u{g}}l{\i}k{\c{c}}{\i}, H.~Luo, A.~Olgun, J.~Park,
  H.~Hassan, M.~Patel, J.~S. Kim, and O.~Mutlu, ``{A Deeper Look into
  RowHammer's Sensitivities: Experiemental Analysis of Real DRAM Chips and
  Implications on Future Attacks and Defenses},'' in \emph{MICRO}, 2021.

\bibitem{mutlu2017rowhammer}
O.~Mutlu, ``{The RowHammer Problem and Other Issues we may Face as Memory
  Becomes Denser},'' in \emph{DATE}, 2017.

\bibitem{kim2014flipping}
Y.~Kim, R.~Daly, J.~Kim, C.~Fallin, J.~H. Lee, D.~Lee, C.~Wilkerson, K.~Lai,
  and O.~Mutlu, ``{Flipping Bits in Memory Without Accessing Them: An
  Experimental Study of DRAM Disturbance Errors},'' in \emph{ISCA}, 2014.

\bibitem{mutlu2019rowhammer}
O.~Mutlu and J.~Kim, ``{RowHammer: A Retrospective},'' in \emph{TCAD}, 2019.

\bibitem{yauglikcci2022understanding}
A.~G. Ya{\u{g}}l{\i}k{\c{c}}{\i}, H.~Luo, G.~F. De~Oliviera, A.~Olgun,
  M.~Patel, J.~Park, H.~Hassan, J.~S. Kim, L.~Orosa, and O.~Mutlu,
  ``{Understanding RowHammer Under Reduced Wordline Voltage: An Experimental
  Study Using Real DRAM Devices},'' in \emph{DSN}, 2022.

\bibitem{mutlu2023fundamentally}
O.~Mutlu, A.~Olgun, and A.~G. Ya{\u{g}}l{\i}kc{\i}, ``{Fundamentally
  Understanding and Solving Rowhammer},'' in \emph{ASP-DAC}, 2023.

\bibitem{cojocar2019exploiting}
L.~Cojocar, K.~Razavi, C.~Giuffrida, and H.~Bos, ``{Exploiting Correcting
  Codes: On The Effectiveness Of ECC Memory Against Rowhammer Attacks},'' in
  \emph{S\&P}, 2019.

\bibitem{yaglikci2021blockhammer}
A.~G. Ya{\u{g}}l{\i}k{\c{c}}{\i}, M.~Patel, J.~S. Kim, R.~Azizibarzoki,
  A.~Olgun, L.~Orosa, H.~Hassan, J.~Park, K.~Kanellopoullos, T.~Shahroodi,
  S.~Ghose, and O.~Mutlu, ``{BlockHammer: Preventing RowHammer at Low Cost by
  Blacklisting Rapidly-Accessed DRAM Rows},'' in \emph{HPCA}, 2021.

\bibitem{park2020graphene}
Y.~Park, W.~Kwon, E.~Lee, T.~J. Ham, J.~H. Ahn, and J.~W. Lee, ``{Graphene:
  Strong yet Lightweight Row Hammer Protection},'' in \emph{MICRO}, 2020.

\bibitem{apple2015about}
{Apple Inc.}, ``{About the Security Content of Mac EFI Security Update
  2015-001},'' \url{https://support.apple.com/en-us/HT204934}, 2015.

\bibitem{brasser2016can}
F.~Brasser, L.~Davi, D.~Gens, C.~Liebchen, and A.-R. Sadeghi, ``{Can't Touch
  This: Practical and Generic Software-Only Defenses Against Rowhammer
  Attacks},'' arXiv, 2016.

\bibitem{konoth2018zebram}
R.~K. Konoth, M.~Oliverio, A.~Tatar, D.~Andriesse, H.~Bos, C.~Giuffrida, and
  K.~Razavi, ``{ZebRAM: Comprehensive and Compatible Software Protection
  Against Rowhammer Attacks},'' in \emph{OSDI}, 2018.

\bibitem{van2018guardion}
V.~van~der Veen, M.~Lindorfer, Y.~Fratantonio, H.~P. Pillai, G.~Vigna,
  C.~Kruegel, H.~Bos, and K.~Razavi, ``{GuardION: Practical Mitigation of
  DMA-Based Rowhammer Attacks on ARM},'' in \emph{DIMVA}, 2018.

\bibitem{aweke2016anvil}
Z.~B. Aweke, S.~F. Yitbarek, R.~Qiao, R.~Das, M.~Hicks, Y.~Oren, and T.~Austin,
  ``{ANVIL: Software-Based Protection Against Next-Generation Rowhammer
  Attacks},'' in \emph{ASPLOS}, 2016.

\bibitem{lee2019twice}
E.~Lee, I.~Kang, S.~Lee, G.~{Edward Suh}, and J.~{Ho Ahn}, ``{TWiCe: Preventing
  Row-Hammering by Exploiting Time Window Counters},'' in \emph{ISCA}, 2019.

\bibitem{seyedzadeh2017cbt}
S.~M. Seyedzadeh, A.~K. Jones, and R.~Melhem, ``{Counter-Based Tree Structure
  for Row Hammering Mitigation in DRAM},'' \emph{CAL}, 2017.

\bibitem{son2017making}
M.~Son, H.~Park, J.~Ahn, and S.~Yoo, ``{Making DRAM Stronger Against Row
  Hammering},'' in \emph{DAC}, 2017.

\bibitem{you2019mrloc}
J.~M. You and J.-S. Yang, ``{MRLoc : Mitigating Row-Hammering Based on Memory
  Locality},'' in \emph{DAC}, 2019.

\bibitem{greenfield2016throttling}
Z.~Greenfield and T.~Levy, ``{Throttling Support for Row-Hammer Counters},''
  2016, {U.S.\ Patent 9,251,885}.

\bibitem{yauglikcci2021security}
A.~G. Ya{\u{g}}l{\i}k{\c{c}}{\i}, J.~S. Kim, F.~Devaux, and O.~Mutlu,
  ``{Security Analysis of the Silver Bullet Technique for RowHammer
  Prevention},'' \emph{arXiv preprint arXiv:2106.07084}, 2021.

\bibitem{kim2022mithril}
M.~J. Kim, J.~Park, Y.~Park, W.~Doh, N.~Kim, T.~J. Ham, J.~W. Lee, and J.~H.
  Ahn, ``{Mithril: Cooperative Row Hammer Protection on Commodity DRAM
  Leveraging Managed Refresh},'' in \emph{HPCA}, 2022.

\bibitem{taouil2021lightroad}
M.~Taouil, C.~Reinbrecht, S.~Hamdioui, and J.~Sep{\'u}lveda, ``{LightRoAD:
  Lightweight Rowhammer Attack Detector},'' in \emph{ISVLSI}, 2021.

\bibitem{devaux2021method}
F.~Devaux and R.~Ayrignac, ``{Method and Circuit for Protecting a DRAM Memory
  Device from the Row Hammer Effect},'' Jan.~5 2021, {US Patent 10,885,966}.

\bibitem{marazzi2023rega}
M.~Marazzi, F.~Solt, P.~Jattke, K.~Takashi, and K.~Razavi, ``{REGA: Scalable
  Rowhammer Mitigation with Refresh-Generating Activations},'' in \emph{S\&P},
  2023.

\bibitem{woo2023scalable}
J.~Woo, G.~Saileshwar, and P.~J. Nair, ``{Scalable and Secure Row-Swap:
  Efficient and Safe Row Hammer Mitigation in Memory Systems},'' in
  \emph{HPCA}, 2023.

\bibitem{wi2023shadow}
M.~Wi, J.~Park, S.~Ko, M.~J. Kim, N.~S. Kim, E.~Lee, and J.~H. Ahn, ``{SHADOW:
  Preventing Row Hammer in DRAM with Intra-Subarray Row Shuffling},'' in
  \emph{HPCA}, 2023.

\bibitem{kim2023ddr5}
W.~Kim, C.~Jung, S.~Yoo, D.~Hong, J.~Hwang, J.~Yoon, O.~Jung, J.~Choi, S.~Hyun
  \emph{et~al.}, ``{A 1.1V 16Gb DDR5 DRAM with Probabilistic-Aggressor
  Tracking, a Refresh Management Function, Per-Row Hammer Tracking, a
  Multi-Step Precharge, and Core-Bias-Voltage Modulation for Security and
  Reliability Enhancement},'' in \emph{ISSCC}, 2023.

\bibitem{softmcsource}
{SoftMC Source Code}, \url{https://github.com/CMU-SAFARI/SoftMC}.

\bibitem{chang2017understanding}
K.~K. Chang, A.~G. Ya{\u{g}}l{\i}k{\c{c}}{\i}, S.~Ghose, A.~Agrawal,
  N.~Chatterjee, A.~Kashyap, D.~Lee, M.~O'Connor, H.~Hassan, and O.~Mutlu,
  ``{Understanding Reduced-Voltage Operation in Modern DRAM Devices:
  Experimental Characterization, Analysis, and Mechanisms},'' in
  \emph{SIGMETRICS}, 2017.

\bibitem{frigo2020trrespass}
P.~Frigo, E.~Vannacci, H.~Hassan, V.~van~der Veen, O.~Mutlu, C.~Giuffrida,
  H.~Bos, and K.~Razavi, ``{TRRespass: Exploiting the Many Sides of Target Row
  Refresh},'' in \emph{IEEE S\&P}, 2020.

\bibitem{gao2019computedram}
F.~Gao, G.~Tziantzioulis, and D.~Wentzlaff, ``{ComputeDRAM: In-Memory Compute
  using Off-the-Shelf DRAMs},'' in \emph{MICRO}, 2019.

\bibitem{kim2019d}
J.~S. Kim, M.~Patel, H.~Hassan, L.~Orosa, and O.~Mutlu, ``{D-RaNGe: Using
  Commodity DRAM Devices to Generate True Random Numbers With Low Latency And
  High Throughput},'' in \emph{HPCA}, 2019.

\bibitem{hassan2021uncovering}
H.~Hassan, Y.~C. Tugrul, J.~S. Kim, V.~Van~der Veen, K.~Razavi, and O.~Mutlu,
  ``{Uncovering In-DRAM RowHammer Protection Mechanisms: A New Methodology,
  Custom RowHammer Patterns, and Implications},'' in \emph{MICRO}, 2021.

\bibitem{ghose2018your}
S.~Ghose, A.~G. Ya{\u{g}}l{\i}k{\c{c}}{\i}, R.~Gupta, D.~Lee, K.~Kudrolli,
  W.~X. Liu, H.~Hassan, K.~K. Chang, N.~Chatterjee, A.~Agrawal, M.~O'Connor,
  and O.~Mutlu, ``{What Your DRAM Power Models Are Not Telling You: Lessons
  from a Detailed Experimental Study},'' \emph{SIGMETRICS}, 2018.

\bibitem{olgun2021quac}
A.~Olgun, M.~Patel, A.~G. Ya{\u{g}}l{\i}k{\c{c}}{\i}, H.~Luo, J.~S. Kim,
  N.~Bostanc{\i}, N.~Vijaykumar, O.~Ergin, and O.~Mutlu, ``{QUAC-TRNG:
  High-Throughput True Random Number Generation Using Quadruple Row Activation
  in Commodity DRAM Chips},'' in \emph{ISCA}, 2021.

\bibitem{orosa2021codic}
L.~Orosa, Y.~Wang, M.~Sadrosadati, J.~S. Kim, M.~Patel, I.~Puddu, H.~Luo,
  K.~Razavi, J.~G{\'o}mez-Luna, H.~Hassan, N.~Mansouri-Ghiasi, S.~Ghose, and
  O.~Mutlu, ``{CODIC: A Low-Cost Substrate for Enabling Custom In-DRAM
  Functionalities and Optimizations},'' \emph{ISCA}, 2021.

\bibitem{talukder2018exploiting}
B.~Talukder, J.~Kerns, B.~Ray, T.~Morris, and M.~T. Rahman, ``{Exploiting DRAM
  Latency Variations for Generating True Random Numbers},'' \emph{ICCE}, 2019.

\bibitem{talukder2019prelatpuf}
B.~B. Talukder, B.~Ray, D.~Forte, and M.~T. Rahman, ``{PreLatPUF: Exploiting
  DRAM Latency Variations For Generating Robust Device Signatures},''
  \emph{IEEE Access}, 2019.

\bibitem{talukder2018ldpuf}
B.~Talukder, B.~Ray, M.~Tehranipoor, D.~Forte, and M.~T. Rahman, ``{LDPUF:
  Exploiting DRAM Latency Variations to Generate Robust Device Signatures},''
  \emph{arXiv preprint arXiv:1808.02584}, 2018.

\bibitem{talukder2020towards}
B.~B. Talukder, V.~Menon, B.~Ray, T.~Neal, and M.~T. Rahman, ``{Towards the
  Avoidance of Counterfeit Memory: Identifying the DRAM Origin},'' in
  \emph{HOST}, 2020.

\bibitem{bepary2022dram}
M.~K. Bepary, B.~M. S.~B. Talukder, and M.~T. Rahman, ``{DRAM Retention
  Behavior with Accelerated Aging in Commercial Chips},'' \emph{Applied
  Sciences}, 2022.

\bibitem{farmani2021rhat}
M.~Farmani, M.~Tehranipoor, and F.~Rahman, ``{RHAT: Efficient RowHammer-Aware
  Test for Modern DRAM Modules},'' in \emph{ETS}, 2021.

\bibitem{yaglikci2022hira}
A.~G. Ya{\u{g}}l{\i}k{\c{c}}{\i}, A.~Olgun, M.~Patel, H.~Luo, H.~Hassan,
  L.~Orosa, O.~Ergin, and O.~Mutlu, ``{HIRA: Hidden Row Activation for Reducing
  Refresh Latency of Off-the-Shelf DRAM Chips},'' in \emph{MICRO}, 2022.

\bibitem{gao2022fracdram}
F.~Gao, G.~Tziantzioulis, and D.~Wentzlaff, ``{FracDRAM: Fractional Values in
  Off-the-Shelf DRAM},'' in \emph{MICRO}, 2022.

\bibitem{lee2014green}
J.~Lee, ``{Green Memory Solution},'' {Investor’s Forum}, {Samsung
  Electronics}, 2014.

\bibitem{micronddr4}
Micron, ``{DDR4 SDRAM Datasheet},'' 2016.

\bibitem{utrrsource}
{U-TRR Source Code}, \url{https://github.com/CMU-SAFARI/U-TRR}.

\bibitem{jedec2020lpddr5}
JEDEC, ``{Low Power Double Data Rate 5 (LPDDR5) SDRAM Specification},''
  \emph{JEDEC Standard JESD209--5A}, 2020.

\bibitem{jattke2022blacksmith}
P.~Jattke, V.~van~der Veen, P.~Frigo, S.~Gunter, and K.~Razavi, ``{Blacksmith:
  Scalable Rowhammering in the Frequency Domain},'' in \emph{S\&P}, 2022.

\bibitem{smdsource}
{Self-Managing DRAM (SMD) Source Code},
  \url{https://github.com/CMU-SAFARI/SelfManagingDRAM}.

\bibitem{hassan2022self}
H.~Hassan, A.~Olgun, A.~G. Ya{\u{g}}l{\i}k{\c{c}}{\i}, H.~Luo, and O.~Mutlu,
  ``{A Case for Self-Managing DRAM Chips: Improving Performance, Efficiency,
  Reliability, and Security via Autonomous in-DRAM Maintenance Operations},''
  \emph{arXiv preprint arXiv:2207.13358}, 2022.

\bibitem{patel2017reaper}
M.~Patel, J.~S. Kim, and O.~Mutlu, ``{The Reach Profiler (REAPER): Enabling the
  Mitigation of DRAM Retention Failures via Profiling at Aggressive
  Conditions},'' in \emph{ISCA}, 2017.

\bibitem{crow_spice_github}
{SAFARI Research Group}, ``{CROW --- GitHub Repository},''
  \url{https://github.com/CMU-SAFARI/CROW}.

\bibitem{hassan2019crow}
H.~Hassan, M.~Patel, J.~S. Kim, A.~G. Ya{\u{g}}l{\i}k{\c{c}}{\i},
  N.~Vijaykumar, N.~M. Ghiasi, S.~Ghose, and O.~Mutlu, ``{CROW: A Low-Cost
  Substrate for Improving DRAM Performance, Energy Efficiency, and
  Reliability},'' in \emph{ISCA}, 2019.

\bibitem{lee2016simultaneous}
D.~Lee, S.~Ghose, G.~Pekhimenko, S.~Khan, and O.~Mutlu, ``{Simultaneous
  Multi-Layer Access: Improving 3D-Stacked Memory Bandwidth at Low Cost},'' in
  \emph{{TACO}}, 2016.

\bibitem{lee2017design}
D.~Lee, S.~Khan, L.~Subramanian, S.~Ghose, R.~Ausavarungnirun, G.~Pekhimenko,
  V.~Seshadri, and O.~Mutlu, ``{Design-Induced Latency Variation in Modern DRAM
  Chips: Characterization, Analysis, and Latency Reduction Mechanisms},'' in
  \emph{SIGMETRICS}, 2017.

\bibitem{lee2015decoupled}
D.~Lee, L.~Subramanian, R.~Ausavarungnirun, J.~Choi, and O.~Mutlu, ``{Decoupled
  Direct Memory Access: Isolating CPU and IO Traffic by Leveraging a
  Dual-Data-Port DRAM},'' in \emph{PACT}, 2015.

\bibitem{seshadri2013rowclone}
V.~Seshadri, Y.~Kim, C.~Fallin, D.~Lee, R.~Ausavarungnirun, G.~Pekhimenko,
  Y.~Luo, O.~Mutlu, P.~B. Gibbons, M.~A. Kozuch, P.~B. Gibbons, and T.~C.
  Mowry, ``{RowClone: Fast and Energy-Efficient In-DRAM Bulk Data Copy and
  Initialization},'' in \emph{MICRO}, 2013.

\bibitem{seshadri2017ambit}
V.~Seshadri, D.~Lee, T.~Mullins, H.~Hassan, A.~Boroumand, J.~Kim, M.~A. Kozuch,
  O.~Mutlu, P.~B. Gibbons, and T.~C. Mowry, ``{Ambit: In-Memory Accelerator for
  Bulk Bitwise Operations Using Commodity DRAM Technology},'' in \emph{MICRO},
  2017.

\bibitem{seshadri2015gather}
V.~Seshadri, T.~Mullins, A.~Boroumand, O.~Mutlu, P.~B. Gibbons, M.~A. Kozuch,
  and T.~C. Mowry, ``{Gather-Scatter DRAM: In-DRAM Address Translation to
  Improve the Spatial Locality of Non-Unit Strided Accesses},'' in
  \emph{MICRO}, 2015.

\bibitem{seshadri2020indram}
V.~Seshadri and O.~Mutlu, ``{In-DRAM Bulk Bitwise Execution Engine},''
  \emph{Advances in Computers}, 2020.

\bibitem{zhang2014half}
T.~Zhang, K.~Chen, C.~Xu, G.~Sun, T.~Wang, and Y.~Xie, ``{Half-DRAM: A
  High-Bandwidth and Low-Power DRAM Architecture from the Rethinking of
  Fine-Grained Activation},'' in \emph{ISCA}, 2014.

\bibitem{wang2020figaro}
Y.~Wang, L.~Orosa, X.~Peng, Y.~Guo, S.~Ghose, M.~Patel, J.~S. Kim, J.~G. Luna,
  M.~Sadrosadati, N.~M. Ghiasi, and O.~Mutlu, ``{FIGARO: Improving System
  Performance via Fine-Grained In-DRAM Data Relocation and Caching},'' in
  \emph{MICRO}, 2020.

\bibitem{cojocar2020are}
L.~Cojocar, J.~Kim, M.~Patel, L.~Tsai, S.~Saroiu, A.~Wolman, and O.~Mutlu,
  ``{Are We Susceptible to Rowhammer? An End-to-End Methodology for Cloud
  Providers},'' in \emph{IEEE S\&P}, 2020.

\bibitem{patel2019understanding}
M.~Patel, J.~S. Kim, H.~Hassan, and O.~Mutlu, ``{Understanding and Modeling
  On-Die Error Correction in Modern DRAM: An Experimental Study Using Real
  Devices},'' in \emph{DSN}, 2019.

\bibitem{zhang2016restore}
X.~Zhang, Y.~Zhang, B.~R. Childers, and J.~Yang, ``{Restore Truncation for
  Performance Improvement in Future DRAM Systems},'' in \emph{HPCA}, 2016.

\bibitem{kim2018dram}
J.~S. Kim, M.~Patel, H.~Hassan, and O.~Mutlu, ``{The DRAM Latency PUF: Quickly
  Evaluating Physical Unclonable Functions by Exploiting the
  Latency-Reliability Tradeoff in Modern Commodity DRAM Devices},'' in
  \emph{HPCA}, 2018.

\bibitem{kim2015ramulator}
Y.~Kim, W.~Yang, and O.~Mutlu, ``{Ramulator: A Fast and Extensible DRAM
  Simulator},'' in \emph{CAL}, 2015.

\bibitem{bostanci2022dr}
F.~Bostanc{\i}, A.~Olgun, L.~Orosa, A.~G. Ya{\u{g}}l{\i}k{\c{c}}{\i}, J.~S.
  Kim, H.~Hassan, O.~Ergin, and O.~Mutlu, ``{DR-STRaNGe: End-to-End System
  Design for DRAM-based True Random Number Generators},'' in \emph{HPCA}, 2022.

\bibitem{hajinazar2021simdram}
N.~Hajinazar, G.~F. Oliveira, S.~Gregorio, J.~Ferreira, N.~M. Ghiasi, M.~Patel,
  M.~Alser, S.~Ghose, J.~G. Luna, and O.~Mutlu, ``{SIMDRAM: An End-to-End
  Framework for Bit-Serial SIMD Computing in DRAM},'' \emph{ASPLOS}, 2021.

\bibitem{patel2020bit}
M.~Patel, J.~Kim, T.~Shahroodi, H.~Hassan, and O.~Mutlu, ``{Bit-Exact ECC
  Recovery (BEER): Determining DRAM On-Die ECC Functions by Exploiting DRAM
  Data Retention Characteristics},'' in \emph{MICRO}, 2020.

\bibitem{davis2001modern}
B.~T. Davis, ``{Modern DRAM architectures},'' Ph.D. dissertation, University of
  Michigan, 2001.

\bibitem{rixner2000memory}
S.~Rixner, W.~J. Dally, U.~J. Kapasi, P.~Mattson, and J.~D. Owens, ``{Memory
  Access Scheduling},'' in \emph{ISCA}, 2000.

\bibitem{keeth2007dram}
B.~Keeth, R.~J. Baker, B.~Johnson, and F.~Lin, \emph{{DRAM Circuit Design:
  Fundamental and High-Speed Topics}}.\hskip 1em plus 0.5em minus 0.4em\relax
  {John Wiley \& Sons}, 2007.

\bibitem{itoh2013vlsi}
K.~Itoh, \emph{{VLSI Memory Chip Design}}.\hskip 1em plus 0.5em minus
  0.4em\relax Springer Science \& Business Media, 2013, vol.~5.

\bibitem{jacob2010memory}
B.~Jacob, S.~Ng, and D.~Wang, \emph{{Memory Systems: Cache, DRAM, Disk}}.\hskip
  1em plus 0.5em minus 0.4em\relax Morgan Kaufmann, 2010.

\bibitem{gong2015clean}
S.-L. Gong, M.~Rhu, J.~Kim, J.~Chung, and M.~Erez, ``{Clean-ECC: High
  Reliability ECC for Adaptive Granularity Memory System},'' in \emph{MICRO},
  2015.

\bibitem{gong2018duo}
S.-L. Gong, J.~Kim, S.~Lym, M.~Sullivan, H.~David, and M.~Erez, ``{DUO:
  Exposing On-Chip Redundancy to Rank-Level ECC for High Reliability},'' in
  \emph{HPCA}, 2018.

\bibitem{jeong2012balancing}
M.~K. Jeong, D.~H. Yoon, D.~Sunwoo, M.~Sullivan, I.~Lee, and M.~Erez,
  ``{Balancing DRAM Locality and Parallelism in Shared Memory CMP Systems},''
  in \emph{HPCA}, 2012.

\bibitem{jeong2012drsim}
M.~K. Jeong, D.~H. Yoon, and M.~Erez, ``{DrSim: A Platform for Flexible {DRAM}
  System Research},'' \url{http://lph.ece.utexas.edu/public/DrSim}, 2012.

\bibitem{kim2015bamboo}
J.~Kim, M.~Sullivan, and M.~Erez, ``{Bamboo ECC: Strong, Safe, and Flexible
  Codes For Reliable Computer Memory},'' in \emph{HPCA}, 2015.

\bibitem{kim2015frugal}
J.~Kim, M.~Sullivan, S.-L. Gong, and M.~Erez, ``{Frugal ECC: Efficient And
  Versatile Memory Error Protection Through Fine-Grained Compression},'' in
  \emph{SC}, 2015.

\bibitem{kim2016all}
J.~Kim, M.~Sullivan, S.~Lym, and M.~Erez, ``{All-Inclusive ECC: Thorough
  End-to-End Protection for Reliable Computer Memory},'' in \emph{ISCA}, 2016.

\bibitem{kim2016relaxfault}
D.~W. Kim and M.~Erez, ``{RelaxFault Memory Repair},'' in \emph{ISCA}, 2016.

\bibitem{yoon2010virtualized}
D.~H. Yoon and M.~Erez, ``{Virtualized and Flexible ECC for Main Memory},'' in
  \emph{ASPLOS}, 2010.

\bibitem{nair2013archshield}
P.~J. Nair, D.-H. Kim, and M.~K. Qureshi, ``{ArchShield: Architectural
  Framework for Assisting DRAM Scaling by Tolerating High Error Rates},'' in
  \emph{ISCA}, 2013.

\bibitem{nair2016xed}
P.~J. Nair, V.~Sridharan, and M.~K. Qureshi, ``{XED: Exposing On-Die Error
  Detection Information for Strong Memory Reliability},'' in \emph{ISCA}, 2016.

\bibitem{saileshwar2022randomized}
G.~Saileshwar, B.~Wang, M.~Qureshi, and P.~J. Nair, ``{Randomized Row-Swap:
  Mitigating Row Hammer by Breaking Spatial Correlation between Aggressor and
  Victim Rows},'' in \emph{ASPLOS}, 2022.

\bibitem{chandrasekar2014exploiting}
K.~Chandrasekar, S.~Goossens, C.~Weis, M.~Koedam, B.~Akesson, N.~Wehn, and
  K.~Goossens, ``{Exploiting Expendable Process-Margins in DRAMs for Run-Time
  Performance Optimization},'' in \emph{DATE}, 2014.

\bibitem{jung2014optimized}
M.~Jung, C.~Weis, N.~Wehn, M.~Sadri, and L.~Benini, ``{Optimized Active and
  Power-Down Mode Refresh Control in 3D-DRAMs},'' in \emph{VLSI-SoC}, 2014.

\bibitem{jung2015dramsys}
M.~Jung, C.~Weis, and N.~Wehn, ``{DRAMSys: A Flexible DRAM Subsystem Design
  Space Exploration Framework},'' \emph{T-SLDM}, 2015.

\bibitem{jung2015omitting}
M.~Jung, {\'E}.~Zulian, D.~M. Mathew, M.~Herrmann, C.~Brugger, C.~Weis, and
  N.~Wehn, ``{Omitting Refresh: A Case Study for Commodity and Wide I/O
  DRAMs},'' in \emph{MEMSYS}, 2015.

\bibitem{jung2016reverse}
M.~Jung, C.~C. Rheinl{\"a}nder, C.~Weis, and N.~Wehn, ``{Reverse Engineering of
  DRAMs: Row Hammer with Crosshair},'' in \emph{MEMSYS}, 2016.

\bibitem{jung2017platform}
M.~Jung, D.~M. Mathew, C.~C. Rheinl{\"a}nder, C.~Weis, and N.~Wehn, ``{A
  Platform to Analyze DDR3 DRAM's Power and Retention Time},'' \emph{IEEE
  Design \& Test}, 2017.

\bibitem{kraft2018improving}
K.~Kraft, C.~Sudarshan, D.~M. Mathew, C.~Weis, N.~Wehn, and M.~Jung,
  ``{Improving the Error Behavior of DRAM by Exploiting its Z-Channel
  Property},'' in \emph{DATE}, 2018.

\bibitem{weis2015retention}
C.~Weis, M.~Jung, P.~Ehses, C.~Santos, P.~Vivet, S.~Goossens, M.~Koedam, and
  N.~Wehn, ``{Retention Time Measurements and Modelling of Bit Error Rates of
  Wide I/O DRAM in MPSoCs},'' in \emph{DATE}, 2015.

\bibitem{weis2017dramspec}
C.~Weis, A.~Mutaal, O.~Naji, M.~Jung, A.~Hansson, and N.~Wehn, ``{DRAMSpec: A
  High-Level DRAM Timing, Power and Area Exploration Tool},'' \emph{IJPP},
  2017.

\bibitem{chandrasekar2012drampower}
K.~Chandrasekar, C.~Weis, Y.~Li, S.~Goossens, M.~Jung, O.~Naji, B.~Akesson,
  N.~Wehn, and K.~Goossens, ``{DRAMPower: Open-Source DRAM Power \& Energy
  Estimation Tool},'' \url{http://www.drampower.info}.

\bibitem{ipek2008self}
E.~Ipek, O.~Mutlu, J.~F. Mart{\'\i}nez, and R.~Caruana, ``{Self-Optimizing
  Memory Controllers: A Reinforcement Learning Approach},'' in \emph{ISCA},
  2008.

\bibitem{kim2010atlas}
Y.~Kim, D.~Han, O.~Mutlu, and M.~Harchol-Balter, ``{ATLAS: A Scalable and
  High-Performance Scheduling Algorithm for Multiple Memory Controllers},'' in
  \emph{HPCA}, 2010.

\bibitem{kim2010thread}
Y.~Kim, M.~Papamichael, O.~Mutlu, and M.~Harchol-Balter, ``{Thread Cluster
  Memory Scheduling: Exploiting Differences in Memory Access Behavior},'' in
  \emph{MICRO}, 2010.

\bibitem{mutlu2007stall}
O.~Mutlu and T.~Moscibroda, ``{Stall-Time Fair Memory Access Scheduling for
  Chip Multiprocessors},'' in \emph{MICRO}, 2007.

\bibitem{mutlu2008parallelism}
O.~Mutlu and T.~Moscibroda, ``{Parallelism-Aware Batch Scheduling: Enhancing
  both Performance and Fairness of Shared DRAM Systems},'' in \emph{ISCA},
  2008.

\bibitem{subramanian2016bliss}
L.~Subramanian, D.~Lee, V.~Seshadri, H.~Rastogi, and O.~Mutlu, ``{BLISS:
  Balancing Performance, Fairness and Complexity in Memory Access
  Scheduling},'' in \emph{TPDS}, 2016.

\bibitem{jedec2016gddr6}
JEDEC, ``{Graphics Double Data Rate (GDDR6) SGRAM Standard},'' \emph{JEDEC
  Standard JESD250C}, 2021.

\bibitem{zhang2015exploiting}
X.~Zhang, Y.~Zhang, B.~R. Childers, and J.~Yang, ``{Exploiting DRAM Restore
  Time Variations In Deep Sub-Micron Scaling},'' in \emph{DATE}, 2015.

\bibitem{jedec2014lpddr4}
JEDEC, ``{Low Power Double Data Rate 4 (LPDDR4) SDRAM Specification},''
  \emph{JEDEC Standard JESD209--4B}, 2014.

\bibitem{chang2017thesis}
K.~K. Chang, ``{Understanding and Improving Latency of DRAM-Based Memory
  Systems},'' Ph.D. dissertation, Carnegie Mellon University, 2017.

\bibitem{choi2015multiple}
J.~Choi, W.~Shin, J.~Jang, J.~Suh, Y.~Kwon, Y.~Moon, and L.-S. Kim, ``{Multiple
  Clone Row DRAM: A Low Latency and Area Optimized DRAM},'' in \emph{ISCA},
  2015.

\bibitem{hennessy2011computer}
J.~L. Hennessy and D.~A. Patterson, \emph{{Computer Architecture: A
  Quantitative Approach}}.\hskip 1em plus 0.5em minus 0.4em\relax Elsevier,
  2011.

\bibitem{nguyen2018nonblocking}
K.~Nguyen, K.~Lyu, X.~Meng, V.~Sridharan, and X.~Jian, ``{Nonblocking Memory
  Refresh},'' in \emph{ISCA}, 2018.

\bibitem{shin2014nuat}
W.~Shin, J.~Yang, J.~Choi, and L.-S. Kim, ``{NUAT: A Non-Uniform Access Time
  Memory Controller},'' in \emph{HPCA}, 2014.

\bibitem{wang2018reducing}
Y.~Wang, A.~Tavakkol, L.~Orosa, S.~Ghose, N.~M. Ghiasi, M.~Patel, J.~S. Kim,
  H.~Hassan, M.~Sadrosadati, and O.~Mutlu, ``{Reducing DRAM Latency Via
  Charge-Level-Aware Look-Ahead Partial Restoration},'' in \emph{MICRO}, 2018.

\bibitem{kim2018solar}
J.~S. Kim, M.~Patel, H.~Hassan, and O.~Mutlu, ``{Solar-DRAM: Reducing DRAM
  Access Latency by Exploiting the Variation in Local Bitlines},'' in
  \emph{ICCD}, 2018.

\bibitem{hidaka1990cache}
H.~Hidaka, Y.~Matsuda, M.~Asakura, and K.~Fujishima, ``{The Cache DRAM
  Architecture: A DRAM with an on-Chip Cache Memory},'' \emph{MICRO}, 1990.

\bibitem{gulur2012multiple}
N.~D. Gulur, R.~Manikantan, M.~Mehendale, and R.~Govindarajan, ``{Multiple
  Sub-Row Buffers in DRAM: Unlocking Performance and Energy Improvement
  Opportunities},'' in \emph{ICS}, 2012.

\bibitem{takemura2007long}
R.~Takemura, K.~Itoh, T.~Sekiguchi, S.~Akiyama, S.~Hanzawa, K.~Kajigaya, and
  T.~Kawahara, ``{Long-Retention-Time, High-Speed DRAM Array with 12-F 2 Twin
  Cell for Sub 1-V Operation},'' \emph{TOE}, 2007.

\bibitem{micron-lpddr4}
{Micron Technology, Inc.}, ``{x64 Mobile LPDDR4 SDRAM Datasheet},''
  \url{https://prod.micron.com/~/media/documents/products/data-sheet/dram/mobile-dram/low-power-dram/lpddr4/272b\_z9am\_qdp\_mobile\_lpddr4.pdf}.

\bibitem{isen2009eskimo}
C.~Isen and L.~John, ``{ESKIMO: Energy Savings Using Semantic Knowledge of
  Inconsequential Memory Occupancy for DRAM Subsystem},'' in \emph{MICRO},
  2009.

\bibitem{kim2003block}
J.~Kim and M.~C. Papaefthymiou, ``{Block-Based Multiperiod Dynamic Memory
  Design for Low Data-Retention Power},'' in \emph{TVLSI}, 2003.

\bibitem{venkatesan2006retention}
R.~K. Venkatesan, S.~Herr, and E.~Rotenberg, ``{Retention-Aware Placement in
  DRAM (RAPID): Software Methods for Quasi-Non-Volatile DRAM},'' in
  \emph{HPCA}, 2006.

\bibitem{riho2014partial}
Y.~Riho and K.~Nakazato, ``{Partial Access Mode: New Method for Reducing Power
  Consumption of Dynamic Random Access Memory},'' \emph{TVLSI}, 2014.

\bibitem{patel2005energy}
K.~Patel, L.~Benini, E.~Macii, and M.~Poncino, ``{Energy-Efficient Value-Based
  Selective Refresh for Embedded DRAMs},'' in \emph{PATMOS}, 2005.

\bibitem{kim2009new}
K.~Kim and J.~Lee, ``{A New Investigation of Data Retention Time in Truly
  Nanoscaled DRAMs},'' in \emph{EDL}, 2009.

\bibitem{kim2015architectural}
Y.~Kim, ``{Architectural Techniques to Enhance DRAM Scaling},'' Ph.D.
  dissertation, Carnegie Mellon University, 2015.

\bibitem{meza2015revisiting}
J.~Meza, Q.~Wu, S.~Kumar, and O.~Mutlu, ``{Revisiting Memory Errors in
  Large-Scale Production Data Centers: Analysis and Modeling of New Trends from
  the Field},'' in \emph{DSN}, 2015.

\bibitem{schroeder2009dram}
B.~Schroeder, E.~Pinheiro, and W.-D. Weber, ``{DRAM Errors in the Wild: a
  Large-Scale Field Study},'' in \emph{SIGMETRICS}, 2009.

\bibitem{sridharan2015memory}
V.~Sridharan, N.~DeBardeleben, S.~Blanchard, K.~B. Ferreira, J.~Stearley,
  J.~Shalf, and S.~Gurumurthi, ``{Memory Errors in Modern Systems: The Good,
  the Bad, and the Ugly},'' in \emph{ASPLOS}, 2015.

\bibitem{sridharan2012study}
V.~Sridharan and D.~Liberty, ``{A Study of DRAM Failures in the Field},'' in
  \emph{SC}, 2012.

\bibitem{weis2015thermal}
C.~Weis, M.~Jung, O.~Naji, C.~Santos, P.~Vivet, and A.~Hansson, ``{Thermal
  Aspects and High-Level Explorations of 3D Stacked DRAMs},'' in \emph{ISVLSI},
  2015.

\bibitem{lin2012secret}
C.~H. Lin, D.-Y. Shen, Y.-J. Chen, C.-L. Yang, and M.~Wang, ``{SECRET:
  Selective Error Correction for Refresh Energy Reduction in DRAMs},'' in
  \emph{ICCD}, 2012.

\bibitem{ohsawa1998optimizing}
T.~Ohsawa, K.~Kai, and K.~Murakami, ``{Optimizing the DRAM Refresh Count for
  Merged DRAM/logic LSIs},'' in \emph{ISLPED}, 1998.

\bibitem{wang2014proactivedram}
J.~Wang, X.~Dong, and Y.~Xie, ``{ProactiveDRAM: A DRAM-Initiated Retention
  Management Scheme},'' in \emph{ICCD}, 2014.

\bibitem{sutar2016d}
S.~Sutar, A.~Raha, and V.~Raghunathan, ``{D-PUF: An Intrinsically
  Reconfigurable DRAM PUF for Device Authentication in Embedded Systems},'' in
  \emph{CASES}, 2016.

\bibitem{tehranipoor2017investigation}
F.~Tehranipoor, N.~Karimian, W.~Yan, and J.~A. Chandy, ``{Investigation of DRAM
  PUFs Reliability Under Device Accelerated Aging Effects},'' in \emph{ISCAS},
  2017.

\bibitem{sutar2018d}
S.~Sutar, A.~Raha, D.~Kulkarni, R.~Shorey, J.~Tew, and V.~Raghunathan,
  ``{D-PUF: An Intrinsically Reconfigurable DRAM PUF for Device Authentication
  and Random Number Generation},'' \emph{TECS}, 2018.

\bibitem{tehranipoor2016robust}
F.~Tehranipoor, W.~Yan, and J.~A. Chandy, ``{Robust Hardware True Random Number
  Generators using DRAM Remanence Effects},'' in \emph{HOST}, 2016.

\bibitem{olgun2021pidram}
A.~Olgun, J.~G. Luna, K.~Kanellopoulos, B.~Salami, H.~Hassan, O.~Ergin, and
  O.~Mutlu, ``{PiDRAM: A Holistic End-to-end FPGA-based Framework for
  Processing-in-DRAM},'' \emph{TACO}, 2022.

\bibitem{patel2021harp}
M.~Patel, G.~F. de~Oliveira, and O.~Mutlu, ``{HARP: Practically and Effectively
  Identifying Uncorrectable Errors in Memory Chips That Use On-Die
  Error-Correcting Codes},'' in \emph{MICRO}, 2021.

\bibitem{mukhanov2020dstress}
L.~Mukhanov, D.~S. Nikolopoulos, and G.~Karakonstantis, ``{DStress: Automatic
  Synthesis of DRAM Reliability Stress Viruses using Genetic Algorithms},'' in
  \emph{MICRO}, 2020.

\bibitem{bacchini2014characterization}
A.~Bacchini, M.~Rovatti, G.~Furano, and M.~Ottavi, ``{Characterization of Data
  Retention Faults in DRAM Devices},'' in \emph{DFT}, 2014.

\bibitem{restle1992dram}
P.~J. Restle, J.~Park, and B.~F. Lloyd, ``{DRAM Variable Retention Time},'' in
  \emph{IEDM}, 1992.

\bibitem{shirley2014copula}
C.~G. Shirley and W.~R. Daasch, ``{Copula Models of Correlation: A DRAM Case
  Study},'' in \emph{TC}, 2014.

\bibitem{kim2011characterization}
H.~Kim, B.~Oh, Y.~Son, K.~Kim, S.-Y. Cha, J.-G. Jeong, S.-J. Hong, and H.~Shin,
  ``{Characterization of the Variable Retention Time in Dynamic Random Access
  Memory},'' \emph{TED}, 2011.

\bibitem{kim2011study}
H.~Kim, B.~Oh, Y.~Son, K.~Kim, S.-Y. Cha, J.-G. Jeong, S.-J. Hong, and H.~Shin,
  ``{Study of Trap Models Related to the Variable Retention Time Phenomenon in
  DRAM},'' \emph{TED}, 2011.

\bibitem{kumar2014detection}
N.~Kumar, ``{Detection of Variable Retention Time in DRAM},'' Master's thesis,
  Portland State University, Portland, Oregon, 2014.

\bibitem{mori2005origin}
Y.~Mori, K.~Ohyu, K.~Okonogi, and R.~i.~Yamada, ``{The Origin of Variable
  Retention Time in DRAM},'' in \emph{IEDM}, 2005.

\bibitem{ohyu2006quantitative}
K.~Ohyu, T.~Umeda, K.~Okonogi, S.~Tsukada, M.~Hidaka, S.~Fujieda, and
  Y.~Mochizuki, ``{Quantitative Identification for the Physical Origin of
  Variable Retention Time: A Vacancy-Oxygen Complex Defect Model},'' in
  \emph{IEDM}, 2006.

\bibitem{micron2017whitepaper}
{Micron Technology Inc.}, ``{ECC Brings Reliability and Power Efficiency to
  Mobile Devices},'' {Micron Technology Inc.}, Tech. Rep., 2017.

\bibitem{cha2017defect}
S.~Cha, O.~Seongil, H.~Shin, S.~Hwang, K.~Park, S.~J. Jang, J.~S. Choi, G.~Y.
  Jin, Y.~H. Son, H.~Cho, J.~H. Ahn, and N.~S. Kim, ``{Defect Analysis and
  Cost-Effective Resilience Architecture for Future DRAM Devices},'' in
  \emph{HPCA}, 2017.

\bibitem{sharifi2017online}
R.~Sharifi and Z.~Navabi, ``{Online Profiling for Cluster-Specific Variable
  Rate Refreshing in High-Density DRAM Systems},'' in \emph{ETS}, 2017.

\bibitem{borucki2008comparison}
L.~Borucki, G.~Schindlbeck, and C.~Slayman, ``{Comparison of Accelerated DRAM
  Soft Error Rates Measured at Component and System Level},'' in \emph{IEEE
  IRPS}, 2008.

\bibitem{may1979alpha}
T.~C. May and M.~H. Woods, ``{Alpha-Particle-Induced Soft Errors in Dynamic
  Memories},'' \emph{TED}, 1979.

\bibitem{guenzer1979single}
C.~Guenzer, E.~Wolicki, and R.~Allas, ``{Single Event Upset of Dynamic RAMs by
  Neutrons and Protons},'' \emph{IEEE Transactions on Nuclear Science}, 1979.

\bibitem{ziegler1996terrestrial}
J.~F. Ziegler, ``{Terrestrial Cosmic Rays},'' \emph{IBM Journal of Research and
  Development}, 1996.

\bibitem{hwang2012cosmic}
A.~A. Hwang, I.~A. Stefanovici, and B.~Schroeder, ``{Cosmic Rays Don't Strike
  Twice: Understanding the Nature of DRAM Errors and the Implications for
  System Design},'' in \emph{SIGPLAN Notices}, 2012.

\bibitem{hamamoto1998retention}
T.~Hamamoto, S.~Sugiura, and S.~Sawada, ``{On the Retention Time Distribution
  of Dynamic Random Access Memory (DRAM)},'' in \emph{TED}, 1998.

\bibitem{hou2013fpga}
C.-S. Hou, J.-F. Li, C.-Y. Lo, D.-M. Kwai, Y.-F. Chou, and C.-W. Wu, ``{An
  FPGA-Based Test Platform for Analyzing Data Retention Time Distribution of
  DRAMs},'' in \emph{VLSI-DAT}, 2013.

\bibitem{kong2008analysis}
W.~Kong, P.~C. Parries, G.~Wang, and S.~S. Iyer, ``{Analysis of Retention Time
  Distribution of Embedded DRAM-A New Method to Characterize Across-Chip
  Threshold Voltage Variation},'' in \emph{ITC}, 2008.

\bibitem{kim2010high}
I.~Kim, S.~Cho, D.~Im, E.~Cho, D.~Kim, G.~Oh, D.~Ahn, S.~Park, S.~Nam, J.~Moon
  \emph{et~al.}, ``{High Performance PRAM Cell Scalable to Sub-20nm Technology
  with Below 4F2 Cell Size, Extendable to DRAM Applications},'' in
  \emph{VLSIT}, 2010.

\bibitem{halderman2008lest}
J.~A. Halderman, S.~D. Schoen, N.~Heninger, W.~Clarkson, W.~Paul, J.~A.
  Calandrino, A.~J. Feldman, J.~Appelbaum, and E.~W. Felten, ``{Lest We
  Remember: Cold-Boot Attacks on Encryption Keys},'' \emph{{USENIX Security}},
  2008.

\bibitem{kim2016ecc}
D.-H. Kim and L.~S. Milor, ``{ECC-ASPIRIN: An ECC-Assisted Post-Package Repair
  Scheme for Aging Errors in DRAMs},'' in \emph{VTS}, 2016.

\bibitem{seaborn2015exploiting}
M.~Seaborn and T.~Dullien, ``{Exploiting the DRAM RowHammer Bug to Gain Kernel
  Privileges},'' \emph{Black Hat}, 2015.

\bibitem{rh_project_zero}
M.~Seaborn and T.~Dullien, ``{Exploiting the DRAM Rowhammer Bug to Gain Kernel
  Privileges},''
  \url{http://googleprojectzero.blogspot.com.tr/2015/03/exploiting-dram-rowhammer-bug-to-gain.html},
  2015.

\bibitem{veen2016drammer}
V.~Van Der~Veen, Y.~Fratantonio, M.~Lindorfer, D.~Gruss, C.~Maurice, G.~Vigna,
  H.~Bos, K.~Razavi, and C.~Giuffrida, ``{Drammer: Deterministic Rowhammer
  Attacks on Mobile Platforms},'' \emph{{CCS}}, 2016.

\bibitem{gruss2016rowhammer}
D.~Gruss, C.~Maurice, and S.~Mangard, ``{Rowhammer.js: A Remote
  Software-Induced Fault Attack in Javascript},'' in \emph{DIMVA}, 2016.

\bibitem{qiao2016new}
R.~Qiao and M.~Seaborn, ``{A New Approach for Rowhammer Attacks},'' in
  \emph{HOST}, 2016.

\bibitem{kwong2020rambleed}
A.~Kwong, D.~Genkin, D.~Gruss, and Y.~Yarom, ``{RAMBleed: Reading Bits in
  Memory Without Accessing Them},'' in \emph{S\&P}, 2020.

\bibitem{pessl2016drama}
P.~Pessl, D.~Gruss, C.~Maurice, M.~Schwarz, and S.~Mangard, ``{DRAMA:
  Exploiting DRAM Addressing for Cross-CPU Attacks},'' in \emph{USENIX
  Security}, 2016.

\bibitem{bhattacharya2016curious}
S.~Bhattacharya and D.~Mukhopadhyay, ``{Curious Case of RowHammer: Flipping
  Secret Exponent Bits using Timing Analysis},'' in \emph{CHES}, 2016.

\bibitem{jang2017sgx}
Y.~Jang, J.~Lee, S.~Lee, and T.~Kim, ``{SGX-Bomb: Locking Down the Processor
  via Rowhammer Attack},'' in \emph{SysTEX}, 2017.

\bibitem{zhang2020pthammer}
Z.~Zhang, Y.~Cheng, D.~Liu, S.~Nepal, Z.~Wang, and Y.~Yarom, ``{PThammer:
  Cross-User-Kernel-Boundary Rowhammer through Implicit Accesses},'' in
  \emph{MICRO}, 2020.

\bibitem{weissman2020jackhammer}
Z.~Weissman, T.~Tiemann, D.~Moghimi, E.~Custodio, T.~Eisenbarth, and B.~Sunar,
  ``{JackHammer: Efficient Rowhammer on Heterogeneous FPGA--CPU Platforms},''
  arXiv:1912.11523 [cs.CR], 2020.

\bibitem{ji2019pinpoint}
S.~Ji, Y.~Ko, S.~Oh, and J.~Kim, ``{Pinpoint Rowhammer: Suppressing Unwanted
  Bit Flips on Rowhammer Attacks},'' in \emph{ASIACCS}, 2019.

\bibitem{gruss2018another}
D.~Gruss, M.~Lipp, M.~Schwarz, D.~Genkin, J.~Juffinger, S.~O'Connell,
  W.~Schoechl, and Y.~Yarom, ``{Another Flip in the Wall of Rowhammer
  Defenses},'' in \emph{IEEE S\&P}, 2018.

\bibitem{bosman2016dedup}
E.~Bosman, K.~Razavi, H.~Bos, and C.~Giuffrida, ``{Dedup Est Machina: Memory
  Deduplication as an Advanced Exploitation Vector},'' 2016.

\bibitem{deridder2021smash}
F.~de~Ridder, P.~Frigo, E.~Vannacci, H.~Bos, C.~Giuffrida, and K.~Razavi,
  ``{SMASH: Synchronized Many-sided Rowhammer Attacks from JavaScript},'' in
  \emph{USENIX Security}, 2021.

\bibitem{frigo2018grand}
P.~Frigo, C.~Giuffrida, H.~Bos, and K.~Razavi, ``{Grand Pwning Unit:
  Accelerating Microarchitectural Attacks with the GPU},'' \emph{{IEEE S\&P}},
  2018.

\bibitem{razavi2016flip}
K.~Razavi, B.~Gras, E.~Bosman, B.~Preneel, C.~Giuffrida, and H.~Bos, ``{Flip
  Feng Shui: Hammering a Needle in the Software Stack},'' in \emph{USENIX
  Sec.}, 2016.

\bibitem{xiao2016one}
Y.~Xiao, X.~Zhang, Y.~Zhang, and M.~Teodorescu, ``{One Bit Flips, One Cloud
  Flops: Cross-VM Row Hammer Attacks and Privilege Escalation},'' in
  \emph{USENIX Sec.}, 2016.

\bibitem{tatar2018defeating}
A.~Tatar, C.~Giuffrida, H.~Bos, and K.~Razavi, ``{Defeating Software
  Mitigations Against Rowhammer: A Surgical Precision Hammer},'' in
  \emph{RAID}, 2018.

\bibitem{lipp2018nethammer}
M.~Lipp, M.~T. Aga, M.~Schwarz, D.~Gruss, C.~Maurice, L.~Raab, and L.~Lamster,
  ``{Nethammer: Inducing Rowhammer Faults Through Network Requests},''
  \emph{arXiv}, 2018.

\bibitem{yao2020deephammer}
F.~Yao, A.~S. Rakin, and D.~Fan, ``{Deephammer: Depleting the Intelligence of
  Deep Neural Networks Through Targeted Chain of Bit Flips},'' in \emph{USENIX
  Security}, 2020.

\bibitem{hong2019terminal}
S.~Hong, P.~Frigo, Y.~Kaya, C.~Giuffrida, and T.~Dumitra\c{s}, ``{Terminal
  Brain Damage: Exposing the Graceless Degradation in Deep Neural Networks
  under Hardware Fault Attacks},'' in \emph{USENIX Security}, 2019.

\bibitem{rh-cisco}
T.~Fridley and O.~Santos, ``{Mitigations Available for the DRAM Row Hammer
  Vulnerability},''
  http://blogs.cisco.com/security/mitigations-available-for-the-dram-row-hammer-vulnerability,
  2015.

\bibitem{lenovo2015row}
{Lenovo Group Ltd.}, ``{Row Hammer Privilege Escalation},''
  \url{https://support.lenovo.com/us/en/ product security/row hammer}, March
  2015.

\bibitem{enterprise2015hp}
{Hewlett-Packard Enterprise}, ``{HP Moonshot Component Pack Version
  2015.05.0},'' 2015.

\bibitem{marazzi2022protrr}
M.~Marazzi, P.~Jattke, S.~Flavien, and K.~Razavi, ``{PROTRR: Principled yet
  Optimal In-DRAM Target Row Refresh},'' in \emph{S\&P}, 2022.

\bibitem{qureshi2022hydra}
M.~Qureshi, A.~Rohan, G.~Saileshwar, and P.~J. Nair, ``{Hydra: Enabling
  Low-Overhead Mitigation of Row-Hammer at Ultra-Low Thresholds via Hybrid
  Tracking},'' in \emph{ISCA}, 2022.

\bibitem{mosalikanti2011high}
P.~Mosalikanti, C.~Mozak, and N.~Kurd, ``{High Performance DDR Architecture in
  Intel Core Processors Using 32nm CMOS High-K Metal-Gate Process},'' in
  \emph{VLSI-DAT}, 2011.

\bibitem{yitbarek2017cold}
S.~F. Yitbarek, M.~T. Aga, R.~Das, and T.~Austin, ``{Cold Boot Attacks are
  Still Hot: Security Analysis of Memory Scramblers in Modern Processors},'' in
  \emph{HPCA}, 2017.

\bibitem{gruhn2013practicability}
M.~Gruhn and T.~M{\"u}ller, ``{On the Practicability of Cold Boot Attacks},''
  in \emph{ARES}, 2013.

\bibitem{bauer2016lest}
J.~Bauer, M.~Gruhn, and F.~C. Freiling, ``{Lest We Forget: Cold-Boot Attacks on
  Scrambled DDR3 Memory},'' \emph{Digital Investigation}, 2016.

\bibitem{simmons2011security}
P.~Simmons, ``{Security Through Amnesia: A Software-Based Solution to the Cold
  Boot Attack on Disk Encryption},'' in \emph{ACSAC}, 2011.

\bibitem{muller2011tresor}
T.~M{\"u}ller, F.~C. Freiling, and A.~Dewald, ``{TRESOR Runs Encryption
  Securely Outside RAM},'' in \emph{USENIX Security}, 2011.

\bibitem{henson2014memory}
M.~Henson and S.~Taylor, ``{Memory Encryption: A Survey of Existing
  Techniques},'' \emph{CSUR}, 2014.

\bibitem{yang2005improving}
J.~Yang, L.~Gao, and Y.~Zhang, ``{Improving Memory Encryption Performance in
  Secure Processors},'' \emph{IEEE TC}, 2005.

\bibitem{seol2019amnesiac}
H.~Seol, M.~Kim, Y.~Kim, T.~Kim, and L.-S. Kim, ``{Amnesiac DRAM: A Proactive
  Defense Mechanism Against Cold Boot Attacks},'' \emph{IEEE Trans. on Comp.},
  2019.

\bibitem{kim2005technology}
K.~Kim, ``{Technology for sub-50nm DRAM and NAND Flash Manufacturing},'' ser.
  IEDM, 2005.

\bibitem{mueller2005challenges}
W.~Mueller, G.~Aichmayr, W.~Bergner, E.~Erben, T.~Hecht, C.~Kapteyn, A.~Kersch,
  S.~Kudelka, F.~Lau, J.~Luetzen \emph{et~al.}, ``{Challenges for the DRAM Cell
  Scaling to 40nm},'' in \emph{IEDM}, 2005.

\bibitem{nakagome1988impact}
Y.~Nakagome, M.~Aoki, S.~Ikenaga, M.~Horiguchi, S.~Kimura, Y.~Kawamoto, and
  K.~Itoh, ``{The Impact of Data-Line Interference Noise on DRAM Scaling},'' in
  \emph{JSSC}, 1988.

\bibitem{sridharan2013feng}
V.~Sridharan, J.~Stearley, N.~DeBardeleben, S.~Blanchard, and S.~Gurumurthi,
  ``{Feng Shui of Supercomputer Memory: Positional Effects in DRAM and SRAM
  Faults},'' in \emph{SC}, 2013.

\bibitem{nassif2000delay}
S.~Nassif, ``{Delay Variability: Sources, Impacts and Trends},'' in
  \emph{ISSCC}, 2000.

\bibitem{srinivasan1994accurate}
G.~Srinivasan, P.~Murley, and H.~Tang, ``{Accurate, Predictive Modeling of Soft
  Error Rate due to Cosmic Rays and Chip Alpha Radiation},'' in \emph{IRPS},
  1994.

\bibitem{hazucha2000impact}
P.~Hazucha and C.~Svensson, ``{Impact of CMOS Technology Scaling on the
  Atmospheric Neutron Soft Error Rate},'' \emph{TNS}, 2000.

\bibitem{lee2010mechanism}
M.~J. Lee and K.~W. Park, ``{A Mechanism for Dependence of Refresh Time on Data
  Pattern in DRAM},'' in \emph{EDL}, 2010.

\bibitem{hidaka1989twisted}
H.~Hidaka, K.~Fujishima, Y.~Matsuda, M.~Asakura, and T.~Yoshihara, ``{Twisted
  Bit-Line Architectures for Multi-Megabit DRAMs},'' \emph{JSSC}, 1989.

\bibitem{advantest}
Advantest, ``{V6000 Memory Platform},''
  \url{https://www.advantest.com/products/ic-test-systems/v6000-memory}.

\bibitem{nickel}
{Nickel Electronics}, ``{DRAM Memory Testing},''
  \url{https://www.nickelelectronics.com/memory-testing/}.

\bibitem{teradyne}
Teradyne, ``{Magnum Memory Test System},''
  \url{http://www.teradyne.com/products/semiconductor-test/magnum}.

\bibitem{futureplus}
{FuturePlus}, ``{FS2800 DDR Detective},''
  \url{http://www.futureplus.com/DDR-Detective-Standalone/summary-2800.html}.

\bibitem{yang2015random}
H.~Yang, S.-H. Kuo, T.-H. Huang, C.-H. Chen, C.~Lin, and M.~C.-T. Chao,
  ``{Random Pattern Generation for Post-Silicon Validation of DDR3 SDRAM},'' in
  \emph{VTS}, 2015.

\bibitem{huang2000fpga}
J.~Huang, C.-K. Ong, K.-T. Cheng, and C.-W. Wu, ``{An FPGA-Based
  Re-Configurable Functional Tester for Memory Chips},'' in \emph{ATS}, 2000.

\bibitem{keezer2015fpga}
D.~Keezer, T.~Chen, T.~Moon, D.~Stonecypher, A.~Chatterjee, H.~Choi, S.~Kim,
  and H.~Yoo, ``{An FPGA-Based ATE Extension Module for Low-Cost Multi-GHz
  Memory Test},'' in \emph{ETS}, 2015.

\bibitem{bojnordi2012pardis}
M.~N. Bojnordi and E.~Ipek, ``{PARDIS: A Programmable Memory Controller for the
  DDRx Interfacing Standards},'' in \emph{ISCA}, 2012.

\bibitem{you1985self}
Y.~You and J.~Hayes, ``{A Self-Testing Dynamic RAM Chip},'' \emph{TED}, 1985.

\bibitem{querbach2014reusable}
B.~Querbach, R.~Khanna, D.~Blankenbeckler, Y.~Zhang, R.~T. Anderson, D.~G.
  Ellis, Z.~T. Schoenborn, S.~Deyati, and P.~Chiang, ``{A Reusable BIST with
  Software Assisted Repair Technology for Improved Memory and IO Debug,
  Validation and Test Time},'' in \emph{ITC}, 2014.

\bibitem{querbach2016architecture}
B.~Querbach, R.~Khanna, S.~Puligundla, D.~Blankenbeckler, J.~Crop, and
  P.~Chiang, ``{Architecture of a Reusable BIST Engine for Detection and Auto
  Correction of Memory Failures and for IO Debug, Validation, Link Training,
  and Power Optimization on 14nm SOC},'' \emph{IEEE D\&T}, 2016.

\bibitem{bernardi2010programmable}
P.~Bernardi, M.~Grosso, M.~S. Reorda, and Y.~Zhang, ``{A Programmable BIST for
  DRAM Testing and Diagnosis},'' in \emph{ITC}, 2010.

\bibitem{yang2015hybrid}
C.~Yang, J.-F. Li, Y.-C. Yu, K.-T. Wu, C.-Y. Lo, C.-H. Chen, J.-S. Lai, D.-M.
  Kwai, and Y.-F. Chou, ``{A Hybrid Built-In Self-Test Scheme for DRAMs},'' in
  \emph{VLSI-DAT}, 2015.

\bibitem{jacobsen2015riffa}
M.~Jacobsen, D.~Richmond, M.~Hogains, and R.~Kastner, ``{{RIFFA} 2.1: A
  Reusable Integration Framework for {FPGA} Accelerators},'' \emph{TRETS},
  2015.

\bibitem{david2011memory}
H.~David, C.~Fallin, E.~Gorbatov, U.~R. Hanebutte, and O.~Mutlu, ``{Memory
  Power Management via Dynamic Voltage/Frequency Scaling},'' in \emph{ICAC},
  2011.

\bibitem{ml605}
{Xilinx}, \emph{{ML605 Hardware User Guide}}, Oct. 2012.

\bibitem{xilpcie}
Xilinx, ``{Virtex-6 FPGA Integrated Block for PCI Express},''
  \url{http://www.xilinx.com/support/documentation/user_guides/v6_pcie_ug517.pdf}.

\bibitem{xilphy}
Xilinx, ``{Virtex-6 FPGA Memory Interface Solutions},''
  \url{http://www.xilinx.com/support/documentation/ip_documentation/mig/v3_92/ug406.pdf}.

\bibitem{tomishima2016dram}
S.~Tomishima, Personal Communication, Dec. 2016.

\bibitem{koop2008performance}
M.~J. Koop, W.~Huang, K.~Gopalakrishnan, and D.~K. Panda, ``{Performance
  Analysis and Evaluation of PCIe 2.0 and Quad-Data Rate Infiniband},'' in
  \emph{HOTI}, 2008.

\bibitem{bittner2014direct}
R.~Bittner, E.~Ruf, and A.~Forin, ``{Direct GPU/FPGA Communication via PCI
  Express},'' \emph{Cluster Computing}, 2014.

\bibitem{kadric2012fpga}
E.~Kadric, N.~Manjikian, and Z.~Zilic, ``{An FPGA Implementation for a
  High-Speed Optical Link with a PCIe Interface},'' in \emph{SoC}, 2012.

\bibitem{ramulatorgithub}
``{Ramulator Source Code},'' \url{https://github.com/CMU-SAFARI/ramulator}.

\bibitem{vc709}
{Xilinx}, \emph{{VC709 FPGA User Guide}}, Aug. 2016.

\bibitem{luo2014characterizing}
Y.~Luo, S.~Govindan, B.~Sharma, M.~Santaniello, J.~Meza, A.~Kansal, J.~Liu,
  B.~Khessib, K.~Vaid, and O.~Mutlu, ``{Characterizing Application Memory Error
  Vulnerability to Optimize Datacenter Cost via Heterogeneous-Reliability
  Memory},'' in \emph{DSN}, 2014.

\bibitem{raoux2008phare}
S.~Raoux, G.~Burr, M.~Breitwisch, C.~Rettner, Y.~Chen, R.~Shelby, M.~Salinga,
  D.~Krebs, S.-H. Chen, H.~Lung, and C.~Lam, ``{Phase-Change Random Access
  Memory: A Scalable Technology},'' \emph{IBM JRD}, 2008.

\bibitem{lee2010phase}
B.~C. Lee, P.~Zhou, J.~Yang, Y.~Zhang, B.~Zhao, E.~Ipek, O.~Mutlu, and
  D.~Burger, ``{Phase-Change Technology and the Future of Main Memory},''
  \emph{IEEE Micro}, 2010.

\bibitem{lee2009architecting}
B.~C. Lee, E.~Ipek, O.~Mutlu, and D.~Burger, ``{Architecting Phase Change
  Memory as a Scalable DRAM Alternative},'' in \emph{ISCA}, 2009.

\bibitem{kawahara2008spram}
T.~Kawahara, R.~Takemura, K.~Miura, J.~Hayakawa, S.~Ikeda, Y.~M. Lee,
  R.~Sasaki, Y.~Goto, K.~Ito, T.~Meguro, F.~Matsukura, H.~Takahashi,
  H.~Matsuoka, and H.~Ohno, ``{2 Mb SPRAM (SPin-Transfer Torque RAM) with
  Bit-by-Bit Bi-Directional Current Write and Parallelizing-Direction Current
  Read},'' \emph{JSSC}, 2008.

\bibitem{kultursay2013evaluating}
E.~K{\"u}lt{\"u}rsay, M.~Kandemir, A.~Sivasubramaniam, and O.~Mutlu,
  ``{Evaluating STT-RAM as an Energy-Efficient Main Memory Alternative},'' in
  \emph{ISPASS}, 2013.

\bibitem{akinaga2010resistive}
H.~Akinaga and H.~Shima, ``{Resistive Random Access Memory (ReRAM) Based on
  Metal Oxides},'' \emph{Proc. IEEE}, 2010.

\bibitem{wong2012metal}
H.-S.~P. Wong, H.-Y. Lee, S.~Yu, Y.-S. Chen, Y.~Wu, P.-S. Chen, B.~Lee, F.~T.
  Chen, and M.-J. Tsai, ``{Metal--Oxide RRAM},'' \emph{Proc. IEEE}, 2012.

\bibitem{micron2016xpoint}
Micron, ``{3D XPoint Memory},''
  \url{http://www.micron.com/about/innovations/3d-xpoint-technology}, 2016.

\bibitem{everspin2021sttmram}
{Everspin Technologies}, ``{16Mb MRAM MR4A16B}.''

\bibitem{subramanian2014blacklisting}
L.~Subramanian, D.~Lee, V.~Seshadri, H.~Rastogi, and O.~Mutlu, ``{The
  Blacklisting Memory Scheduler: Achieving High Performance And Fairness At Low
  Cost},'' in \emph{ICCD}, 2014.

\bibitem{muralidhara2011reducing}
S.~P. Muralidhara, L.~Subramanian, O.~Mutlu, M.~Kandemir, and T.~Moscibroda,
  ``{Reducing Memory Interference in Multicore Systems via Application-Aware
  Memory Channel Partitioning},'' in \emph{MICRO}, 2011.

\bibitem{ebrahimi2011parallel}
E.~Ebrahimi, R.~Miftakhutdinov, C.~Fallin, C.~J. Lee, J.~A. Joao, O.~Mutlu, and
  Y.~N. Patt, ``{Parallel Application Memory Scheduling},'' in \emph{MICRO},
  2011.

\bibitem{deng2011memscale}
Q.~Deng, D.~Meisner, L.~Ramos, T.~F. Wenisch, and R.~Bianchini, ``{MemScale:
  Active Low-Power Modes for Main Memory},'' in \emph{ASPLOS}, 2011.

\bibitem{cai2011fpga}
Y.~Cai, E.~F. Haratsch, M.~McCartney, and K.~Mai, ``{{FPGA}-Based Solid-State
  Drive Prototyping Platform},'' in \emph{FCCM}, 2011.

\bibitem{luo2016enabling}
Y.~Luo, S.~Ghose, Y.~Cai, E.~F. Haratsch, and O.~Mutlu, ``{Enabling Accurate
  and Practical Online Flash Channel Modeling for Modern {MLC NAND} Flash
  Memory},'' in \emph{JSAC}, 2016.

\bibitem{cai2015read}
Y.~Cai, Y.~Luo, S.~Ghose, and O.~Mutlu, ``{Read Disturb Errors in MLC NAND
  Flash Memory: Characterization, Mitigation, and Recovery},'' in \emph{DSN},
  2015.

\bibitem{luo2015warm}
Y.~Luo, Y.~Cai, S.~Ghose, J.~Choi, and O.~Mutlu, ``{{WARM}: Improving NAND
  Flash Memory Lifetime with Write-Hotness Aware Retention Management},'' in
  \emph{MSST}, 2015.

\bibitem{cai2015data}
Y.~Cai, Y.~Luo, E.~F. Haratsch, K.~Mai, and O.~Mutlu, ``{Data Retention in MLC
  NAND Flash Memory: Characterization, Optimization, and Recovery},'' in
  \emph{HPCA}, 2015.

\bibitem{cai2014neighbor}
Y.~Cai, G.~Yalcin, O.~Mutlu, E.~F. Haratsch, O.~Unsal, A.~Cristal, and K.~Mai,
  ``{Neighbor-Cell Assisted Error Correction for MLC NAND Flash Memories},'' in
  \emph{SIGMETRICS}, 2014.

\bibitem{cai2013program}
Y.~Cai, O.~Mutlu, E.~F. Haratsch, and K.~Mai, ``{Program Interference in {MLC
  NAND} Flash Memory: Characterization, Modeling, and Mitigation},'' in
  \emph{ICCD}, 2013.

\bibitem{cai2013error}
Y.~Cai, G.~Yalcin, O.~Mutlu, E.~F. Haratsch, A.~Cristal, O.~S. Unsal, and
  K.~Mai, ``{Error Analysis and Retention-Aware Error Management for {NAND}
  Flash Memory},'' in \emph{ITJ}, 2013.

\bibitem{cai2013threshold}
Y.~Cai, E.~F. Haratsch, O.~Mutlu, and K.~Mai, ``{Threshold Voltage Distribution
  in {MLC NAND} Flash Memory: Characterization, Analysis, and Modeling},'' in
  \emph{DATE}, 2013.

\bibitem{cai2012flash}
Y.~Cai, G.~Yalcin, O.~Mutlu, E.~F. Haratsch, A.~Cristal, O.~S. Unsal, and
  K.~Mai, ``{Flash Correct-And-Refresh: Retention-Aware Error Management for
  Increased Flash Memory Lifetime},'' in \emph{ICCD}, 2012.

\bibitem{cai2012error}
Y.~Cai, E.~F. Haratsch, O.~Mutlu, and K.~Mai, ``{Error Patterns in {MLC NAND}
  Flash Memory: Measurement, Characterization, and Analysis},'' in \emph{DATE},
  2012.

\bibitem{cai2017vulnerabilities}
Y.~Cai, S.~Ghose, Y.~Luo, K.~Mai, O.~Mutlu, and E.~F. Haratsch,
  ``{Vulnerabilities in MLC NAND Flash Memory Programming: Experimental
  Analysis, Exploits, and Mitigation Techniques},'' in \emph{HPCA}, 2017.

\bibitem{yang2019trap}
T.~Yang and X.-W. Lin, ``{Trap-Assisted DRAM Row Hammer Effect},'' \emph{EDL},
  2019.

\bibitem{walker2021dram}
A.~J. Walker, S.~Lee, and D.~Beery, ``{On DRAM Rowhammer and the Physics of
  Insecurity},'' \emph{TED}, 2021.

\bibitem{ryu2017overcoming}
S.-W. Ryu, K.~Min, J.~Shin, H.~Kwon, D.~Nam, T.~Oh, T.-S. Jang, M.~Yoo, Y.~Kim,
  and S.~Hong, ``{Overcoming the Reliability Limitation in the Ultimately
  Scaled DRAM using Silicon Migration Technique by Hydrogen Annealing},'' in
  \emph{IEDM}, 2017.

\bibitem{park2016experiments}
K.~Park, C.~Lim, D.~Yun, and S.~Baeg, ``{Experiments and Root Cause Analysis
  for Active-Precharge Hammering Fault In DDR3 SDRAM Under 3$\times$ Nm
  Technology},'' \emph{Microelectronics Reliability}, 2016.

\bibitem{yang2016suppression}
C.~M. Yang, C.~K. Wei, Y.~J. Chang, T.~C. Wu, H.~P. Chen, and C.~S. Lai,
  ``{Suppression of Row Hammer Effect by Doping Profile Modification in
  Saddle-Fin Array Devices for Sub-30-nm DRAM Technology},'' \emph{IEEE
  Transactions on Device and Materials Reliability}, 2016.

\bibitem{yang2017scanning}
C.-M. Yang, C.-K. Wei, H.-P. Chen, J.-S. Luo, Y.~J. Chang, T.-C. Wu, and C.-S.
  Lai, ``{Scanning Spreading Resistance Microscopy for Doping Profile in
  Saddle-Fin Devices},'' \emph{TNANO}, 2017.

\bibitem{gautam2019row}
S.~Gautam, S.~Manhas, A.~Kumar, M.~Pakala, and E.~Yieh, ``{Row Hammering
  Mitigation Using Metal Nanowire in Saddle Fin DRAM},'' \emph{TED}, 2019.

\bibitem{jiang2021quantifying}
Y.~Jiang, H.~Zhu, D.~Sullivan, X.~Guo, X.~Zhang, and Y.~Jin, ``{Quantifying
  Rowhammer Vulnerability for DRAM Security},'' in \emph{DAC}, 2021.

\bibitem{micronddr4trr}
Micron, ``{8Gb: x4, x8, x16 DDR4 SDRAM Features - Excessive Row Activation},''
  2020.

\bibitem{scarfone2008guide}
K.~Scarfone, W.~Jansen, and M.~Tracy, ``{Guide to General Server Security},''
  \emph{NIST Special Publication}, 2008.

\bibitem{anderson2020security}
R.~Anderson, \emph{{Security Engineering: A Guide to Building Dependable
  Distributed Systems}}.\hskip 1em plus 0.5em minus 0.4em\relax John Wiley \&
  Sons, 2020.

\bibitem{saltzer1975protection}
J.~H. Saltzer and M.~D. Schroeder, ``{The Protection of Information in Computer
  Systems},'' \emph{Proceedings of the IEEE}, 1975.

\bibitem{ddr4operationhynix}
{SK Hynix}, ``{DDR4 SDRAM Device Operation},''
  \url{https://pdf.directindustry.com/pdf/hynix/ddr4-sdram-device-operation/34497-773768.html}.

\bibitem{son2015cidra}
Y.~H. Son, S.~Lee, O.~Seongil, S.~Kwon, N.~S. Kim, and J.~H. Ahn, ``{CiDRA: A
  cache-Inspired DRAM resilience architecture},'' in \emph{HPCA}, 2015.

\bibitem{horiguchi2011nanoscale}
M.~Horiguchi and K.~Itoh, \emph{{Nanoscale Memory Repair}}.\hskip 1em plus
  0.5em minus 0.4em\relax Springer SBM, 2011.

\bibitem{ddr4}
JEDEC, ``{Double Data Rate 4 (DDR4) SDRAM Standard},'' 2012.

\bibitem{dell1997white}
T.~J. Dell, ``{A White Paper on the Benefits of Chipkill-Correct ECC for PC
  Server Main Memory},'' \emph{{IBM Microelectronics Division}}, 1997.

\bibitem{gong2017dram}
S.-L. Gong, J.~Kim, and M.~Erez, ``{DRAM Scaling Error Evaluation Model Using
  Various Retention Time},'' in \emph{DSN-W}, 2017.

\bibitem{costello2004ecc}
S.~Lin and D.~J. Costello, \emph{{Error Control Coding: Fundamentals and
  Applications}}, 2004.

\bibitem{hamming1950error}
R.~W. Hamming, ``{Error Detecting and Error Correcting Codes},'' in \emph{Bell
  Labs Technical Journal}, 1950.

\bibitem{oh2014a}
T.-Y. Oh, H.~Chung, J.-Y. Park, K.-W. Lee, S.~Oh, S.-Y. Doo, H.-J. Kim, C.~Lee,
  H.-R. Kim, J.-H. Lee \emph{et~al.}, ``{A 3.2 Gbps/Pin 8 Gbit 1.0 V LPDDR4
  SDRAM with Integrated ECC Engine for Sub-1 V DRAM Core Operation},''
  \emph{JSSC}, 2014.

\bibitem{kwak2017a}
N.~Kwak, S.-H. Kim, K.~H. Lee, C.-K. Baek, M.~S. Jang, Y.~Joo, S.-H. Lee, W.~Y.
  Lee, E.~Lee, D.~Han \emph{et~al.}, ``{A 4.8 Gb/s/pin 2Gb LPDDR4 SDRAM with
  Sub-100$\mu$A Self-Refresh Current for IoT Applications},'' in \emph{ISSCC},
  2017.

\bibitem{kwon2017an}
H.-J. Kwon, E.~Seo, C.-Y. Lee, Y.-H. Seo, G.-H. Han, H.-R. Kim, J.-H. Lee,
  M.-S. Jang, S.-G. Do, S.-H. Cho \emph{et~al.}, ``{An Extremely
  Low-Standby-Power 3.733 Gb/s/pin 2Gb LPDDR4 SDRAM for Wearable Devices},'' in
  \emph{ISSCC}, 2017.

\bibitem{im2016im}
{Intelligent Memory}, ``{IM ECC DRAM with Integrated Error Correcting Code},''
  2016, {Product Brief}.

\bibitem{jeong2020pair}
S.~Jeong, S.~Kang, and J.-S. Yang, ``{PAIR: Pin-aligned In-DRAM ECC
  architecture using expandability of Reed-Solomon code},'' in \emph{DAC},
  2020.

\bibitem{amd2009bkdg}
AMD, ``{BKDG for AMD NPT Family 0Fh Processors},'' 2009.

\bibitem{reed1960polynomial}
I.~S. Reed and G.~Solomon, ``{Polynomial Codes Over Certain Finite Fields},''
  \emph{SIAM}, 1960.

\bibitem{huffman2003fundamentals}
W.~C. Huffman and V.~Pless, \emph{{Fundamentals of Error-Correcting
  Codes}}.\hskip 1em plus 0.5em minus 0.4em\relax {Cambridge University Press},
  2003.

\bibitem{aga2017good}
M.~T. Aga, Z.~B. Aweke, and T.~Austin, ``{When Good Protections go Bad:
  Exploiting anti-DoS Measures to Accelerate Rowhammer Attacks},'' in
  \emph{HOST}, 2017.

\bibitem{rowhammergithub}
{SAFARI Research Group}, ``{RowHammer --- GitHub Repository},''
  \url{https://github.com/CMU-SAFARI/rowhammer}.

\bibitem{google2021halfdouble}
S.~Qazi, Y.~Kim, N.~Boichat, E.~Shiu, and M.~Nissler, ``{Introducing
  Half-Double: New Hammering Rechnique for DRAM Rowhammer Bug},''
  \url{http://googleprojectzero.blogspot.com.tr/2015/03/exploiting-dram-rowhammer-bug-to-gain.html},
  2021.

\bibitem{awasthi2012efficient}
M.~Awasthi, M.~Shevgoor, K.~Sudan, B.~Rajendran, R.~Balasubramonian, and
  V.~Srinivasan, ``{Efficient Scrub Mechanisms for Error-Prone Emerging
  Memories},'' in \emph{HPCA}, 2012.

\bibitem{hong2010memory}
S.~Hong, ``{Memory Technology Trend and Future Challenges},'' in \emph{IEDM},
  2010.

\bibitem{lee2016technology}
S.-H. Lee, ``{Technology Scaling Challenges and Opportunities of Memory
  Devices},'' in \emph{IEDM}, 2016.

\bibitem{park2015technology}
S.-K. Park, ``{Technology Scaling Challenge and Future Prospects of DRAM and
  NAND Flash Memory},'' in \emph{IMW}, 2015.

\bibitem{mukherjee2004cache}
S.~S. Mukherjee, J.~Emer, T.~Fossum, and S.~K. Reinhardt, ``{Cache Scrubbing in
  Microprocessors: Myth or Necessity?}'' in \emph{SDC}, 2004.

\bibitem{saleh1990reliability}
A.~M. Saleh, J.~J. Serrano, and J.~H. Patel, ``{Reliability of Scrubbing
  Recovery-Techniques for Memory Systems},'' \emph{TR}, 1990.

\bibitem{siddiqua2017lifetime}
T.~Siddiqua, V.~Sridharan, S.~E. Raasch, N.~DeBardeleben, K.~B. Ferreira,
  S.~Levy, E.~Baseman, and Q.~Guan, ``{Lifetime Memory Reliability Data from
  the Field},'' in \emph{DFT}, 2017.

\bibitem{micron2019whitepaper}
R.~Rooney and N.~Koyle, ``{Micron DDR5 SDRAM: New Features},'' {Micron
  Technology Inc.}, Tech. Rep., 2019.

\bibitem{kim2000dynamic}
J.~Kim and M.~C. Papaefthymiou, ``{Dynamic Memory Design for Low Data-Retention
  Power},'' in \emph{PATMOS}, 2000.

\bibitem{kwon2021reducing}
H.~Kwon, K.~Kim, D.~Jeon, and K.-S. Chung, ``{Reducing Refresh Overhead with
  In-DRAM Error Correction Codes},'' in \emph{ISOCC}, 2021.

\bibitem{park2016statistical}
K.~Park, D.~Yun, and S.~Baeg, ``{Statistical Distributions of Row-Hammering
  Induced Failures in DDR3 Components},'' \emph{Microelectronics Reliability},
  2016.

\bibitem{bloom1970space}
B.~H. Bloom, ``{Space/Time Trade-offs in Hash Coding with Allowable Errors},''
  \emph{Communications of the ACM}, 1970.

\bibitem{fan1998summary}
L.~Fan, P.~Cao, J.~Almeida, and A.~Z. Broder, ``{Summary Cache: A Scalable
  Wide-Area Web Cache Sharing Protocol},'' \emph{SIGCOMM}, 1998.

\bibitem{pontarelli2016improving}
S.~Pontarelli, P.~Reviriego, and J.~A. Maestro, ``{Improving Counting Bloom
  Filter Performance with Fingerprints},'' \emph{Information Processing
  Letters}, 2016.

\bibitem{misra1982finding}
J.~Misra and D.~Gries, ``{Finding Repeated Elements},'' \emph{{Science of
  Computer Programming}}, 1982.

\bibitem{drampowergithub}
``{DRAMPower Source Code},'' https://github.com/ravenrd/DRAMPower.

\bibitem{chandrasekar2011improved}
K.~Chandrasekar, B.~Akesson, and K.~Goossens, ``{Improved Power Modeling of DDR
  SDRAMs},'' in \emph{DSD}, 2011.

\bibitem{luk2005pin}
C.-K. Luk, R.~Cohn, R.~Muth, H.~Patil, A.~Klauser, G.~Lowney, S.~Wallace, V.~J.
  Reddi, and K.~Hazelwood, ``{Pin: Building Customized Program Analysis Tools
  with Dynamic Instrumentation},'' in \emph{{PLDI}}, 2005.

\bibitem{spec2006}
``{Standard Performance Evaluation Corporation},'' http://www.spec.org/cpu2006.

\bibitem{tpc}
{Transaction Processing Performance Council}, ``{TPC Benchmarks},''
  \url{http://www.tpc.org/}.

\bibitem{stream}
J.~D. McCalpin, ``{STREAM: Sustainable Memory Bandwidth in High Performance
  Computers},'' \url{https://www.cs.virginia.edu/stream/}.

\bibitem{fritts2005mediabench}
J.~E. Fritts, F.~W. Steiling, and J.~A. Tucek, ``{Mediabench II Video:
  Expediting the Next Generation of Video Systems Research},'' in
  \emph{Electronic Imaging}, 2005.

\bibitem{eyerman2008system}
S.~Eyerman and L.~Eeckhout, ``{System-Level Performance Metrics for
  Multiprogram Workloads},'' in \emph{IEEE Micro}, 2008.

\bibitem{snavely2000symbiotic}
A.~Snavely and D.~M. Tullsen, ``{Symbiotic Jobscheduling for a Simultaneous
  Mutlithreading Processor},'' in \emph{ASPLOS}, 2000.

\bibitem{michaud2012demystifying}
P.~Michaud, ``{Demystifying Multicore Throughput Metrics},'' \emph{CAL}, 2012.

\bibitem{muralimanohar2009cacti}
N.~Muralimanohar, R.~Balasubramonian, and N.~P. Jouppi, ``{CACTI 6.0: A Tool to
  Model Large Caches},'' HP Laboratories, Tech. Rep. HPL-2009-85, 2009.

\bibitem{udipi2011combining}
A.~N. Udipi, N.~Muralimanohar, R.~Balasubramonian, A.~Davis, and N.~P. Jouppi,
  ``{Combining Memory and a Controller with Photonics Through 3D-Stacking to
  Enable Scalable and Energy-Efficient Systems},'' in \emph{ISCA}, 2011.

\bibitem{ham2013disintegrated}
T.~J. Ham, B.~K. Chelepalli, N.~Xue, and B.~C. Lee, ``{Disintegrated Control
  for Energy-Efficient and Heterogeneous Memory Systems},'' in \emph{HPCA},
  2013.

\bibitem{fang2011memory}
K.~Fang, L.~Chen, Z.~Zhang, and Z.~Zhu, ``{Memory Architecture for Integrating
  Emerging Memory Technologies},'' in \emph{PACT}, 2011.

\bibitem{cooper2012buffer}
E.~Cooper-Balis, P.~Rosenfeld, and B.~Jacob, ``{Buffer-on-Board Memory
  Systems},'' in \emph{ISCA}, 2012.

\bibitem{seyedzadeh2018cbt}
S.~M. {Seyedzadeh}, A.~K. {Jones}, and R.~{Melhem}, ``Mitigating wordline
  crosstalk using adaptive trees of counters,'' in \emph{ISCA}, 2018.

\bibitem{herath2015these}
N.~Herath and {Anders Fogh}, ``These are {{Not Your Grand Daddy}}'s {{CPU
  Performance Counters}},'' in \emph{Black Hat Briefings}, 2015.

\bibitem{udipi2010rethinking}
A.~N. Udipi, N.~Muralimanohar, N.~Chatterjee, R.~Balasubramonian, A.~Davis, and
  N.~P. Jouppi, ``{Rethinking DRAM Design and Organization for
  Energy-Constrained Multi-Cores},'' in \emph{ISCA}, 2010.

\bibitem{ghasempour2015armor}
M.~Ghasempour, M.~Lujan, and J.~Garside, ``{{ARMOR}: A Run-Time Memory Hot-Row
  Detector},'' 2015.

\bibitem{seyedzadeh2017mitigating}
S.~M. Seyedzadeh, D.~Kline~Jr, A.~K. Jones, and R.~Melhem, ``{Mitigating
  Bitline Crosstalk Noise in DRAM Memories},'' in \emph{ISMS}, 2017.

\bibitem{rambus}
{Rambus Inc.}, ``{{DRAM} Power Model},'' 2016, http://www.rambus.com/energy/.

\bibitem{vogelsang2010understanding}
T.~Vogelsang, ``{Understanding the Energy Consumption of Dynamic Random Access
  Memories},'' in \emph{MICRO}, 2010.

\bibitem{zhaoptm}
W.~Zhao and Y.~Cao, ``{New Generation of Predictive Technology Model for Sub-45
  nm Early Design Exploration},'' \emph{TED}, 2006.

\bibitem{ptmweb}
{Arizona State Univ., NIMO Group}, ``{Predictive Technology Model},''
  \url{http://ptm.asu.edu/}, 2012.

\bibitem{park2013regularities}
H.~Park, S.~Baek, J.~Choi, D.~Lee, and S.~H. Noh, ``{Regularities Considered
  Harmful: Forcing Randomness to Memory Accesses to Reduce Row Buffer Conflicts
  for Multi-Core, Multi-Bank Systems},'' in \emph{ASPLOS}, 2013.

\bibitem{zuravleff1997controller}
W.~K. Zuravleff and T.~Robinson, ``{Controller for a Synchronous DRAM that
  Maximizes Throughput by Allowing Memory Requests and Commands to be Issued
  Out of Order},'' 1997, {US} Patent 5,630,096.

\bibitem{mutlu2007memory}
T.~Moscibroda and O.~Mutlu, ``{Memory Performance Attacks: Denial of Memory
  Service in Multi-Core Systems},'' in \emph{USENIX Security}, 2007.

\bibitem{simpoint}
G.~Hamerly, E.~Perelman, J.~Lau, and B.~Calder, ``{SimPoint 3.0: Faster and
  More Flexible Program Phase Analysis},'' \emph{JILP}, 2005.

\bibitem{iacobovici2004effective}
S.~Iacobovici, L.~Spracklen, S.~Kadambi, Y.~Chou, and S.~G. Abraham,
  ``{Effective Stream-Based and Execution-Based Data Prefetching},'' in
  \emph{ICS}, 2004.

\bibitem{lu2015improving}
S.-L. Lu, Y.-C. Lin, and C.-L. Yang, ``{Improving DRAM Latency with Dynamic
  Asymmetric Subarray},'' in \emph{MICRO}, 2015.

\bibitem{rixner2004memory}
S.~Rixner, ``{Memory Controller Optimizations for Web Servers},'' in
  \emph{MICRO}, 2004.

\bibitem{herrero2012thread}
E.~Herrero, J.~Gonzalez, R.~Canal, and D.~Tullsen, ``{Thread Row Buffers:
  Improving Memory Performance Isolation and Throughput in Multiprogrammed
  Environments},'' \emph{TC}, 2012.

\bibitem{olgun2022dram}
A.~Olgun, H.~Hassan, A.~G. Ya{\u{g}}l{\i}k{\c{c}}{\i}, Y.~C. Tu{\u{g}}rul,
  L.~Orosa, H.~Luo, M.~Patel, O.~Ergin, and O.~Mutlu, ``{DRAM Bender: An
  Extensible and Versatile FPGA-based Infrastructure to Test State-of-the-art
  DRAM Chips},'' \emph{arXiv preprint}, 2022.

\bibitem{drambendergithub}
``{DRAM Bender Source Code},'' \url{https://github.com/CMU-SAFARI/dram-bender}.

\bibitem{tannu2017cryogenic}
S.~S. Tannu, D.~M. Carmean, and M.~K. Qureshi, ``{Cryogenic-DRAM based Memory
  System for Scalable Quantum Computers: A Feasibility Study},'' in
  \emph{MEMSYS}, 2017.

\bibitem{wang2018dram}
F.~Wang, T.~Vogelsang, B.~Haukness, and S.~C. Magee, ``{DRAM Retention at
  Cryogenic Temperatures},'' in \emph{IMW}, 2018.

\bibitem{ware2017superconducting}
F.~Ware, L.~Gopalakrishnan, E.~Linstadt, S.~A. McKee, T.~Vogelsang, K.~L.
  Wright, C.~Hampel, and G.~Bronner, ``{Do Superconducting Processors Really
  Need Cryogenic Memories? The Case for Cold DRAM},'' in \emph{MEMSYS}, 2017.

\bibitem{kelly2019some}
T.~Kelly, P.~Fernandez, T.~Vogelsang, S.~A. McKee, L.~Gopalakrishnan, S.~Magee,
  K.~Padgett, D.~Barrow, J.~Rizza, D.~Doidge, K.~L. Wright, C.~Hampel, and
  G.~Bronner, ``{Some Like It Cold: Initial Testing Results for Cryogenic
  Computing Components},'' in \emph{Journal of Physics: Conference Series},
  2019.

\bibitem{lee2021cryoguard}
G.-H. Lee, S.~Na, I.~Byun, D.~Min, and J.~Kim, ``{CryoGuard: A Near
  Refresh-Free Robust DRAM Design for Cryogenic Computing},'' in \emph{ISCA},
  2021.

\bibitem{lee2019cryogenic}
G.-h. Lee, D.~Min, I.~Byun, and J.~Kim, ``{Cryogenic Computer Architecture
  Modeling with Memory-Side Case Studies},'' in \emph{ISCA}, 2019.

\bibitem{garzon2021gain}
E.~Garz{\'o}n, Y.~Greenblatt, O.~Harel, M.~Lanuzza, and A.~Teman, ``{Gain-Cell
  Embedded DRAM Under Cryogenic Operation—A First Study},'' \emph{IEEE
  TVLSI}, 2021.

\bibitem{gurumurthi2021hbm3}
S.~Gurumurthi, K.~Lee, M.~Jang, V.~Sridharan, A.~Nygren, Y.~Ryu, K.~Sohn,
  T.~Kim, and H.~Chung, ``{HBM3: Enabling Memory Resilience at Scale},''
  \emph{CAL}, 2021.

\bibitem{qureshi2009scalable}
M.~K. Qureshi, V.~Srinivasan, and J.~A. Rivers, ``{Scalable High Performance
  Main Memory System Using Phase-change Memory Technology},'' in \emph{ISCA},
  2009.

\bibitem{seong2013tri}
N.~H. Seong, S.~Yeo, and H.-H.~S. Lee, ``{Tri-Level-Cell Phase Change Memory:
  Toward an Efficient and Reliable Memory System},'' in \emph{ISCA}, 2013.

\bibitem{wong2010phase}
H.-S.~P. Wong, S.~Raoux, S.~Kim, J.~Liang, J.~P. Reifenberg, B.~Rajendran,
  M.~Asheghi, and K.~E. Goodson, ``{Phase Change Memory},'' \emph{Proc. IEEE},
  2010.

\bibitem{hamdioui2017test}
S.~Hamdioui, P.~Pouyan, H.~Li, Y.~Wang, A.~Raychowdhur, and I.~Yoon, ``{Test
  and Reliability of Emerging Non-Volatile Memories},'' in \emph{ATS}, 2017.

\bibitem{qureshi2011pay}
M.~K. Qureshi, ``{Pay-As-You-Go: Low-Overhead Hard-Error Correction for Phase
  Change Memories},'' in \emph{MICRO}, 2011.

\bibitem{tavana2017remap}
M.~K. Tavana, A.~K. Ziabari, M.~Arjomand, M.~Kandemir, C.~Das, and D.~Kaeli,
  ``{REMAP: A Reliability/Endurance Mechanism for Advancing PCM},'' in
  \emph{MEMSYS}, 2017.

\bibitem{zhang2012memory}
Z.~Zhang, W.~Xiao, N.~Park, and D.~J. Lilja, ``{Memory Module-Level Testing And
  Error Behaviors For Phase Change Memory},'' in \emph{ICCD}, 2012.

\bibitem{huai2008spin}
Y.~Huai, ``{Spin-Transfer Torque MRAM (STT-MRAM): Challenges and Prospects},''
  \emph{AAPPS bulletin}, 2008.

\bibitem{apalkov2013spin}
D.~Apalkov, A.~Khvalkovskiy, S.~Watts, V.~Nikitin, X.~Tang, D.~Lottis, K.~Moon,
  X.~Luo, E.~Chen, A.~Ong, A.~Driskill-Smith, and M.~Krounbi, ``{Spin-Transfer
  Torque Magnetic Random Access Memory (STT-MRAM)},'' \emph{JETC}, 2013.

\bibitem{chun2012scaling}
K.~C. Chun, H.~Zhao, J.~D. Harms, T.-H. Kim, J.-P. Wang, and C.~H. Kim, ``{A
  Scaling Roadmap and Performance Evaluation of In-Plane and Perpendicular MTJ
  Based STT-MRAMs for High-Density Cache Memory},'' \emph{JSSC}, 2012.

\bibitem{akram2021performance}
S.~Akram, ``{Performance Evaluation of Intel Optane Memory for Managed
  Workloads},'' \emph{TACO}, 2021.

\bibitem{shanbhag2020large}
A.~Shanbhag, N.~Tatbul, D.~Cohen, and S.~Madden, ``{Large-Scale in-Memory
  Analytics on Intel Optane DC Persistent Memory},'' in \emph{Proceedings of
  the 16th International Workshop on Data Management on New Hardware}, 2020.

\bibitem{ahn2016scalable}
J.~Ahn, S.~Hong, S.~Yoo, O.~Mutlu, and K.~Choi, ``{A Scalable
  Processing-In-Memory Accelerator for Parallel Graph Processing},''
  \emph{ISCA}, 2016.

\bibitem{ahn2015pim}
J.~Ahn, S.~Yoo, O.~Mutlu, and K.~Choi, ``{PIM-Enabled Instructions: A
  Low-Overhead, Locality-Aware Processing-in-Memory Architecture},'' in
  \emph{ISCA}, 2015.

\bibitem{fromm1997energy}
R.~Fromm, S.~Perissakis, N.~Cardwell, C.~Kozyrakis, B.~McGaughy, D.~Patterson,
  T.~Anderson, and K.~Yelick, ``{The Energy Efficiency of IRAM
  Architectures},'' in \emph{ISCA}, 1997.

\bibitem{li2016pinatubo}
S.~Li, C.~Xu, Q.~Zou, J.~Zhao, Y.~Lu, and Y.~Xie, ``{Pinatubo: A
  Processing-in-Memory Architecture for Bulk Bitwise Operations in Emerging
  non-Volatile Memories},'' in \emph{DAC}, 2016.

\bibitem{mutlu2019processing}
O.~Mutlu, S.~Ghose, J.~G{\'o}mez-Luna, and R.~Ausavarungnirun, ``{Processing
  Data Where It Makes Sense: Enabling in-Memory Computation},''
  \emph{Microprocessors and Microsystems}, 2019.

\bibitem{shin2018mcdram}
H.~Shin, D.~Kim, E.~Park, S.~Park, Y.~Park, and S.~Yoo, ``{McDRAM: Low Latency
  and Energy-Efficient Matrix Computations in DRAM},'' \emph{TCAD}, 2018.

\bibitem{ferreira2022pluto}
J.~Ferreira, G.~Falcao, J.~G{\'o}mez-Luna, M.~Alser, G.~F. Oliveira, J.~Kim,
  M.~Sadrosadati, L.~Orosa, T.~Shahroodi, A.~Nori, and O.~Mutlu, ``{pLUTo:
  Enabling Massively Parallel Computation in DRAM via Lookup Tables},'' in
  \emph{MICRO}, 2022.

\bibitem{deng2018dracc}
Q.~Deng, L.~Jiang, Y.~Zhang, M.~Zhang, and J.~Yang, ``{DrAcc: A DRAM based
  Accelerator for Accurate CNN Inference},'' in \emph{DAC}, 2018.

\bibitem{li2017drisa}
S.~Li, D.~Niu, K.~T. Malladi, H.~Zheng, B.~Brennan, and Y.~Xie, ``{DRISA: A
  Dram-based Reconfigurable in-Situ Accelerator},'' in \emph{MICRO}, 2017.

\bibitem{xin2020elp2im}
X.~Xin, Y.~Zhang, and J.~Yang, ``{ELP2IM: Efficient and Low Power Bitwise
  Operation Processing in DRAM},'' in \emph{HPCA}, 2020.

\bibitem{akin2015data}
B.~Akin, F.~Franchetti, and J.~C. Hoe, ``{Data Reorganization in Memory Using
  3D-Stacked DRAM},'' in \emph{ISCA}, 2016.

\bibitem{augusta2015jafar}
A.~Augusta and S.~Idreos, ``{JAFAR: Near-Data Processing for Databases},'' in
  \emph{SIGMOD}, 2015.

\bibitem{farmahini2015nda}
A.~Farmahini-Farahani, J.~H. Ahn, K.~Morrow, and N.~S. Kim, ``{NDA: Near-DRAM
  Acceleration Architecture Leveraging Commodity DRAM Devices and Standard
  Memory Modules},'' in \emph{HPCA}, 2015.

\bibitem{gao2016hrl}
M.~Gao and C.~Kozyrakis, ``{HRL: Efficient and Flexible Reconfigurable Logic
  for Near-Data Processing},'' in \emph{HPCA}, 2016.

\bibitem{gao2017tetris}
M.~Gao, J.~Pu, X.~Yang, M.~Horowitz, and C.~Kozyrakis, ``{TETRIS: Scalable and
  Efficient Neural Network Acceleration with 3D Memory},'' in \emph{ASPLOS},
  2017.

\bibitem{hsieh2016accelerating}
K.~Hsieh, S.~Khan, N.~Vijaykumar, K.~K. Chang, A.~Boroumand, S.~Ghose, and
  O.~Mutlu, ``{Accelerating Pointer Chasing in 3D-Stacked Memory: Challenges,
  Mechanisms, Evaluation},'' in \emph{ICCD}, 2016.

\bibitem{kim2016neurocube}
D.~Kim, J.~Kung, S.~Chai, S.~Yalamanchili, and S.~Mukhopadhyay, ``{Neurocube: A
  Programmable Digital Neuromorphic Architecture with High-Density 3D
  Memory},'' in \emph{ISCA}, 2016.

\bibitem{nai2017graphpim}
L.~Nai, R.~Hadidi, J.~Sim, H.~Kim, P.~Kumar, and H.~Kim, ``{GraphPIM: Enabling
  Instruction-Level PIM Offloading in Graph Computing Frameworks},'' in
  \emph{HPCA}, 2017.

\bibitem{zhang2014top}
D.~Zhang, N.~Jayasena, A.~Lyashevsky, J.~L. Greathouse, L.~Xu, and
  M.~Ignatowski, ``{TOP-PIM: Throughput-Oriented Programmable Processing in
  Memory},'' in \emph{HPDC}, 2014.

\bibitem{zhu2013accelerating}
Q.~Zhu, T.~Graf, H.~E. Sumbul, L.~Pileggi, and F.~Franchetti, ``{Accelerating
  Sparse Matrix-Matrix Multiplication with 3D-Stacked Logic-in-Memory
  Hardware},'' in \emph{HPEC}, 2013.

\bibitem{mutlu2021primer}
O.~Mutlu, S.~Ghose, J.~Gomez-Luna, and R.~Ausavarungnirun, ``{A Modern Primer
  on Processing in Memory},'' in \emph{{arXiv}}, 2020.

\bibitem{cho2020chonda}
B.~Y. Cho, Y.~Kwon, S.~Lym, and M.~Erez, ``{Near Data Acceleration with
  Concurrent Host Access},'' in \emph{ISCA}, 2020.

\bibitem{oliveira2022accelerating}
G.~F. Oliveira, J.~G{\'o}mez-Luna, S.~Ghose, A.~Boroumand, and O.~Mutlu,
  ``{Accelerating Neural Network Inference with Processing-in-DRAM: From the
  Edge to the Cloud},'' \emph{IEEE Micro}, 2022.

\bibitem{eccsimgithub}
``{EINSim Source Code},'' \url{https://github.com/CMU-SAFARI/EINSim}.

\bibitem{beergithub}
``{BEER Source Code},'' \url{https://github.com/CMU-SAFARI/BEER}.

\bibitem{clrdramgithub}
``{CLR-DRAM Source Code},'' \url{https://github.com/CMU-SAFARI/CLRDRAM}.

\bibitem{olgun2022sectored}
A.~Olgun, F.~Bostanci, G.~F. Oliveira, Y.~C. Tugrul, R.~Bera, A.~G. Yaglikci,
  H.~Hassan, O.~Ergin, and O.~Mutlu, ``{Sectored DRAM: An Energy-Efficient
  High-Throughput and Practical Fine-Grained DRAM Architecture},'' \emph{arXiv
  preprint arXiv:2207.13795}, 2022.

\bibitem{pidramgithub}
``{PiDRAM Source Code},'' \url{https://github.com/CMU-SAFARI/PiDRAM}.

\bibitem{olgun2022pidram}
A.~Olgun, J.~G. Luna, K.~Kanellopoulos, B.~Salami, H.~Hassan, O.~Ergin, and
  O.~Mutlu, ``{PiDRAM: A Holistic End-to-end FPGA-based Framework for
  Processing-in-DRAM},'' \emph{TACO}, 2022.

\bibitem{blockhammergithub}
``{BlockHammer Source Code},'' \url{https://github.com/CMU-SAFARI/BlockHammer}.

\bibitem{boroumand2019conda}
A.~Boroumand, S.~Ghose, M.~Patel, H.~Hassan, B.~Lucia, R.~Ausavarungnirun,
  K.~Hsieh, N.~Hajinazar, K.~T. Malladi, H.~Zheng, and O.~Mutlu, ``{CoNDA:
  Efficient Cache Coherence Support For Near-Data Accelerators},'' in
  \emph{ISCA}, 2019.

\bibitem{boroumand2016lazypim}
A.~Boroumand, S.~Ghose, M.~Patel, H.~Hassan, B.~Lucia, K.~Hsieh, K.~T. Malladi,
  H.~Zheng, and O.~Mutlu, ``{LazyPIM: An Efficient Cache Coherence Mechanism
  For Processing-In-Memory},'' \emph{CAL}, 2016.

\bibitem{vijaykumar2022metasys}
N.~Vijaykumar, A.~Olgun, K.~Kanellopoulos, F.~N. Bostanci, H.~Hassan, M.~Lotfi,
  P.~B. Gibbons, and O.~Mutlu, ``{MetaSys: A Practical Open-source Metadata
  Management System to Implement and Evaluate Cross-layer Optimizations},''
  \emph{TACO}, 2022.

\bibitem{metasysgithub}
``{MetaSys Source Code},'' \url{https://github.com/CMU-SAFARI/MetaSys}.

\bibitem{kim2018grim}
J.~S. Kim, D.~S. Cali, H.~Xin, D.~Lee, S.~Ghose, M.~Alser, H.~Hassan, O.~Ergin,
  C.~Alkan, and O.~Mutlu, ``{GRIM-Filter: Fast seed location filtering in DNA
  read mapping using processing-in-memory technologies},'' \emph{BMC Genomics},
  2018.

\bibitem{alser2017gatekeeper}
M.~Alser, H.~Hassan, H.~Xin, O.~Ergin, O.~Mutlu, and C.~Alkan, ``{GateKeeper: A
  New Hardware Architecture for Accelerating Pre-Alignment in DNA Short Read
  Mapping},'' \emph{Bioinformatics}, 2017.

\bibitem{gatekeepergithub}
``{GateKeeper Source Code},''
  \url{https://github.com/BilkentCompGen/GateKeeper}.

\bibitem{alser2019shouji}
M.~Alser, H.~Hassan, A.~Kumar, O.~Mutlu, and C.~Alkan, ``{Shouji: A Fast and
  Efficient Pre-Alignment Filter for Sequence Alignment},''
  \emph{Bioinformatics}, 2019.

\bibitem{shoujigithub}
``{Shouji Source Code},'' \url{https://github.com/CMU-SAFARI/Shouji}.

\end{thebibliography}


\begin{thebibliography}{100}
\providecommand{\url}[1]{#1}
\csname url@samestyle\endcsname
\providecommand{\newblock}{\relax}
\providecommand{\bibinfo}[2]{#2}
\providecommand{\BIBentrySTDinterwordspacing}{\spaceskip=0pt\relax}
\providecommand{\BIBentryALTinterwordstretchfactor}{4}
\providecommand{\BIBentryALTinterwordspacing}{\spaceskip=\fontdimen2\font plus
\BIBentryALTinterwordstretchfactor\fontdimen3\font minus
  \fontdimen4\font\relax}
\providecommand{\BIBforeignlanguage}[2]{{%
\expandafter\ifx\csname l@#1\endcsname\relax
\typeout{** WARNING: IEEEtranS.bst: No hyphenation pattern has been}%
\typeout{** loaded for the language `#1'. Using the pattern for}%
\typeout{** the default language instead.}%
\else
\language=\csname l@#1\endcsname
\fi
#2}}
\providecommand{\BIBdecl}{\relax}
\BIBdecl

\bibitem{keeth2007dram}
\emph{DRAM Circuit Design: Fundamental and High-Speed Topics}.

\bibitem{drampowergithub}
``{DRAMPower Source Code},'' https://github.com/tukl-msd/DRAMPower.

\bibitem{rh-apple}
{Apple Inc.}, ``{About the Security Content of Mac EFI Security Update
  2015-001},'' \url{https://support.apple.com/en-us/HT204934}, 2015.

\bibitem{awasthi2012efficient}
M.~Awasthi, M.~Shevgoor, K.~Sudan, B.~Rajendran, R.~Balasubramonian, and
  V.~Srinivasan, ``{Efficient Scrub Mechanisms for Error-Prone Emerging
  Memories},'' in \emph{HPCA}, 2012.

\bibitem{aweke2016anvil}
Z.~B. Aweke, S.~F. Yitbarek, R.~Qiao, R.~Das, M.~Hicks, Y.~Oren, and T.~Austin,
  ``{ANVIL: Software-Based Protection Against Next-Generation Rowhammer
  Attacks},'' in \emph{ASPLOS}, 2016.

\bibitem{baek2014refresh}
S.~Baek, S.~Cho, and R.~Melhem, ``{Refresh Now and Then},'' in \emph{TC}, 2014.

\bibitem{bhati2013coordinated}
I.~Bhati, Z.~Chishti, and B.~Jacob, ``{Coordinated Refresh: Energy Efficient
  Techniques for DRAM Refresh Scheduling},'' in \emph{ISPLED}, 2013.

\bibitem{bloom1970space}
B.~H. Bloom, ``{Space/Time Trade-offs in Hash Coding with Allowable Errors},''
  \emph{Communications of the ACM}, 1970.

\bibitem{brasser2016can}
F.~Brasser, L.~Davi, D.~Gens, C.~Liebchen, and A.-R. Sadeghi, ``{Can't Touch
  This: Practical and Generic Software-only Defenses Against RowHammer
  Attacks},'' \emph{USENIX Security}, 2017.

\bibitem{cha2017defect}
S.~Cha, O.~Seongil, H.~Shin, S.~Hwang, K.~Park, S.~J. Jang, J.~S. Choi, G.~Y.
  Jin, Y.~H. Son, H.~Cho, J.~H. Ahn, and N.~S. Kim, ``{Defect Analysis and
  Cost-Effective Resilience Architecture for Future DRAM Devices},'' in
  \emph{HPCA}, 2017.

\bibitem{chandrasekar2011improved}
K.~Chandrasekar, B.~Akesson, and K.~Goossens, ``{Improved Power Modeling of DDR
  SDRAMs},'' in \emph{DSD}, 2011.

\bibitem{chang2014improving}
K.~K. Chang, D.~Lee, Z.~Chishti, A.~R. Alameldeen, C.~Wilkerson, Y.~Kim, and
  O.~Mutlu, ``{Improving DRAM Performance by Parallelizing Refreshes with
  Accesses},'' in \emph{HPCA}, 2014.

\bibitem{cojocar2019exploiting}
L.~Cojocar, K.~Razavi, C.~Giuffrida, and H.~Bos, ``{Exploiting Correcting
  Codes: On the Effectiveness of ECC Memory Against RowHammer Attacks},'' in
  \emph{S\&P}, 2019.

\bibitem{cooper2012buffer}
E.~Cooper-Balis, P.~Rosenfeld, and B.~Jacob, ``{Buffer-on-Board Memory
  Systems},'' in \emph{ISCA}, 2012.

\bibitem{cui2014dtail}
Z.~Cui, S.~A. McKee, Z.~Zha, Y.~Bao, and M.~Chen, ``{DTail: A Flexible Approach
  to DRAM Refresh Management},'' in \emph{SC}, 2014.

\bibitem{deridder2021smash}
F.~de~Ridder, P.~Frigo, E.~Vannacci, H.~Bos, C.~Giuffrida, and K.~Razavi,
  ``{SMASH: Synchronized Many-sided Rowhammer Attacks from JavaScript},'' in
  \emph{USENIX Security}, 2021.

\bibitem{dell1997white}
T.~J. Dell, ``{A White Paper on the Benefits of Chipkill-Correct ECC for PC
  Server Main Memory},'' \emph{IBM Microelectronics Division}, 1997.

\bibitem{emma2008rethinking}
P.~G. Emma, W.~R. Reohr, and M.~Meterelliyoz, ``{Rethinking Refresh: Increasing
  Availability and Reducing Power in DRAM for Cache Applications},'' in
  \emph{MICRO}, 2008.

\bibitem{eyerman2008system}
S.~Eyerman and L.~Eeckhout, ``{System-level Performance Metrics for
  Multiprogram Workloads},'' in \emph{IEEE Micro}, 2008.

\bibitem{fan1998summary}
L.~Fan, P.~Cao, J.~Almeida, and A.~Z. Broder, ``{Summary Cache: A Scalable
  Wide-Area Web Cache Sharing Protocol},'' \emph{SIGCOMM}, 1998.

\bibitem{fang2011memory}
K.~Fang, L.~Chen, Z.~Zhang, and Z.~Zhu, ``{Memory Architecture for Integrating
  Emerging Memory Technologies},'' in \emph{PACT}, 2011.

\bibitem{frigo2020trrespass}
P.~Frigo, E.~Vannacci, H.~Hassan, V.~van~der Veen, O.~Mutlu, C.~Giuffrida,
  H.~Bos, and K.~Razavi, ``{TRRespass: Exploiting the Many Sides of Target Row
  Refresh},'' in \emph{SP}, 2020.

\bibitem{fritts2005mediabench}
J.~E. Fritts, F.~W. Steiling, and J.~A. Tucek, ``{Mediabench II Video:
  Expediting the Next Generation of Video Systems Research},'' in
  \emph{Electronic Imaging}, 2005.

\bibitem{gautam2019row}
S.~Gautam, S.~Manhas, A.~Kumar, M.~Pakala, and E.~Yieh, ``{Row Hammering
  Mitigation Using Metal Nanowire in Saddle Fin DRAM},'' \emph{TED}, 2019.

\bibitem{ghosh2007smart}
M.~Ghosh and H.-H.~S. Lee, ``{Smart Refresh: An Enhanced Memory Controller
  Design for Reducing Energy in Conventional and 3D Die-stacked DRAMs},'' in
  \emph{MICRO}, 2007.

\bibitem{gong2018duo}
S.-L. Gong, J.~Kim, S.~Lym, M.~Sullivan, H.~David, and M.~Erez, ``{DUO:
  Exposing on-chip Redundancy to Rank-level ECC for High Reliability},'' in
  \emph{HPCA}, 2018.

\bibitem{greenfield2016throttling}
Z.~Greenfield and L.~Tomer, ``{Throttling Support for Row-Hammer Counters},''
  2016, {US Patent 9,251,885}.

\bibitem{ham2013disintegrated}
T.~J. Ham, B.~K. Chelepalli, N.~Xue, and B.~C. Lee, ``{Disintegrated Control
  for Energy-Efficient and Heterogeneous Memory Systems},'' in \emph{HPCA},
  2013.

\bibitem{hamming1950error}
R.~W. Hamming, ``{Error Detecting and Error Correcting Codes},'' in \emph{Bell
  Labs Technical Journal}, 1950.

\bibitem{hassan2019crow}
H.~Hassan, M.~Patel, J.~S. Kim, A.~G. Ya\u{g}l{\i}k\c{c}{\i}, N.~Vijaykumar,
  N.~M. Ghiasi, S.~Ghose, and O.~Mutlu, ``{CROW: A Low-Cost Substrate for
  Improving DRAM Performance, Energy Efficiency, and Reliability},'' in
  \emph{ISCA}, 2019.

\bibitem{hassan2021uncovering}
H.~Hassan, Y.~C. Tugrul, J.~S. Kim, V.~Van~der Veen, K.~Razavi, and O.~Mutlu,
  ``{Uncovering In-DRAM RowHammer Protection Mechanisms: A New Methodology,
  Custom RowHammer Patterns, and Implications},'' in \emph{MICRO}, 2021.

\bibitem{hassan2017softmc}
H.~Hassan, N.~Vijaykumar, S.~Khan, S.~Ghose, K.~Chang, G.~Pekhimenko, D.~Lee,
  O.~Ergin, and O.~Mutlu, ``{SoftMC: A Flexible and Practical Open-Source
  Infrastructure for Enabling Experimental DRAM Studies},'' in \emph{HPCA},
  2017.

\bibitem{herath2015these}
N.~Herath and {Anders Fogh}, ``These are {{Not Your Grand Daddy}}'s {{CPU
  Performance Counters}},'' in \emph{Black Hat Briefings}, 2015.

\bibitem{hong2010memory}
S.~Hong, ``{Memory Technology Trend and Future Challenges},'' in \emph{IEDM},
  2010.

\bibitem{ddr4operationhynix}
S.~Hynix, ``{DDR4 SDRAM Device Operation}.''

\bibitem{isen2009eskimo}
C.~Isen and L.~John, ``{ESKIMO: Energy Savings Using Semantic Knowledge of
  Inconsequential Memory Occupancy for DRAM Subsystem},'' in \emph{MICRO},
  2009.

\bibitem{jacob2010memory}
B.~Jacob, D.~Wang, and S.~Ng, \emph{{Memory Systems: Cache, DRAM, Disk}}.\hskip
  1em plus 0.5em minus 0.4em\relax Morgan Kaufmann, 2010.

\bibitem{jattke2022blacksmith}
P.~Jattke, V.~van~der Veen, P.~Frigo, S.~Gunter, and K.~Razavi, ``{Blacksmith:
  Scalable Rowhammering in the Frequency Domain},'' in \emph{S\&P}, 2022.

\bibitem{ddr4}
JEDEC, ``{Double Data Rate 4 (DDR4) SDRAM Standard},'' 2012.

\bibitem{2014lpddr4}
JEDEC, ``{Low Power Double Data Rate 4 (LPDDR4) SDRAM Specification},'' 2014.

\bibitem{lpddr5}
JEDEC, ``{LPDDR5 SDRAM Specification - JESD209-5},'' 2019.

\bibitem{jedec2021ddr5}
JEDEC, ``{DDR5 SDRAM Specification - JESD79-5A},'' 2021.

\bibitem{jiang2021quantifying}
Y.~Jiang, H.~Zhu, D.~Sullivan, X.~Guo, X.~Zhang, and Y.~Jin, ``{Quantifying
  Rowhammer Vulnerability for DRAM Security},'' in \emph{DAC}, 2021.

\bibitem{jung2015omitting}
M.~Jung, {\'E}.~Zulian, D.~M. Mathew, M.~Herrmann, C.~Brugger, C.~Weis, and
  N.~Wehn, ``{Omitting Refresh: A Case Study for Commodity and Wide I/O
  DRAMs},'' in \emph{MEMSYS}, 2015.

\bibitem{kang2014co}
U.~Kang, H.~S. Yu, C.~Park, H.~Zheng, J.~Halbert, K.~Bains, S.~Jang, and J.~S.
  Choi, ``{Co-Architecting Controllers and DRAM to Enhance DRAM Process
  Scaling},'' in \emph{The Memory Forum}, 2014.

\bibitem{khan2014efficacy}
S.~Khan, D.~Lee, Y.~Kim, A.~R. Alameldeen, C.~Wilkerson, and O.~Mutlu, ``{The
  Efficacy of Error Mitigation Techniques for DRAM Retention Failures: A
  Comparative Experimental Study},'' in \emph{SIGMETRICS}, 2014.

\bibitem{khan2016parbor}
S.~Khan, D.~Lee, and O.~Mutlu, ``{PARBOR: An Efficient System-Level Technique
  to Detect Data-Dependent Failures in DRAM},'' in \emph{DSN}, 2016.

\bibitem{khan2017detecting}
S.~Khan, C.~Wilkerson, Z.~Wang, A.~R. Alameldeen, D.~Lee, and O.~Mutlu,
  ``{Detecting and Mitigating Data-Dependent DRAM Failures by Exploiting
  Current Memory Content},'' in \emph{MICRO}, 2017.

\bibitem{kim2020revisiting}
J.~S. Kim, M.~Patel, A.~G. Yaglikci, H.~Hassan, R.~Azizi, L.~Orosa, and
  O.~Mutlu, ``{Revisiting RowHammer: An Experimental Analysis of Modern DRAM
  Devices and Mitigation Techniques},'' in \emph{ISCA}, 2020.

\bibitem{kim2000dynamic}
J.~Kim and M.~C. Papaefthymiou, ``{Dynamic Memory Design for Low Data-Retention
  Power},'' in \emph{PATMOS}, 2000.

\bibitem{kim2003block}
J.~Kim and M.~C. Papaefthymiou, ``{Block-Based Multiperiod Dynamic Memory
  Design for Low Data-Retention Power},'' in \emph{TVLSI}, 2003.

\bibitem{kim2009new}
K.~Kim and J.~Lee, ``{A New Investigation of Data Retention Time in Truly
  Nanoscaled DRAMs},'' in \emph{EDL}, 2009.

\bibitem{kim2021mithril}
M.~J. Kim, J.~Park, Y.~Park, W.~Doh, N.~Kim, T.~J. Ham, J.~W. Lee, and J.~H.
  Ahn, ``{Mithril: Cooperative Row Hammer Protection on Commodity DRAM
  Leveraging Managed Refresh},'' \emph{arXiv preprint arXiv:2108.06703}, 2021.

\bibitem{kim2022mithril}
M.~J. Kim, J.~Park, Y.~Park, W.~Doh, N.~Kim, T.~J. Ham, J.~W. Lee, and J.~H.
  Ahn, ``Mithril: Cooperative row hammer protection on commodity dram
  leveraging managed refresh,'' in \emph{HPCA}, 2022.

\bibitem{kim2020charge}
S.~Kim, W.~Kwak, C.~Kim, D.~Baek, and J.~Huh, ``{Charge-Aware DRAM Refresh
  Reduction with Value Transformation},'' in \emph{HPCA}, 2020.

\bibitem{kim2014flipping}
Y.~Kim, R.~Daly, J.~Kim, C.~Fallin, J.~H. Lee, D.~Lee, C.~Wilkerson, K.~Lai,
  and O.~Mutlu, ``{Flipping Bits in Memory Without Accessing Them: An
  Experimental Study of DRAM Disturbance Errors},'' in \emph{ISCA}, 2014.

\bibitem{kim2015ramulator}
Y.~Kim, W.~Yang, and O.~Mutlu, ``{Ramulator: A Fast and Extensible DRAM
  Simulator},'' in \emph{CAL}, 2015.

\bibitem{konoth2018zebram}
R.~K. Konoth, M.~Oliverio, A.~Tatar, D.~Andriesse, H.~Bos, C.~Giuffrida, and
  K.~Razavi, ``{ZebRAM: Comprehensive and Compatible Software Protection
  Against Rowhammer Attacks},'' in \emph{OSDI}, 2018.

\bibitem{kwon2021reducing}
H.~Kwon, K.~Kim, D.~Jeon, and K.-S. Chung, ``{Reducing Refresh Overhead with
  In-DRAM Error Correction Codes},'' in \emph{ISOCC}, 2021.

\bibitem{lee2015adaptive}
D.~Lee, Y.~Kim, G.~Pekhimenko, S.~Khan, V.~Seshadri, K.~Chang, and O.~Mutlu,
  ``{Adaptive-Latency DRAM: Optimizing DRAM Timing for the Common-Case},'' in
  \emph{HPCA}, 2015.

\bibitem{lee2017design}
D.~Lee, S.~Khan, L.~Subramanian, S.~Ghose, R.~Ausavarungnirun, G.~Pekhimenko,
  V.~Seshadri, and O.~Mutlu, ``{Design-Induced Latency Variation in Modern DRAM
  Chips: Characterization, Analysis, and Latency Reduction Mechanisms},'' in
  \emph{SIGMETRICS}, 2017.

\bibitem{lee2019twice}
E.~Lee, I.~Kang, S.~Lee, G.~{Edward Suh}, and J.~{Ho Ahn}, ``{TWiCe: Preventing
  Row-Hammering by Exploiting Time Window Counters},'' in \emph{ISCA}, 2019.

\bibitem{lee2018twice}
E.~Lee, S.~Lee, G.~E. Suh, and J.~H. Ahn, ``{TWiCe: Time Window Counter Based
  Row Refresh to Prevent Row-Hammering},'' \emph{CAL}, 2018.

\bibitem{lee2016technology}
S.-H. Lee, ``{Technology Scaling Challenges and Opportunities of Memory
  Devices},'' in \emph{IEDM}, 2016.

\bibitem{li2011dram}
Y.~Li, H.~Schneider, F.~Schnabel, R.~Thewes, and D.~Schmitt-Landsiedel, ``{DRAM
  Yield Analysis and Optimization by a Statistical Design Approach},'' in
  \emph{CSI}, 2011.

\bibitem{liu2013experimental}
J.~Liu, B.~Jaiyen, Y.~Kim, C.~Wilkerson, and O.~Mutlu, ``{An Experimental Study
  of Data Retention Behavior in Modern DRAM Devices: Implications for Retention
  Time Profiling Mechanisms},'' in \emph{ISCA}, 2013.

\bibitem{liu2012raidr}
J.~Liu, B.~Jaiyen, R.~Veras, and O.~Mutlu, ``{RAIDR: Retention-Aware
  Intelligent DRAM Refresh},'' in \emph{ISCA}, 2012.

\bibitem{liu2012flikker}
S.~Liu, K.~Pattabiraman, T.~Moscibroda, and B.~G. Zorn, ``{Flikker: Saving DRAM
  Refresh-Power Through Critical Data Partitioning},'' in \emph{ASPLOS}, 2012.

\bibitem{luk2005pin}
C.-K. Luk, R.~Cohn, R.~Muth, H.~Patil, A.~Klauser, G.~Lowney, S.~Wallace, V.~J.
  Reddi, and K.~Hazelwood, ``{Pin: Building Customized Program Analysis Tools
  with Dynamic Instrumentation},'' in \emph{PLDI}, 2005.

\bibitem{luo2014characterizing}
Y.~Luo, S.~Govindan, B.~Sharma, M.~Santaniello, J.~Meza, A.~Kansal, J.~Liu,
  B.~Khessib, K.~Vaid, and O.~Mutlu, ``{Characterizing Application Memory Error
  Vulnerability to Optimize Datacenter Cost via Heterogeneous-reliability
  Memory},'' in \emph{DSN}, 2014.

\bibitem{mandelman2002challenges}
J.~A. Mandelman, R.~H. Dennard, G.~B. Bronner, J.~K. DeBrosse, R.~Divakaruni,
  Y.~Li, and C.~J. Radens, ``{Challenges and Future Directions for the Scaling
  of Dynamic Random-access Memory (DRAM)},'' in \emph{IBM JRD}, 2002.

\bibitem{marazzi2022protrr}
M.~Marazzi, P.~Jattke, S.~Flavien, and K.~Razavi, ``{PROTRR: Principled yet
  Optimal In-DRAM Target Row Refresh},'' in \emph{S\&P}, 2022.

\bibitem{stream}
J.~D. McCalpin, ``{STREAM: Sustainable Memory Bandwidth in High Performance
  Computers},'' \url{https://www.cs.virginia.edu/stream/}.

\bibitem{meza2015revisiting}
J.~Meza, Q.~Wu, S.~Kumar, and O.~Mutlu, ``{Revisiting Memory Errors in
  Large-scale Production Data Centers: Analysis and Modeling of New Trends from
  the Field},'' in \emph{DSN}, 2015.

\bibitem{michaud2012demystifying}
P.~Michaud, ``{Demystifying Multicore Throughput Metrics},'' \emph{CAL}, 2012.

\bibitem{micronddr4}
Micron, ``{DDDR4 SDRAM Datasheet},'' 2016.

\bibitem{misra1982finding}
J.~Misra and D.~Gries, ``{Finding Repeated Elements},'' \emph{{Science of
  Computer Programming}}, 1982.

\bibitem{mukherjee2004cache}
S.~S. Mukherjee, J.~Emer, T.~Fossum, and S.~K. Reinhardt, ``{Cache Scrubbing in
  Microprocessors: Myth or Necessity?}'' in \emph{SDC}, 2004.

\bibitem{mukundan2013understanding}
J.~Mukundan, H.~Hunter, K.-h. Kim, J.~Stuecheli, and J.~F. Mart{\'\i}nez,
  ``{Understanding and Mitigating Refresh Overheads in High-density DDR4 DRAM
  Systems},'' in \emph{ISCA}, 2013.

\bibitem{muralimanohar2009cacti}
N.~Muralimanohar, R.~Balasubramonian, and N.~P. Jouppi, ``{CACTI 6.0: A Tool to
  Model Large Caches},'' HP Laboratories, Tech. Rep. HPL-2009-85, 2009.

\bibitem{mutlu2013memory}
O.~Mutlu, ``{Memory Scaling: A Systems Architecture Perspective},'' in
  \emph{IMW}, 2013.

\bibitem{mutlu2019rowhammer}
O.~Mutlu and J.~S. Kim, ``{RowHammer: A Retrospective},'' \emph{TCAD}, 2019.

\bibitem{mutlu2007stall}
O.~Mutlu and T.~Moscibroda, ``{Stall-Time Fair Memory Access Scheduling for
  Chip Multiprocessors},'' in \emph{MICRO}, 2007.

\bibitem{mutlu2014research}
O.~Mutlu and L.~Subramanian, ``{Research Problems and Opportunities in Memory
  Systems},'' in \emph{SUPERFRI}, 2014.

\bibitem{nair2013case}
P.~Nair, C.-C. Chou, and M.~K. Qureshi, ``{A Case for Refresh Pausing in DRAM
  Memory Systems},'' in \emph{HPCA}, 2013.

\bibitem{nair2014refresh}
P.~J. Nair, C.-C. Chou, and M.~K. Qureshi, ``{Refresh Pausing in DRAM Memory
  Systems},'' in \emph{TACO}, 2014.

\bibitem{nair2013archshield}
P.~J. Nair, D.-H. Kim, and M.~K. Qureshi, ``{ArchShield: Architectural
  Framework for Assisting DRAM Scaling by Tolerating High Error Rates},'' in
  \emph{ISCA}, 2013.

\bibitem{nair2016xed}
P.~J. Nair, V.~Sridharan, and M.~K. Qureshi, ``{XED: Exposing On-Die Error
  Detection Information for Strong Memory Reliability},'' in \emph{ISCA}, 2016.

\bibitem{nguyen2018nonblocking}
K.~Nguyen, K.~Lyu, X.~Meng, V.~Sridharan, and X.~Jian, ``{Nonblocking Memory
  Refresh},'' in \emph{ISCA}, 2018.

\bibitem{park2016experiments}
K.~Park, C.~Lim, D.~Yun, and S.~Baeg, ``{Experiments and Root Cause Analysis
  for Active-Precharge Hammering Fault in DDR3 SDRAM under 3$\times$ nm
  Technology},'' in \emph{Microelectronics Reliability}, 2016.

\bibitem{park2016statistical}
K.~Park, D.~Yun, and S.~Baeg, ``{Statistical Distributions of Row-Hammering
  Induced Failures in DDR3 Components},'' in \emph{Microelectronics
  Reliability}, 2016.

\bibitem{park2015technology}
S.-K. Park, ``{Technology Scaling Challenge and Future Prospects of DRAM and
  NAND Flash Memory},'' in \emph{IMW}, 2015.

\bibitem{park2020graphene}
Y.~Park, W.~Kwon, E.~Lee, T.~J. Ham, J.~H. Ahn, and J.~Lee, ``Graphene: Strong
  yet lightweight row hammer protection,'' in \emph{MICRO}, 2020.

\bibitem{patel2005energy}
K.~Patel, L.~Benini, E.~Macii, and M.~Poncino, ``{Energy-Efficient Value-based
  Selective Refresh for Embedded DRAMs},'' in \emph{PATMOS}, 2005.

\bibitem{patel2021harp}
M.~Patel, G.~F. de~Oliveira, and O.~Mutlu, ``{HARP: Practically and Effectively
  Identifying Uncorrectable Errors in Memory Chips That Use On-Die
  Error-Correcting Codes},'' in \emph{MICRO}, 2021.

\bibitem{patel2020bit}
M.~Patel, J.~Kim, T.-M. Shahroodi, H.~Hassan, and O.~Mutlu, ``{Bit-Exact ECC
  Recovery (BEER): Determining DRAM On-Die ECC Functions by Exploiting DRAM
  Data Retention Characteristics},'' in \emph{MICRO}, 2020.

\bibitem{patel2019understanding}
M.~Patel, J.~S. Kim, H.~Hassan, and O.~Mutlu, ``{Understanding and Modeling
  On-Die Error Correction in Modern DRAM: An Experimental Study Using Real
  Devices},'' in \emph{DSN}, 2019.

\bibitem{patel2017reaper}
M.~Patel, J.~S. Kim, and O.~Mutlu, ``{The Reach Profiler (REAPER): Enabling the
  Mitigation of DRAM Retention Failures via Profiling at Aggressive
  Conditions},'' in \emph{ISCA}, 2017.

\bibitem{pontarelli2016improving}
S.~Pontarelli, P.~Reviriego, and J.~A. Maestro, ``{Improving Counting Bloom
  Filter Performance with Fingerprints},'' \emph{Information Processing
  Letters}, 2016.

\bibitem{qureshi2015avatar}
M.~K. Qureshi, D.~Kim, S.~Khan, P.~J. Nair, and O.~Mutlu, ``{AVATAR: A
  Variable-Retention-Time (VRT) Aware Refresh for DRAM Systems},'' in
  \emph{DSN}, 2015.

\bibitem{riho2014partial}
Y.~Riho and K.~Nakazato, ``{Partial Access Mode: New Method for Reducing Power
  Consumption of Dynamic Random Access Memory},'' 2014.

\bibitem{rooney2019micron}
R.~Rooney and N.~Koyle, ``{Micron DDR5 SDRAM: New Features},'' \emph{Micron
  Technology Inc., Tech. Rep}, 2019.

\bibitem{ryu2017overcoming}
S.-W. Ryu, K.~Min, J.~Shin, H.~Kwon, D.~Nam, T.~Oh, T.-S. Jang, M.~Yoo, Y.~Kim,
  and S.~Hong, ``{Overcoming the Reliability Limitation in the Ultimately
  Scaled DRAM using Silicon Migration Technique by Hydrogen Annealing},'' in
  \emph{IEDM}, 2017.

\bibitem{ramulatorgithub}
{SAFARI Research Group}, ``{Ramulator Source Code},''
  https://github.com/CMU-SAFARI/ramulator.

\bibitem{saileshwar2022randomized}
G.~Saileshwar, B.~Wang, M.~Qureshi, and P.~J. Nair, ``{Randomized Row-Swap:
  Mitigating Row Hammer by Breaking Spatial Correlation between Aggressor and
  Victim Rows},'' in \emph{ASPLOS}, 2022.

\bibitem{saino2000impact}
K.~Saino, S.~Horiba, S.~Uchiyama, Y.~Takaishi, M.~Takenaka, T.~Uchida,
  Y.~Takada, K.~Koyama, H.~Miyake, and C.~Hu, ``{Impact of Gate-induced Drain
  Leakage Current on the Tail Distribution of DRAM Data Retention Time},'' in
  \emph{IEDM}, 2000.

\bibitem{saleh1990reliability}
A.~M. Saleh, J.~J. Serrano, and J.~H. Patel, ``{Reliability of Scrubbing
  Recovery-Techniques for Memory Systems},'' \emph{IEEE Transactions on
  Reliability}, 1990.

\bibitem{schroeder2009dram}
B.~Schroeder, E.~Pinheiro, and W.-D. Weber, ``{DRAM Errors in the Wild: a
  Large-scale Field Study},'' in \emph{SIGMETRICS}, 2009.

\bibitem{seyedzadeh2018cbt}
S.~M. {Seyedzadeh}, A.~K. {Jones}, and R.~{Melhem}, ``{Mitigating Wordline
  Crosstalk Using Adaptive Trees of Counters},'' in \emph{ISCA}, 2018.

\bibitem{seyedzadeh2017counter}
S.~M. Seyedzadeh, A.~K. Jones, and R.~Melhem, ``{Counter-based Tree Structure
  for Row Hammering Mitigation in DRAM},'' \emph{CAL}, 2017.

\bibitem{sharifi2017online}
R.~Sharifi and Z.~Navabi, ``{Online Profiling for Cluster-Specific Variable
  Rate Refreshing in High-Density DRAM Systems},'' in \emph{ETS}, 2017.

\bibitem{siddiqua2017lifetime}
T.~Siddiqua, V.~Sridharan, S.~E. Raasch, N.~DeBardeleben, K.~B. Ferreira,
  S.~Levy, E.~Baseman, and Q.~Guan, ``{Lifetime Memory Reliability Data from
  the Field},'' in \emph{DFT}, 2017.

\bibitem{snavely2000symbiotic}
A.~Snavely and D.~M. Tullsen, ``{Symbiotic Jobscheduling for a Simultaneous
  Mutlithreading Processor},'' in \emph{ASPLOS}, 2000.

\bibitem{son2017making}
M.~Son, H.~Park, J.~Ahn, and S.~Yoo, ``{Making DRAM Stronger Against Row
  Hammering},'' in \emph{DAC}, 2017.

\bibitem{spec2006}
{Standard Performance Evaluation Corp.}, ``{SPEC CPU@2006},'' 2006,
  http://www.spec.org/cpu2006.

\bibitem{stuecheli2010elastic}
J.~Stuecheli, D.~Kaseridis, H.~C. Hunter, and L.~K. John, ``{Elastic Refresh:
  Techniques to Mitigate Refresh Penalties in High Density Memory},'' in
  \emph{MICRO}, 2010.

\bibitem{taouil2021lightroad}
M.~Taouil, C.~Reinbrecht, S.~Hamdioui, and J.~Sep{\'u}lveda, ``{LightRoAD:
  Lightweight Rowhammer Attack Detector},'' in \emph{ISVLSI}, 2021.

\bibitem{tpc}
{Transaction Processing Performance Council}, ``{TPC Benchmarks},''
  \url{http://www.tpc.org/}.

\bibitem{udipi2011combining}
A.~N. Udipi, N.~Muralimanohar, R.~Balasubramonian, A.~Davis, and N.~P. Jouppi,
  ``{Combining Memory and a Controller with Photonics Through 3D-Stacking to
  Enable Scalable and Energy-Efficient Systems},'' in \emph{ISCA}, 2011.

\bibitem{van2018guardion}
V.~van~der Veen, M.~Lindorfer, Y.~Fratantonio, H.~P. Pillai, G.~Vigna,
  C.~Kruegel, H.~Bos, and K.~Razavi, ``{GuardION: Practical Mitigation of
  DMA-Based Rowhammer Attacks on ARM},'' in \emph{DIMVA}, 2018.

\bibitem{venkatesan2006retention}
R.~K. Venkatesan, S.~Herr, and E.~Rotenberg, ``{Retention-Aware Placement in
  DRAM (RAPID): Software Methods for Quasi-Non-Volatile DRAM},'' in
  \emph{HPCA}, 2006.

\bibitem{walker2021dram}
A.~J. Walker, S.~Lee, and D.~Beery, ``{On DRAM Rowhammer and the Physics of
  Insecurity},'' \emph{TED}, 2021.

\bibitem{yaglikci2021blockhammer}
A.~G. Ya{\u{g}}lik{\c{c}}i, M.~Patel, J.~Kim, R.~Azizi, A.~Olgun, L.~Orosa,
  H.~Hassan, J.~Park, K.~Kanellopoulos, T.~Shahroodi, S.~Ghose, and O.~Mutlu,
  ``Blockhammer: Preventing rowhammer at low cost by blacklisting
  rapidly-accessed dram rows,'' \emph{ArXiv}, 2021.

\bibitem{yauglikcci2021security}
A.~G. Ya{\u{g}}l{\i}k{\c{c}}{\i}, J.~S. Kim, F.~Devaux, and O.~Mutlu,
  ``{Security Analysis of the Silver Bullet Technique for RowHammer
  Prevention},'' \emph{arXiv preprint arXiv:2106.07084}, 2021.

\bibitem{yang2016suppression}
C.-M. Yang, C.-K. Wei, Y.~J. Chang, T.-C. Wu, H.-P. Chen, and C.-S. Lai,
  ``{Suppression of Row Hammer Effect by Doping Profile Modification in
  Saddle-Fin Array Devices for sub-30-nm DRAM Technology},'' \emph{TDMR}, 2016.

\bibitem{yang2019trap}
T.~Yang and X.-W. Lin, ``{Trap-Assisted DRAM Row Hammer Effect},'' \emph{EDL},
  2019.

\bibitem{you2019mrloc}
J.~M. You and J.-S. Yang, ``{MRLoc: Mitigating Row-Hammering Based on Memory
  Locality},'' in \emph{DAC}, 2019.

\end{thebibliography}
\end{singlespace}

\bookmarksetup{startatroot}
\end{document}